\documentclass[10pt]{article}

\usepackage[english]{babel}
\usepackage{amsthm}
\usepackage{amssymb}
\usepackage{mathtools}

\newtheorem{theorem}{Theorem}[section]
\newtheorem{lemma}[theorem]{Lemma}
\newtheorem{proposition}[theorem]{Proposition}
\newtheorem{corollary}[theorem]{Corollary}
\newtheorem{remark}[theorem]{Remark}
\newtheorem{definition}[theorem]{Definition}
\makeatletter

\@addtoreset{equation}{section}
\usepackage{color}
\usepackage{graphicx}

\makeatother
\usepackage{newlfont}
\usepackage{pdfpages}
\def \n {\noindent}
\begin{document}

\title{{\color{blue}{\bf {\small {\large Spectral analysis of some non-normal operators arising  in  Reggeon field theory}}}}}
\author{{\small {\bf Abdelkader INTISSAR}}}
\date{{\small }}
 \maketitle

\begin{center}
Equipe d'Analyse spectrale, Facult\'e des Sciences et Techniques\\ Universit\'e de Cort\'e, 20250 Cort\'e, France \\
\quad\\
Le Prador\\
129,  rue commandant Rolland\\
13008 Marseille France\\
{\color{blue}abdelkader.intissar@orange.fr}\\
\end{center}

\begin{abstract}

\n In this work, we present a complete spectral study of a family of non-normal operators arising in Reggeon field theory. This family of operators is an original example who permit us to discover the recent theory of physical requirement of space-time reflection symmetry (PT symmetry) without losing any of the essential physical features of quantum mechanics {\color{blue} [Bender]}.\\
 
\n Early studies of Reggeon field theory, in the late 1970's led a number of investigators to observe that model cubic quantum-mechanical Hamiltonians might have real eigenvalues {\color{blue} [Bower et al]}. \\

\n The study of this family of operators, permit us to discover some fine results of Spectral Theory and Functional Analysis in particular the results connected with completeness of elementary solutions of mathematical physics problems. \\

\n We use knowledge  basics of holomorphic functions of one complex variable and the properties of Hilbert spaces. In this work the  knowledge of spectral theory and the functional analysis of standard level is required (see {\color{blue}[Kato]} , {\color{blue}[Ghohberg2 et al]} and {\color{blue}[Markus]}). It is also requested to know the basic properties of semigroup theory {see for example \color{blue}([Pazy]}).\\

\n {\bf Keywords :} Spectral theory ; Gribov operators ; semigroups ; Non-self-adjoint operators ; Bargmann space ; Reggeon field theory.
\end{abstract}
\newpage
\tableofcontents
\newpage

\section{ Introduction}

\n For self-adjoint and more generally normal operators $T$ on some complex Hilbert space $\mathcal{H}$,  we have a nice theory, including the spectral theorem and the wellknown and important resolvent estimate,\\
\begin{equation}
\displaystyle{\mid\mid (H - \lambda I)^{-1} \mid\mid \leq \frac{1}{(dist (\lambda, \sigma(T))}}
\end{equation}
\n where $\sigma(H)$ denotes the spectrum of $T$. The spectral theorem also gives very nice control over functions of self-adjoint operators, so for instance if $H$ is
self-adjoint with spectrum contained in the half interval $\displaystyle{[\lambda_{0}, +\infty[}$, then \\
\begin{equation}
\displaystyle{\mid\mid e^{-tH} \mid\mid \leq e^{-\lambda_{0}t} , t \geq 0 }
\end{equation}
\n However, non-normal operators appear frequently in different problems;\\

 \n Convection-diffusion problems, Theory of elasticity, transport equation, damped wave equations, linearized operators in fluid dynamics, Reggeon field theory etc. \\
 
 \n Then typically,  $\displaystyle{\mid\mid (H - \lambda I)^{-1} \mid\mid}$ may not exist or may be very large even when $\lambda$ is far from the spectrum and this implies mathematical difficulties:\\

\n - When studying the existence of eigenvalues and their distribution,\\  

\n - When studying functions of the operator, like $\displaystyle{e^{-tH}}$, \\

\n - When studying the operators $H$ in the form $\displaystyle{H = H_{0} + i H_{1} }$; $i^{2} = -1$ where  $H_{0}$ is self adjoint and  $H_{1}$  is a strong perturbation of $H_{0}$  with $\sigma(H_{1}) = \mathbb{C}$,\\

\n - When the domains of the adjoint and anti-adjoint parts of $H$ are not included in one another.\\

\n When choosing the ambient Hilbert space, like in problems for scattering poles or  in problems for Reggeon field  theory.\\

\n  This is a source of mathematical and numerical difficulties.\\

\n This text is devoted to some old and recent general results for non-self-adjoint operators \\
\begin{equation}
\displaystyle{H_{\mu, \lambda} = \mu \mathcal {N} + i\lambda (\mathcal{N} A + A^{*}\mathcal{N})}
\end{equation}
\begin{equation}
\displaystyle{H_{\lambda', \mu, \lambda} =  \lambda'A^{*^{2}}A^{2} + \mu \mathcal {N} + i\lambda (\mathcal{N} A + A^{*}\mathcal{N})}
\end{equation}
\begin{equation}
\displaystyle{H_{ \lambda'', \mu, \lambda} =   \lambda''A^{*^{3}}A^{3} + \mu \mathcal {N} + i\lambda (\mathcal{N} A + A^{*}\mathcal{N})}
\end{equation}

\n arising in Reggeon field theory  where \\

\n - $\mu$, $\lambda$; $\lambda'$ and $\lambda''$ are real parameters and $i^{2} = -1$.\\

\n  - $A^{*}$ and $A$, are the creation and annihilation operators. These operators satisfy the commutation relation:\\
\begin{equation}
[A, A^{*}] = I
\end{equation}
\n and  $\mathcal{N} = A^{*}A$ is Harmonic oscillator.\\

\n $H_{\mu, \lambda}$  can be defined  in Bargmann space as the closure of its restriction on the polynomials. Then the first mathematical difficulty when studying the minimal domain and the maximal domain of  our operator is to show their equality,\\

\n $H_{\mu, \lambda}$  acts as Hamiltonian of Reggeon field theory. In this theory, it is the simplest Hamiltonian that can describe a field with interaction and it is plausible that it describes the dominant interaction between two hadrons.\\

\n In the above, the word <<simple>> only applies to the algebraic expression of the operator. If we look at the family a little more generally,\\
\begin{equation}
\displaystyle{H_{\lambda', \mu, \lambda} = \lambda' A^{*^{2}}A^{2} +  H_{\mu, \lambda}}
\end{equation}
\n For $\lambda' \neq 0$,  the study of $H_{\lambda', \mu, \lambda}$ is easier than the study of $H_{\mu, \lambda}$ and permit us  by taking the limit $\lambda' = 0$ to confirm  that model cubic quantum-mechanical Hamiltonian  has the eigenvalues and which are real.\\

\n For $\lambda'' \neq 0$,  the study of $H_{\lambda'', \mu, \lambda}$ is easier than the study of $H_{\lambda', \mu, \lambda}$ and permit us to give a regularized  trace formula of $H_{\lambda'', \mu, \lambda}$ and  a regularized  trace formula of its semigroup. \\

\n {\bf {\color{blue}Here is the plan of this work}}.\\

\n {\color{blue}{\bf In Section 2 }} we give an overview of Reggeon Field Theory (RFT) and the operators which caracterize this theory for one site. This Reggeon field theory (RFT) is an attempt to predict the high-energy behaviour of soft pro- cesses; the RFT can be derived from the assumed softness of hadronic interactions at low transverse momenta, which seems to be well established experimentally in hadron-hadron and hadron-nucleus interaction. \\

\n The Reggeon Field Theory (RFT) was invented by Vladimir Naumovich Gribov who was  one of the creators of modern theoretical physics. The concepts and methods that Gribov has developed in the second half of the 20th century became cornerstones of the physics of high energy hadron interactions (relativistic theory of complex angular momenta, a notion of the vacuum pole-Pomeron, effective Reggeon field theory), condensed matter physics (critical phenomena), neutrino oscillations, and nuclear physics.\\

\n Originally, RFT was formulated by Gribov in 1969 as a field theory (or as quantum mechanics in zero transverse dimensions) of pomerons. The basic degrees of freedom in this formulation are the Gribov fields  $\psi$ and $\psi^{+}$ that create and annihilate the pomeron. The action defining the theory with triple pomeron
couplings only (MRFT) is defined in the following way:\\
\begin{equation}
\displaystyle{\mathcal{S} = \int dy\{\psi^{+}\partial_{y}\psi  - \mu \psi^{+}\psi  + i \lambda \psi^{+}( \psi + \psi^{+})\psi\}}
\end{equation}
\n where $\mu$ is the bare intercept of the pomeron and $\lambda$ is the coupling of the triple pomeron interaction and $i^{2} = -1$.\\

\n After redefinition of the Gribov fields  by simple harmonic oscillator annihilation and creation operators :\:

\n $\psi \longrightarrow A$  \quad $\overline{\psi} \longrightarrow A^{*}$\\

\n Then, the Hamiltonian of the problem is given by:\\
\begin{equation}
\displaystyle{ H_{\mu, \lambda} = - \mu A^{*}A + i\lambda A^{*}(A + A^{*})A}
\end{equation}

\n The commutation relation of the pomeron annihilation and creation operators, $[\psi^{+}, \psi] = 1 $ implies $[A^{*}, A]= I$\\

\n We choose like ambient Hilbert space the Bargmann space \\
\begin{equation}
\displaystyle{  \mathbb{B }= \{\varphi : \mathbb{C} \longrightarrow \mathbb{C}  \, entire \, ; \int_{\mathbb{C} }\mid \varphi(z)\mid^{2}e^{-\mid z \mid^{2}}dx dy < \infty \}}
\end{equation}

\n to  give a complete spectral  analysis of $H_{\mu, \lambda}$ called Gribov-Intissar operators. \\

\n This naming of these operators was suggested by a referee of Advance in Mathematics  (CHINA) who reviewed an article of  author titled ``On Spectral Approximation of Unbounded Gribov-Intissar Operators in Bargmann Space'', Advances in mathematics (China), Vol. 46. N0. 1, Jan., 2017  13-33, {\color{blue}doi: 10.11845/sxjz.2015026b}.\\

\n  In Bargmann representation $\displaystyle{A = \frac{\partial}{\partial z}}$  and  $A^{*} = z$  where $z = x + iy ; (x, y) \in \mathbb{R}^{2}$ and the Hamiltonian takes the form:\\
\begin{equation}
\displaystyle{H_{\mu, \lambda}  = i \lambda z \frac{\partial^{2}}{\partial z^{2}} + (i\lambda z^{2} - \mu z)\frac{\partial}{\partial z}}
\end{equation} 
\n The interacting Pomeron system whose state is given by  $u(t, z)$ evolves according to a Schrodinger equation \\
\begin{equation}
\displaystyle{\frac{\partial u(t, z)}{\partial t} = H_{\mu, \lambda}u(t,z)}
\end{equation}
\n The complete definition of the problem requires that the domain of $H_{\mu; \lambda}$ and the boundary conditions are specified.\\

\n The asymptotic behaviour of reggeon quantum mechanics (i.e., reggeon field theory without transverse dimensions) has been analyzed in a series of papers {\color{blue}[Alessandrini et al]}, {\color{blue}[Jengo]}, \, {\color{blue}[Bronzan et al]} and {\color{blue}[Ciafaloni et al]}.\\

\n However, White {\color{blue}[White]} has tried to cast doubt on the conclusion of refs. {\color{blue}[Amati et al]} and {\color{blue}[Ciafaloni et al]}, as the mathematics used in ref. {\color{blue}[Ciafaloni et al]} is not beyond criticism (although it is quite respectable according to 'physicists' standards).\\

\n It was during this period that the professor Michel Lebellac of University of Nice submitted the suject to Professor Martin Zerner and that the latter in turn brought the subject to the author for a mathematical study. We present in the next sections the evolution of significant mathematical results from $1980$ to the present.\\

\n Also, we note that  Grassberger has given an interesting interpretation of RFT; following Feynman, he assumes that hadron cross sections at high energy are governed by the interactions of wee partons. If a hadron is boosted, the wee partons can split and recombine, they can be converted into hard partons, and thus are lost from the point of view of the collision, or they can recombine to give hard partons.\\

\n  And with  Sundermeyer  they show that  Reggeon field theory with a quartic coupling in addition to the standard cubic one is mathematically equivalent to a chemical process where a radical can undergo diffusion, absorption, recombination, and autocatalytic production. Physically, these ``radicals'' are wee partons.\\

\n {\color{blue}{\bf In Section 3}},   we do not assume any prior knowledge of holomorphic function spaces. But we will have an idea of their usefulness as soon as we discover the properties of the Bargmann space which will be our ambient space for our operators throughout the work.\\

\n The Bargmann-Segal space has originally been introduced by V. Bargmann, see  {\color{blue}[Bargmann 1]}, {\color{blue}[Bargmann 2]} and {\color{blue}[Bargmann 3]} and E.
Segal, see  {\color{blue}[Segal 1]},  {\color{blue}[Segal 2]} and  {\color{blue}[Segal 3]}.\\

\n Despite the bountiful results on this space, we present some original results on it that we will applied to the study of our operator in particular to give an explicit inversion of $H_{\mu, \lambda}, \mu > 0$  on the axis of negative imaginaries ($z = - iy ; y \geq 0$) and to study an extension of integral operator associated to $H_{\mu, \lambda} ^ {- 1} $ at a Hilbet-Shmidt operator on $\displaystyle{L_{2}(]-\infty, 0], \theta(y)dy)}$ with a suitable choice of the function  $\theta$ to prove the existence of the eigenvalues of $H_{\mu, \lambda} $.\\

\n Notice that if $\mu = 0$, $H_{0, \lambda}$ is chaotic. A complete spectral analysis  of this operator is given in {\color{blue}[Intissar 11]}(1998) and the study of its chaoticity is is given in {\color{blue}[Intissar 2 et al]} (2017).\\

\n In similar way, we show that the integral operator $H_{\lambda', \mu, \lambda} ^ {- 1}, \lambda' \neq 0 $ can be extended toi Hilbet-Shmidt operator on $\displaystyle{L_{2}(]-i\rho', 0], r(y)dy)}$; $\rho' = \frac{\lambda}{\lambda'}, \lambda' \neq 0$ with a suitable choice of the function  $r(y)$. these techniques allow us to show that :\\

\n (i)  The spectral radius of $\displaystyle{ H_{\lambda', \mu, \lambda} ^ {- 1} }$ converges to spectral radius of $\displaystyle{ H_{\mu, \lambda} ^ {- 1} }$ as $\lambda' \longrightarrow 0$.\\

\n (ii) Each of the above spectral radius can be extended analytically over $\mathbb{R}$ with respect to the parameter $\mu$.\\

\n Notice also that it is shown in this section that $\displaystyle{ H_{\mu, \lambda} ^ {- 1} }$ is the class of Carlamann $\displaystyle{\mathcal{C}_{1+ \epsilon}}$ of order $ 1 + \epsilon $ \quad $\forall \,\, \epsilon > 0$. \\

\n where  we define  the class of Carleman $\mathcal{C}_{p}$ of order $p$ as follow :\\

\n Let $K$ be a compact operator in the Hilbert space $\mathcal{H}$. $K$ is said to belong to the Carleman-class $\mathcal{C}_{p}, p > 0$, with order $p$, if the
series $\displaystyle{\sum_{n=1}^{+\infty}s_{n}^{p}(K)}$. converges, where $s_{n}(K)$. are the eigenvalues of the operator $\displaystyle{\sqrt{K^{*}K}}$.\\

\n  In the particular case $p = 2$, $\mathcal{C}_{2}$ is exactly the Hilbert-Schmidt class and for $p = 1$, $\mathcal{C}_{1}$ is said class of nuclear operators or trace operators.
For a systematic treatment of the operators of Carleman-class, we refer to the Gohberg and Krein's book {\color {blue}[Gohberg et al]}.\\

\n From a mathematical point of view, one of the results which greatly contributed to meeting the desires of physicists is the work of professors T.Ando and M.Zerner (1984) on the extension of the small eigenvalue of $H_{\mu, \lambda} $ over any $\mathbb{R}$ with respect to the parameter $ \mu $ {\color {blue} [Ando et al]}. \\

\n We have regularized $H_{\mu, \lambda}$  by adding a perturbation of the form $\displaystyle {\lambda' A ^ {* ^ {2}} A ^ {2} }$ which greatly facilitates the study of the operator: \\
\begin{equation}
 \displaystyle{H _ {\lambda ', \mu, \lambda} = \lambda' A ^ {* ^ {2}} A ^ {2} + \mu A ^ {*} A + i \lambda A ^ {*} ( A + A ^ {*}) A }
\end{equation}
\n or \\
\begin{equation}
\displaystyle{H _ {\lambda ', \mu, \lambda} = (\lambda'z^{2} + i\lambda z)\frac{\partial^{2}}{\partial z^{2}} + (i \lambda z^{2} + \mu z) \frac{\partial}{\partial z}}
\end{equation}
\n  For $\lambda' \neq 0$, $ H_{\lambda', \mu, \lambda}$ has a rich set of spectral properties desired by physicists  and which allows us to deduce the results of Ando-Zerner when $\lambda'$ approaches zero. \\

\n {\color{blue}{\bf In Section 4}}, for the study of eigenvalue problem associated to $ H_{\lambda', \mu, \lambda}$, $\lambda' \neq 0$, we begin by to study  the singularity  at $\infty$  of \\
\begin{equation}
\displaystyle{ H_{\lambda',\mu, \lambda}\varphi(z) = \sigma(\lambda', \mu, \lambda) \varphi(z)}
\end{equation}
\n i.e. We consider the equation \\
\begin{equation}
\displaystyle{(\lambda'z^{2} + i\lambda z)\varphi^{''}(z) + (i \lambda z^{2} + \mu z)\varphi^{'}(z) = \sigma(\lambda', \mu, \lambda) \varphi(z)  \, at \,\quad \infty}
\end{equation}
\n This equation  belongs to family of the bi-confluent  Heun equations of class $(0, 1, 1_{4})$  which has three regular singular points $ z_{0} = 0$ and $z_{1} = -i \rho'$ where $\displaystyle{\rho' = \frac{\lambda}{\lambda'}}$; $\lambda' \neq 0$ and one irregular singular point $z_{\infty} = \infty$. \\

\n If $ \lambda' = 0$,  it has one regular singular points $ z_{0} = 0$ and one irregular singular point $z_{\infty} = \infty$. \\

\n As  $\displaystyle{ \mathbb{B}_{0} = \{ \varphi \in \mathbb{B}; \varphi (0) = 0 \}}$  is invariant under the  actions of $ H_{\lambda', \mu, \lambda}$ and $H_{\mu, \lambda}$ respectively then it is naturl to take $\varphi(0) = 0$ and at $\infty $  we  will prove that a solution $\varphi$ of (0.10)  is in Bargmann space.\\

\n \n  Let $\psi \in \mathbb{B}_{0}$, we consider the equation  $H_{\lambda',\mu,\lambda}\varphi = \psi$  (respectively  $H_{\mu,\lambda}\varphi = \psi$) which can be written under following form :\\
\begin{equation}
\displaystyle{(\lambda'z^{2} + i\lambda z)\varphi''(z) + (i\lambda z^{2}  + \mu z)\varphi'(z) = \psi(z)}
\end{equation}
\n respectively :\\
\begin{equation}
\displaystyle{ i\lambda z\varphi''(z) + (i\lambda z^{2}  + \mu z)\varphi'(z) = \psi(z)}
\end{equation}
\n Now if $\psi \in \mathbb{B}_{0}$ and choosing the straight line connecting $-i\rho' , z \in \mathbb{C}$ parametrized  by \\
\begin{equation}
\displaystyle{\gamma : [0, 1] \rightarrow \mathbb{C}, \gamma(t) = -i\rho' + t(z + i\rho') \quad \quad (\gamma(0) = -i\rho' , \gamma(1) = z)}
\end{equation}
\n Then we define $\displaystyle{\int_{-i\rho'}^{z}\psi(\xi)d\xi:=\int_{\gamma}\psi(\xi)d\xi:= \int_{0}^{1}}\psi(\gamma(t))\gamma'(t)dt$ to deduce that our  equation can be transformed to following integral equation :\\
\begin{equation}
\displaystyle{\varphi(z) = \frac{1}{\lambda'}\int_{-i\rho'}^{z}e^{-i\rho'\eta}(\eta + i\rho')^{-(\delta +1)} [\int_{-i\rho'}^{\eta}e^{i\rho'\xi}(\xi + i\rho')^{\delta}\frac{\psi(\xi)}{\xi}d\xi]d\eta}
\end{equation}
\n In the same way we get on $]-i\infty, 0]$ an integral  equation with $\lambda' = 0$\\

\n However, the integral representation of $\varphi(z)$ in  the equations is hard to study in $\mathbb{C}$ for some existing results on  eigenvalues and eigenfunctions of our operators.\\

\n To overcome this difficulty, the study of integral equations associated to our equations are restricted on negative imaginary axis by applying an original result on Bargmann space  which say :\\

\n If $\varphi \in \mathbb{B}$ then its restriction on $x + i\mathbb{R}$ is square integrable function with measure $\displaystyle{e^{-|y|^{2}} dy, y \in  \mathbb{R}}$ for {\bf all} fixed $x \in  \mathbb{R}$.\\

\n Observe the difference of above result  between the version for {\bf almost} $x \in \mathbb{R}$ which  follows easily from Fubini theorem and the version  {\bf all} fixed $x \in  \mathbb{R}$  which was showed  for the first time by the author  in 1987 in {\color{blue}[Intissar2]} which has been reproduced at the end of the section above.\\

\n Therefore led to put $u(y) = \varphi(-iy)$ ; $y \in [0, + \infty[$ to study the following problems :\\
\begin{equation}
\displaystyle{-\lambda y u_{\mu, \lambda}^{''}(y) + (\lambda y^{2} + \mu y)u_{\mu, \lambda}^{'}(y) = \sigma(\mu, \lambda)u_{\mu, \lambda}(y)}
\end{equation}
\begin{equation}
\displaystyle{(\lambda' y^{2} - \lambda y)u_{\lambda', \mu, \lambda}^{"}(y) + (\lambda y^{2} + \mu y)u_{\mu, \lambda}^{'}(y) = \sigma(\lambda', \mu, \lambda)u_{\lambda', \mu, \lambda}(y)}
\end{equation}
\n Let  $\displaystyle{\rho'  = \frac{\lambda}{\lambda'}; \lambda' \neq 0 \, > 0,  y \in [0, \rho']}$ and $\displaystyle{L_{2}((0, \rho'), r(y)dy)}$ be the Hilbert space of square integrable functions on $[0, \rho']$ with inner product\\
\begin{equation}
\displaystyle{< u, v > = \int_{0}^{\rho'}u(y)\overline{v(y)}r(y)dy}
\end{equation}
\n with $r(y)$ called weight function.\\

 \n In this section we restrict (0.10) on the axis of negative imaginaries ($z = - iy ; y \geq 0$) that  gives the following equation :\\
\begin{equation}
\displaystyle{-y (\lambda - \lambda' y)u^{"}(y) + y(\lambda y +\mu y)u'(y) = \sigma(\lambda', \mu, \lambda)u(y)}
\end{equation}
\n with $u(y) = \varphi(-iy)$ ; $y \geq 0$ and we construct the fonction $r(y)$ such that $H_{\lambda', \mu, \lambda}$ is symmetric on $\displaystyle{L_{2}((0, \rho'), r(y)dy)}$.\\
 
 \n Precisely, we get :\\
 
 \n If $\displaystyle{\rho'(\rho + \rho')  \geq 1}$  where  $\displaystyle{\rho = \frac{\mu}{\lambda}; \lambda \neq 0 }$ then $H_{\lambda', \mu, \lambda}$ is symmetric on $\displaystyle{L_{2}((0, \rho'), r(y)dy)}$ and takes the following form\\
\begin{equation}
\displaystyle{H_{\lambda', \mu, \lambda}u(y) = \frac{1}{r(y)}\frac{\partial}{\partial y}(p(y)u'(y))},  where \displaystyle{p(y) = \lambda' y(\rho' - y) r(y)}
\end{equation}
\n {\color{blue}{\bf In Section 5}}, we recall a list of fundamental spectral properties of $H_{\lambda',\mu, \lambda}$ that we have presented in above sections and we give an explicit inverse of $H_{\lambda', \mu, \lambda.}$ and  $H_{\mu, \lambda.}$ respectively.\\

\n {\color{black}$\bullet$} Precisely, for $ y \in [0, \rho^{'}]$, $\displaystyle{\rho' = \frac{\lambda}{\lambda'}}$ and $\displaystyle{\rho = \frac{\mu}{\lambda}}$, we get:\\ 
\begin{equation}
\displaystyle{H_{\lambda', \mu, \lambda}^{-1} \varphi(-iy) = \int_{0}^{\rho'}\mathcal{N}_{\lambda', \mu,\lambda}(y, s)\varphi(-is)ds}
\end{equation}
\n where $\displaystyle{\mathcal{N}_{\mu,\lambda}(y, s) = \frac{1}{\lambda s}e^{\rho' s}(1 -  \frac{s}{\rho'})^{\delta}\int_{0}^{min(y, s)}e^{-\rho' u }(1 - \frac{u}{\rho'})^{-(\rho + 1)} du}$\\

\n and\\

\n {\color{black}$\bullet$} For $ y \in [0, +\infty[$, we get:\\
\begin{equation}
\displaystyle{H_{\mu, \lambda}^{-1} \varphi(-iy) = \int_{0}^{\infty}\mathcal{N}_{\mu,\lambda}(y, s)\varphi(-is)ds}
\end{equation}
\n where  $\displaystyle{\mathcal{N}_{\mu, \lambda}(y, s) = \frac{1}{s}e^{-\frac{s^{2}}{2} - \frac{\mu}{\lambda}s}\int_{0}^{min(y, s)}e^{\frac{u^{2}}{2} +\frac{\mu}{\lambda}u}du}$\\

\n {\color{blue}{\bf In Section 6}}, we shall consider a solution $u$ of the operational differential equation \\
\begin{equation}
\displaystyle{u'(t) =  H_{ \lambda'', \lambda', \mu , \lambda}u(t) }  for t > 0
\end{equation}
\n Where \\
\n  $\displaystyle{(\lambda'', \lambda', \mu, \lambda) \in \mathbb{R}^{4}}$\\

\begin{equation}
\displaystyle{H_{ \lambda'', \, \lambda', \, \mu, \, \lambda} = \lambda'' G + \lambda' S + \mu \mathcal{N} + i\lambda H_{I} ; \quad  i^{2} = -1}
\end{equation}
\n with\\
\begin{equation}
\displaystyle{G = A^{*^{3}}A^{3}}
\end{equation}
\begin{equation}
\displaystyle{S = A^{*^{2}}A^{2}}
\end{equation}
\begin{equation}
\displaystyle{\mathcal{N} = A^{*}A}
\end{equation}
\begin{equation}
\displaystyle{H_{I} = A^{*}(A + A^{*})A }
\end{equation}

\n to establish the following results\\

\n {\bf $\bullet$} For $\lambda'' > 0$ , we prove that the system of generalized eigenvectors of the operator  $H_{ \lambda'', \lambda', \mu , \lambda}$ is an unconditional basis in Bargmann space $\mathbb{B}$.\\

\n {\bf $\bullet$}   For $\lambda'' > 0$, we give a generalized diagonalization in the Abel sense of the semigroups associated to $H_{ \lambda'', \lambda', \mu , \lambda}$ and to its square root, respectively.\\

\n {\bf $\bullet$}   For $\lambda'' = 0$ , $\lambda' > 0$ and $\displaystyle{\lambda'^{2} \leq \mu \lambda' + \lambda^{2}}$, we prove  that the solution of Cauchy's problem :\\
\begin{equation}
\displaystyle{u'(t) = - H_{ 0, \lambda', \mu , \lambda}u(t) } ; t > 0 \quad and \quad u(0) = \phi \in \mathbb{B}
\end{equation}
\n can be expanded  in the following form :\\

\n $\displaystyle{u(t) = \sum_{k=1}^{+\infty}\frac{< \phi, \phi_{k}^{*} >}{\phi_{k}, \phi_{k}^{*} >}e^{-\sigma_{k}t}\phi_{k}}$.\\

\n where \\

\n $\displaystyle{H_{ 0, \lambda', \mu , \lambda}\phi_{k} = \sigma_{k} \phi_{k}}$ and \n $\displaystyle{H_{ 0, \lambda', \mu , \lambda}^{*}\phi_{k}^{*} = \sigma_{k} \phi_{k}^{*}}$.\\

\n {\bf $\bullet$}   For $\lambda'' = 0$ and $\lambda' = 0$ , we consider the Cauchy problem :\\
\begin{equation}
\displaystyle{ u'(t) = - H_{ 0, 0, \mu , \lambda}u(t) , t > 0} \quad ; \quad  u(0) = \phi \in \mathbb{B}
\end{equation}

\n For $\mu > 0$, the semigroup $e^{-tH_{ 0, 0, \mu , \lambda}}$ associated to $H_{ 0, 0, \mu , \lambda}$ is compact in particular we have\\
\begin{equation}
\sigma (e^{-tH_{ 0, 0, \mu , \lambda}}) = e^{-t\sigma(H_{ 0, 0, \mu , \lambda})} \cup \{0\}
\end{equation}
\n  Let $\sigma_{0}$ et $\sigma_{1}$ are respectively the smallest and the second eigenvalue of the operator $H_{ 0, 0, \mu , \lambda}$. \\

\n Then \\
\begin{equation}
 (i) \mid\mid e^{-tH_{ 0, 0, \mu , \lambda}} \mid\mid = e^{-\sigma_{0}t} + O(e^{-\epsilon t})\, \forall \epsilon < \sigma_{1}
\end{equation}
\begin{equation}
 (ii) < e^{-tH_{ 0, 0, \mu , \lambda}}\phi, \psi > = <\phi, \phi_{0} > <\phi_{0} , \psi >e^{-\sigma_{0}t} + O(e^{-\sigma_{0}t})\, \forall \phi \in \mathbb{B}, \forall \psi \in \mathbb{B}
\end{equation}
\n where $\phi_{0}$  is the eigenfunction of  $H_{ 0, 0, \mu , \lambda}$ associated to $\sigma_{0}$.\\

\n {\bf $\bullet$} Notice that $H_{I}$ does not generates a semigroup in Bargmann space, which implies that we cannot apply the usual Trotter formula to the operator $H_{ 0, 0, \mu , \lambda}$.\\

\n Nevertheless  in 2005,  we have given in {\color{blue}[Intissar 13]} a ``generalized Trotter product formula'' for this operator $H_{ 0, 0, \mu , \lambda}$.\\

\n {\color{blue}{\bf In Section 7}},   we consider the operator :\\
\begin{equation}
\displaystyle{H_{\lambda'', \mu, \lambda} = \lambda''G + H_{\mu, \lambda}}
\end{equation}
\n where $\displaystyle{G = A^{*^{3}}A^{3}}$ and $\lambda''$ is called  the magic coupling of Pomeron by the specialists of high energies.\\

\n AS $\displaystyle{H_{\lambda'', \mu, \lambda}}$ is a  ordinary differential operator, we present a regularized trace formula for this family operators.\\

\n Notice that the theory of regularized traces of ordinary differential operators has a long history. First, the trace formulas for the Sturm-Liouville operator with the Dirichlet boundary conditions and sufficiently smooth potential $q(x)$ were established in [{\color{blue}[Dikii 1]},  {\color{blue}[Gelfand et al]}].\\

\n Afterwards these investigations were continued in many directions, for instance, the trace formulas for the Sturm-Liouville operator with periodic or antiperiodic
boundary conditions were obtained in [{\color{blue}[Lax]}, {\color{blue}[Sansuc et al}]\\

\n and for regular but not strongly regular ones {\color{blue}[Naimark]}\\

\n similar formulas were found in {\color{blue}[Makin]}. \\

\n A method for calculating trace formulas for general problems involving ordinary differential equations on a finite interval was proposed in {\color{blue}[Lidskii et al]}.\\

 \n The bibliography on the subject is very extensive and we refer to the list of the works in [{\color{blue}[Levitan et al]}, {\color{blue}[Sadovnichi et al]} (2006)].\\

\n The trace formulas can be used for approximate calculation of the first eigenvalues of an operator.\\

\n we use the results of  {\color{blue}[Sadovnichi et al]} (2002)] to give the number of corrections sufficient for the existence of finite formula of the trace of  $\displaystyle{H_{\lambda'', \mu, \lambda}}$.\\

\n Precisely we prove the following statement.\\

\n  Let $ \mathbb{B}$ be the Bargmann space, $\displaystyle{ H_{\lambda'', \mu, \lambda} = \lambda{''}G + H_{\mu,\lambda}}$ acting on $\mathbb{B}$ where $ G = A^{*^{3}}A^{3}$ and $ H_{\mu,\lambda} = \mu A^{*}A + i\lambda A^{*}( A + A^{*})A$ , $A$ and $A^{*}$ are the standard Bose annihilation and creation operators.\\

\n Then there exists an increasing sequence of radius $r_{m}$ such that $r_{m} \rightarrow \infty$ as $ m \rightarrow \infty$  and\\
\begin{equation}
\displaystyle{\lim\limits_{m \rightarrow \infty} (\sum_{n=0}^{m}(\sigma_{n} - \lambda{''}\lambda_{n}) + \frac{1}{2i\pi}\int_{\gamma_{m}} Tr[\sum_{k=1}^{4}\frac{(-1)^{k-1}}{k}[H_{\mu,\lambda}(\lambda{''}G - \sigma I)^{-1}]^{k}]d\sigma) = 0}
\end{equation}
\n  Where\\

 \n - $\sigma_{n}$ are the eigenvalues of the operator $ H_{\lambda'', \mu, \lambda}  = \lambda{''}G + H_{\mu,\lambda}$\\
 
\n  - $\lambda_{n} = n(n-1)(n-2)$ are the eigenvalues of the operator $ G $\\

 \n - $ (\lambda{''}G - \sigma I)^{-1}$ is the resolvent of the operator $\lambda{''}G$\\
 
\n  and\\
 
\n  - $\gamma_{m}$ is the circle of radius $r_{m}$ centered at zero in complex plane.\\

\n {\bf {\color{blue} In Section 8}}, We show that  $\displaystyle{- H_{\lambda'', \mu, \lambda}}$  generates a semigroup $e^{-t\mathbb{H}_{\lambda''}}$ of Carleman class $\mathcal{C}_{p}$ for all $ p > 0$ and all $ t > 0$. In particular, we go to derive an asymptotic expansion of the trace of $e^{-t\mathbb{H}_{\lambda''}}$ as $t\rightarrow 0^{+}$. \\ 
 
 \n Precisely we prove the following statement.\\
 \begin{equation}
 \mid\mid e^{-tH_{\lambda''}} - e^{-t\lambda''G} \mid\mid_{1} = t\mid\mid e^{-t\lambda''G} H_{\mu,\lambda}\mid\mid_{1} + \mid\mid (\lambda''G)^{\delta} e^{-\frac{t}{3}\lambda''G }\mid\mid_{1}O(t^{2})
 \end{equation}
\n $\displaystyle{\forall \, \delta \geq \frac{1}{2}}$.\\

\section{Presentation of non-normal operators arising in Reggeon field theory}

\subsection{A short introduction of Reggeon Field Theory (RFT) }

\n In this section we present a short introduction of Reggeon Field Theory (RFT) without transverse dimensions and the operators which characterize this theory.\\

\n The Reggeon field theory (see Poghosyan in {\color{blue} [Poghosyan]} for an elementary introduction to this theory) is one of the attempts to understand strong interactions that is, the interactions between, among other less stable particles, protons and neutrons. For Gribov's reggeon calculus, we can see  {\color{blue} [Gribov1]}, {\color{blue}[Gribov2]}, {\color{blue}  [Intissar1 et al]} and {\color{blue} [Baker et al.]}\\

\n Many experiments in physics (atomic and nuclear physics, solid state physics and high energy physics) consist of directing a beam of particles {\bf(1)} onto a target of particles {\bf(2)}, and studying the resulting collisions. This is possible since the various particles constituting the final state of the system (state after the collision) are
detected and their characteristics (direction of emission, energy, etc.) are measured.\\

\n Aim of such experiments: Determine the interactions that occur between the various particles entering the collision. The collisions give rise to reactions:\\
$$\displaystyle{{\color{red}(1) + (2) \quad {\color{blue}\longrightarrow} \quad  (3) + (4) + (5) + . . . .}}$$\\
\n In quantum mechanics, we can only speak of probabilities of the possible states coming out of the collision. Among the reactions possible under given conditions, scattering
reactions are defined as those in which the final state and the initial state are composed of the same particles (1) and (2). In an elastic scattering reaction, none of the particles? internal states change during the collision.\\
\n The experimental quantity (see {\color{blue}[Roy]} and  {\color{blue}[Levin1, Levin2]}) that is wanted to be computed is the total cross-section that is, the proportion of particles that are destroyed to create new ones. \\
\subsection{The representation of Hamiltonian of Reggeon Field Theory in Bargmann space }

\n In Bargmann representation {\color{blue}[Barg1]} and {\color{blue}[Barg2]}, the Reggeon's field theory in zero transverse dimensions is characterized by Gribov's operator {\color{blue}[Gribov1]}:\\
\begin{equation}
H_{\mu, \lambda} = \mu A^{*}A + i\lambda A^{*}(A + A^{*})A
\end{equation}
\n where $\mu$ and $\lambda$ are real parameters , ($\mu$ is Pemeron's intercept, $\lambda$ is  triple coupling of Pemeron and $i^{2} = -1$.\\

\n $A^{*}$ and $A$, are the creation and annihilation operators represented in Bargmann representation by $\displaystyle{A = \frac{\partial}{\partial z}}$  and  $A^{*} = z$  where $z = x + iy ; (x, y) \in \mathbb{R}^{2}$. These operators satisfy the commutation relation:\\
\begin{equation}
[A, A^{*}] = I
\end{equation}
\n The operator $H_{\mu, \lambda}$ can be regularized by  adding the term $\displaystyle{\lambda'A^{*^{2}}A^{2}}$ to (2.1)  to get:\\
\begin{equation}
H_{\lambda', \mu, \lambda} = \lambda' A^{*^{2}}A^{2}  + \mu A^{*}A + i\lambda A^{*}(A + A^{*})A
\end{equation}
\n where $\lambda'$ is quartic Reggeon coupling\\

\n The operator $H_{\lambda', \mu, \lambda}$ can be regularized by adding the term $\displaystyle{\lambda'' A^{*^{3}}A^{3}}$ to (2.3) to get:\\
\begin{equation}
H_{\lambda'', \lambda', \mu, \lambda} =\lambda'' A^{*^{3}}A^{3} +  \lambda' A^{*^{2}}A^{2}  + \mu A^{*}A + i\lambda A^{*}(A + A^{*})A
\end{equation}
\n where $\lambda''$ is magic Reggeon coupling\\

\n In Bargmann representation :\\
\begin{equation}
\displaystyle{  \mathbb{B }= \{\varphi : \mathbb{C} \longrightarrow \mathbb{C}  \, entire \, ; \int_{\mathbb{C} }\mid \varphi(z)\mid^{2}e^{-\mid z \mid^{2}}dx dy < \infty \}}
\end{equation}
\n the operators $H_{\mu, \lambda}$, $H_{\lambda',\mu, \lambda}$ and $H_{\lambda'', \lambda',\mu, \lambda}$ takes respectively  the  following form:\\
\begin{equation}
\displaystyle{H_{\mu, \lambda} = i\lambda z\frac{\partial^{2}}{\partial z^{2}} +  (i\lambda z^{2} + \mu z)\frac{\partial}{\partial z}}
\end{equation}
\n acting on its {\bf maximal domain} $\displaystyle{D(H_{\mu, \lambda}) = \{\varphi \in  \mathbb{B} ; H_{\mu, \lambda} \varphi \in  \mathbb{B}\}}$.\\
\begin{equation}
\displaystyle{H_{\lambda', \mu, \lambda} = (\lambda' z^{2} + i\lambda z)\frac{\partial^{2}}{\partial z^{2}} +  (i\lambda z^{2} + \mu z)\frac{\partial}{\partial z}}
\end{equation}
\n acting on its {\bf maximal domain} $\displaystyle{D(H_{\lambda', \mu, \lambda}) = \{\varphi \in  \mathbb{B} ; H_{\lambda' , \mu, \lambda} \varphi \in  \mathbb{B}\}}$.\\
\begin{equation}
\displaystyle{H_{\lambda'', \lambda', \mu, \lambda} = \lambda'' z^{3}\frac{\partial^{3}}{\partial z^{3}} +  (\lambda' z^{2} + i\lambda z)\frac{\partial^{2}}{\partial z^{2}} +  (i\lambda z^{2} + \mu z)\frac{\partial}{\partial z}}
\end{equation}
\n acting on its {\bf maximal domain} $\displaystyle{D(H_{\lambda'', \lambda', \mu, \lambda}) = \{\varphi \in  \mathbb{B} ; H_{\lambda'', \lambda' , \mu, \lambda} \varphi \in  \mathbb{B}\}}$.\\

\n Now, we suppose that $ \lambda'' > 0$, $\lambda' > 0$, $\mu > 0$ and $\lambda \neq 0$, hence  we consider the evolution problems associated respectively to $H_{\mu, \lambda}$, $H_{\lambda', \mu, \lambda}$ and $H_{\lambda'', \lambda', \mu, \lambda}$ with their eigenvalue problems:\\
\begin{equation}
\left \{ \begin{array} [c] {l}\displaystyle{ \frac{\partial u(t, z)}{\partial t} = - H_{\mu, \lambda} u(t, z); \quad t > 0, z \in \mathbb{C}} \\ \quad\\
\displaystyle{ u(0, z) =\phi(z) ; \,  \phi \in \mathbb{B} } \\
\end{array} \right.
\end{equation}
\n with associated eigenvalue problem.\\
\begin{equation}
\left \{ \begin{array} [c] {l}\displaystyle{H_{\mu, \lambda}\varphi_{\mu, \lambda} = \sigma(\mu, \lambda)\varphi_{\mu, \lambda}}\\ \quad \\
\varphi_{\mu, \lambda} \in D(H_{\mu, \lambda}), \sigma(\mu, \lambda) \in \mathbb{C}\\
\end{array} \right .
\end{equation}

\n In (2.9) the physical literature does not put the minus sign. But it very different from a real Schrodinger equation for two reasons at least :\\

\n (i) $t$ is not time but rapidity (the logarithm of collision energy).\\

\n (ii) The $i$ factor is missing and anyway $H_{\mu, \lambda}$ (called three pomerons coupling) is neither self-adjoint nor skew-adjoint. As will be seen, this is the major source of mathematical difficulties.\\

\n (iii) At first glance it is not obvious that imaginary time can be used instead of real time. To motivate the use of imaginary time we can consult  {\color{blue}[Eppens]} (section 4.3, pp: 32-34) entitled ``Real time vs. Imaginary time '' where the imaginary time Schrodinger equation is used to simplify the numerical simulation of quantum adiabatic computation.\\

\n The perturbation of $H_{\mu, \lambda}$ by $\lambda'A^{*^{2}}A^{2}$ (four Pomeron) with the parameter $\lambda' $ (called four coupling of Pomeron) allows us to regularize   $H_{\mu, \lambda}$, then we also consider the following problem:\\
\begin{equation}
\left \{ \begin{array} [c] {l}\displaystyle{ \frac{\partial u(t, z)}{\partial t} = - H_{\lambda' , \mu, \lambda} u(t, z); \quad t > 0, z \in \mathbb{C}} \\ \quad\\
\displaystyle{ u(0, z) =\phi(z) ; \,  \phi \in \mathbb{B} } \\
\end{array} \right. 
\end{equation}
\n with associated eigenvalue problem.\\
\begin{equation}
\left \{ \begin{array} [c] {l}\displaystyle{H_{\lambda',\mu, \lambda}\varphi_{\lambda', \mu, \lambda} = \sigma(\lambda', \mu, \lambda)\varphi_{\lambda', \mu, \lambda}}\\ \quad \\
\varphi_{\lambda', \mu, \lambda} \in D(H_{\lambda', \mu, \lambda}), \, \sigma(\lambda', \mu, \lambda) \in \mathbb{C}\\
\end{array} \right . 
\end{equation}
\n The perturbation of $H_{\lambda', \mu, \lambda}$ by $\lambda''A^{*^{3}}A^{3}$ where the parameter $\lambda'' $ is called by the specialists of high energies ``magic coupling of Pomeron'' allows us to regularize   $H_{\lambda', \mu, \lambda}$, then we also consider the following problem:\\
\begin{equation}
\left \{ \begin{array} [c] {l}\displaystyle{ \frac{\partial u(t, z)}{\partial t} = - H_{\lambda'', \lambda' , \mu, \lambda} u(t, z); \quad t > 0, z \in \mathbb{C}} \\ \quad\\
\displaystyle{ u(0, z) =\phi(z) ; \,  \phi \in \mathbb{B} } \\
\end{array} \right.
\end{equation}
\n with associated eigenvalue problem.\\
\begin{equation}
\left \{ \begin{array} [c] {l}\displaystyle{H_{\lambda'', \lambda',\mu, \lambda}\varphi_{\lambda'', \lambda', \mu, \lambda} = \sigma(\lambda'', \lambda', \mu, \lambda)\varphi_{\lambda'', \lambda', \mu, \lambda}}\\ \quad \\
\varphi_{\lambda', \mu, \lambda} \in D(H_{\lambda', \mu, \lambda}), \, \sigma(\lambda'', \lambda', \mu, \lambda) \in \mathbb{C}\\
\end{array} \right .
\end{equation}
\n And the abstract elliptic problem on $\mathbb{B}$ of Dirichlet type,\\
\begin{equation}
\left \{ \begin{array} [c] {l}\displaystyle{ \frac{\partial^{2} u(t, z)}{\partial t^{2}} = - H_{\lambda'', \lambda' , \mu, \lambda} u(t, z); \quad t > 0, z \in \mathbb{C}} \\ \quad\\
\displaystyle{ u(0, z) =\phi(z) ; \,  \phi \in D(H_{\lambda'', \lambda' , \mu, \lambda}) } \\
\end{array} \right. 
\end{equation}
\n i.e. to consider the evolution problem associated to $\displaystyle{H_{\lambda'', \lambda' , \mu, \lambda}^{\frac{1}{2}}}$ and to show that Then the solution $u(t)$ can  be expanded in $\mathbb{B}$ into the series :\\

 \n $\displaystyle{u(t) = \sum_{k=1}^{+\infty}P_{k}e^{-tH_{\lambda'', \lambda' , \mu, \lambda}^{\frac{1}{2}}} = \sum_{k=1}^{+\infty}e^{-tH_{\lambda'', \lambda' , \mu, \lambda}^{\frac{1}{2}}}P_{k}}$ fot $t >0$.\\
 
 \n where the $P_{k}$'s are the projector operators given by $\displaystyle{P_{k} = \frac{1}{2i\pi}\int_{C_{k}}^{ } (H_{\lambda'', \lambda' , \mu, \lambda} - \sigma I)^{-1}d\sigma}$  with the $C_{k}$'s as small circles centered at the eigenvalues $\{\sigma_{k}\}$ of the operator $H_{\lambda'', \lambda' , \mu, \lambda}$ in the Bargmann space $\mathbb{B}$.\\

\n The complete definition of the above problems requires that the domains respectively  of $H_{\mu, \lambda}$, $H_{\lambda',\mu, \lambda}$ and $H_{\lambda'', \lambda',\mu, \lambda}$ and the boundary conditions be specified.\\

\n The asymptotic behaviour of Reggeon quantum mechanics (i.e., Reggeon field theory without transverse dimensions) has been analyzed by certain specialists of high energies in a series of papers see {\color{blue}[Alessandrini et al]}, {\color{blue}[Jengo]}, {\color{blue}[Bronzan et al]} and {\color{blue}[Ciafaloni et al]}.\\

\n Around the 1980s a controversy raised by Professor R; White with his physic colleagues on certain results established in certain articles by Professors Amati, Ciafaloni and Le bellac without being demonstrated mathematically. \\

\n We recall that the Reggeon field theory (RFT) is an attempt to predict the high-energy behaviour of soft processes; the RFT can be derived from the assumed softness of hadronic interactions at low transverse momenta, which seems to be well established experimentally in hadron-hadron and hadron-nucleus interactions. Unfortunately, nobody has been able up to now to derive the RFT from the fashionable theory of strong interactions, quantum chromodynamics. For this reason, the RFT is at present somewhat out of fashion, but it must be remembered that it is the most elaborate model of soft processes, and may well prove to be a valuable tool for the analysis of forthcoming data from new colliders. We think
that it is still an important task to understand its spectral properties in particular the the Reggeon field theory for $n$ sites which have been began by the author in $1987$ in {\color{blue}[Intissar 2]}.\\

\n The most important question is that of the asymptotic behaviour of total cross sections. Since it seems that the intercept of the bare Pomeron $\alpha := 1 -\mu$ is above one, we have to deal with the problem of a "supercritical" RFT. A solution to the problem has been given in ref.  {\color{blue}[Amati et al]}, and one of its ingredients is the solution of the RFT in zero transverse dimension. In that case, the hamiltonian is given by (2.1),\\

\n The properties of $H_{\mu, \lambda}$ have been studied in ref.  {\color{blue}[Amati et al]}, and this study has been refined in ref.  {\color{blue}[Ciafaloni et al]}.\\

\n  However around the 1980s, White  {\color{blue}[White]} has tried to cast doubt on the conclusion of refs. {\color{blue} [Amati et al] } and {\color{blue}[Ciafaloni et al]}, as the mathematics used in ref.  {\color{blue}[Ciafaloni et al ]} is not beyond criticism (although it is quite respectable according to physicists' standards).\\

\n It was during this period that the professor Michel Lebellac of University of Nice submitted the subject to Professor Martin Zerner and that the latter in turn brought the subject to the author for a mathematical study.\\

\n  We present in the next sections the evolution of significant mathematical results from 1980 to present \\

\n Ciafaloni-Onofri in {\color{blue}[Ciafaloni-Onofri]} have confirmed the results of ref.  {\color{blue}[Ciafaloni et al]} by adding a term $\displaystyle{ \lambda'A^{*^{2}}A^{2}}$ to (2.1) and taking the limit $\lambda' = 0$. However, it is worthwhile to have a completely rigorous study of the hamiltonian (2.1), and this work sums up such a mathematic study which confirms the results of ref. {\color{blue}[Ciafaloni et al]}. Then there cannot be any doubt about the high-energy behaviour of a supercritical RFT.\\

\n The mathematical difficulties of this problem come of course from the non-self-adjointness of $H_{\lambda'', \lambda' , \mu, \lambda}$; $\lambda \neq 0$. Notice that this non-self-adjointness is a rather wild one; the
word ``wild'' meaning here that the domains of the adjoint and anti-adjoint parts are not included in one another, nor is the domain of their commutator. To make
this more precise we want to point out two unpleasant properties of non-self-adjoint operators. First recall that a complex number $z$ is said to be in the spectrum of an operator $V$ if $V - zI$ has no inverse and it is an eigenvalue of V if there is some non-zero vector $u$ such that $Vu = zu$. Now if $V$ is compact, zero is in the spectrum and all other points of the spectrum are eigenvalues. This is a rather powerful result which we will use; but it should not fool us into thinking that all compact operators have eigenvalues (not to speak of a basis of eigenvectors!).\\

\n For instance look at the operator $V$ defined by $\displaystyle{V(u(y) = \int_{0}^{y}u(s)ds}$\\

\n It is a compact operator in $L^{2}(0,1)$ (and in fact in any reasonable Banach space of functions); still it has {\bf no} eigenvalue whatsoever. The second property has to do with the relationship between the spectrum and the asymptotic behaviour of $\displaystyle{e^{ -tH_{\mu, \lambda}}}$, $\displaystyle{e^{ -tH_{ \lambda', \mu, \lambda}}} $ and $\displaystyle{e^{ -tH_{ \lambda'',  \lambda', \mu, \lambda}}}$. \\

\n It is possible to construct an operator $V$ with the two following properties:\\

\n - All points of the spectrum of $V$ are pure imaginary eigenvalues.\\

\n - There are vectors $u$ such that the norm of $\displaystyle{e^{t V}u}$ tends to $\infty $ with $t$.\\

\section{First spectral properties of the Hamiltonian $\displaystyle{ H_{\lambda',\mu, \lambda}}$ of  Reggeon field theory}

\subsection{Some original results on Bargmann space}

\n Let be $O(\mathbb{C})$ the set of entire functions, $\mathbb{M}$ is the operator of multiplication by $z = x + iy ; (x, y) \in \mathbb{R}^{2}$ and $i^{2} = -1$ (i.e $\mathbb{M}\varphi = z\varphi, \varphi \in O(\mathbb{C})$) and  $\mathbb{D}$ is the operator of derivative in order $z$ (i.e   $\mathbb{D}\varphi = \frac{\partial}{\partial z}\varphi, \varphi \in O(\mathbb{C})$). then we have:\\
\begin{lemma}

\n If we request that $\mathbb{M}$ is adjoint to $\mathbb{D}$ in $O(\mathbb{C})$ with inner product :\\
 $\displaystyle{  < \varphi, \psi > = \int_{\mathbb{C}} \varphi(z) \overline{\psi(z)}\rho(z, \bar{z})dxdy}$ where  $\rho(z, \bar{z})$ is a  measure.\\
 Then $\displaystyle{ \rho(z, \bar{z}) = e^{-\mid z\mid^{2}}}$ (Gaussian measure).\\
\end{lemma}
\n {\bf Proof}\\

\n The requirement that $\mathbb{M} = \mathbb{D}^{*}$ gives :\\

\n $\displaystyle{\int_{\mathbb{C}}\frac{\partial}{\partial z }[\varphi(z)]\overline{\psi(z)}\rho((z, \bar{z})dxdy =  \int_{\mathbb{C}}\varphi(z)\overline{z\psi(z)}\rho((z, \bar{z})dxdy}$\\

\n As \\

\n $\displaystyle{\frac{\partial}{\partial z }[\varphi(z)]\overline{\psi(z)}\rho((z, \bar{z}) = \frac{\partial}{\partial z }[\varphi(z)\overline{\psi(z)}\rho((z, \bar{z})] - \varphi(z)\frac{\partial}{\partial z }[\overline{\psi(z)}]\rho((z, \bar{z}) - \varphi(z)\overline{\psi(z)}\frac{\partial}{\partial z }[\rho((z, \bar{z})]}$\\

\n then\\

\n $\displaystyle{\int_{\mathbb{C}}\frac{\partial}{\partial z }[\varphi(z)]\overline{\psi(z)}\rho((z, \bar{z})dxdy = \int_{\mathbb{C}}\frac{\partial}{\partial z }[\varphi(z)\overline{\psi(z)}\rho((z, \bar{z})]dxdy - \int_{\mathbb{C}}\varphi(z)\frac{\partial}{\partial z }[\overline{\psi(z)}]\rho((z, \bar{z})dxdy }$\\

\n $\displaystyle{- \int_{\mathbb{C}}\varphi(z)\overline{\psi(z)}\frac{\partial}{\partial z }[\rho((z, \bar{z})]dxdy}$\\

\n In the right hand side, the first term of the integrand vanishes if we assume that the inner product between $\varphi$ and $\psi$  is finite, so that $\varphi \overline{\psi}\rho \longrightarrow 0$  sufficiently fast  as $\mid z \mid \longrightarrow \infty$. The second term also vanishes,because  $\psi$ is holomorphic, so that  $\overline{\psi}$ is anti-holomorphic and hence $\displaystyle{\frac{\partial}{\partial z}\overline{\psi} = 0}$.This gives:\\
\begin{equation}
\displaystyle{ \int_{\mathbb{C}}\varphi(z)\overline{z\psi(z)}\rho((z, \bar{z})dxdy + \int_{\mathbb{C}}\varphi(z)\frac{\partial}{\partial z }[\overline{\psi(z)}]\rho((z, \bar{z})dxdy = 0}
\end{equation}
\n which is solved for arbitrary $\varphi$ and $\psi$ if \\
\begin{equation}
\displaystyle{\bar{z}\rho((z, \bar{z}) + \frac{\partial}{\partial z }[\rho((z, \bar{z})] = 0}
\end{equation}
\n giving \\
\begin{equation}
\displaystyle{ \rho((z, \bar{z}) = Ce^{-\mid z\mid^{2}}}
\end{equation}
.\hfill { } $\square$

\n The constant $C$ is chosen to be $\displaystyle{\frac{1}{\pi}}$, so that the norm of the constant function $\varphi(z) = 1$ is one. This explains why the space of holomorphic functions equipped with a Gaussian measure gives Bargmann space:\\
\begin{equation}
\displaystyle{ \mathbb{B }= \{\varphi : \mathbb{C} \longrightarrow \mathbb{C}  \, entire \, ; \int_{\mathbb{C} }\mid \varphi(z)\mid^{2}e^{-\mid z \mid^{2}}dx dy < \infty \}}
\end{equation}
\n Let 
\begin{equation}
\displaystyle{ \mathbb{B}_{s} = \{(a_{n})_{n=0}^{\infty} \in \mathbb{C} ; \sum_{n=0}^{\infty}n!\mid a_{n}\mid^{2} < \infty \}}
\end{equation}
\n $\mathbb{B}_{s}$ is also a Hilbert space. The scalar product on $\mathbb{B}_{s}$ is defined by\\
 \begin{equation}
 \displaystyle{ < (a_{n}) , (b_{n} >_{s} = \sum_{n=0}^{\infty}n!a_{n}\overline{b_{n}}}
 \end{equation}
\n  and the associated norm is denoted by $\mid\mid . \mid\mid_{s}$.\\

\n It is well known that  $\mathbb{B}$ is related to $\mathbb{B}_{s} $ by unitary transform of  $\mathbb{B}$ onto $\mathbb{B}_{s} $, given by the following transform :\\
\begin{equation}
\displaystyle{ I: \mathbb{B} \longrightarrow \mathbb{B}_{s} , \varphi(z) = \sum_{n=0}^{\infty}a_{n}z^{n} \longrightarrow I(\varphi) = (\frac{1}{n!}\varphi^{(n)}(0))_{n=0}^{\infty} = (a_{n})_{n=0}^{\infty}}
\end{equation}
\begin{lemma}

\n As the operator of the derivative on Bargmann space $\mathbb{B}$ is unbounded then we have the following result :\\

\n $\displaystyle{\varphi \in \mathbb{B}\, \Longleftrightarrow \,   z \longrightarrow \frac{\varphi'(z) - \varphi'(0)}{z} \in \mathbb{B}}$\\
\end{lemma}
\n {\bf Proof}\\

\n $\Longrightarrow$\\

\n For an entire function  $\displaystyle{\phi(z) = }\displaystyle{\sum_{k=0}^{+\infty}a_{k}z^{k}}$  on $\mathbb{C}$ we have:\\

\n $\displaystyle{\phi'(z) = \displaystyle{\sum_{k=0}^{+\infty}(k+1)a_{k+1}z^{k}}}$, $\displaystyle{\phi'(0) = a_{1}}$  and  $\displaystyle{\frac{\phi'(z) - \phi'(0)}{z} =  \sum_{k=0}^{+\infty}(k+2)a_{k+2}z^{k}}$\\ 
 
\n By using the isometry between Bargmann space $\mathbb{B}$ and $\mathbb{B}_{s}$, we deduce that:\\

\n $\displaystyle{\mid\mid \frac{\varphi'(z) - \varphi'(0)}{z} \mid\mid^{2} =  \displaystyle{\sum_{k=0}^{+\infty}k!(k+2)^{2}\mid a_{k+2}\mid^{2}}}$\\
i.e.,\\
\begin{equation}
\displaystyle{\mid\mid \frac{\varphi'(z) - \varphi'(0)}{z} \mid\mid^{2} =  \displaystyle{\sum_{k=0}^{+\infty}\frac{k+2}{k+1}(k+2)!\mid a_{k+2}\mid^{2}}}
\end{equation}
\n As $\displaystyle{\frac{k+2}{k+1} \leq 2}$, then   $\displaystyle{\mid\mid \frac{\varphi'(z) - \varphi'(0)}{z} \mid\mid^{2}  \leq 2\sum_{k=0}^{+\infty}(k+2)!\mid a_{k+2}\mid^{2}}$,
 it follows that if \\$\displaystyle{\varphi \in \mathbb{B} \,\, then \,   z \longrightarrow \frac{\varphi'(z) - \varphi'(0)}{z} \in \mathbb{B}}$\\
 
\n $\Longleftarrow$\\

\n For the reciprocity, we remark that  $ \displaystyle{\frac{k+2}{k+1} > 1 }$, and from (3.8) we deduce that $\displaystyle{ \mid\mid \varphi \mid\mid \leq \mid\mid \frac{\varphi'(z) - \varphi'(0)}{z} \mid\mid}$, it follows that if $\displaystyle{ z \longrightarrow \frac{\varphi'(z) - \varphi'(0)}{z} \in \mathbb{B} \,\, then \,  \varphi \in \mathbb{B} }$. \hfill { } $\square$\\
\begin{theorem}

\n The Segal-Bargmann transform $\displaystyle{B : L_{2}(\mathbb{R}) \longrightarrow \mathbb{B} : f  \longrightarrow \varphi}$ defined by\\

\n $\displaystyle{B[f](z) = \varphi(z) = \int_{\mathbb{R}}\mathcal{A}(z, u)f(u)du}$ \hfill { } ($\star$)\\

\n with \\

\n $\displaystyle{\mathcal{A}(z, u) = e^{-\frac{u^{2}}{2} + \sqrt{2}uz -\frac{z^{2}}{2}}}$  \hfill { } ($\star\star$)\\

\n is surjective isometry.\\
\end{theorem}
\n {\bf Proof}\\

\n It is well Known that the proof of this theorem consists of three steps. First by showing that for any $f \in L_{2}(\mathbb{R})$ the image $Bf$ is a holomorphic function. Secondly by introducing the complete orthogonal sequences $\{f_{k}\} \in L_{2}(\mathbb{R})$ and $\{z^{k}\} \in \mathbb{B}$ and showing that $\mathcal{B}f_{k} = z^{k}$ . Lastly by gathering all the data obtained from above steps, we can conclude that $B$ maps into $\mathbb{B}$ isometrically and surjectively. \hfill { } $\square$\\ 

\n The Segal-Bargmann transform was introduced (depently) by I. Segal and V. Bargmann near 1960 and also by F.A Berezin in the same time give a generalization of this transformation). \\
 \begin{theorem}

\n Let $\mathbb{B}$ be Bargmann space,  if $\varphi \in \mathbb{B}$ then its restriction on  $x + i\mathbb{R}$ is square integrable function with measure $\displaystyle{e^{-\mid y \mid^{2}}dy , y \in \mathbb{R}}$ for all  fixed $x \in \mathbb{R}$.\\
\end{theorem}
\n {\bf Proof}\\

\n Bargmann has built an isometry between the space $\mathbb{B}$ and $L_{2}(\mathbb{R})$ so that,  $\forall \,\varphi \in \mathbb{B}$ is uniquely represented by $f \in  L_{2}(\mathbb{R})$ by means of the following integral:\\
\begin{equation}
\displaystyle{\varphi (z) = c\int_{\mathbb{R}}e^{- \frac{z^{2}}{2} - \frac{q^{2}}{2} +\sqrt{2}zq } f(q)dq  }
\end{equation}
\begin{equation}
\displaystyle{\mid\mid \varphi \mid\mid = \mid\mid f \mid\mid_{L_{2}( \mathbb{R})}}
\end{equation}

\n Now, we put $\displaystyle{g_{z}(q) = e^{- \frac{q^{2}}{2} +\sqrt{2}zq }}$, which is function belonging to $L_{2}(\mathbb{R})$. Then we can write (1.2.9) in the following form :\\
\begin{equation}
\displaystyle{e^{\frac{z^{2}}{2}}\phi (z) = }c\displaystyle{\int_{\mathbb{R}}g_{z}(q)f(q)dq  }
\end{equation}

\n As $g_{z}(q)$ and $f$ are in  $L_{2}(\mathbb{R})$, we can apply the Parseval's identity at (3.11) to get :\\
\begin{equation}
\displaystyle{e^{\frac{z^{2}}{2}}\varphi (z) = c_{1}}\displaystyle{\int_{\mathbb{R}}\hat{g}_{z}(p)\hat{f}(p)dp}
\end{equation}

\n As $\displaystyle{g_{z}(q)}$ is a Gaussian's function,  one knows how to calculate its Fourier transform:\\

\n $\displaystyle{\hat{g}_{z}(p) = \int_{\mathbb{R}}e^{ipq}g_{z}(q)dq =\int_{\mathbb{R}}e^{- \frac{q^{2}}{2} + (\sqrt{2}z - ip)q} dq = (2\pi)^{\frac{1}{2}}e^{\frac{1}{2}(\sqrt{2}z - ip)^{2}}}$\\

 \n i.e.\\
\begin{equation}
\displaystyle{\hat{g}_{z}(p) = (2\pi)^{\frac{1}{2}}e^{\frac{1}{2}(\sqrt{2}z - ip)^{2}}}
\end{equation}

 \n Let $z = x + iy$, then for all fixed $x \in \mathbb{R}$, the function $\hat{g}_{z}(p)$ can be written under following form:\\
\begin{equation}
\displaystyle{\hat{g}_{z}(p) = (2\pi)^{\frac{1}{2}}e^{\frac{1}{2}(\sqrt{2}x - i(p - \sqrt{2}y)^{2}}}
\end{equation}

\n  If we put $\displaystyle{h_{x}(p) = e^{x^{2}}e^{-\frac{1}{2}(p^{2} + i\frac{\sqrt{2}}{2}px)}}$ then $\displaystyle{e^{\frac{z^{2}}{2}}\varphi (z)}$ is the convolution product of $h_{x}$ with $\hat{f}$ evaluated in $\sqrt{2}y$, i.e.,\\
\begin{equation}
\displaystyle{e^{\frac{(x + iy)^{2}}{2}}\varphi (x + iy) = C_{2} h_{x}*\hat{f}(\sqrt{2}y) }
\end{equation}

\n It follows that $\displaystyle{y \rightarrow e^{\frac{(x + iy)^{2}}{2}}\varphi (x + iy)}$ is in  $L_{2}(\mathbb{R})$ for all $x \in \mathbb{R}$\\.

\n By applying the Young's inequality we deduce that :\\
\begin{equation}
\mid\mid e^{\frac{(x + iy)^{2}}{2}}\varphi (x + iy)\mid\mid^{2}_{L_{2}(\mathbb{R})} \leq \mid\mid h_{x}\mid\mid^{2}_{L_{1}((\mathbb{R}^{N})}.\mid\mid \hat{f}\mid\mid^{2}_{L_{2}(\mathbb{R})}
\end{equation}

\n where $C_{1}, C_{2}$ and $C$ are constants.\\

\n As $\mid\mid h_{x}\mid\mid^{2}_{L_{1}(\mathbb{R})} = C_{3}e^{2\mid x \mid^{2}}$ et $\mid\mid \hat{f}\mid\mid^{2}_{L_{2}(\mathbb{R})} = C_{4}\mid\mid f \mid\mid^{2}_{L_{2}(\mathbb{R})}$, then we deduce that :\\
\begin{equation}
\displaystyle{\int_{\mathbb{R}}e^{-\mid y \mid^{2}}\mid \varphi (x + iy)\mid^{2}dy \leq C_{5}e^{-\mid x \mid^{2}}\mid\mid \varphi \mid\mid^{2}}
\end{equation}

\n where $C_{5}$ depends of  all the previous constants.\hfill { } $\square$\\

\n   (3.17) is valid on $\mathbb{C}^{n}$ by taking :\\
\begin{equation}
\displaystyle{ \mathbb{B }= \{\varphi : \mathbb{C}^{n} \longrightarrow \mathbb{C}  \, entire \, ; \int_{\mathbb{C}^{n} }\mid \varphi(z)\mid^{2}e^{-\mid z \mid^{2}}dx dy < \infty \}}
\end{equation}

\n where \\

\n $\displaystyle{x = (x_{1}, ..., x_{j}, ...,x_{n })\in \mathbb{R}^{n},  y = (y_{1}, ..., y_{j}, ...,y_{n })\in \mathbb{R}^{n}}$ and \\

\n $\displaystyle{ z = (z_{1}, ..., z_{j}, ...,z_{n })\in \mathbb{C}^{n}; z_{j} = x_{j} + iy_{j}}$\\
\subsection{Representation of $H_{\lambda', \mu, \lambda}$ in $\displaystyle{\mathbb{B}_{s} = \{ (a_{n})_{n=0}^{\infty} \in \mathbb{C} ; \sum_{n=0}^{\infty} n! \mid a_{n} \mid^{2} < \infty\}}$} 
\n The Segal-Bargmann transform is a very well-known operator acting from the space $L_{2}(\mathbb{R})$ and the Segal-Bargmann-Fock space which is unitary and allows identifying these two spaces see theorem 3.3.\\

\n In this under-section, we  built an isometry $I$ between the space $\mathbb{B}$ and \\

\n  $\displaystyle{\mathbb{B}_{s} = \{ (a_{n})_{n=0}^{\infty} \in \mathbb{C} ; \sum_{n=0}^{\infty} n! \mid a_{n} \mid^{2} < \infty\}}$ to represent the annihilation and creation operators in this representation $\displaystyle{\mathbb{B}_{s}}$.\\

\n Now  we recall that $\mathbb{B}$ is defined as follow \\
\begin{equation}
\displaystyle{ \mathbb{B }= \{\varphi : \mathbb{C} \longrightarrow \mathbb{C}  \, entire \, ; \int_{\mathbb{C} }\mid \varphi(z)\mid^{2}e^{-\mid z \mid^{2}}dx dy < \infty \}}
\end{equation}

\n The scalar product on $\mathbb{B}$ is defined by\\
\begin{equation}
<\phi,\psi> = \displaystyle {\int_{\mathbb{C}}}\displaystyle{\phi(z)\overline{\psi(z)}e^{-\mid z\mid^{2}}dxdy}
\end{equation}

\n and the associated norm is denoted by $\mid\mid . \mid\mid $.\\

\n $\mathbb{B}$ is closed in $L_{2}(\mathbb{C}, d\mu(z))$ where the measure $d\mu(z) = e^{-\mid z\mid^{2}}dxdy$ and it is closed related to $L_{2}(\mathbb{R} )$ by an unitary transform of $L_{2}(\mathbb{R} )$ onto $\mathbb{B}$ given in {\color{blue}[Bargmann1]} by the following integral transform\\
\begin{equation}
\phi(z) = \displaystyle{\int_{\mathbb{R} }\,e^{-\frac{1}{2}(z^{2} + u^{2}) +\sqrt{2}uz}f(u)du }
\end{equation}

\n if $f$ $\in L_{2}(\mathbb{R} )$ the integral converges absolutely.\\
\begin{lemma}

\n Let $\phi $ be in $\mathbb{B}$ with $\phi(z) = \displaystyle{\sum_{n=0}^{\infty}a_{n}z^{n}; (a_{n})_{n=0}^{\infty} \in \mathbb{C} }$ then \\ $\displaystyle{\mid\mid \phi \mid\mid^{2} = \pi \sum_{n=0}^{\infty}n!\mid a_{n}\mid^{2}}$\\
\end{lemma}
\n {\bf Proof}\\

\n Let $\phi$ be in $B$ and we put $z = re^{i\theta}$ with $r \in [0, +\infty[$ and $\theta \in [0, 2\pi]$ then\\ $\phi(z) = \phi(re^{i\theta}) = \displaystyle{\sum_{n=0}^{+\infty}a_{n}r^{n}e^{in\theta}}$\\

\n We write $\mid\mid \phi \mid\mid^{2}$ and $\mid \phi(re^{i\theta})\mid^{2}$ as following:\\

\n $\mid\mid \phi \mid\mid^{2} =\displaystyle{\int_{\mathbb{C}}e^{-\mid z\mid^{2}}\mid\phi(z)\mid^{2}dzd\bar{z}} =\displaystyle{ _{_{_{\sigma \rightarrow +\infty}}}\!\!\!\!\!\!\!\!\!\!Lim \int_{0}^{\sigma}\int_{0}^{2\pi}e^{-r^{2}}\mid \phi(re^{i\theta})\mid^{2}rdrd\theta}$\\

\n and\\

\n $\mid \phi(re^{i\theta})\mid^{2} = \displaystyle{\sum_{n=0}^{+\infty}\displaystyle{\sum_{p+q =n}a_{p}\bar{a}_{q}r^{p+q}e^{i(p-q)\theta}}}$\\

\n By Beppo-Levi and Fubini theorems, we obtain that\\

\n $\displaystyle{\int_{\mathbb{C}}e^{-\mid z\mid^{2}}\mid\phi(z)\mid^{2}dzd\bar{z}}$ =
$\displaystyle{ _{_{_{\sigma \rightarrow +\infty}}}\!\!\!\!\!\!\!\!\!\!lim \int_{0}^{\sigma}e^{-r^{2}}}$
$\displaystyle{\sum_{n=0}^{+\infty}}$ $\displaystyle{\sum_{p+q=n}a_{p}\bar{a}_{q}r^{p+q+1}}$
$\displaystyle{\int_{0}^{2\pi}e^{i(p-q)\theta}drd\theta}$\\

\n As $\displaystyle{\int_{0}^{2\pi}e^{i(p-q)\theta}d\theta} = 2\pi\delta_{pq}$ where $\delta_{pq}$ is Kronoecker's symbol then we get\\

\n $\displaystyle{\int_{\mathbb{C}}e^{-\mid z\mid^{2}}\mid\phi(z)\mid^{2}dzd\bar{z}} = \displaystyle{ _{_{_{\sigma \rightarrow +\infty}}}\!\!\!\!\!\!\!\!\!\!lim \int_{0}^{\sigma}e^{-r^{2}}\displaystyle{\sum_{n=0}^{+\infty}\displaystyle{\sum_{2p=n}a_{p}\bar{a}_{p}r^{2p+1}}}}2\pi dr$\\

$= $
$2\pi\displaystyle{_{_{_{\sigma \rightarrow +\infty}}}\!\!\!\!\!\!\!\!\!\!lim \int_{0}^{\sigma}e^{-r^{2}}}$
$\displaystyle{\sum_{n=0}^{+\infty}a_{n}\bar{a}_{n}r^{2n+1}}dr$\\

\n Again by applying Beppo-Levi's theorem, we obtain\\

\n $\displaystyle{\int_{\mathbb{C}}e^{-\mid z\mid^{2}}\mid\phi(z)\mid^{2}dzd\bar{z}}$ = $\displaystyle{\sum_{n=0}^{+\infty}\mid a_{n}\mid^{2}\displaystyle{ _{_{_{\sigma \rightarrow +\infty}}}\!\!\!\!\!\!\!\!\!\!lim \int_{0}^{\sigma}e^{-r^{2}}r^{2n+1}2\pi dr}}$\\

\n But \\

\n $\displaystyle{_{_{_{\sigma \rightarrow +\infty}}}\!\!\!\!\!\!\!\!\!\!lim \int_{0}^{\sigma}e^{-r^{2}}r^{2n+1}2\pi dr}$
= $\pi n!$\\

\n Therefore\\

\n $\displaystyle{\int_{\mathbb{C}}e^{-\mid z\mid^{2}}\mid\phi(z)\mid^{2}dzd\bar{z}} = \pi\displaystyle{\sum_{n=0}^{+\infty}n! \mid a_{n}\mid^{2} < +\infty}$.\hfill { } $\square$\\

\n Now let
\begin{equation}
\displaystyle{ B_{s} = \{(a_{n})_{n=0}^{\infty} \in \mathbb{C} \, such\, that\, \sum_{n=0}^{\infty}n!\mid a_{n}\mid^{2} < \infty \}}
\end{equation}

\n $\mathbb{B}_{s}$ is also a Hilbert space. The scalar product on $\mathbb{B}_{s}$ is defined by\\
\begin{equation}
< (a_{n})_{n=0}^{\infty}, (b_{n})_{n=0}^{\infty}> = \displaystyle{\sum_{n=0}^{\infty}n!a_{n}\overline{b_{n}}}
\end{equation}
and the associated norm is denoted by $\mid\mid . \mid\mid_{s} $.\\

\n $\mathbb{B}$ related to $\mathbb{B}_{s}$ by an unitary transform of $\mathbb{B}$ onto $\mathbb{B}_{s}$, given  by the following transform {\bf I}\\
\begin{equation}
{\bold I} : B \mapsto \mathbb{B}_{s} ; I(\phi) = (\frac{1}{n!}\phi^{(n)}(0))_{0}^{\infty} = (a_{n})_{n=0}^{\infty}
\end{equation}
\n Let\\
\begin{equation}
B_{s}^{0} = \{(a_{n})_{n=0}^{\infty} \in B_{s} \, such \, that \, a_{0} = 0\} 
\end{equation}

\n This space is associated to Bargmann space\\
\begin{equation}
\mathbb{B}_{0} = \{\phi \in \mathbb{B}\, such\,\, that \, \phi(0) = 0\}
\end{equation}

\n The realization in $\mathbb{B}_{s}$ of operators $\displaystyle{A\phi(z) = \frac{d}{dz}\phi(z)}$ with domain $\displaystyle{D(A) = \{\phi \in \mathbb{B}; \frac{d}{dz}\phi \in \mathbb{B}\}}$

\n and $\displaystyle{A^{*}\phi(z) = z\phi(z)}$ with domain $\displaystyle{D(A^{*}) = \{\phi \in \mathbb{B}; z\phi \in \mathbb{B}\}}$\\

\n is given by\\
\begin{equation}
\left\{\begin{array}[c]{l}A(a_{n})_{n=0}^{\infty}=((n+1)a_{n+1})_{n=0}^{\infty}\\ \quad \\ with \quad domain:\\

\n D(A) = \{(a_{n})_{n=0}^{\infty} \in \mathbb{B}_{s}; \displaystyle{\sum_{n=0}^{\infty}n!n\mid a_{n}
\mid^{2}} < \infty\} \\ \end{array}\right.
\end{equation}
\n and\\
\begin{equation}
\left\{\begin{array}[c]{l}A^{*}(a_{n})_{n=0}^{\infty}=(a_{n-1})_{n=0}^{\infty}\quad;\quad a_{-1} = 0 \\ \quad \\ with \quad domain: \\
D(A^{*}) = \{(a_{n})_{n=0}^{\infty} \in \mathbb{B}_{s};\displaystyle{\sum_{n=0}^{\infty}n!\mid a_{n-1}\mid^{2}} < \infty\} \\ \end{array}\right. 
\end{equation}
\n In $\mathbb{B}_{s}$ realization, the first result on the domains of $A$ and $A^{*}$ is\\
\begin{lemma}

\n i) $D(A)$ is dense in $\mathbb{B}_{s}$.\\

\n ii) $D(A) = D(A^{*})$\\

\n iii) $D(A) \hookrightarrow \mathbb{B}_{s}$ is compact.\\
\end{lemma}
\n {\bf Proof }\\

\n i) and ii) is obvious.\\

\n iii) Let $C_{s}$ be the unit ball of $D(A)$ equipped with graph norm:\\

\n $C_{s} =\{(a_{n})_{n=0}^{\infty} \in D(A)$ such that $\mid\mid(a_{n})_{n=0}^{\infty}\mid\mid_{s}^{2} + \mid\mid A((a_{n})_{n=0}^{\infty})\mid\mid_{s}^{2} \leq 1\}$ i.e. \\

\n $C_{s} =\{(a_{n})_{n=0}^{\infty} \in D(A)$ such that $\displaystyle {\mid a_{0}\mid^{2} +\sum_{n=1}^{\infty}(n+1)!\mid a_{n}\mid^{2} \leq 1} \}$\\

\n To prove that $C_{s}$ is relatively compact for the norm associated to $\mathbb{B}_{s}$, we observe that $\mathbb{B}_{s}$ is complete space then
it suffices to show that $C_{s}$ can be covering by finite number of unit balls of radius $\frac{1}{p}$ with arbitrary $p\in \mathbb{N}-\{0\}$ \\

\n Writing $\displaystyle {\sum_{n=1}^{\infty}(n+1)!\mid a_{n}\mid^{2}}$ as follows\\

 \n  $\displaystyle {\sum_{n=1}^{\infty}(n+1)!\mid a_{n}\mid^{2} = 2!\mid a_{1}\mid^{2} + 3!\mid a_{2}\mid^{2}+4!\mid a_{3}\mid^{2}+ ............} $\\

 \n  $ = $ $1!\mid a_{1}\mid^{2} + 2!\mid a_{2}\mid^{2}+3!\mid a_{3}\mid^{2}+ ............................+$\\

\n   $\quad$ $1!\mid a_{1}\mid^{2} + 2!\mid a_{2}\mid^{2}+3!\mid a_{3}\mid^{2}+ ............................+$\\

\n  $\quad$ $0 \quad \quad \quad + 2!\mid a_{2}\mid^{2}+3!\mid a_{3}\mid^{2}+ .............................+$\\

\n   $\quad$ $0 \quad \quad \quad +  0 \quad \quad \quad +  3!\mid a_{3}\mid^{2}+ ............................+$\\

\n   $\quad$ $............................................................................... +$\\

\n  $\quad$ $................................................................................+$\\

\n $ = 2 \displaystyle {\sum_{n=1}^{\infty}n!\mid a_{n}\mid^{2}}$ + $\displaystyle {\sum_{n=2}^{\infty}n!\mid a_{n}\mid^{2}}$ + ........ + $\displaystyle{\sum_{n=p}^{\infty}n!\mid a_{n}\mid^{2}}$ + .......\\

\n then we deduce that $(p+1)\displaystyle{\sum_{n=p}^{\infty}n!\mid a_{n}\mid^{2}} \leq 1$
i.e. $(0,0, ......, 0,a_{p}, a_{p+1}, ......)$ is in $C_{s}^{0}(0,\frac{1}{p})$ the ball
with radius $\frac{1}{p}$ around the origin of $\mathbb{B}_{s}$.\\

\n As the set $\{(a_{1}, a_{2}, .....,a_{p-1}) \in \mathbb{C}^{p-1}$ ; $\displaystyle {\sum_{n=1}^{p-1}n!\mid a_{n}\mid^{2}}\leq 1\}$ is compact then they exist $m$ balls $K(x_{i},\frac{1}{p})$ with center $x_{i}\in \mathbb{C}^{p-1}$ of radius $\frac{1}{p}$ such that
$\{(a_{1}, a_{2}, .....,a_{p-1} \in \mathbb{C}^{p-1}$;$\displaystyle {\sum_{n=1}^{p-1}n!\mid a_{n}\mid^{2}}\leq 1\} = \bigcup_{i=1}^{m} K(x_{i},\frac{1}{p})$\\

\n then we get\\

\n $C_{s}\subset \displaystyle {\bigcup_{i=1}^{m} }K(x_{i},\frac{1}{p})\bigcup C_{s}^{0}(0,\frac{1}{p})$
and we deduce the property iii)..\hfill { } $\square$ \\

 \n We can also prove iii) by applying a following classic proposition\\
\begin{proposition}
{\color{blue} [Fr\'echet]}\\

\n  Let $K \subset \mathbb{B}_{s}$ such that\\

\n  i) $K$ is bounded and closed.\\

\n  ii) $\forall \varepsilon > 0$, $\exists N_{\varepsilon} > 0$ such that $\displaystyle{\sum_{n= N_{\varepsilon}}^{\infty}n!\mid a_{n}\mid^{2} \leq \varepsilon \quad \forall \{a_{n}\}_{n=1}^{\infty}} \in K$\\

\n then $ K $ is compact.\hfill { } $\square$\\
\end{proposition}
\n In $\mathbb{B}$ representation or $\mathbb{B}_{s}$  representation we have the following fundamental lemma \\
\begin{lemma}

\n Let $e_{1}=^{t}(0,1,0,.........)$ and $S(a_{n})_{n=0}^{\infty}=(a_{n+1})_{n=0}^{\infty}$ then we have\\

\n i) In $\mathbb{B}$, $\displaystyle{\phi \in \mathbb{B}  \Leftrightarrow z \rightarrow \frac{\phi^{'}(z) - \phi^{'}(0)}{z} \in \mathbb{B}}$\\

\n ii) In $\mathbb{B}_{s}$, $\displaystyle{(a_{n})_{n=0}^{\infty}\in \mathbb{B}_{s} \Leftrightarrow S(a(a_{n})_{n=0}^{\infty} -a_{1}e_{1}) \in \mathbb{B}_{s} \Leftrightarrow  \sum_{n=0}^{\infty}n!(n +2)^{2}\mid a_{n+2}\mid^{2}< \infty}$\\
\end{lemma}
\n {\bf Proof }\\

\n i) For $\phi(z) = \displaystyle {\sum_{n=0}^{\infty}a_{n}z^{n}}$ then $\phi^{'}(z) = \displaystyle {\sum_{n=0}^{\infty}(n+1)a_{n+1}z^{n}}$ and $\frac{\phi^{'}(z) - \phi^{'}(0)}{z} = \displaystyle {\sum_{n=0}^{\infty}(n + 2)a_{n+2}z^{n}}$\\

\n By using the isometry  between Bargmann space $\mathbb{B}$ and $\mathbb{B}_{s}$  we deduce that \\

\n $\mid\mid \frac{\phi^{'}(z) - \phi^{'}(0)}{z}\mid\mid^{2} = \displaystyle {\sum_{n=0}^{\infty}n!(n+2)^{2}\mid a_{n+2}\mid^{2}}$, i.e.\\

\n  $\mid\mid \frac{\phi^{'}(z) - \phi^{'}(0)}{z}\mid\mid^{2} = \displaystyle {\sum_{n=0}^{\infty}\frac{n+2}{n+1}(n+2)!\mid a_{n+2}\mid^{2}}$ $\hfill { }  (*)$\\

\n Now, as $\frac{n+2}{n+1} < 2$ then $\mid\mid \frac{\phi^{'}(z) - \phi^{'}(0)}{z}\mid\mid^{2} \leq \displaystyle {2\sum_{n=0}^{\infty}(n+2)!\mid a_{n+2}\mid^{2}}$ \\

\n if $\phi \in \mathbb{B}$ then $z\rightarrow \frac{\phi^{'}(z) - \phi^{'}(0)}{z} \in \mathbb{B}$\\

\n For the reciprocity, we remak that $\frac{n+2}{n+1} > 1$ and from (*) we deduce that
$\mid\mid \phi \mid\mid \leq  \mid\mid \frac{\phi^{'}(z) - \phi^{'}(0)}{z}\mid\mid$\\

\n ii) is a simple version of i) in $\mathbb{B}_{s}$.\hfill { } $\square$\\

\n Now starting by a list of notations and remarks\\

\n In $\mathbb{B}$ representation the operator  $ H := H_{\mu,\lambda}$ is defined by\\
\begin{equation}
 \left\{\begin{array}[c]{l}H\phi(z) = \quad i\lambda z\phi^{''}(z)  + (i\lambda z^{2} + \mu z)\phi^{'}(z)\\ \quad \\ with \quad maximal \quad domain: \\\quad \\
D(H_{max}) = \{\phi \in \mathbb{B} \quad such \quad that \quad  H\phi \in \mathbb{B}\}\\ \end{array}\right. 
\end{equation}
\begin{remark}

\n i) Let $\mathcal{P}$ be space of polynomials then it is dense in $\mathbb{B}$.\\

\n  ii) Let  $H_{\mid_{\mathcal{P}}}$ be the restriction of $H_{\mu, \lambda}$ to polynomials space,  we can define $H_{min}$ as the closure of operator $H_{\mid_{\mathcal{P}}}$ in Bargmann space:\\
\begin{equation}
\left\{\begin{array}[c]{l}H_{min}\phi= i\lambda z\phi^{''}  + (i\lambda z^{2} + \mu z)\phi^{'} \\ \quad \\
with  \quad minimal \quad domain: \\ \quad \\
D(H_{min}) = \{\phi \in \mathbb{B}; \, \exists p_{n} \in \mathcal{P}\, and \, \psi \in \mathbb{B} ; p_{n}\rightarrow \phi \, and\, Hp_{n}\rightarrow \psi\}\\ \end{array}\right.
\end{equation}
\n iii) \begin{equation}
H_{min}^{*} = H_{\mu,-\lambda} \,  \,  with \, \, domain  \, \, D(H_{min}^{*}) = D(H_{max})
\end{equation}
\n iv) An orthonormal basis of $B$ is given by :
\begin{equation}
\displaystyle{ e_{n}(z) = \frac{z^n}{\sqrt{n!}}; n = 0, 1, ....}
\end{equation}
\n v)  The action of the operators $A$, $A^{*}$ and $H$ on the basis $\displaystyle{e_{n}(z)= \frac{z^{n}}{\sqrt{n!}} n = 0, 1, .....}$ is:\\

\n {\color{blue}$\bullet$} $A(e_{n}) = \sqrt{n}e_{n-1}$ with the convention $e_{-1}=0$\\

\n {\color{blue}$\bullet$} $A^{*}(e_{n}) = \sqrt{n+1}e_{n+1}$\\

\n  and\\

\n {\color{blue}$\bullet$} $H_{\mu, \lambda}(e_{n}) =  i\lambda(n-1)\sqrt{n}e_{n-1} + \mu ne_{n} + i\lambda n\sqrt{n+1}e_{n+1}$\\

\n vi) let $l^{2}(\mathbb{N})$ = $\displaystyle{\{ (\phi_{n})_{n=0}^{\infty}\in \mathbb{C}\quad such \quad that\quad \sum_{n=0}^{\infty}\mid\phi_{n}\mid^{2} < +\infty\}}$ with the inner product:\\
\begin{equation}
\displaystyle{< (\phi_{n})_{n=0}^{\infty}, (\psi_{n})_{n=0}^{\infty} > = \sum_{n=0}^{\infty}\phi_{n}\bar{\psi}_{n}}
\end{equation}
\n In the representation $l^{2}(\mathbb{N})$ where the coefficients $\phi_{n}$ define an entire function \\
$\phi(z) = \displaystyle{\sum_{n=0}^{\infty}\phi_{n}e_{n}(z)}$ in Bargmann space we have\\
- \begin{equation}
\left\{\begin{array}[c]{l}(A\phi)_{n}=\sqrt{n}\phi_{n-1}, \phi_{-1}= 0\\ \quad \\ with \quad domain: \\ \quad \\ D(A) = \{(\phi)_{n} \in l^{2}(\mathbb{N}); \displaystyle{\sum_{n=0}^{\infty}n\mid \phi_{n}\mid^{2} < \infty} \}\\ \end{array}\right. 
\end{equation}
-\begin{equation}
 \left\{\begin{array}[c]{l}(A^{*}\phi)_{n} =\sqrt{n+1}\phi_{n+1}\\ \quad \\ with \quad domain: \\ \quad \\ D(A^{*})=\{(\phi)_{n} \in l^{2}(\mathbb{N});\displaystyle {\sum_{n=0}^{\infty}n\mid \phi_{n}\mid^{2} < \infty }\} \\ \end{array}\right. 
\end{equation}
-\begin{equation}
\left\{\begin{array}[c]{l} (H_{\mu, \lambda}\phi)_{n} = i\lambda (n-1)\sqrt{n}\phi_{n-1} + \mu n\phi_{n} + i\lambda n\sqrt{n+1}\phi_{n+1}\\ \quad \\ with \quad domain: \\ \quad \\ D(H_{\mu, \lambda}) = \{\phi \in l^{2}(\mathbb{N}); H\phi \in l^{2}(I\!\!N)\}\end{array}\right. 
\end{equation}
\n vi) $\mathbb{B} = \mathbb{B}_{0}\bigoplus\{constants\}$  and zero is eigenvalue of  $H_{\mu, \lambda}$ without interest.\\

\n vii) We define $l_{0}^{2}(\mathbb{N}) = \{\phi =(\phi_{n})_{n=0}^{\infty} \in l^{2}(\mathbb{N}); \phi_{0} = 0\}$ and the operators $A$ and $A^{*}$ by (3.34) and (3.35) and \\

\n  $H_{\mu, \lambda} = \mu A^{*}A + i\lambda A^{*}(A + A^{*})A$ with domain $D(H_{\mu, \lambda}) = \{\phi \in l_{0}^{2}(\mathbb{N}); H\phi \in l_{0}^{2}(\mathbb{N})\}$\\

\n In the representation $l_{0}^{2}(\mathbb{N})$ where the coefficients $\phi_{n}$ define an entire function\\ $\phi(z) = \displaystyle{\sum_{n=1}^{\infty}\phi_{n}e_{n}(z)}$ in Bargmann space $B_{0}$, we study, in the present paper, a class of Jacobi-Gribov matrices with unbounded entries:\\

\begin{equation}
\left \{ \begin{array}{c} (H\phi)_{n} = \alpha_{n-1}\phi_{n-1} + q_{n}\phi_{n} +\alpha_{n}\phi_{n+1}, n\geq 2 \\
\quad\\
with\,\, the \,\, initial \,\, condition\\
\quad\\(H\phi)_{1} = q_{1}\phi_{1} + \alpha_{1}\phi_{2},\\
\end{array} \right.
\end{equation}
\n where\\

\n $q_{n} = \mu n$, and $\alpha_{n} = i\lambda n\sqrt{n+1}$, ( $\mu$ and $\lambda$ are real numbers and $i^{2} = -1$).\\

\n We will write from now the tridiagonal Jacobi-Gribov matrix as\\

\n $H = $ $\left(\begin{array}{c c c c c c c c} \mu & i\lambda \sqrt{2} & 0 & \cdots\\i\lambda \sqrt{2} & 2\mu & i\lambda 2\sqrt{3} &0 & .\\0 &i\lambda 2\sqrt{3} & 3\mu & i\lambda 3\sqrt{4}& 0 &.\\\vdots & 0 & i\lambda 3\sqrt{4} & 4\mu & * & 0 & .\\\vdots & \ddots & 0 & * & * & * & \ddots & .\\\vdots & \ddots & \ddots & 0 & * & * & * & \ddots\\
    \end{array}\right).$ \\

\n i.e. \\
\begin{equation}
\left\{\begin{array}[c]{l}H = (h_{m,n})_{m,n=1}^{\infty}\quad with \quad the \quad elements:\\\quad \\h_{nn} = \mu n,h_{n,n+1}=h_{n+1,n}= i\lambda n\sqrt{n+1}; n = 1,2,...\\ \quad \\and \\ \quad \\h_{mn} = 0 \quad for \quad \mid m - n \mid > 1 \\ \end{array}\right.
\end{equation}

\n The above ``Jacobi-Gribov'' matrix determines two linear operators in $l_{0}^{2}(\mathbb{N})$ by the formal matrix product $H\phi$. the first operator is defined in the linear manifold of vectors in $l_{0}^{2}(\mathbb{N})$ with finite support related to the set $\mathcal{P}$ of polynomials in Bargmann space, this operator is densely defined and closable. Let $H_{min}$ be its closure.\\

\n  The second operator $H_{max}$ has the domain $D(H_{max}) =\{ \phi = (\phi_{n})_{n=0}^{\infty} \in l_{0}^{2}(\mathbb{N}); H\phi \in l_{0}^{2}(\mathbb{N})\}$\\

\n The  family of infinite ``Jacobi-Gribov'' matrices is tridiagonal of the form $J + i\lambda G$, where the matrix $ J$ is diagonal with entries $J_{nn} := q_{n}= \mu n$ , and the matrix $G$ is off-diagonal, with nonzero entries $G_{n,n+1} = G_{n+1,n} := G_{n} = n\sqrt{n+1}$.\\

\n The Jacobi operators with zero main diagonal and $G_{n}\sim n^{\alpha}$ with $\frac{1}{2} \leq \alpha\leq 1$ has been studied in  {\color{blue}[Janas3 et al]}, 
 {\color{blue}[Janas4 et al]} and   {\color{blue}[Janas5 et al]}.\\

\n The spectral analysis of Jacobi operators corresponding to matrices with rapidly growing weights has been treated in few occasions see  {\color{blue}[Janas2 et al]},   {\color{blue}[Moszynski] ]} and {\color{blue}[Silva]}.\\

\n In contrast, literature on Jacobi matrices with slowly growing weights is more abundant (see for example {\color{blue}[Tur]} and references therein) :\\

\n In  {\color{blue}[Janas1 et al]} Janas and Malejkib have studied  the asymptotic behaviour of the point spectrum for special classes of Jacobi matrices with analytic models where $J_{n} = \delta n^{\alpha}$ and $G_{n} = \gamma n^{\beta}(1 + \Delta_{n}) , \Delta_{n}$ goes to zero as $n$ goes to infinity, with $\alpha - \beta$ equals, respectively, to $1, \frac{1}{2}, 0 $ . In particular  their analytic model of example 4 with $\alpha - \beta = \frac{1}{2}$ uses the Bargmann space.\\

\n The spectral analysis of ``Jacobi-Gribov'' operator as analytical model is based on the above representations.\\
\end{remark}
\begin{remark}

\n Let the infinite matrix $ ^{t\!}H $ be obtained from $H$ by transposing of the elements and the infinite matrix $ H^{\bot}$ be obtained from $H$ by transposing and by taking complex conjugates of the elements. Then\\

\n i) $ H $ is symmetric complex matrix i.e. $ H = ^{t\!}H$.\\

\n ii) $ H \neq  H^{\bot}$ \\

\n iii) $ H_{min}^{*} = H_{max}$ and $ H_{max}^{*} = H_{min}$, in particular the maximal operator $H_{max}$ is a closed extension of $H_{min}$.\\

\n iv) As $G_{n} = O(n^{\alpha})$ with $\alpha = \frac{3}{2} > 1$ then the standard perturbation theory is not applicable.\\
\end{remark}
\n For a systematic treatment of complex Jacobi matrices, we refer to  {\color{blue}[Beckermann]} and the references therein.\\
\subsection{Some properties of domains of annihilation and creation operators and of domains of $H_{\lambda', \mu, \lambda}$ in $\mathbb{B}_{s}$}

\n For an operator $T$ acting on Hilbert space $\mathcal{H}$, we denote by $D(T)$; $R(T)$ or $\Im m(T)$ (French notation); $N(T)$, $\sigma (T)$ and $\rho(T)$, its domain of definition, its range, its kernel, its spectrum, and its resolvent set.\\
\begin{theorem}
\n In $l_{0}^{2}(\mathbb{N})$ we have:\\

\n 1) $\mid\mid A\phi\mid\mid \geq \mid\mid \phi\mid\mid \quad \forall \phi \in D(A)$\\

\n 2) $ < G\phi , \phi > = \displaystyle{\sum_{n=1}^{\infty}n\sqrt{n+1}[\phi_{n}\overline{\phi_{n+1}} + \phi_{n+1}\overline{\phi_{n}}]  \quad \forall \phi \in D(G_{max})}$\\

\n  3) For $ \mu \neq 0$, $\mid\mid H_{min}\phi\mid\mid \geq \mid \mu \mid\mid\mid \phi\mid\mid \quad \forall \phi \in D(H_{min})$\\

\n 4) $D(H_{min}) \hookrightarrow D(A)$ is continuous.\\

\n 5) For $ \mu \neq 0$, \quad $ 0 \in \rho(H_{min}) $\\

\n 6) The spectrum of $H_{min}$ is discrete.\\
\end{theorem}
\n {\bf Proof }\\

\n  1) We see that $\mid\mid A\phi\mid\mid^{2} = \displaystyle{\sum_{n=1}^{\infty}(n+1)\mid \phi_{n}\mid^{2} \, \forall \phi =\{\phi_{n}\}_{n=1}^{\infty} \, \in D(A)}$\\
  then\\ $\mid\mid A\phi\mid\mid \geq \mid\mid \phi\mid\mid \, \forall \phi \in D(A)$\\

\n  2) This property is obvious\\

\n  3) For $\lambda \in \mathbb{R}$ we observe that
 $\mid \Re e < H_{min}\phi, \phi>\mid = \mid \mu \mid\mid\mid A\phi\mid\mid^{2}$\\

\n  and by using the property 1) we deduce that \\

$\mid \Re e < H_{min}\phi, \phi>\mid$ $\geq$ $\mid\mu\mid.\mid\mid \phi\mid\mid^{2} \quad \forall \phi \in D(H_{min})$\\
 then \\
 
  $\mid\mid H_{min}\phi\mid\mid$ $\geq$ $\mid\mu\mid.\mid\mid \phi\mid\mid \quad \forall \phi \in D(H_{min})$.\\

\n  From this calculations, we deduce that  $N(H_{min}) = \{0\}$  $\Im m (H_{min})$ is closed and the property 4) i.e. $D(H_{min}) \hookrightarrow D(A)$ is continuous.\\

\n  Now by using the property 4) and above lemma, we will deduce that\\

\n  $D(H_{min}) \hookrightarrow l_{0}^{2}(\mathbb{N})$ is compact. \hfill { }  (**)\\

\n 5) We begin by showing that $Im(H_{min})$ is dense in $l_{0}^{2}(\mathbb{N})$\\

\n  In fact $ < H_{min}\phi,\psi> = 0 \Leftrightarrow H_{max}^{\perp}\psi = 0$ and in Bargmann representation, we have\\
\begin{equation}
-i\lambda z\psi^{''}(z) + (\mu z - i\lambda z^{2}) \psi^{'}(z) = 0 
\end{equation}
\n  For $\lambda \neq 0$ then \\
\begin{equation}
\displaystyle{\psi^{'}(z) = e^{-\frac{1}{2}z^{2} + i\frac{\mu}{\lambda}z}}
\end{equation}
 \n Now we show that the norma-lizability requirement for $ \psi(z) $ is not verified in some direction of the $z$-plane.\\

\n  In fact we will use the fundamental lemma 3.8 for $ z\rightarrow -i\infty$\\

\n  Let $ z = \Re e z + i\Im m z$, then\\

\n  $e^{-\mid z \mid^{2}}\mid \frac{\psi^{'}(z)}{z}\mid^{2}$ = $e^{-\mid z \mid^{2}}\mid \frac{e^{-\frac{1}{2}z^{2} + i\frac{\mu}{\lambda}z}}{z}\mid^{2}$ = $\frac{1}{\mid z \mid^{2}}e^{-2(\Re z)^{2}}e^{-2\frac{\mu}{\lambda}\Im z}$\\

\n  For $ \mu $ and $\lambda $ given, there exists an direction of the z-plane where the function \\
 $\psi (z) = \displaystyle{\int_{0}^{z} e^{-\frac{1}{2}\xi^{2} + i\frac{\mu}{\lambda}\xi}d\xi}$ cannot be considered in Bargmann space ( for example $\displaystyle{\frac{\mu}{\lambda} > 0}$ and $\Im m z < 0$).\\

\n  By using the property 3), we deduce that $H_{min}^{-1}$ exists, i.e. $ 0 \in \rho(H_{min}) $.\\

\n  6)  As $H_{min}^{-1}$ exists, we can write it in the form\\

\n  $H_{min}^{-1}$ = $i_{2}oi_{1}oH_{min}^{-1}$ where $i_{1}:D(H_{min}) \hookrightarrow D(A)$ is continuous(see the property 4) and \\

\n $\displaystyle{ i_{2}:D(A) \hookrightarrow l_{0}^{2}(\mathbb{N})}$ is compact , i.e. $H_{min}^{-1}$ is compact.\\

\n  By Chapter III, Theorem 6.29 {\color{blue} [Kato]}, we deduce that the spectrum of $H_{min}$ consists entirely \\

\n of isolated eigenvalues with finite multiplicities, and $(H_{min} - \sigma I)^{-1}$ is compact $\forall \sigma \in \rho(H_{min})$.\hfill { } $\square$\\

\n The first interesting result on this ``Jacobi-Gribov'' matrix is the following theorem \\
\begin{theorem}

\n For $\mu \neq 0 $, the ``Jacobi-Gribov'' matrix is proper, i.e $H_{min}$  = $H_{max}$ and its spectrum $\sigma(H_{max})$ is discrete\\
\end{theorem}
\n {\bf Proof }\\

\n i) $ D(H_{min})\subset D(H_{max})$ is obvious.\\

\n  ii) $ D(H_{max})\subset D(H_{min})$.\\

\n In fact let $\psi = (\psi_{n})_{n=1}^{\infty} \in D(H_{max})$, according that $H_{min}$ is invertible by using the last proposition then there exists $ \phi = (\phi_{n})_{n=1}^{\infty} \in D(H_{min})$ such that $ H_{min}\phi = \psi$.\\

\n In particular there exists a sequence of polynomials $q_{m}$ such that  $q_{m} \rightarrow \psi ; m \rightarrow \infty$ then $H_{min}^{-1}q_{m} \rightarrow H_{min}^{-1}\psi = \phi$ but $H_{min}^{-1}q_{m}$ can be not  polynomials.\\
We proceed as follows to resolve this difficulty\\

\n We consider the polynomials $p_{m}$ such that $\mid\mid\mid p_{m} - H_{min}^{-1}q_{m} \mid\mid\mid \leq \frac{1}{m}$ where $\mid\mid\mid . \mid\mid\mid$ is graph norm.\\

\n Then\\
\begin{equation}
\displaystyle{\mid\mid p_{m} - \phi \mid\mid \leq \mid\mid p_{m} - H_{min}^{-1}q_{m} \mid\mid + \mid\mid H_{min}^{-1}q_{m} - \phi \mid\mid}
\end{equation}

\n In particular we obtain \\

\n $\mid\mid p_{m} - \phi \mid\mid \leq \frac{1}{m} + \mid\mid H_{min}^{-1}q_{m} - \phi \mid\mid$, i.e. $p_{m}\rightarrow \phi$ as $ m \rightarrow \infty$\\

\n Now to show that $H_{max}p_{m}\rightarrow H_{max}\phi$ as $ m \rightarrow \infty$, we observe that \\

\n $\mid\mid H_{max} p_{m} - H_{max}\phi \mid\mid \leq \mid\mid H_{max}p_{m} - q_{m}\mid\mid  + \mid\mid q_{m} - H_{max}\phi \mid\mid \leq \mid\mid\mid H_{max}\mid\mid\mid(\frac{1}{m} + \mid\mid H_{min}^{-1}q_{m} - \phi \mid\mid)$.\\

\n where $\mid\mid\mid H_{max}\mid\mid\mid $ denote the graph norm of operator $H_{max}$.\\

\n  i.e.\\

\n $\displaystyle{H_{max} p_{m} \longrightarrow H_{max} \phi}$  as $m \longrightarrow +\infty$.\hfill { } $\square$\\
 \begin{theorem}

\n  (i) Let $\mathcal{P}_{0}$ = $\{p \in \mathcal{P}; p(0) = 0\}$ Then $H[\mathcal{P}]$ is dense in $B_{0}$.\\

\n (ii)  For $\mu = 0 $, the Jacobi-Gribov matrix is not proper, i.e $H_{min}$  $\neq $ $H_{max}$ and  $\sigma(H_{max}) =  \mathbb{C} $. \\
\end{theorem}
\n {\bf Proof }\\

\n (i) Let $\psi \in B_{0}$, as $ 0 \in \rho(H_{min})$ and $H_{min} = H_{max}: = H$ then there exists $\phi \in D(H):= D(H_{max}) = D(H_{min})$ such that
$ \psi = H\phi$, hence there exists $p_{m} \in \mathcal{P}_{0}$ ;$ p_{m} \rightarrow \phi$ and $\psi_{1} \in B_{0}$ ;$Hp_{m} \rightarrow \psi_{1}$ when $ m \rightarrow \infty$.
As $Hp_{m} \rightarrow H\phi$ when $ m \rightarrow \infty$ of distributions sense, we deduce that $\psi = \psi_{1}$.\\

\n (ii) For $\mu = 0$ the boundary conditions at infinity had used in a description of all maximal dissipative extensions of $H_{min}$ and the characteristic functions of these dissipative extensions had computed. Completeness theorems had obtained for the system of generalized eigenvectors.\\.\hfill { } $\square$\\

\n To end this section, we observe that  our operator can be  generalized  to :\\

\n  $\displaystyle{ \mathbb{H}_{\lambda',  \mu, \lambda} = \lambda' A^{*^{p+1}} A^{p+1} + \mu A^{*^{p}}A^{p} + i\lambda A^{*^{p}}(A + A^{*})A^{p}\, (p = 0, 1, 2 ...)}$ \hfill { } ((i))\\

\n and  If $p = 0$, $\mu= 0$ and $\lambda \in i\mathbb{R}$ then $\displaystyle{ \mathbb{H}_{\lambda',  \mu, \lambda}}$ is the displaced harmonic oscillator of the following form:\\

\n $\displaystyle{\mathbb{H}_{\omega} = \omega A^{*}A + \lambda (A + A^{*}) = \omega (A^{*} + \frac{\lambda}{\omega})(A +  \frac{\lambda}{\omega}) - \frac{\lambda^{2}}{\omega}}.$ \hfill { }  ((ii))\\

\n As it is well known that this Hamiltonian is trivially solved by introducing the operators :\\

\n $\displaystyle{B = A + \frac{\lambda}{\omega}}$ and $\displaystyle{B^{*} = A^{*} + \frac{\lambda}{\omega}}$ which are unitarily equivalent to the $A$ and $A^{*}$.\\
\n  As $\displaystyle{[B, B^{*}] = \mathbb{I}}$ and $\displaystyle{\mathbb{H}_{\omega} = \omega B^{*}B  - \frac{\lambda^{2}}{\omega}}$ then the energy spectrum is given by :\\

\n $\displaystyle{\sigma_{n} = n\omega -   \frac{\lambda^{2}}{\omega}}$ \hfill { }  ((iii))\\

\n In terms of Bargmann realization $\displaystyle{A \longrightarrow \frac{d}{dz}}$, $\displaystyle{A^{*} \longrightarrow z}$ the eigenvalue equation \\

\n $\displaystyle{\mathbb{H}_{\omega}\phi = \sigma \phi}$ \hfill { }  ((iv))\\

\n becomes :\\

\n $\displaystyle{[z\frac{d}{dz} + \frac{\lambda}{\omega}(z + \frac{d}{dz}) - \frac{\sigma}{\omega}]\phi(z) = 0}$  \hfill { }  ((v))\\

\n By using the power series method, we choose $\displaystyle{\phi(z) = \sum_{n=0}^{\infty}a_{n}(z +  \frac{\lambda}{\omega})^{n + s}}$ to get the indicial equation :\\

\n $\displaystyle{s = \frac{\sigma}{\omega} + \frac{\lambda^{2}}{\omega^{2}}}$\hfill { } ((vi))\\

\n and the recurrence relation for the $a_{n}$'s would read\\

\n $\displaystyle{a_{n} = -\frac{\lambda}{n\omega}a_{n-1}  = \frac{(-\lambda)^{n}}{(\omega)^{n}\Gamma(n +1)}a_{0}} $ \hfill { } ((vii))\\

\n The series $\displaystyle{\sum_{n}a_{n}z^{n}}$ thus converges for all $z$, and

\n  $\displaystyle{\phi(z) = a_{0}(z + \frac{\lambda}{\omega})^{\frac{\sigma}{\omega} + \frac{\lambda^{2}}{\omega^{2}}}\sum_{n=0}^{\infty}\frac{(-\lambda)^{n}}{\omega^{n}\Gamma(n + 1)}(z + \frac{\lambda}{\omega})^{n}}$ \hfill { }  ((viii))\\

\n From the requirement that  $\phi(z)$ be entire we obtain $\displaystyle{\frac{\sigma}{\omega} + \frac{\lambda^{2}}{\omega^{2}} = n \in \mathbb{N}}$\\

\n The  $\phi_{n}(z)$ associated to $\displaystyle{\sigma_{n} = n \omega - \frac{\lambda^{2}}{\omega}}$ are in domain of $\mathbb{H}_{\omega}$.\\

\subsection{On existence of eigenvalues of $\displaystyle{H_{\mu, \lambda} \, and \, H_{\lambda', \mu, \lambda} ; \lambda' \neq 0}$}
\n {\color{black}{ \bf (1) Explicit inversion of $H_{\mu,\lambda}$ on $]-\infty, \, 0]$}}\\
\n We consider the eigenvalue problem:\\

\n $ \left \{ \begin{array} {c}  \displaystyle{ i\, \lambda z \frac{\partial^{2}\varphi_{\mu, \lambda}}{\partial z^{2}}(z) + (i\lambda z^{2} + \mu z)\frac{\partial \varphi_{\mu, \lambda}}{\partial z}(z) = \sigma(\mu, \lambda)\varphi_{\mu, \lambda} (z)}\\
\quad\\
\varphi \in D(H_{\mu, \lambda})\quad \quad \quad \quad \quad \quad  \quad \quad \quad \quad \quad  \quad \quad \quad \quad \quad  \quad \quad \quad \quad   \\
\quad\\
\sigma(\mu, \lambda) \in \mathbb{C}  \quad \quad \quad \quad \quad  \quad \quad \quad \quad \quad  \quad \quad \quad \quad \quad  \quad \quad \quad \quad \quad \\
\end{array} \right. $ \hfill { } (*)\\

\n Denoting $\displaystyle{\varphi_{\mu, \lambda}(z) = \varphi(z),  \sigma(\mu, \lambda) = \lambda\sigma, \frac{\partial}{\partial z}\varphi (z) = \varphi^{'}(z)\, and \, \frac{\partial^{2}}{\partial z^{2}}\varphi (z) = \varphi^{"}(z) }$ then the eigenvalue problem associated to $H_{\mu, \lambda}; \lambda \neq 0$ can be written as follows \\
\begin{equation}
\displaystyle{\varphi^{"}(z) + p(z) \varphi^{'}(z) + q(z) \varphi(z) = 0}; \displaystyle{ p(z) = (z - i\rho)}\, and\, \displaystyle{ q(z) =  \frac{i\sigma}{z}}
\end{equation}
\n We restrict the study of the equation (3.42) on the lines parallel to the imaginary axis :\\

\n Let $z = a + iy$ and $\varphi(a + iy) = \varphi_{a}(y)$ where $a$ is fixed. Then we get\\
\begin{equation}
\left \{ \begin{array} [c] {l} \displaystyle{\varphi_{a}^{"}(y) - (y - \gamma) \varphi_{a}^{'}(y) - \frac{i\sigma}{a + iy}\varphi_{a}(y) = 0}\\
\quad\\
\gamma = \rho + ia\\
\end{array} \right . 
\end{equation}
\n We put \\
\begin{equation}
\displaystyle{\varphi_{a}^{'}(y) = K_{a}(y)e^{\frac{1}{2}(y - \gamma)^{2}}}
\end{equation}
\n and substitute it in (3.43) then we get\\
\begin{equation}
\displaystyle{K_{a}^{'}(y) = i\sigma e^{-\frac{1}{2}(y - \gamma)^{2}}\frac{\varphi_{a}(y)}{a + iy}}
\end{equation}
\n We choose a primitive representation of the function $\displaystyle{ y \longrightarrow K_{a}^{'}(y)}$ as follows\\
\begin{equation}
\displaystyle{K_{a}(y) = i\sigma \int_{-\infty}^{y}e^{-\frac{1}{2}(u - \gamma)^{2}}\frac{\varphi_{a}(u)}{a + iu}du + c_{a} \, ; y \in ]-\infty, 0]}
\end{equation}
\n where $c_{a}$ is a constant.\\
\n Consequently, equation (3.44) can be written as follows\\
\begin{equation}
\displaystyle{\varphi_{a}^{'}(y) = i\sigma e^{\frac{1}{2}(y - \gamma)^{2}} \int_{-\infty}^{y}e^{-\frac{1}{2}(u - \gamma)^{2}}\frac{\varphi_{a}(u)}{a + iu}du  + c_{a}e^{\frac{1}{2}(y - \gamma)^{2}} }
\end{equation}
\begin{lemma}
\n For $\rho > 0$ the representation (3.45) has a sense.\\
\end{lemma}
\n {\bf Proof}\\

\n For $\varphi \in \mathbb{B}$ we have $\displaystyle{\int_{\mathbb{R}^{2}}e^{-(x^{2} + y^{2})}\mid \varphi(x + iy) \mid^{2} dxdy < + \infty}$ then we apply the Fubini theorem to deduce that  \\
\begin{equation}
\displaystyle{\int_{\mathbb{R}}e^{- y^{2}}\mid \varphi_{x}(y) \mid^{2} dy < + \infty}
\end{equation}
\n Now, as  $\displaystyle{ \int_{-\infty}^{y}e^{-\frac{1}{2}(u - \gamma)^{2}}\frac{\varphi_{a}(u)}{a + iu}du = e^{-\frac{1}{2}\gamma^{2}}\int_{-\infty}^{y}e^{-\frac{1}{2}u^{2}}\varphi_{a}(u)\frac{e^{\gamma u}}{a + iu}du }$ and we look for $\varphi$ in Bargmann space $\mathbb{B}$, we have on the one hand $\displaystyle{u \longrightarrow  e^{-\frac{1}{2}u^{2}}\varphi_{a}(u)}$ is square integrable and on the other hand for $\rho > 0$ the function $\displaystyle{u \longrightarrow  \frac{e^{\gamma u}}{ a + iu}}$ is also square integrable on $]-\infty, 0]$. Consequently, $\displaystyle{ \int_{-\infty}^{y}e^{-\frac{1}{2}(u - \gamma)^{2}}\frac{\varphi_{a}(u)}{a + iu}du}$ has a sense. \hfill { } $\square$ \\
\begin{remark}

\n (i) If $\rho < 0$, we choose a primitive representation of the function $\displaystyle{ y \longrightarrow K_{a}^{'}(y)}$ as follows\\
\begin{equation}
\displaystyle{K_{a}(y) = i\sigma \int_{y}^{+\infty}e^{-\frac{1}{2}(u - \gamma)^{2}}\frac{\varphi_{a}(u)}{a + iu}du + c_{a} }, c_{a} \, is \, constant 
\end{equation}

\n This representation  has  a sense.\\

\n (ii) From (3.43) and (3.45), we deduce that\\
\begin{equation}
\displaystyle{\varphi_{a}^{'}(y) = i\sigma e^{\frac{1}{2}(y - \gamma)^{2}} \int_{y}^{+\infty}e^{-\frac{1}{2}(u - \gamma)^{2}}\frac{\varphi_{a}(u)}{a + iu}du + c_{a}e^{\frac{1}{2}(u - \gamma)^{2}}}
\end{equation}
\end{remark}
\begin{theorem}

\n (i) In the representation (3.45) we have $c_{a} = 0$\\

\n (ii)  \begin{equation}
\displaystyle{Lim \, e^{-\frac{1}{2}(y - \gamma)^{2}}\varphi_{a}^{'}(y) = 0}\, as \, y \longrightarrow -\infty\n (boundary condition)
\end{equation}
\end{theorem}
\n {\bf Proof}\\

\n (i) By multiplying the two members of the equation (3.49) by $\displaystyle{e^{-\frac{1}{2}(u - \gamma)^{2}}}$, we get\\
\begin{equation}
\displaystyle{e^{-\frac{1}{2}(u - \gamma)^{2}}\varphi_{a}^{'}(y) = i\sigma  \int_{y}^{+\infty}e^{-\frac{1}{2}(u - \gamma)^{2}}\frac{\varphi_{a}(u)}{a + iu}du + c_{a}}
\end{equation}

\n As the function $\displaystyle{u \longrightarrow e^{-\frac{1}{2}(u - \gamma)^{2}}\frac{\varphi_{a}(u)}{a + iu}}$ is integrable we deduce that \\
\begin{equation}
\displaystyle{Lim \, e^{-\frac{1}{2}(y - \gamma)^{2}}\varphi_{a}^{'}(y) = c_{a}}\, as \,y \longrightarrow - \infty
\end{equation}

\n Now, if $c_{a} \neq 0$ then we get that \\
\begin{equation}
\displaystyle{\varphi_{a}^{'}(y) \thicksim c_{a}e^{\frac{1}{2}(y - \gamma)^{2}}}\, and   \,\displaystyle{\frac{e^{-\frac{1}{2}y^{2}}}{y}\varphi_{a}^{'}(y) \thicksim c_{a}\frac{e^{- \rho y}}{y}e^{\frac{\gamma^{2}}{2} - ia}}
\end{equation}

\n We consider the space\\
\begin{equation}
\displaystyle{\mathbb{B}_{a} = \{ \varphi_{a} : \mathbb{R} \longrightarrow \mathbb{C} entire ; \int_{\mathbb{R}}e^{-\frac{1}{2}y^{2}}\mid \varphi_{a}(y)\mid^{2}dy < +\infty\}}
\end{equation}

\n Then \\

\n ($\alpha$) From theorem 3.4, we deduce  that $\forall \, \varphi \in \mathbb{B}$ then $\varphi_{a} \in \mathbb{B}_{a}$\\

\n ($\beta$) From lemma 3.2, we deduce  that $\forall \, \varphi_{a} \in \mathbb{B}_{a}$ then the function \\$\displaystyle{y \longrightarrow  \frac{\varphi_{a}^{'}(y) - \varphi_{a}^{'}(0)}{y} \in \mathbb{B}_{a}}$

\n As in (3.53) the function $\displaystyle{y \longrightarrow \frac{e^{- \rho y}}{y}e^{\frac{\gamma^{2}}{2} - ia}}$ is not square integrable then $\varphi_{a} \notin \mathbb{B}_{a}$\\

\n which is contradictory with the property ($\beta$) and therefore $c_{a} = 0$.\\

\n (ii) we apply (i) to (3.53).\hfill { } $\square$\\
\begin{remark}

\n By applying the above theorem, the last equation  can be written as follows\\
\begin{equation}
\displaystyle{\varphi_{a}^{'}(y) = i\sigma e^{\frac{1}{2}(y - \gamma)^{2}} \int_{-\infty}^{y}e^{-\frac{1}{2}(u - \gamma)^{2}}\frac{\varphi_{a}(u)}{a + iu}du; \, y \in \,]-\infty, 0] }
\end{equation}
\end{remark}
\n {\bf {\color{black} (2) Justification for crossing to the limit in the equation (3.56) as $a \longrightarrow 0$}}\\
\begin{proposition}

\n If $a \longrightarrow 0$ in (3.56) then we get \\
\begin{equation}
\displaystyle{\varphi_{0}^{'}(y) = \sigma e^{\frac{1}{2}(y - \rho)^{2}} \int_{-\infty}^{y}e^{-\frac{1}{2}(u - \rho)^{2}}\frac{\varphi_{0}(u)}{u}du; \, y \in \,]-\infty, 0] }
\end{equation}
\end{proposition}
\n {\bf Proof}\\
\n As the functions $\displaystyle{a \longrightarrow \varphi_{a}^{'}(y)}$ and $\displaystyle{a \longrightarrow \frac{e^{-\frac{1}{2}(u - \rho - ia)^{2}}}{a + iu}\varphi_{a}(u)}$ are continuous  then 
\begin{equation}
\displaystyle{\lim\limits_{a \rightarrow 0} \varphi_{a}^{'}(y) = \varphi_{0}^{'}(y)} \, and \, \displaystyle{\lim\limits_{a \rightarrow 0}\frac{e^{-\frac{1}{2}(u - \rho - ia)^{2}}}{a + iu}\varphi_{a}(u) =  \frac{e^{-\frac{1}{2}(u - \rho)^{2}}}{iu}\varphi_{0}(u)}
\end{equation}

\n We have to show that $\displaystyle{u \longrightarrow \frac{e^{-\frac{1}{2}(u - \rho)^{2}}}{iu}\varphi_{0}(u)}$ is integrable. For this purpose we consider the following function\\
\begin{equation}
\displaystyle{f_{a}(u) = \frac{e^{-\frac{1}{2}(u - \rho - ia)^{2}}}{a + iu}\varphi_{a}(u)}
\end{equation}

 \n  As for $\epsilon > 0$ enough small  we have $\varphi_{a}(a + iu) \thicksim a +iu ; u \in ]-\epsilon, 0]$, we deduce that \\
 \begin{equation}
 \displaystyle{\mid f_{a}(u)\mid \leq e^{\rho u}}
 \end{equation}
  
\n Now, for $u \in ]-\infty , -\epsilon[$, we observe that \\
 \begin{equation}
 \displaystyle{\mid f_{a}(u)\mid \leq e^{-\frac{\rho^{2}}{2}}\frac{e^{\rho u}}{u}}
 \end{equation}

\n As the function\\
\begin{equation}
g(u) = \left \{ \begin{array}[c] {l} \displaystyle{e^{-\frac{\rho^{2}}{2}}\frac{e^{\rho u}}{u}}; u \in ]-\infty , -\epsilon[\\
\quad\\
\displaystyle{e^{\rho u}}; u \in ]-\epsilon, 0]\\
\end{array} \right .
\end{equation}

\n is integrable and the the sequence $f_{a}(u)$ converges pointwise to  $\displaystyle{\frac{e^{-\frac{1}{2}(u - \rho^{2}}}{iu}\varphi_{0}(u)}$ as $a \longrightarrow 0$ then by applying the Lebesgue's Dominated Convergence Theorem, we deduce that\\
 \begin{equation}
 \displaystyle{\lim\limits_{a \rightarrow 0} \int_{-\infty}^{y}\frac{e^{-\frac{1}{2}(u - \rho - ia)^{2}}}{a + iu}\varphi_{a}(u)du = \int_{-\infty}^{y}\frac{e^{-\frac{1}{2}(u - \rho^{2}}}{iu}\varphi_{0}(u)}
 \end{equation}

\n Consequently, we get\\
 \begin{equation}
 \displaystyle{\varphi_{0}^{'}(y) = \sigma e^{\frac{1}{2}(y - \rho)^{2}} \int_{-\infty}^{y}e^{-\frac{1}{2}(u - \rho)^{2}}\frac{\varphi_{0}(u)}{u}du; \, y \in \,]-\infty, 0] }
 \end{equation}
.\hfill { } $\square$\\
\begin{theorem}
\n (i) An operator integral associated to $H_{\mu, \lambda}$ is given as follows\\
\begin{equation}
\displaystyle{K\psi(y) = \int_{-\infty}^{0}\mathcal{N}(y, s)\psi(s)ds}
\end{equation}

\n where\\
\n ($\alpha$) {\color{red}$\rhd$}
\begin{equation}
\displaystyle{\psi(y) = \frac{\varphi_{0}^{'}(y)}{y}e^{-\frac{y^{2}}{2}}\theta(y)}
\end{equation}

\n with\\

\n $\theta(y) = \left \{ \begin{array}[c] {l}\,  y ; \quad y \in [-1, 0]\\
\quad\\
- 1; \quad y \in ]-\infty, -1]\\
\end{array} \right .$ \\
\quad\\
\n ($\beta$) {\color{red}$\rhd$}
\begin{equation}
\displaystyle{\mathcal{N}(y, s) = e^{-\rho y}\frac{\theta(y)}{y}.\frac{s}{\theta(s)}e^{\frac{s^{2}}{2}}\int_{-\infty}^{min(y, s)}e^{-\frac{1}{2}(u - \rho)^{2}}{u}du}
\end{equation}

\n where\\

\n $min(y, s) = \left \{ \begin{array}[c] {l} y ; \quad y \leq s\\
\quad\\
s; \quad  s \leq y \\
\end{array} \right .$ \\
\quad\\
\n ($\gamma$) {\color{red}$\rhd$}
\begin{equation}
\displaystyle{\psi \in L_{2}(]-\infty, 0], \theta(y)dy)}
\end{equation}

\n (ii) For $\psi \in \mathbb{B}_{0}$, an explicit inverse of $H_{\mu, \lambda}$ restricted on imaginary axis ; $y \in [0,  +\infty[$ is given by\\
\begin{equation}
\displaystyle{H_{\mu, \lambda}^{-1} \psi(-iy) = \int_{0}^{\infty}\mathcal{N}_{\mu,\lambda}(y, s)\psi(-is)ds}
\end{equation}

\n where  $\displaystyle{\mathcal{N}_{\mu,\lambda}(y, s) = \frac{1}{\lambda s}e^{-\frac{s^{2}}{2} - \frac{\mu}{\lambda}s}\int_{0}^{min(y, s)}e^{\frac{u^{2}}{2} +\frac{\mu}{\lambda}u}du}$\\
\end{theorem}
\n {\bf Proof}\\

\n (i) We substitue $\displaystyle{\varphi_{0}(u) = -\int_{u}^{0} \varphi_{0}^{'}(s)ds}$ in (3.64) to obtain\\
\begin{equation}
\displaystyle{\varphi_{0}^{'}(y) = -\sigma e^{\frac{1}{2}(y - \rho)^{2}} \int_{-\infty}^{y}\frac{e^{-\frac{1}{2}(u - \rho)^{2}}}{u}\int_{u}^{0}\varphi_{0}^{'}(s)ds; \, y \in \,]-\infty, 0] }
\end{equation}

\n By applying the Fubini theorem we get\\
\begin{equation}
\displaystyle{\int_{-\infty}^{y}\frac{e^{-\frac{1}{2}(u - \rho)^{2}}}{u}\int_{u}^{0}\varphi_{0}^{'}(s)ds = \int_{-\infty}^{0}[\int_{-\infty}^{min(y, s)}\frac{e^{-\frac{1}{2}(u - \rho)^{2}}}{u}du] \varphi_{0}^{'}(s)ds}
\end{equation}

\n and\\
\begin{equation}
\displaystyle{\varphi_{0}^{'}(y) = -\sigma e^{\frac{1}{2}(y - \rho)^{2}}\int_{-\infty}^{0}[\int_{-\infty}^{min(y, s)}} \displaystyle{\frac{e^{-\frac{1}{2}(u - \rho)^2}}{u}du] \varphi_{0}^{'}(s)ds}
\end{equation}
\n or\\
\begin{equation}
\displaystyle{e^{-\frac{y^{2}}{2}}\varphi_{0}^{'}(y)} \displaystyle{ = -\sigma e^{\frac{\rho^{2}}{2}}\int_{-\infty}^{0}[ e^{\rho y} \int_{-\infty}^{min(y, s)}\frac{e^{-\frac{1}{2}(u - \rho)^{2}}}{u}du] \varphi_{0}^{'}(s)ds}
\end{equation}

\n Now, if we put :\\
\begin{equation}
\displaystyle{\psi(y) = \frac{\varphi_{0}^{'}(y)}{y}e^{-\frac{y^{2}}{2}}\theta(y)}
\end{equation} 

\n with $\theta(y) = \left \{ \begin{array}[c] {l}\,  y ; \quad y \in [-1, 0]\\
\quad\\
- 1; \quad y \in ]-\infty, -1]\\
\end{array} \right .$ \\
\quad\\

\n and if we substitute $\psi$ in (3.73) then we get \\
\begin{equation}
\displaystyle{\psi(y) = -\sigma e^{\frac{\rho^{2}}{2}}\int_{-\infty}^{0}\mathcal{N}(y, s) \psi(s)ds }
\end{equation}

\n Where $\mathcal{N}(y, s)$ is given by (3.67).\\

\n (ii) Let $\varphi \in \mathbb{B}_{0}$ and  $\psi \in \mathbb{B}_{0}$, we consider the equation $H_{\mu, \lambda}\varphi = \psi (z)$ i.e.\\
\begin{equation}
\displaystyle{i\lambda z \varphi^{"}(z)  + (i\lambda z^{2} + \mu z)\varphi'(z) = \psi(z)}
\end{equation}

\n Let $ z = -iy$, $u(y) = \varphi(-iy)$ and $ f(y) = \psi(-iy)$ with $y \in [0, \infty[$ then (3.76) can be written in the following form:\\
\begin{equation}
\displaystyle{ - \lambda y u''(y) + (\lambda y^{2} + \mu y) u'(y) = f(y)}
\end{equation}

\n or\\
\begin{equation}
\displaystyle{ - u''(y) + (y + \rho ) u'(y) = \frac{f(y)}{\lambda y}}
\end{equation}

\n with help theorem 3.4, Lemma 3.2 and a similar technique used in proof of (i) of this theorem we get (3.69).\hfill { } $\square$\\

\n {\bf {\color{black} (3) The extension of integral operator  associated to the restriction of $H_{\mu, \lambda}$  on negative  imaginary axis}} {\color{black}{\bf to Hilbert-Scmidt operator on}} {\color{black} $\displaystyle{L_{2}(]-\infty, 0], \theta(y)dy)}$}\\
\begin{proposition}

\n Le $ \mathcal{N}(y, s)$ = $ e^{-\rho y} \frac{\theta(y)}{y} \frac{s}{\theta(s)}e^{\frac{1}{2}s^{2}}$ $\displaystyle{\int_{-\infty}^{min(y,s)}}$ $\frac{e^{-\frac{1}{2}(u - \rho)^{2}}}{u}du$ be the kernel of the equation:\\
\begin{equation}
\displaystyle{-\lambda y\varphi''(iy) + (\lambda y^{2} + \mu y)\varphi'(iy)  = \psi(iy)  , y < 0}
\end{equation}

\n where  $\displaystyle{\rho = \frac{\mu}{\lambda}}$ then:\\

\n (i) $ \mathcal{N}(y, s)$ is Hilbert-Shmidt i.e. $\displaystyle{\mathcal{N}(y, s)}$ belongs to \\

$\displaystyle{L^{2}(]-\infty, 0]\times]-\infty, 0], \theta(y)\theta(s)dyds)}$.\\

\n (ii) $\displaystyle{\mathcal{N}(y, s)}$ belongs to $\displaystyle{L^{2}(]-\infty, 0]\times]-\infty, 0], \theta(y)\theta(s)dyds)}$.\\
\end{proposition}
\n {\bf \underline{Proof} }\\

\n (i) To apply the theorem of dominated convergence, we will increase $K(y, s)$ by a function $\tilde{K}(y, s) \in L^{2}(]-\infty, 0]\times]-\infty, 0], \theta(y)\theta(s)dyds)$\\

\n For this we pose $min(y, s) = m$  then we have :\\
\begin{equation}
\displaystyle{\mid \frac{e^{-\frac{1}{2}(u - \rho)^{2}}}{u}\mid \leq e^{-\frac{1}{2}\rho^{2}}\frac{1}{m(m - \rho)}\mid (\rho - u) e^{-\frac{1}{2}u^{2} + \rho u}\mid}
\end{equation}

\n The function $ u \longrightarrow \displaystyle{\frac{1}{u(u - \rho)}} $ is bounded for $-\infty < u \leq m < 0$\\

\n Hence a first increase:\\
\begin{equation}
\mid \mathcal{N}(y, s) \mid \leq \displaystyle{e^{-\frac{1}{2}\rho^{2}}\mid \frac{e^{-\rho y}}{y}\frac{\theta(y)}{\theta(s)}se^{\frac{1}{2}s^{2}}\frac{e^{-\frac{1}{2}m^{2} + \rho^{m}}}{m(m - \rho)}\mid}
\end{equation}

\n If $y \leq s$, we have: \\

\n $ m = y$ et $\displaystyle{e^{\frac{1}{2}s^{2} - \frac{1}{2}y^{2}} < 1}.$\\

\n Then :\\
\begin{equation}
\mid \mathcal{N}(y, s) \mid \leq \displaystyle{e^{-\frac{1}{2}\rho^{2}}\mid \frac{\theta(y)}{y}\frac{s}{\theta(s)}\frac{1}{y(y - \rho)}}\mid
\end{equation}

\n If $s \leq y$, we have: \\

\n $ m = s$ et $e^{\rho (s - y)} < 1.$\\

\n Therefore\\
\begin{equation}
\mid \mathcal{N}(y, s) \mid \leq e^{-\frac{1}{2}\rho^{2}}\mid \frac{\theta(y)}{y}\frac{s}{\theta(s)}\frac{1}{s(s - \rho)}\mid
\end{equation}

\n Finally we put :\\
\begin{equation}
\tilde{\mathcal{N}}(y, s) = \left\{
  \begin{array}{ c }
e^{-\frac{1}{2}\rho^{2}}\mid \frac{\theta(y)}{y}\frac{s}{\theta(s)}\frac{1}{y(y - \rho)} \quad pour \quad y \leq s \\
e^{-\frac{1}{2}\rho^{2}}\mid \frac{\theta(y)}{y}\frac{s}{\theta(s)}\frac{1}{s(s - \rho)} \quad pour \quad s \leq y \\
\end{array} \right .
\end{equation}

\n to get :\\
\begin{equation}
\mid \mathcal{N}(y, s) \mid \leq \mid \tilde{N}(y, s) \mid
\end{equation}

\n it remains to show that $ \tilde{N}(y, s) \in L^{2}(]-\infty, 0]\times]-\infty, 0], \theta(y)\theta(s)dyds)$\\

\n Let $\Delta = ]-\infty, 0]\times]-\infty, 0] -]-1, 0]\times]-1, 0], \Delta_{1} = \{(y,s) \in \Delta ; y \leq s\}$ et \\

\n $\Delta_{2} = \{(y,s) \in \Delta ; s \leq y\}$. Then\\

\n $\displaystyle{\int_{-\infty}^{0}\theta(y)dy}\displaystyle{\int_{-\infty}^{0}\theta(s)\mid \tilde{N}(y, s) \mid^{2} ds} = $\\

\n $\displaystyle{\int_{\Delta_{1}}\theta(y)\theta(s)\mid \frac{\theta(y)}{y}\frac{s}{\theta(s)}\frac{1}{y(y - \rho)}\mid^{2}dyds} + \displaystyle{\int_{\Delta_{2}}\theta(y)\theta(s)\mid \frac{\theta(y)}{y}\frac{s}{\theta(s)}\frac{1}{s(s - \rho)}\mid^{2}dyds}$\\

\n As $\rho > 0$  and the function  $\displaystyle{ u \rightarrow \frac{1}{(u - \rho)^{2}}}$  is integrable at the origin, we deduce that:\\
\begin{equation}
\displaystyle{\int_{-\infty}^{0}\theta(y)dy}\displaystyle{\int_{-\infty}^{0}\theta(s)\mid \tilde{N}(y, s) \mid^{2} ds < +\infty}
\end{equation}

\n then $\mathcal{N}(y, s)$ is a kernel of Hilbert-Shmidt.\\

\n We note that the functions :\\

\n $ y \rightarrow \displaystyle{\int_{-\infty}^{y}\frac{e^{-\frac{1}{2}(u - \rho)^{2}}}{u}du} ;\quad y \leq s$\\

\n and \\

\n $s \rightarrow \displaystyle{\int_{-\infty}^{s}\frac{e^{-\frac{1}{2}(u - \rho)^{2}}}{u}du} ;\quad s \leq y$\\

\n are decreasing and tend to zero respectively when$ y \rightarrow -\infty $ and $ s \rightarrow -\infty $,we deduce that:\\

\n $-\mathcal{N}(y, s)$ non negative.\hfill { } $\square$\\
\begin{remark}

\n  The operator $\mathbb{K}\psi(iy) = \displaystyle{\int_{-\infty}^{0} \mathcal{N}(y, s)\psi(is)ds}$ is:\\

\n (i)  Hilbert-Shmidt operator.\\

\n (ii) The spectrum of $\mathbb{K}$ is not empty and discrete.\\
\end{remark}

\n It follows that we can give an explicit form of inverse of $H_{0, \mu,\lambda : H_{\mu, \lambda}}$ on negative imaginary axis; $z = -iy $ with $y > 0$.\\

\n  $H_{0,\mu,\lambda}^{-1}$ can be written as integral operator in following form :\\
\begin{equation}
\displaystyle{H_{0,\mu,\lambda}^{-1}\psi(-iy) = \int_{0}^{+\infty}N_{0,\mu,\lambda}(y, s)\psi(-is)ds}
\end{equation}

\n where\\
\begin{equation}
\displaystyle{N_{0,\mu,\lambda}(y, s) = \frac{1}{\lambda s}e^{\frac{-s^{2}}{2} - \frac{\mu}{\lambda}s}\int_{0}^{min(y, s)}e^{\frac{u^{2}}{2} + \frac{\mu}{\lambda}u}du}
\end{equation}

\n Let $\displaystyle{L_{2}([0, +\infty[, e^{-x^{2} -2\omega x}dx)}$ be the space of square integrable  functions with respect to the measure $e^{-x^{2} -2\omega x}dx$  then  we have the following result :\\
\begin{proposition}

\n (i) $\forall \, \mu \in \mathbb{C}; \mathcal{R}e \mu \geq \omega$  then $H_{0,\mu,\lambda}^{-1}$ can be extended to Hilbert-Schmidt operator of $\displaystyle{L_{2}([0, +\infty[, e^{-y^{2} -2\omega y}dy)}$ to $\displaystyle{L_{2}([0, +\infty[, e^{-y^{2} -2\omega y}dy)}$ .\\

\n (ii) For $\mu > 0$, let $\sigma(0,\mu)$ be the smallest eigenvalue of $H_{0,\mu,\lambda}$ then $\sigma (0,\mu)$\\ is positive, increasing and analytic function on the whole real line with respect to $\mu$.\\
\end{proposition}

\n {\bf Proof}\\

\n (i) From the above proposition, we deduce (i).\\

\n (ii) is the main result of Ando-Zerner (see proposition 1 of  {\color{blue}[Ando et al]} in  Commun. Math. Phys. 93, (1984), p:123-139).\hfill { } $\square$ \\
\begin{proposition}
 (continuity of $H_{\mu, \lambda}^{-1}$ with respect to $\lambda$)\\

\n Let $\mu > 0$ and $\displaystyle{f: \mathbb{R} \longrightarrow \mathcal{B}_{0} ; \lambda \mapsto f(\lambda) = H_{\mu, \lambda}^{-1}}$, then $f$ is continuous.\\
\end{proposition}

\n {\bf Proof}\\

\n As  for any $\lambda \in \mathbb{R}$ and $\mu > 0$ we have $\displaystyle{\mu \mid\mid \varphi \mid\mid \leq \mid\mid H_{\mu, \lambda}\varphi \mid\mid}$  on $\mathcal{B}_{0}$ then \\

\n $\displaystyle{\mid\mid H_{\mu, \lambda}^{-1}\psi \mid\mid \leq \frac{1}{\mu}\mid\mid \psi \mid\mid}$ $\forall\, \psi \in \mathcal{B}_{0}$. \hfill { } (*)\\

\n Now, for $\lambda_{0} \in \mathbb{R}$ we have :\\

\n $\displaystyle{H_{\mu, \lambda}^{-1} - H_{\mu, \lambda_{0}}^{-1} = i(\lambda_{0} - \lambda)H_{\mu, \lambda}^{-1} A^{*}(A + A^{*})H_{\mu, \lambda_{0}}^{-1} }$.\\

\n By applying (*) we get \\

\n $\displaystyle{\mid\mid H_{\mu, \lambda}^{-1} - H_{\mu, \lambda_{0}}^{-1}\mid\mid \leq  \frac{1}{\mu}\mid\mid A^{*}(A + A^{*})H_{\mu, \lambda_{0}}^{-1}\mid\mid. \mid \lambda_{0} - \lambda \mid}$.\\

\n and as \\

\n $\displaystyle{\mid\mid A^{*}(A + A^{*})H_{\mu, \lambda_{0}}^{-1}\mid\mid = \frac{1}{\mid \lambda_{0}\mid }( 1 + \mu \mid\mid A^{*}AH_{\mu, \lambda_{0}}^{-1}\mid\mid}$.\\

\n From the density of the polynomial set $\mathcal{P}_{0}$ in $\mathbb{B}_{0}$, the equality of minimal domain with maximal domain of $\displaystyle{H_{\mu, \lambda}}$  and $\displaystyle{[A^{*}A]^{-1}(\mathcal{P}_{0}) = \mathcal{P}_{0}}$, we deduce the following fundamental inequality :\\

\n $\displaystyle{ \forall\, \epsilon > 0 , (\mu - \epsilon)\mid\mid A^{*}A\varphi \mid\mid \leq \mid\mid H_{\mu, \lambda}\varphi \mid\mid  + \frac{\mid \lambda\mid^{2}}{4\epsilon}\mid\mid \varphi \mid\mid  \, \forall\, \varphi \in D(H_{\mu, \lambda})}$ \hfill { } (**)\\

\n We choose $\displaystyle{\epsilon = \frac{\mu}{2}}$ to get :\\

\n  $\displaystyle{\frac{1}{\mid \lambda_{0}\mid }( 1 + \mu \mid\mid A^{*}AH_{\mu, \lambda_{0}}^{-1}\mid\mid) \leq \frac{1}{\mid\lambda_{0}\mid}(3 + \frac{\mid\lambda_{0}\mid^{2}}{\mu})}$.\\
 
\n  Consequently, we get:\\

\n $\displaystyle{\mid\mid H_{\mu, \lambda}^{-1} - H_{\mu, \lambda_{0}}^{-1}\mid\mid  \leq \frac{1}{\mu} \frac{1}{\mid\lambda_{0}\mid}(3 + \frac{\mid\lambda_{0}\mid^{2}}{\mu})\mid \lambda_{0} - \lambda\mid}$\\

\n i.e the continuity of  $H_{\mu, \lambda}^{-1}$ in $\lambda_{0} \neq 0$.\\

\n Now for $\lambda_{0} = 0$, we have:\\

\n $\displaystyle{\mid\mid H_{\mu, \lambda}^{-1}\psi - \frac{1}{\mu}[A^{*}A]^{-1}\psi \mid\mid \leq \mid\mid H_{\mu, \lambda_{0}}^{-1}\mid\mid. \mid\mid \psi - \frac{1}{\mu}H_{\mu, \lambda}[A^{*}A]^{-1}\psi \mid\mid \forall\, [A^{*}A]^{-1}\psi \in D(H_{\mu, \lambda})}$ \\i.e. $\forall\, \psi \in A^{*}A[D(H_{\mu, \lambda})]$.\\

\n Hence we use the fundamental inequality (**) to get:\\

\n $\displaystyle{\mid\mid H_{\mu, \lambda}^{-1}\psi - \frac{1}{\mu}[A^{*}A]^{-1}\psi \mid\mid \leq \frac{\mid \lambda \mid}{\mu^{2}}\mid\mid A^{*}(A + A^{*})A[A^{*}A]^{-1}\psi\mid\mid}$ which converges to zero as $\lambda \longrightarrow 0$ $\forall\, \psi \in A^{*}A[D(H_{\mu, \lambda})]$. \\

\n Now, as $\displaystyle{A^{*}A[D(H_{\mu, \lambda})]}$ is dense in $\mathbb{B}_{0}$, we deduce the continuity of $f$ at zero.\hfill { } $\square$\\
\subsection{what class Carleman operators belong the inverse of $H_{\mu, \lambda}$ ? }

\n {\bf {\color{black}$\bigstar$} {\color{black}On the factorization of $H_{\mu, \lambda}^{-1}$}}\\

\n Which bounded linear operator on a Hilbert space can be factored as the product of finitely many normal operators ?  \n What is the answer if is normal operators is replaced by involutions ? partial isometrics ? or other classes of familiar operators ? Just as in the case of the factorization of integers, polynomials, or other objects in mathematics, such operator factorization problems seem to arise very naturally in the course of study of operators.\\

\n  In the cases when they are solved, the solutions are usually neat and elegant; otherwise, they pose interesting and challenging questions whose solutions may lead us to a deeper understanding of the nature of the operators under consideration.\\

\n For the products of two Hermitian operators on finite-dimensional spaces, we have the following classical characterization  \\
\begin{theorem}
 
\n The following statements are equivalent for a finite matrix $T$: \\

\n  (1) $T$ is the product of two Hermitian matrices; \\

 \n  (2) $T$ is the product of two Hermitian matrices, one of which is nonsingular; \\

\n  (3) $T$ is similar to a matrix with real entries; \\

\n  (4) $T$ is similar to $T^{*}$. \\
\end{theorem}
\n{\bf Proof}\\

\n The proof is an easy exercise in linear algebra. In the context of infinite- dimensional spaces, the implication $(1) \Longrightarrow (4)$ is no longer true.  As a consolation, it was conjectured that $T$ is similar to $T^{*}$ if $T$ is the product of two Hermitian Fredholm operators. This seems still to be open. \\

\n As for the implication $(4)  \Longrightarrow  (1)$ despite the abundance of supporting special cases, its validity is still unconfirmed even under the extra assumption that $T$ is invertible.\hfill { } $\square$ \\
\begin{remark}

\n The following conditions are equivalent for an operator $T$ on an infinite-dimensional space:\\

\n (1) $T$ is a commutator;\\

\n (2) $T$ is not the sum of nonzero scalar and a compact operator.\\
\end{remark}
\n In this context, the products of commutators seem not to have been considered. It would also be interesting to characterize products of a special kind of commutators: self-commutators, operators of the form $T^{*}T - TT^{*}$. \\
\begin{remark}

\n Any compact operator on an infinite-dimensional space is the product of finitely many nonnegative operators.\\
\end{remark}
\n For our operator $H_{\mu, \lambda}$, we will answered to following question: Its inverse $H_{\mu, \lambda}^{-1}$ can be factorized as the product of two operators  of Carleman's classes ?\\

\n Let $(\mu, \lambda) \in \mathbb{R}^{2}$ and  $H_{\mu, \lambda} = \mu H_{0} + i\lambda H_{1}$ with $H_{0} = A^{*}A$ and $H_{1} =  A^{*}( A + A^{*})A$ then  we can  to write $H_{0}^{-1}H_{1}$ but not $H_{1}H_{0}^{-1}$ because the domain of $H_{0}$  and the domain of $H_{1}$ are not included  in each other, nor is the domain of their commutator.\\

\n Observing that $H_{0}^{-1}(\mathcal{P}) = \mathcal{P} = H_{0}(\mathcal{P})$ then $H_{1}H_{0}^{-1}$ is well defined on the polynomial space $\mathcal{P}$ and we have:\\

\n $\displaystyle{\forall\, p \in \mathcal{P}, H_{\mu, \lambda} H_{0}^{-1}p = \mu p + i\lambda H_{1} H_{0}^{-1}p = \mu p + i\lambda A^{*}A^{2}H_{0}^{-1}p + i\lambda A^{*^{2}}AH_{0}^{-1}p}$\\

\n $\displaystyle{ = \mu p + i\lambda (AA^{*} - I)AH_{0}^{-1}p + i\lambda A^{*}p = \mu p + i\lambda(A + A^{*})p - i\lambda AH_{0}^{-1}p}$.\\

\n In particular, we get:\\

\n $\displaystyle{\mid \Re e < H_{\mu, \lambda} H_{0}^{-1}p, p > \mid \geq \mu \mid\mid p  \mid\mid^{2} -  \mid \lambda \mid \mid \Im m < AH_{0}^{-1}p, p > \mid  \forall \, p \in \mathcal{P}}$ \hfill { }  ($\bigstar_{1}$)\\

\n and\\

\n $\displaystyle{\mid\mid H_{\mu, \lambda} H_{0}^{-1}p\mid\mid.\mid\mid p \mid\mid \geq \mu \mid\mid p  \mid\mid^{2} -  \mid \lambda \mid \mid\mid AH_{0}^{-1}p \mid\mid. \mid \mid p \mid \mid \forall \, p \in \mathcal{P}}$  \hfill { }  ($\bigstar_{2}$)\\

\n Let $\displaystyle{p(z) = \sum_{k= 1}^{n}a_{k}\frac{z^{k}}{\sqrt{k!}}}$ then from  ($\bigstar_{1}$),  ($\bigstar_{2}$) and :\\

\n (i) $\displaystyle{AH_{0}^{-1}p(z) = \sum_{k= 1}^{n}a_{k}AH_{0}^{-1}\frac{z^{k}}{\sqrt{k!}} = \sum_{k= 1}^{n}\frac{a_{k}}{\sqrt{k}}\frac{z^{k-1}}{\sqrt{(k-1)!}};}$ \\

\n (ii) $\displaystyle{\mid\mid A\varphi\mid\mid^{2} = < A^{*}A\varphi, \varphi > = < H_{0}\varphi, \varphi >; \varphi \in D(H_{0})}$,\\

\n  we deduce that\\

\n $\displaystyle{\mid\mid AH_{0}^{-1}p\mid\mid^{2} \leq \mid\mid p \mid\mid^{2} \forall \, p \in \mathcal{P};}$ \hfill { }  ($\bigstar_{3}$)\\

 \n $\displaystyle{\mid\mid A\varphi\mid\mid \leq \mid\mid H_{0}\varphi\mid\mid^{\frac{1}{2}}\mid\mid \varphi\mid\mid^{\frac{1}{2}}, \forall\, \varphi \in D(H_{0});}$ \hfill { }  ($\bigstar_{4}$)\\
 
 \n $\displaystyle{\mid\mid AH_{0}^{-1}p \mid\mid \leq \mid\mid \varphi \mid\mid^{\frac{1}{2}}\mid\mid H_{0}^{-1}p \mid\mid^{\frac{1}{2}}\leq \frac{\epsilon}{\mid \lambda \mid} \mid\mid p \mid\mid  + \frac{\mid \lambda \mid }{\mid 4\epsilon\mid} \mid\mid H_{0}^{-1}p \mid\mid.}$ \hfill { }  ($\bigstar_{5}$)\\
 
 \n $\displaystyle{(\mu - \epsilon)\mid\mid p \mid\mid \leq \mid\mid H_{\mu, \lambda}H_{0}^{-1}p \mid\mid +  \frac{\mid \lambda \mid^{2}}{4\epsilon} \mid\mid H_{0}^{-1}p \mid\mid  \forall \, p \in \mathcal{P} }$ \hfill { }  ($\bigstar_{6}$)\\
 
 \n Now, as $\mathcal{P}$ is dense in $\mathcal{B}_{0}$ and the maximal domain of $H_{\mu, \lambda}$ coincides with its minimal domain, we deduce the following fundamental inequality:\\
 
 \n $\displaystyle{\forall \epsilon > 0, (\mu - \epsilon)\mid\mid H_{0}H_{\mu, \lambda}^{-1}\psi \mid\mid\, \, \leq \, \, (1 +  \frac{\mid \lambda \mid^{2}}{4\mid \mu\epsilon\mid}) \mid\mid \psi \mid\mid  \forall \, \psi \in \mathcal{B}_{0} }$ \hfill { }  ($\bigstar_{7}$)\\
 
\n  We summarize the results of above inequalities  as follows :\\
\begin{theorem}

\n (i) $A$ is $\frac{1}{2}$-subordinated to $H_{0}$,\\

\n (ii) $D(H_{\mu, \lambda}) \subset D(H_{0})$,\\

\n (iii) $H_{0}$ is subordinated to $H_{\mu, \lambda}$,\\

\n (iv) the operator $T = H_{0}H_{\mu, \lambda}^{-1}$ is bounded on Bargmann space $\mathcal{B}_{0}$ and $H_{\mu, \lambda}^{-1}$ is factored as product of $H_{0}^{-1}$ with $T$,\\

\n (v) the operators $H_{0}^{-1}$ and $H_{\mu, \lambda}^{-1}$ are of Carleman class of order $1 + \epsilon$, $\forall \, \epsilon > 0$.\\
\end{theorem}
\n {\bf  Open question}\\

\n Is there exist a strictly positive bounded operator $\mathbb{T}$ acting on Bargmann space such that  $\mathbb{T}H_{\mu, \lambda}^{-1} = H_{\mu, \lambda}^{-1^{*}}\mathbb{T}$ to apply the followiong Krein's theorem:\\
\begin{theorem}
(see {\color{blue}[Krein]})\\

\n Let $\mathcal{H}$ be  an  infinite  dimensional,  separable,  complex  Hilbert  space  and  let $\mathcal{C}_{p} = \mathcal{C}_{p}(\mathcal{H})$ be the Schatten-von Neumann class of compact operators acting on $\mathcal{H}$ and let $\mathbb{M} \in \mathcal{C}_{p}(\mathcal{H})$ be a linear bounded operator with the property that there exists a strictly positive bounded operator $\mathbb{A}$ such that $\mathbb{A}\mathbb{M} = \mathbb{M}^{*}\mathbb{A}$. Then the spectrum of $\mathbb{M}$ is real and for every non-zero eigenvalue $\lambda$, if $(\mathbb{M} -\lambda I)^{m}f = 0$ for some $m > 1$, then $(\mathbb{M} -\lambda I)f = 0$.\\
Moreover, the eigenvectors of $\mathbb{M}^{*}$, including the null vectors, span $\mathcal{H}$.\hfill { } \\
\end{theorem}
\n Although complex symmetric matrices (complex matrices coinciding with their transposes) are as ubiquitous as Hermitian matrices (those coinciding with their
complex adjoint), the first class is less well-known than the second. The matrix associated to $H_{\mu, \lambda}$ belongs to first class.Takagi , was  the first who had remarked that the antilinear eigenvalue problem $T\varphi = \sigma\bar{\varphi}$, where $T$ is a complex symmetric matrix and $\bar{\varphi}$ denotes complex conjugation, entry by entry, of the vector $\varphi$, solves a fundamental interpolation problem for bounded analytic functions in the disk. He had remarked there that the largest such positive skew-eigenvalue $\sigma$ coincides with the operator norm of $T$. Later on, this observation was extended to bounded linear operators with certain symmetries. About half a century ago, Glazman laid the foundations of the theory of unbounded complex symmetric operators. We will tested the Glazman's theory on our operator $H_{\mu, \lambda}$.\\

\n We begin by recalling a few definitions and facts about complex symmetric operators.\\
\begin{definition}

\n We consider a separable Hilbert space $\mathcal{H}$ which carries a conjugation operator $\displaystyle{\mathcal{C} : \mathcal{H} \longrightarrow \mathcal{H}}$ (an antilinear operator satisfying the conditions $\displaystyle{\mathcal{C}^{2} = I}$ and $\displaystyle{ < \mathcal{C}\varphi, \mathcal{C}\psi > = < \varphi, \psi >  \, \forall \, (\varphi, \psi) \in \mathcal{H}\times \mathcal{H} }$.\\

\n For a fixed $\mathcal{C}$, we define the transpose $^{t}T$ of a bounded linear operator $T$ to be  $^{t}T =  \mathcal{C}T^{*} \mathcal{C}$ where $ T^{*}$ is adjoint of $T$.\\

\n  We say that $T$ is $\mathcal{C}$-symmetric if $T = ^{t}T$ (equivalently, if $\displaystyle{\mathcal{C}T = T\mathcal{C}}$). More generally, we say that T is complex symmetric if there exists a $\mathcal{C}$ such that $T$ is $\mathcal{C}$-symmetric.\\

\n We say that a densely defined, closed graph operator is $\mathcal{C}$-symmetric if $\displaystyle{T \subset  \mathcal{C}T^{*} \mathcal{C}}$ and $\mathcal{C}$-selfadjoint if $\displaystyle{T =  \mathcal{C}T^{*} \mathcal{C}}$.\\
\end{definition}
\n By applying the Glazman's theory on our operator $H_{\mu, \lambda}$ then we obtain the following theorem:\\
\begin{theorem}

\n Let  $\mu > 0$ and $\displaystyle{H_{\mu, \lambda} : D(H_{\mu, \lambda}) \longrightarrow \mathcal{B}_{0}}$ then\\

\n (i) $H_{\mu, \lambda}$ is an unbounded $\mathcal{C}$-selfadjoint operator.\\

\n (ii) There exists an orthonormal basis $(\varphi_{n})_{n=0}$ of $\mathcal{B}_{0}$ consisting of solutions of the antilinear eigenvalue problem $\displaystyle{H_{\mu, \lambda} \varphi_{n} = \sigma{_n}\mathcal{C}\varphi_{n}}$ where $(\sigma{_n})_{n=0}$ is an increasing sequence of positive numbers tending to $\infty$.\\

\n (iii)  $\displaystyle{\mid\mid H_{\mu, \lambda}^{-1}\mid\mid = \frac{1}{inf_{n}\sigma_{n}}}$ \\
\end{theorem}
\n {\bf Proof}\\

\n (i) As $ H_{\mu, \lambda} =  H_{\mu, -\lambda}^{*}$ then $ H_{\mu, \lambda}$ is $\mathcal{C}$-symmetric and as $ H_{\mu, \lambda}^{-1}$ exists then by applying  a criterion which goes back to Zhikhar : << if a $\mathcal{C}$-symmetric operator $T$ with dense domain $D(T)$ $\displaystyle{ T: D(T) \subset \mathcal{H} \longrightarrow  \mathcal{H}}$  satisfies $\displaystyle{ (T - z I)D(T) =   \mathcal{H}}$ for some complex number $z$, then $T$ is  $\mathcal{C}$-selfadjoint >>, we deduce that $H_{\mu, \lambda}$ is an unbounded $\mathcal{C}$-selfadjoint operator.\\

\n (i) and (ii) is a direct consequence of results established by Garcia and Putinar in {\color{blue}[Garcia et al]}. \hfill { } $\square$\\
\subsection{Explicit inversion of $H_{\lambda',\mu,\lambda}$ on $[0, -i\frac{\lambda}{\lambda'}]$ and convergence of spectral radius of $H_{\lambda',\mu, \lambda}^{-1}$ to spectral radius of $H_{\mu, \lambda}^{-1}$ as  $\lambda' \longrightarrow 0$}

\n Using the same above technique used to get the kernel of $H_{\mu,\lambda}^{-1}$ on $]-\infty, 0]$, we  explicit the inverse of  $H_{\lambda',\mu,\lambda}$ on $[0, -i\frac{\lambda}{\lambda'}]$. \\

 \n Let $\psi \in \mathbb{B}_{0}$ and $\phi \in D(H_{\lambda',\mu,\lambda})$, we consider the equation  $H_{\lambda',\mu,\lambda}\phi = \psi$ which can be written under following form :\\
\begin{equation}
\displaystyle{(\lambda'z^{2} + i\lambda z)\phi''(z) + (i\lambda z^{2}  + \mu z)\phi'(z) = \psi(z)}
\end{equation}

 \n Choosing the straight line connecting $-i\rho' , z \in \mathbb{C}$ parametrized  by \\
\begin{equation}
\displaystyle{\gamma : [0, 1] \rightarrow \mathbb{C}, \gamma(t) = -i\rho' + t(z + i\rho') \quad \quad (\gamma(0) = -i\rho' , \gamma(1) = z)}
\end{equation}

\n  If we define $\displaystyle{\int_{-i\rho'}^{z}\psi(\xi)d\xi:=\int_{\gamma}\psi(\xi)d\xi:= \int_{0}^{1}}\psi(\gamma(t))\gamma'(t)dt$ then our equation can be transformed to following integral equation :\\
\begin{equation}
\displaystyle{\phi(z) = \frac{1}{\lambda'}\int_{-i\rho'}^{z}e^{-i\rho'\eta}(\eta + i\rho')^{-(\delta +1)} [\int_{-i\rho'}^{\eta}e^{i\rho'\xi}(\xi + i\rho')^{\delta}\frac{\psi(\xi)}{\xi}d\xi]d\eta}
\end{equation}

\n However, the integral representation of $\phi(z)$ in this last equation  is hard to study in $\mathbb{C}$ for some existing results on eigenvalues and eigenfunctions of our operators. To overcome this difficulty, the study of integral operator is restricted on negative imaginary axis by setting :\\

\n $ z = -iy$, $u(y) = \phi(-iy)$ and $ f(y) = \psi(-iy)$ with $y \in [0, \rho']$ then (3.89) can be written in the following form:\\
\begin{equation}
\displaystyle{(\lambda'y^{2} - \lambda y)u''(y) + (\lambda y^{2} + \mu y) u'(y) = f(y)}
\end{equation}

\n As $\psi \in \mathbb{B}_{0}$ then $f(0) = 0$ and if we put $\delta = \rho'(\rho + \rho') -1$, $\eta = -is$ and $\xi = y_{1}$ with $s \in [0, y]$ and $y_{1} \in [s, \rho']$ then (3.91) can be transformed to  following integral equation\\
\begin{equation}
\displaystyle{\phi(-iy) = \frac{1}{\lambda'}\int_{0}^{y}e^{-\rho's}(\rho' - s)^{-(\delta +1)}\int_{s}^{\rho'}e^{\rho'y_{1}}(\rho' - y_{1})^{\delta}\frac{\psi(-iy_{1})}{y_{1}}dy_{1}ds}
\end{equation}

\n then we get\\
\begin{equation}
\displaystyle{H_{\lambda',\mu,\lambda}^{-1}\psi(-iy) = \int_{0}^{\rho'}N_{\lambda',\mu,\lambda}(y, y_{1})\psi(-iy_{1})dy_{1}}
\end{equation}

\n with \\
\begin{equation}
\displaystyle{N_{\lambda',\mu,\lambda}(y, y_{1}) = \frac{1}{\lambda'y_{1}} e^{\rho'y_{1}}(\rho' - y_{1})^{\delta}\int_{0}^{min(y,y_{1})}e^{-\rho's}( \rho' - s)^{-(\delta +1)}ds}
\end{equation}

\n or\\
\begin{equation}
\displaystyle{N_{\lambda',\mu,\lambda}(y, y_{1}) = \frac{1}{\lambda y_{1}} e^{\rho'y_{1}}(1 - \frac{y_{1}}{\rho'})^{\delta}\int_{0}^{min(y,y_{1})}e^{-\rho's}(1 - \frac{s}{\rho'})^{-(\delta +1)}ds}
\end{equation}

\n Let $ \Upsilon = min(y,y_{1})$ and $ \displaystyle{\Theta(\Upsilon) = \int_{0}^{\Upsilon}e^{-\rho's}(1 - \frac{s}{\rho'})^{-(\delta +1)}ds}$ \\

\n then we get the following elementary properties on the function  $\Theta(\Upsilon)$  near to points zero and $\rho'$ respectively:\\

\n (i) $\displaystyle{\Theta(\Upsilon) \equiv \Upsilon}$ near zero.\\

\n (ii) $\displaystyle{\Theta(\Upsilon) \equiv (\rho' -\Upsilon)^{-\delta}}$ near $\rho'$.\\
\begin{theorem}

\n Let $L_{2}([0, \rho'], r(y)dy)$ be  the space of square integrable  functions with respect to the weight  $r(y) = e^{2\rho'y}(1 - \frac{y}{\rho'})^{2\delta}; \rho' = \frac{\lambda}{\lambda'}$ and $\delta = \rho'(\rho + \rho') - 1$.\\
Then for $\lambda' > 0$, $\mu > 0$ and $\delta \geq 0$  $H_{\lambda',\mu,\lambda}^{-1}$ can be extended to Hilbert-Schmidt operator of $L_{2}([0, \rho'], r(y)dy)$ to $L_{2}([0, \rho'], r(y)dy)$.\\
\end{theorem}
\n {\bf Proof (see  follow)}\\

\n {\bf The integral operator}  $H_{\lambda',\mu, \lambda}^{-1}$ {\bf on} $\displaystyle{[0, -i\frac{\lambda}{\lambda'}]}$  {\bf is extended to Hilbert-Scmidt operator on } $\displaystyle{L_{2}([0, -i\frac{\lambda}{\lambda'}], r(y)dy)}$\\

\n We wiil give an explicit inversion of $H_{\lambda',\mu,\lambda}$ on $[0, -i\frac{\lambda}{\lambda'}]$.\\

\n \n Setting $\rho' = \frac{\lambda}{\lambda'}$, $\rho = \frac{\mu}{\lambda}$ and $\delta = \rho'(\rho + \rho') -1$ for $\lambda' \neq 0$ and $\lambda \neq 0$.\\

\n We begin by to explicit the inverse of  $H_{\lambda',\mu,\lambda}$ on $[0, -i\frac{\lambda}{\lambda'}]$. \\

\n  Let $\psi \in \mathbb{B}_{0}$ and $\phi \in D(H_{\lambda',\mu,\lambda}$, we consider the equation  $H_{\lambda',\mu,\lambda}\phi = \psi$ which can be written under following form :\\

\n $\displaystyle{(\lambda'z^{2} + i\lambda z)\phi''(z) = (i\lambda z^{2}  + \mu z)\phi'(z) = \psi(z)}$ \hfill { } {\bf (B.1)}\\

\n  Let $\psi \in \mathbb{B}_{0}$ and choosing the straight line connecting $-i\rho' , z \in \mathbb{C}$ parametrized  by \\

\n  $\displaystyle{\gamma : [0, 1] \rightarrow \mathbb{C}, \gamma(t) = -i\rho' + t(z + i\rho') \quad \quad (\gamma(0) = -i\rho' , \gamma(1) = z)}$\\

 \n If we define $\displaystyle{\int_{-i\rho'}^{z}\psi(\xi)d\xi:=\int_{\gamma}\psi(\xi)d\xi:= \int_{0}^{1}}\psi(\gamma(t))\gamma'(t)dt$ then the equation (B.1) can be transformed to following integral equation :\\

\n $\displaystyle{\phi(z) = \frac{1}{\lambda'}\int_{-i\rho'}^{z}e^{-i\rho'\eta}(\eta + i\rho')^{-(\delta +1)} [\int_{-i\rho'}^{\eta}e^{i\rho'\xi}(\xi + i\rho')^{\delta}\frac{\psi(\xi)}{\xi}d\xi]d\eta}$ \hfill { } {\bf (B.2)}\\

\n However, the integral representation of $\phi(z)$ in this last equation (B.2) is hard to study in $\mathbb{C}$ for some existing results on eigenvalues and eigenfunctions of our operators. To overcome this difficulty, the study of (B.2) is restricted on negative imaginary axis by setting :\\

\n $ z = -iy$, $u(y) = \phi(-iy)$ and $ f(y) = \psi(-iy)$ with $y \in [0, \rho']$ then (B.1) can be written in the following form:\\

\n $\displaystyle{(\lambda'y^{2} - \lambda y)u''(y) + (\lambda y^{2} + \mu y) u'(y) = f(y)}$ \hfill { } {\bf (B.3)}\\

\n As $\psi \in \mathbb{B}_{0}$ then $f(0) = 0$ and if we put $\eta = -is$ and $\xi = y_{1}$ with , $s \in [0, y]$ and $y_{1} \in [s, \rho']$ then (B.3) can be transformed to  following integral equation\\

\n $\displaystyle{\phi(-iy) = \frac{1}{\lambda'}\int_{0}^{y}e^{-\rho's}(\rho' - s)^{-(\delta +1)}\int_{s}^{\rho'}e^{\rho'y_{1}}(\rho' - y_{1})^{\delta}\frac{\psi(-iy_{1})}{y_{1}}dy_{1}ds}$ \hfill { } {\bf (B.4)}\\

\n then we get\\

\n $\displaystyle{K_{\lambda',\mu,\lambda}\psi(-iy) = \int_{0}^{\rho'}N_{\lambda',\mu,\lambda}(y, y_{1})\psi(-iy_{1})dy_{1}}$ \hfill { } {\bf (B.5)}\\

\n with \\

\n $\displaystyle{N_{\lambda',\mu,\lambda}(y, y_{1}) = \frac{1}{\lambda'y_{1}} e^{\rho'y_{1}}(\rho' - y_{1})^{\delta}\int_{0}^{min(y,y_{1})}e^{-\rho's}( \rho' - s)^{-(\delta +1)}ds}$ \hfill { } {\bf (B.6)}\\

\n or\\

\n $\displaystyle{N_{\lambda',\mu,\lambda}(y, y_{1}) = \frac{1}{\lambda y_{1}} e^{\rho'y_{1}}(1 - \frac{y_{1}}{\rho'})^{\delta}\int_{0}^{min(y,y_{1})}e^{-\rho's}(1 - \frac{s}{\rho'})^{-(\delta +1)}ds}$ \hfill { } {\bf (B.7)}\\

\n Let $ \Upsilon = min(y,y_{1})$ and $ \displaystyle{\Theta(\Upsilon) = \int_{0}^{\Upsilon}e^{-\rho's}(1 - \frac{s}{\rho'})^{-(\delta +1)}ds}$ \\

\n then we get the following elementary properties on the function  $\Theta(\Upsilon)$  near to points zero and $\rho'$ respectively:\\

\n  Let $ \Upsilon = min(y,y_{1})$ et $ \displaystyle{\Theta(\Upsilon) = \int_{0}^{\Upsilon}e^{-\rho's}(1 - \frac{s}{\rho'})^{-(\delta +1)}ds}$ then

\n (i) $\displaystyle{\Theta(\Upsilon) \equiv \Upsilon}$ near zero.\\

\n (ii) $\displaystyle{\Theta(\Upsilon) \sim (\rho' -\Upsilon)^{-\delta}}$ near $\rho'$.\\

\n To give an extension of $K_{\lambda',\mu,\lambda}$ to $L_{2}([0, \rho'], r(y)dy)$, we consider

\n Let $L_{2}([0, \rho'], r(y)dy)$ be  the space of square integrable  functions with respect to the weight  $r(y) = e^{2\rho'y}(1 - \frac{y}{\rho'})^{2\delta}; \rho' = \frac{\lambda}{\lambda'}$ and $\delta = \rho'(\rho + \rho') - 1$.\\

\n By using the above lemma, we deduce the following theorem:\\
\begin{theorem}

\n For $\lambda' > 0$, $\mu > 0$ and $\delta \geq 0$  $K_{\lambda',\mu,\lambda}$ can be extended to Hilbert-Schmidt operator of  $L_{2}([0, \rho'], r(y)dy)$ to $L_{2}([0, \rho'], r(y)dy)$.\\
\end{theorem}
\n {\bf Proof}\\

\n  Let $\displaystyle{r(y) = e^{2\rho'y}(1 - \frac{y}{\rho'})^{2\delta}; \delta = \rho'(\rho + \rho') - 1}$ then \\

\n $\displaystyle{K_{\lambda',\mu,\lambda}\psi(-iy) = \int_{0}^{\rho'}N_{\lambda',\mu,\lambda}(y, y_{1})\psi(-iy_{1})dy_{1}}$\\

\n with \\

\n $\displaystyle{N_{\lambda',\mu,\lambda}(y, y_{1}) = \frac{1}{\lambda y_{1}} e^{\rho'y_{1}}(1 - \frac{y_{1}}{\rho'})^{\delta}\int_{0}^{min(y,y_{1})}e^{-\rho's}(1 - \frac{s}{\rho'})^{-(\delta +1)}ds}$ \\

\n can be written under the following form :\\

\n $\displaystyle{K_{\lambda',\mu,\lambda}\psi(-iy) = \int_{0}^{\rho'}\tilde{N}_{\lambda',\mu,\lambda}(y, y_{1})\psi(-iy_{1})r(y_{1})dy_{1}}$ \hfill { } {\bf (B.8)}\\

\n with \\

\n $\displaystyle{\tilde{N}_{\lambda',\mu,\lambda}(y, y_{1}) = \frac{1}{\lambda y_{1}\sqrt{r(y_{1})}} \int_{0}^{min(y,y_{1})}e^{-\rho's}(1 - \frac{s}{\rho'})^{-(\delta +1)}ds}$ \hfill { } {\bf (B.9)}\\

\n or\\

\n $\displaystyle{\tilde{N}_{\lambda',\mu,\lambda}(y, y_{1}) = \frac{1}{\lambda y_{1}\sqrt{r(y_{1})}} \Theta(min(y,y_{1}))}$ \hfill { } {\bf (B.10)}\\

\n Now we consider the following integral: \\

\n $\displaystyle{\mathbb{I} = \int_{0}^{\rho'}\int_{0}^{\rho'}\tilde{N}_{\lambda',\mu,\lambda}^{2}(y, y_{1})r(y)r(y_{1})dydy_{1}}$ \hfill { } {\bf (B.11)}\\

\n or \\

\n $\displaystyle{\mathbb{I} = \int_{0}^{\rho'}\int_{0}^{\rho'}\frac{1}{\lambda^{2} y_{1}^{2}} \Theta^{2}(min(y,y_{1}))r(y)dydy_{1}}$ \hfill { } {\bf (B.12)}\\

\n By applying Fubini theoren then (B.10) can be written as :\\

\n $\displaystyle{\mathbb{I} = \int_{0}^{\rho'}\int_{0}^{y_{1}}\frac{1}{\lambda^{2} y_{1}^{2}} \Theta^{2}(min(y,y_{1}))r(y)dydy_{1} + \int_{0}^{\rho'}\int_{y_{1}}^{\rho'}\frac{1}{\lambda^{2} y_{1}^{2}} \Theta^{2}(min(y,y_{1}))r(y)dydy_{1}}$ \\

\n and we have:\\

\n $\displaystyle{\mathbb{I} = \int_{0}^{\rho'}\int_{0}^{y_{1}}\frac{1}{\lambda^{2} y_{1}^{2}} \Theta^{2}(y)r(y)dydy_{1} + \int_{0}^{\rho'}\int_{y_{1}}^{\rho'}\frac{1}{\lambda^{2} y_{1}^{2}} \Theta^{2}(y_{1})r(y)dydy_{1} = \mathbb{I}_{1} + \mathbb{I}_{2}}$ \\

\n with\\

\n $\displaystyle{\mathbb{I}_{1} = \int_{0}^{\rho'}\frac{1}{\lambda^{2} y_{1}^{2}}\int_{0}^{y_{1}} \Theta^{2}(y)r(y)dydy_{1}}$  \hfill { } {\bf (B.13)}\\

\n and\\

\n $\displaystyle{\mathbb{I}_{2} = \int_{0}^{\rho'}\frac{\Theta^{2}(y_{1})}{\lambda^{2} y_{1}^{2}}\int_{y_{1}}^{\rho'}r(y)dydy_{1}}$ \hfill { } {\bf (B.14)}\\

\n Now we remark that \\

\n (i) For $y_{1}$ near zero we have $\displaystyle{\int_{0}^{y_{1}} \Theta^{2}(y)r(y)dy \equiv y_{1}^{3}}$ \\

\n  and\\

\n (ii) For $y_{1}$ near $\rho'$, the function $ y_{1} \rightarrow \displaystyle{\int_{0}^{y_{1}} \Theta^{2}(y)r(y)dy}$ is continuous.\\

\n then from (i) and (ii), we deduce that the first integral (B.11) converges. In similar way, we verify that the second integral (B.12) converges and consequently the operator $K_{\lambda',\mu,\lambda}$ is Hilbert-Schmidt on $L_{2}([0, \rho'], r(y)dy)$. \hfill { } $\square$\\

\n We summary some above results in the following theorem :\\
\begin{theorem}

\n i) Let $\displaystyle{N_{\rho'}(y, y_{1}) = \frac{1}{\lambda y_{1}\sqrt{r(y_{1})}} \int_{0}^{min(y,y_{1})}e^{-\rho's}(1 - \frac{s}{\rho'})^{-(\delta +1)}ds}$ then\\

\n the function $\rho' \rightarrow N_{\rho'}$ is decreasing.\\

\n ii) Let $\displaystyle{N_{\rho'}(y, y_{1}) = \frac{1}{\lambda  y_{1}}e^{\frac{- y_{1}^{2}}{2} - \frac{\mu}{\lambda} y_{1}}\int_{0}^{min(y,  y_{1})}e^{\frac{s^{2}}{2} + \frac{\mu}{\lambda}s}ds}$ then\\

\n $\displaystyle{Lim \quad \int_{0}^{\rho'}N_{\rho'}(y, y_{1})\psi(-iy_{1})dy_{1} = \int_{0}^{+\infty}N_{\mu,\lambda}(y, y_{1})\psi(-iy_{1})dy_{1}}$ as $\rho' \longrightarrow +\infty$\\

\n iii) The integral operator $K_{\rho'}$ can be extended to Hilbert-Schmidt operator of \\$\displaystyle{L_{2}([0, +\infty[, r_{\infty}(y)dy)}$ to $\displaystyle{L_{2}([0, +\infty[, r_{\infty}(y)dy)}$.\\

\n iv) On $\displaystyle{L_{2}([0, +\infty[, r_{\infty}(y)dy)}$, the integral operator $K_{\rho'}$ converges to  integral operator $K_{0,\mu,\lambda}$ as $\rho'$ goes to infinity with respect to Hilbert-Schmidt norm.\\

\n v) Let $\Omega(\lambda',\mu,\lambda)$ be the spectral radius of $K_{\lambda',\mu,\lambda}$  of kernel $N_{\lambda',\mu,\lambda}$ and let $\Omega(0,\mu,\lambda)$ be the spectral radius of $K_{0,\mu,\lambda}$ of kernel $N_{0,\mu,\lambda}$ then\\

\n $Lim \quad \Omega(\lambda',\mu,\lambda) = \Omega(0,\mu,\lambda)$ as $\lambda' \rightarrow 0$ or $\rho' \rightarrow +\infty$.\\

\n vi) The functions $\mu \rightarrow \tilde{N}_{\lambda',\mu,\lambda}$ and $\mu \rightarrow \Omega(\lambda',\mu,\lambda)$ are creasing with respect to $\mu$.\\

\n vii) The function $\mu \rightarrow \Omega(\lambda',\mu,\lambda)$ is analytic on the whole real line with respect to $\mu$ and the function $\mu \rightarrow \Omega(0,\mu,\lambda)$ is also analytic on the whole real line with respect to $\mu$.\\
\end{theorem}
\n {\bf Proof}\\

\n i) By observing that the map $\displaystyle{\rho' \rightarrow e^{\frac{a}{\rho'}}}$ is decreasing where $a$ is an positive constant, we deduce that the map $\rho' \rightarrow  N_{\rho'}$ is also decreasing.\\

\n ii) We obtain this property from an above lemma.\\

\n iii) If we consider the integral \\

\n $\displaystyle{\int_{0}^{+\infty}\int_{0}^{+\infty}N_{\rho'}^{2}\frac{r_{\infty}(y)}{r_{\infty}(y_{1})}dydy_{1}}$\\

\n then we have:\\

\n $\displaystyle{\int_{0}^{+\infty}\int_{0}^{+\infty}N_{\rho'}^{2}\frac{r_{\infty}(y)}{r_{\infty}(y_{1})}dydy_{1}}$ = $\displaystyle{\int_{0}^{\rho'}\int_{0}^{\rho'}\tilde{N}_{\lambda', \mu, ,\lambda}^{2}e^{-(y^{2} - y_{1}^{2}) - 2\rho(y - y_{1})}dydy_{1}}$\\

\n $\displaystyle{\leq C_{\rho'}\int_{0}^{\rho'}\int_{0}^{\rho'}\tilde{N}_{\lambda', \mu, ,\lambda}^{2}dydy_{1}}$ where $C_{\rho'}$ is a constant because the function $\displaystyle{e^{-(y^{2} - y_{1}^{2}) - 2\rho(y - y_{1})}}$ is bounded.\\

\n As $\displaystyle{\int_{0}^{\rho'}\int_{0}^{\rho'}\tilde{N}_{\lambda', \mu, ,\lambda}^{2}dydy_{1} < +\infty} $ then we deduce the property iii) of this theorem.\\

\n iv) Let $(\mathbb{X},\Sigma, \tau) $ be a measure space with positive measure $\tau$. An version of the classical monotone convergence theorem of Beppo Levi formulated in terms of functions of the Lebesgue space $L_{2}(X,\Sigma, \tau)$ reads as follows: If $(f_{n }: n \geq 1)$ is an decreasing sequence (that is, $f_{n}(x) \geq f_{n+1}(x)$ for every $n$ and almost every $x$ in $\mathbb{X}$) of nonnegative,square integrable functions on $\mathbb{X}$ such that $\displaystyle{\int_{X}\mid f_{1}(x)\mid^{2}d\tau < +\infty}$\\

\n Then\\

\n $f_{n}$ converges to some square integrable function $f$ both almost everywhere and in $\displaystyle{L_{2}}$ norm as $n \rightarrow+\infty$.

\n By applying this theorem, (ii) the above lemma , (i) and (ii) of theorem 3.5 we deduce that:\\

\n $\displaystyle{lim\int_{0}^{+\infty}\int_{0}^{+\infty}G_{\rho'}^{2}\frac{r_{\infty}(y)}{r_{\infty}(y_{1})}dydy_{1}}$ $= \displaystyle{\int_{0}^{+\infty}\int_{0}^{+\infty}N_{\mu,\lambda}^{2}\frac{r_{\infty}(y)}{r_{\infty}(y_{1})}dydy_{1}}$ as $\rho' \rightarrow+\infty$\\

\n and\\

\n $K_{\rho'}$ converges to $K(\mu, \lambda)$ as $\rho'$ goes to infinity in Hilbert-Schmidt norm on $\displaystyle{L_{2}([0, +\infty[, r_{\infty}(y)dy)}$.\\

\n v) Let $\Omega(\lambda',\mu,\lambda)$ be the spectral radius of $K_{\lambda',\mu,\lambda}$  of kernel $N_{\lambda',\mu,\lambda}$ and let $\Omega(0,\mu,\lambda)$ be the spectral radius of $K_{0,\mu,\lambda}$ of kernel $N_{0,\mu,\lambda}$ then\\

\n As $N_{\lambda',\mu,\lambda}$ is positive decreasing sequence of kernels which converges to the positive kernel $N_{0,\mu,\lambda}$ as goes to infinity we deduce that :\\

\n $\Omega(\mu,\lambda) \leq Lim \quad \Omega(\lambda',\mu,\lambda) $ as $\rho' \rightarrow + \infty$.\\

\n  Thus it suffices to show that\\

\n  $\displaystyle{Lim \quad \Omega(\lambda',\mu,\lambda)\leq \Omega(0,\mu,\lambda)}$ as $\rho' \rightarrow + \infty$.\\

\n  From the above property iv) we deduce that\\

\n $\forall \quad n \in \mathbb{N}$, $\displaystyle{\mid\mid K_{\lambda',\mu,\lambda}^{n}\mid\mid^{\frac{1}{n}}}$ $\rightarrow$  $\displaystyle{\mid\mid K_{0,\mu, \lambda}^{n}\mid\mid^{\frac{1}{n}}}$ as $\rho' \rightarrow + \infty$.\\

\n i.e.\\

\n $\displaystyle{\forall \quad \epsilon > 0, \exists\quad \rho_{1}; \forall \quad \rho' \geq \rho_{1}}$ we have $\displaystyle{\mid\mid K_{0,\mu, \lambda}^{n}\mid\mid^{\frac{1}{n}} - \epsilon \leq \mid\mid K_{\lambda',\mu,\lambda}^{n}\mid\mid^{\frac{1}{n}} \leq \mid\mid K_{0,\mu, \lambda}^{n}\mid\mid^{\frac{1}{n}} + \epsilon}$.\\

\n set $\rho_{1} = \frac{\mu}{\lambda_{1}^{'}}$ then we get\\

\n $\displaystyle{\mid\mid K_{\lambda_{1}^{'},\mu,\lambda}^{n}\mid\mid^{\frac{1}{n}}\leq \mid\mid K_{0,\mu, \lambda}^{n}\mid\mid^{\frac{1}{n}} + \epsilon }$.\\

\n In particular, we get\\
\begin{equation}
\displaystyle{\Omega (\lambda_{1}^{'},\mu,\lambda)\leq \Omega(0,\mu, \lambda) + \epsilon }
\end{equation}

\n Now as \\

\n $\displaystyle{\Omega (\lambda',\mu, \lambda) = lim \quad \mid\mid K_{\lambda',\mu, \lambda}^{n}\mid\mid^{\frac{1}{n}}}$ as $n \rightarrow +\infty$\\

\n i.e.\\

\n $\displaystyle{\forall \quad \epsilon > 0, \exists\quad n_{0}\in \mathbb{N}; \forall \quad n \geq n_{0}}$ we have $\displaystyle{\Omega (\lambda',\mu, \lambda) - \epsilon \leq \mid\mid K_{\lambda',\mu, \lambda}^{n}\mid\mid^{\frac{1}{n}} \leq \Omega_{(\lambda',\mu, \lambda)} + \epsilon}$\\

\n By using the above right inequality and the decreasing of the map $\rho' \rightarrow N_{\rho'}$ , we deduce that:\\
\begin{equation}
\displaystyle{\mid\mid K_{\lambda',\mu, \lambda}^{n}\mid\mid^{\frac{1}{n}} \leq \Omega(\lambda',\mu, \lambda) + \epsilon \leq  \Omega(\lambda_{1}^{'},\mu, \lambda) + \epsilon}
\end{equation}

\n and by (3.97) and (3.98) and for all $\epsilon > 0$, we get:\\

\n $\displaystyle{lim \quad\Omega_{\rho'}}$ $\leq$ $\displaystyle{\Omega(0,\mu, \lambda) + 2\epsilon}$ as $\rho'\rightarrow +\infty $\\

\n Then \\

\n $\displaystyle{lim \quad\Omega(\lambda^{'},\mu, \lambda) = \Omega(0,\mu,\lambda)}$ as $\lambda' \rightarrow 0$.\\

\n vi) Let $\displaystyle{\tilde{N}_{\lambda',\mu,\lambda}(y, y_{1}) = \frac{1}{\lambda y_{1}\sqrt{r(y_{1})}} \int_{0}^{min(y,y_{1})}e^{-\rho's}(1 - \frac{s}{\rho'})^{-(\delta +1)}ds}$ where\\ $\delta = \rho'(\rho + \rho') - 1 = \rho'^{2} + \rho\rho' -1$\\

\n then  $\displaystyle{\tilde{N}_{\lambda',\mu,\lambda}(y, y_{1})}$ can be written as \\
\begin{equation}
\displaystyle{\tilde{N}_{\lambda',\mu,\lambda}(y, y_{1}) = \frac{1}{\lambda y_{1}}e^{\rho'y_{1}}(1 - \frac{y_{1}}{\rho'})^{\rho'^{2} -1} \int_{0}^{min(y,y_{1})}e^{-\rho's}(1 - \frac{s}{\rho'})^{-\rho'^{2}}[\frac{\rho' - y_{1}}{\rho' - s}]^{\rho\rho'} ds}
\end{equation}

\n  As $0 \leq \leq y_{1}$, the map $\displaystyle{\mu \rightarrow [\frac{\rho' - y_{1}}{\rho' - s}]^{\rho\rho'}}$ is creasing with respect to $\mu$ and we get vi).\\

\n vii) Let $\displaystyle{\delta_{\beta} = \rho'(\frac{\beta}{\lambda} + \rho') - 1}$ and $\displaystyle{L_{2,\beta}([0, \rho'], r_{\beta}(y)dy)}$ be the space of square integrable  functions with respect to the measure $\displaystyle{r_{\beta}(y) = e^{2\rho'y}(1 - \frac{y}{\rho'})^{2\delta_{\beta}}}$.\\

\n To use the results of chapter VII of Kato's book {\color{blue}[Kato]} on the operators depending of a parameter, we shall consider the map:\\
\begin{equation}
\displaystyle{\mu \rightarrow <\phi, K_{\lambda', \mu, \lambda}\psi > = \int_{0}^{\rho'}\phi(-iy)\int_{0}^{\rho'}\tilde{N}_{\lambda', \mu, \lambda}(y,y_{1})\bar{\psi}(-iy_{1})dy_{1}r_{\beta}(y)dy}
\end{equation}

\n with $\displaystyle{\phi \in L_{2,\beta}([0, \rho'], r_{\beta}(y)dy)}$ and $\displaystyle{\psi \in L_{2,\beta}([0, \rho'], r_{\beta}(y)dy)}$\\

\n We being by showing that the map defined by (3.100) is continuous with respect to $\mu$ :\\

\n Let $\displaystyle{\tilde{N}_{\lambda', \beta}}$ be the expression of  $\displaystyle{\tilde{N}_{\lambda', \mu, \lambda}}$  where we have replaced  $\delta $ by $\delta_{\beta}$.\\

\n Now, we set $\displaystyle{f_{\mu}(y, y_{1}) = e^{2\rho'y}(1 - \frac{y}{\rho'})^{2\delta_{\beta}}\tilde{N}_{\lambda', \mu, \lambda}(y, y_{1})\phi(-iy)\bar{\psi}(-iy_{1})}$ then for $\mu \geq \beta$ we have:\\

\n $\displaystyle{\mid f_{\mu}(y, y_{1})\mid = e^{2\rho'y}(1 - \frac{y}{\rho'})^{2\delta_{\beta}}\tilde{N}_{\lambda', \beta}(y, y_{1})\mid \phi(-iy)\mid \mid\bar{\psi}(-iy_{1})\mid}$ \\

\n then the continuity of (3.100) with respect to $\mu$ follows by applying the bounded convergence theorem.\\

\n Now to show that the map defined by (3.100) is analytic with respect to  $\mu$,  it suffices to observe that for all closed curve we have \\
\begin{equation}
\displaystyle{\int_{\gamma}\tilde{N}_{\lambda', \mu, \lambda}d\mu = 0}
\end{equation}

\n and to apply the following theorem: \\
\begin{theorem}
 (Reed-Simon  {\color{blue}[Reed et al]}, theorem XII.8) \\

\n Let $\mathbb{E}$ be a complex Banach space and $L(\mathbb{E})$ be the set of bounded operators on $\mathbb{E}$ with respect to the operator norm.\\

\n  Let $\mathbb{T}$ be an analytic function $(\mu \in \mathbb{C} \rightarrow \mathbb{T}(\mu) \in L(\mathbb{E}))$ and $\sigma_{0}$  be a simple eigenvalue of $T(\mu_{0})$.Then\\

\n  There exists an analytic function $\sigma$ of $\mu$  for $\mu$ near $\mu_{0}$ such that  $\sigma(\mu)$ is simple eigenvalue of $\mathbb{T}(\mu)$ and $\sigma(\mu_{0}) = \sigma_{0}$.\hfill { } $\square$\\
\end{theorem}
\n We conclude this section by noting that for $\lambda' \neq 0$, the operator $H_{\lambda', \mu, \lambda}$  has a rich set of spectral properties desired by physicists and for $\lambda' = 0$,  the results of Ando-Zerner  were a major contribution in the spectral study of the operator $H_{0, \mu, \lambda}$, however the density of its eigenvectors in Bargmann space is always open question.\\

\section{Study of the singularity of $\displaystyle{ H_{\lambda',\mu, \lambda}\varphi(z) = \sigma(\lambda', \mu, \lambda) \varphi(z)\, at \, \infty}$}
\subsection{On bi-confluent Heun equation associated to eigenvalue problem of  $\displaystyle{ H_{\lambda',\mu, \lambda}}$}

\n We consider the equation \\

\n   $\displaystyle{(\lambda'z^{2} + i\lambda z)\varphi^{''}(z) + (i \lambda z^{2} + \mu z)\varphi^{'}(z) = \sigma(\lambda', \mu, \lambda) \varphi(z)}$ \quad \quad \, at \, $\infty$ \hfill { } ($\star$)\\

\n In our differential equation ($\star$), if we put $\psi(z) = \varphi^{'}(z)$ then we get\\
\begin{equation}
\displaystyle{\psi^{'}(z) = \frac{\sigma(\lambda', \mu, \lambda)}{z(\lambda' z + i\lambda)}\varphi(z) - \frac{ i \lambda + \mu}{\lambda' z + i\lambda}\psi(z)}
\end{equation}

\n Now let $\phi(z) = \left (\begin{array} [c] {l} \varphi(z)\\
\quad\\
\psi(z)\\
\end{array} \right )$ then \\

\begin{equation}
\phi^{'}(z) = \left (\begin{matrix} \displaystyle{0}&\displaystyle{1}\\
\quad\\
\displaystyle{ \frac{\sigma(\lambda', \mu, \lambda)}{z(\lambda' z + i\lambda)}}&\displaystyle{- \frac{ i \lambda + \mu}{\lambda' z + i\lambda}}\\
\end{matrix} \right )\left (\begin{array} [c] {l} \varphi(z)\\
\quad\\
\psi(z)\\
\end{array} \right )
\end{equation}

\n If we put $\displaystyle{z = re^{i\theta}; r \in [0, +\infty[, \theta \in [0, 2\pi]}$ and $W(r) = \left (\begin{array} [c] {l} \varphi(re^{i\theta})\\
\quad\\
\psi(re^{i\theta})\\
\end{array} \right )$, we deduce that\\
\begin{equation}
\displaystyle{ W'(r) =e^{-i\theta}\mathbb{M}(re^{i\theta})W(r)}
\end{equation}

\n where  $\mathbb{M}(re^{i\theta})= \left (\begin{matrix} \displaystyle{0}&\displaystyle{1}\\
\quad\\
\displaystyle{ \frac{\sigma(\lambda', \mu, \lambda)}{re^{i\theta}(\lambda' re^{i\theta} + i\lambda)}}&\displaystyle{- \frac{ i \lambda + \mu}{\lambda' re^{i\theta} + i\lambda}}\\
\end{matrix} \right )$\\
\quad\\

\n In $\mathbb{C}^{2}$ we denote the norm of $W$ by  $\mid\mid W\mid\mid$ and the norm of matrix $\mathbb{M}$ by $\mid\mid\mid \mathbb{M}\mid\mid\mid$.\\

\n As  $\displaystyle{\mid\mid\mathbb{M}(re^{i\theta})W(re^{i\theta})\mid\mid \,\leq \,\mid\mid\mid \mathbb{M}\mid\mid\mid\,\mid\mid W\mid\mid}$, then we deduce that\\
\begin{equation}
\displaystyle{\mid\mid\mathbb{M}^{'}(r)\leq \mid\mid\mid \mathbb{M}\mid\mid\mid\,\mid\mid W(r)\mid\mid}
\end{equation}

\n Let $R > \rho'$ then the norm $\mid\mid\mid \mathbb{M}\mid\mid\mid$ is bounded outside the disk of radius $R$ and by applying the Gronwal Lemma, we deduce the following inequality\\
\begin{equation}
\displaystyle{ \forall \, r \in [R, + \infty[, \mid\mid \phi(z) \mid\mid \leq sup_{\theta \in [0, 2\pi]}\mid\mid W(Re^{i\theta}) \mid\mid e^{\mid\mid\mid \mathbb{M}\mid\mid\mid(\mid z\mid - R)}}
\end{equation}

\n In particular, we get\\
\begin{equation}
\displaystyle{\mid \varphi(z) \mid \leq sup_{\theta \in [0, 2\pi]}\mid\mid W(Re^{i\theta}) \mid\mid e^{\mid\mid\mid \mathbb{M}\mid\mid\mid(\mid z\mid - R)}}
\end{equation}
\n Therefore  $\varphi$ belongs to Bargmann space $\mathbb{B}_{0} $\\
\subsection{On eigenvalues of $H_{\lambda', \mu, \lambda}$ acting on $\displaystyle{L_{2}((0, \rho'), r(y)dy)}$}
\n Let  $\rho' > 0, y \in [0, \rho']$ and $\displaystyle{L_{2}((0, \rho'), r(y)dy)}$ be the Hilbert space of square integrable functions on $[0, \rho']$ with inner
product\\
\begin{equation}
\displaystyle{< u, v > = \int_{0}^{\rho'}u(y)\overline{v(y)}r(y)dy}
\end{equation}
\n with $r(y)$ called weight function.\\

\n We will construct the fonction $r(y)$ such that $H_{\lambda', \mu, \lambda}$ is symmetric on $\displaystyle{L_{2}((0, \rho'), r(y)dy)}$.\\

\n For $\varphi \in \mathbb{B}_{0} $, consider  $u(y) = \varphi(-iy)$ and the eigenvalue problem associated to $H_{\lambda', \mu, \lambda}$:\\
\begin{equation}
\displaystyle{-y (\lambda - \lambda' y)u^{"}(y) + y(\lambda y +\mu y)u'(y) = \sigma(\lambda', \mu, \lambda)u(y)}
\end{equation}
By multiplying the differential equation (4.8) by $r(y)\overline{u(y)}$ and integrating it on $[0, \rho']$ to get \\
\begin{equation} 
\left \{ \begin{array}{c} \displaystyle{\int_{0}^{\rho'}[-y (\lambda - \lambda' y)u^{"}(y)+y(\lambda y +\mu y)u'(y)] \overline{u(y)}r(y)dy =}\\
\displaystyle{\sigma(\lambda', \mu, \lambda)\int_{0}^{\rho'}\mid u(y)\mid^{2}r(y)dy \quad \quad \quad \quad \quad \quad \quad \quad \quad }
\end{array} \right.
 \end{equation}

\n Integrating the first term  of integral equation by parts to get\\

\n $\displaystyle{[s(y)\overline{u(y)}u'(y)(\lambda' y - \lambda)]_{0}^{\rho'} - \int_{0}^{\rho'}s(y)(\lambda' y - \lambda)\mid u'(y)\mid^{2}dy}$ \\

\n + $\displaystyle{\int_{0}^{\rho'}[(\lambda y + \mu)s(y) - \frac{\partial}{\partial y}((\lambda' y - \lambda)s(y))]\mid u(y)\mid^{2}dy}$\\
\begin{equation}
\displaystyle{= \sigma(\lambda', \mu, \lambda)\int_{0}^{\rho'}\mid u(y)\mid^{2}r(y)dy}
\end{equation}

\n where $s(y) = yr(y)$\\

\n We choose $s(y)$ such that \\
\begin{equation}
\displaystyle{ \frac{\partial}{\partial y}((\lambda' y - \lambda)s(y)) = (\lambda y + \mu)s(y)}
\end{equation}

\n Then \\
\begin{equation}
\displaystyle{ s(y) = ce^{\rho' y}\mid y - \rho'\mid^{\rho'(\rho + \rho') - 1}}\, and \,\displaystyle{ r(y) = \frac{s(y)}{y}} , c \, is \, const.
\end{equation}
\begin{theorem}

\n if $\displaystyle{\rho'(\rho + \rho')  \geq 1}$ then $H_{\lambda', \mu, \lambda}$ is symmetric on $\displaystyle{L_{2}((0, \rho'), r(y)dy)}$.\\
\end{theorem}
\n {\bf Proof}\\

\n if $\displaystyle{\rho'(\rho + \rho')  \geq 1}$ then the function $s(y)$ is bounded on $[0, \rho']$ this implies that \\
\begin{equation}
\displaystyle{[s(y)\overline{u(y)}u'(y)(\lambda' y - \lambda)]_{0}^{\rho'} = 0}
\end{equation}

\n Consequently\\
\n if $\displaystyle{\rho'(\rho + \rho')  \geq 1}$ then the function $s(y)$ is bounded on $[0, \rho']$ this implies that \\
\begin{equation}
\displaystyle{ - \int_{0}^{\rho'}s(y)(\lambda' y - \lambda)\mid u'(y)\mid^{2}dy} = \displaystyle{\sigma(\lambda', \mu, \lambda)\int_{0}^{\rho'}\mid u(y)\mid^{2}r(y)dy}
\end{equation}

\n i.e $H_{\lambda', \mu, \lambda}$ is symmetric on $\displaystyle{L_{2}((0, \rho'), r(y)dy)}$.\hfill { } $\square$\\
\begin{corollary}

\n (i)  if $\displaystyle{\rho'(\rho + \rho')  \geq 1}$ then the the eigenvalues of $H_{\lambda', \mu, \lambda}$ are real.\\

\n (ii) On $\displaystyle{L_{2}((0, \rho'), r(y)dy)}$, the operator $H_{\lambda', \mu, \lambda}$ takes the following form\\
\n if $\displaystyle{\rho'(\rho + \rho')  \geq 1}$ then the function $s(y)$ is bounded on $[0, \rho']$ this implies that \\
\begin{equation}
\displaystyle{H_{\lambda', \mu, \lambda}u(y) = \frac{1}{r(y)}\frac{\partial}{\partial y}(p(y)u'(y))}
\end{equation}

\n where $\displaystyle{p(y) = \lambda' y(\rho' - y) r(y)}$.\\
\end{corollary}
\n {\bf Proof}\\

\n (i) It is a simple consequence of the above theorem and the fact that the space of the restrictions of the elements of the domain of the operator $ H_{\lambda', \mu, \lambda}$   as operator acting on Bargmann's space $\mathbb{B}_{0}$ is included in \\

\n $\displaystyle{\mathbb{B}_{2} = \{ u \in L_{2}((0, \rho'), r(y)dy); H_{\lambda', \mu, \lambda}u \in L_{2}((0, \rho'), r(y)dy); u(0) = 0\}}$\\

\n (ii) It is a simple consequence.\hfill { } $\square$\\
\begin{proposition}

\n if $\displaystyle{\rho'(\rho + \rho')  \geq 1}$ then the eigenvalues of $H_{\lambda', \mu, \lambda}$ acting on $\displaystyle{L_{2}((0, \rho'), r(y)dy)}$ are simple.\\
\end{proposition}
\n {\bf Proof}\\

\n Let $u, v \in \displaystyle{\mathbb{B}_{2} }$ then $\displaystyle{vH_{\lambda', \mu, \lambda}u = \frac{v}{r(y)}\frac{\partial}{\partial y}(p(y)u'(y))}$, $\displaystyle{uH_{\lambda', \mu, \lambda}v = \frac{u}{r(y)}\frac{\partial}{\partial y}(p(y)v'(y))}$\\
\n and \\
\n if $\displaystyle{\rho'(\rho + \rho')  \geq 1}$ then the function $s(y)$ is bounded on $[0, \rho']$ this implies that \\
\begin{equation}
\displaystyle{r(y)(vH_{\lambda', \mu, \lambda}u - uH_{\lambda', \mu, \lambda}v = \frac{\partial}{\partial y}[p(y)(v(y)u'(y) - u(y)v'(y)]}
\end{equation}

\n If $\displaystyle{H_{\lambda', \mu, \lambda}u(y)  = \sigma(\lambda', \mu, \lambda)u(y) ; u \neq 0}$ and  $\displaystyle{H_{\lambda', \mu, \lambda}v(y)  = \sigma(\lambda', \mu, \lambda)v(y) ; v \neq 0}$  then \\

\n $\displaystyle{\frac{\partial}{\partial y}[p(y)(v(y)u'(y) - u(y)v'(y)] = 0}$, in particular $p(y)(v(y)u'(y) - u(y)v'(y)$ is a constant.\\

\n  As $p(0)((v(0)u'(0) - u(0)v'(0)) = 0$ then  $v(y)u'(y) - u(y)v'(y) = 0 \, \forall \, y \in [0, \rho']$ and $ v $ is proportional to $u$ i.e. the geometric multiplicity  of $\sigma(\lambda', \mu, \lambda)$ is one. \\

\n Now, let $\displaystyle{Ker(H_{\lambda', \mu, \lambda}  - \sigma(\lambda', \mu, \lambda)I ) = \{u \in  \mathbb{B}_{2}; H_{\lambda', \mu, \lambda}u  = \sigma(\lambda', \mu, \lambda)u ; u \neq 0\}}$, showing that there are no Jordan blocks corresponding to  $\sigma(\lambda', \mu, \lambda)$ i.e;\\

\n $\displaystyle{Ker(H_{\lambda', \mu, \lambda}  - \sigma(\lambda', \mu, \lambda)I) = Ker(H_{\lambda', \mu, \lambda}  - \sigma(\lambda', \mu, \lambda)I)^{2} }$ $, \displaystyle{\forall\,  \sigma(\lambda', \mu, \lambda) \in \sigma(H_{\lambda', \mu, \lambda})}$\\

\n where $\displaystyle{\sigma(H_{\lambda', \mu, \lambda})}$ denotes the spectrum of $\displaystyle{H_{\lambda', \mu, \lambda}}$.\\

\n Let $\displaystyle{u \in Ker(H_{\lambda', \mu, \lambda}  - \sigma(\lambda', \mu, \lambda)I)}$ and $\displaystyle{v \in Ker(H_{\lambda', \mu, \lambda}  - \sigma(\lambda', \mu, \lambda)I)^{2} }$  such that if $\displaystyle{\rho'(\rho + \rho')  \geq 1}$ then the function $s(y)$ is bounded on $[0, \rho']$ this implies that \\
\begin{equation}
\displaystyle{\frac{1}{r(y)}\frac{\partial}{\partial y}(p(y)u'(y))u(y) = \sigma(\lambda', \mu, \lambda)u(y), \quad y \in [0, \rho']}
\end{equation}

\n and\\
\n if $\displaystyle{\rho'(\rho + \rho')  \geq 1}$ then the function $s(y)$ is bounded on $[0, \rho']$ this implies that \\
\begin{equation}
\displaystyle{\frac{1}{r(y)}\frac{\partial}{\partial y}(p(y)v'(y))v(y) = \sigma(\lambda', \mu, \lambda)v(y) + u(y), \quad y \in [0, \rho']}
\end{equation}

\n Multiplyng  the differential equation by $r(y)\overline{v(y)}$ and  by $r(y)\overline{v(y)}$, integrating  by parts the obtained equations on $[0, \rho']$ to get\\
\n if $\displaystyle{\rho'(\rho + \rho')  \geq 1}$ then the function $s(y)$ is bounded on $[0, \rho']$ this implies that \\
\begin{equation}
\displaystyle{\int_{0}^{\rho'}p(y)u'(y)\overline{v}'(y)dy = \sigma(\lambda', \mu, \lambda)\int_{0}^{\rho'}r(y)u(y)\overline{v(y)}dy}
\end{equation}

\n and\\

\n if $\displaystyle{\rho'(\rho + \rho')  \geq 1}$ then the function $s(y)$ is bounded on $[0, \rho']$ this implies that \\
\n \begin{equation}
\left \{ \begin{array} {c} \displaystyle{\int_{0}^{\rho'}p(y)v'(y)\overline{u}'(y)dy = \sigma(\lambda', \mu, \lambda)\int_{0}^{\rho'}r(y)v(y)\overline{u(y)}dy \quad +}\\ \displaystyle{ \int_{0}^{\rho'}r(y)\mid u(y)\mid^{2}dy \quad \quad  \quad \quad  \quad \quad  \quad \quad  \quad  \quad \quad \quad  \quad \quad} \end{array} \right.
\end{equation}

\n Now (4.19) - (4.20) gives \\

\begin{equation}
\left \{ \begin{array} {c}  \displaystyle{\int_{0}^{\rho'}p(y)[ u'(y)\overline{v}'(y) -  v'(y)\overline{u}'(y)]dy = \quad \quad }\\
\displaystyle{\sigma(\lambda', \mu, \lambda)\int_{0}^{\rho'}r(y)[u(y)\overline{v(y)} - v(y)\overline{u(y)}]dy}\\
\displaystyle{ - \int_{0}^{\rho'}r(y)\mid u(y)\mid^{2}dy}    \quad  \quad  \quad  \quad \quad \quad  \quad \quad \end{array} \right.
\end{equation}

\n In particular we have\\

\begin{equation}
\left \{ \begin{array} {c} \displaystyle{2i\int_{0}^{\rho'}p(y)\Im m(u'(y)\overline{v}'(y))dy = }\\

\displaystyle{2i\sigma(\lambda', \mu, \lambda)\int_{0}^{\rho'}r(y)\Im m(u(y)\overline{v(y)}) - v(y)dy}\\

\displaystyle{ - \int_{0}^{\rho'}r(y)\mid u(y)\mid^{2}dy} \end{array} \right.
\end{equation}

\n As $p$ is real function and $ \sigma(\lambda', \mu, \lambda) \in \mathbb{R}$ then $\displaystyle{\int_{0}^{\rho'}r(y)\mid u(y)\mid^{2}dy = 0}$ which is contradictory with $u$ is eigenfunction associated to  $ \sigma(\lambda', \mu, \lambda)$.\hfill { } $\square$\\

\n We end these first spectral properties on $H_{\lambda', \mu, \lambda}$ by the following observations\\

\n (1) The hamiltonian $H_{\lambda', \mu, \lambda}$ acting on Bargmann space (which is not selfadjoint) has a real spectrum.\\ 

\n (2) The PT-symmetry condition (space-time reflection) is not sufficient to provide that the spectrum  of  $H_{\lambda', \mu, \lambda}$ is real, there are some PT-symmetric hamiltonian with polynomial potentials producing non real eigenvalues, see  {\color{blue}[Delabeare et al]}.\\

\n (3) The method used by  {\color{blue}[Dorey et al]} to give a rigorous proof of reality and positivity of the eigenvalues of some non-self-adjoint hamiltonian is not applicable to $H_{\lambda', \mu, \lambda}$ and even less to  $H_{\mu, \lambda}$\\

\section{Review some fondamental spectral properties of $H_{\lambda', \mu, \lambda}$}

\n We now turn to review  the fondamental spectral properties of $H_{\lambda', \mu, \lambda}$ that we will presented in above sections.\\

\n {\color{black}$\bullet$} We have restricted $H_{\lambda', \mu, \lambda}$ to the subspace of Bargmann space consisting of the functions that vanish at $0$. Thus we get rid of the vacuum state.\\

\n  A central role is played by the following inequality (for $\mu > 0$): \\
\begin{equation}
\displaystyle{\Re e < H_{\lambda, \mu} \varphi , \varphi > = \mu \mid\mid A\varphi \mid\mid^{2} \geq  \mid\mid \varphi \mid\mid^{2}, \, \forall\,   \varphi \in D(H_{\lambda, \mu})}
\end{equation}

\n which obviously holds for polynomials ($\Re e$ means the real part).\\

\n  {\color{black}$\bullet$} Now if $\varphi$ is in Bargmann space, $H_{\lambda, \mu}\varphi$ can be defined as an analytic function, so that it makes sense to take as domain of $H_{\lambda, \mu}$ the set of functions $\varphi$ such that $H_{\lambda, \mu}\varphi$ still is in Bargmann space. It turns out that so defined, $H_{\lambda, \mu}$
is the closure of $H_{\lambda, \mu_{\mid_{pol}}}$, has a compact inverse and $e^{-t H_{\lambda, \mu}}$  can be defined for positive $t$ (see  another section of this work for detailed proofs).\\

\n  {\color{black}$\bullet$} Above, a rigorous mathematical analysis of domains of $H_{\mu, \lambda}$ is given , in particular, we have :\\

\begin{equation}
\displaystyle{D(H_{\mu, \lambda}) = D(H_{\mu, \lambda}^{min})}
\end{equation}

\n   where $H_{\mu, \lambda}^{min}$ is the restriction of the closure of $H_{\mu, \lambda}$ on the polynomials.\\ 

\n  {\color{black}$\bullet$} By using the following inequality :\\

\n $\displaystyle{\forall \, \epsilon > 0 , \exists \, C_{\epsilon, \lambda', \mu,  \lambda} > 0 ; \mid\mid H_{\mu, \lambda} \varphi \mid\mid^{2} \leq \epsilon \mid\mid A^{*^{2}}A^{2}\varphi \mid\mid  +  C_{\epsilon, \lambda', \mu,  \lambda} \mid\mid \varphi  \mid\mid \, \forall \, \varphi \in D(A^{*^{2}}A^{2})}$\\

\n we can  show   that for $\lambda' \neq 0$,  we have \\
\begin{equation}
 \displaystyle{D(H_{\lambda', \mu, \lambda}) = D(A^{*^{2}}A^{2})}
\end{equation}

\n {\color{black}$\bullet$} Note that the domain of the adjoint and anti-adjoint parts of $H_{\mu, \lambda}$ are not included  in each other, nor is the domain of their commutator.\\

\n {\color{black}$\bullet$} The restriction of the closure of the anti-part $H_{I} = A^{*}(A + A^{*})A$ of $H_{\mu, \lambda}$  or of $H_{\lambda', \mu, \lambda}$ on the polynomials is symmetric but it is not self-adjoint, $D(H_{I}) \neq  D(H_{I}^{min})$ (where $H_{I}^{min}$  is the restriction of the closure of $H_{I}$ on the polynomials) and  that $H_{I}$ is chaotic in Devaney's sense. In particular, the spectrum  of $H_{I}$ is all $\mathbb{C}$.\\

\n {\color{black}$\bullet$} When we want to proof that the eigenvalues of $H_{\mu, \lambda}$ are real we define :\\

\n $\displaystyle{H_{\lambda'}: = H_{\lambda', \mu, \lambda} = \lambda' A^{*^{2}}A^{2} + H_{\mu, \lambda}}$\\

\n For positive $\lambda'$ and $\mu > 0$ we have,\\
\begin{equation}
\displaystyle{\Re e < H_{\lambda'}\varphi , \varphi > \geq \mu \mid\mid A \varphi \mid\mid^{2}, \, \forall\,   \varphi \in D(H_{\lambda'})}
\end{equation}

\n and $H_{\lambda'}$, which is also invertible satisfies:\\
\begin{equation}
\displaystyle{\mid\mid AH_{\lambda'}^{-1} \psi \mid\mid \leq \frac{1}{\mu}\mid\mid \psi\mid\mid}
\end{equation}

\n Now the set of $\phi$ satisfying $\displaystyle{\mid\mid A\phi \mid\mid^{2} \leq 1}$ is a compact set and we see that all $H_{\lambda'}$, map the unit ball into one compact set.\\

\n This enables us to prove that $\displaystyle{lim\, H_{\lambda'}^{-1} = H_{\mu, \lambda}^{-1}\, as \, \lambda' \longrightarrow 0}$ in the sense of the norm of operators and this implies the convergence of the eigenvalues.\\

\n (v) The operator $H_{\mu, \lambda}$  respectively $H_{\lambda',\mu, \lambda}$ has the form $\displaystyle{p(z)\frac{\partial^{2}}{\partial z^{2}} + q(z))\frac{\partial}{\partial z} }$ with $p(z) = i\lambda z$ (respectively $p(z) = \lambda'z^{2} + i\lambda z$) and $q(z) =  i\lambda z^{2} + \mu z$ of degree $2$ and $H_{\mu, \lambda}$  and $H_{\lambda',\mu, \lambda}$ are of Heun operator type respectively.\\

\n (vi) The eigenvalue problem associated to $H_{\mu, \lambda}$  (respectively $H_{\lambda',\mu, \lambda}$) does not satisfy the classical ordinary differential equation of the form:\\
\begin{equation}
\displaystyle{\sigma(z)\frac{\partial^{2}\varphi}{\partial z^{2}} + \tau(z))\frac{\partial \varphi}{\partial z}  = \alpha \varphi}
\end{equation}

\n where $\sigma(z)$ is a polynomial of degree  at most two, $\tau(z)$ is a polynomial of the degree exactly one and $\alpha$ is a constant.\\

\n (vii) If $\lambda \neq  0$, the eigenvalue problem associated to $H_{\mu, \lambda}$ can be written as follows :\\
\begin{equation}
\displaystyle{z\frac{\partial^{2}\varphi}{\partial z^{2}} +( z^{2} - i\rho z)\frac{\partial \varphi}{\partial z}  = - i\frac{\alpha}{\lambda} \varphi; \rho = \frac{\mu}{\lambda}}
\end{equation}

\n Let $ z = i\sqrt{2}\xi$ and $\varphi(\sqrt{2}\xi) = \psi(\xi)$ then (5.7) can be transformed to :\\
\begin{equation}
\displaystyle{\xi\frac{\partial^{2}\psi}{\partial \xi^{2}} +(- 2\xi^{2}  + \rho\sqrt{2} \xi)\frac{\partial \psi}{\partial\xi} - \frac{\alpha\sqrt{2}}{\lambda} \psi  = 0}
\end{equation}

\n (viii) If $\lambda' \neq  0$ and let $z = -i\rho'\xi$ and $\varphi(z) = \psi(\xi)$ then  the eigenvalue problem associated to $H_{\lambda', \mu, \lambda}$ can be transformed to follows :\\
\begin{equation}
\displaystyle{\xi(\xi - 1)\frac{\partial^{2}\psi}{\partial \xi^{2}} +(\rho'^{2}\xi^{2} + \rho'' \xi)\frac{\partial \psi}{\partial \psi}  = \frac{\alpha}{\lambda'} \psi; \rho' = \frac{\lambda}{\lambda'}, \rho'' = \frac{\mu}{\lambda'}}
\end{equation}

\n (ix)   Our equation  belongs to family of the bi-confluent  Heun equations of class $(0, 1, 1_{4})$ \\
\begin{equation}
\displaystyle{H_{\alpha, \beta, \gamma, \delta}u(z) =  zu''(z) + (-2z^{2} - \beta z  + 1 + \alpha)u'(z) + [(\gamma - \alpha - 2)z -\delta] u(z) = 0}
\end{equation}
\n where  $\displaystyle{ \delta = \frac{1}{2}(\hat{\delta} + \beta(1+\alpha))}$\\

\n  which have one  regular singular points $z_{0} = 0$ and one irregular singular point $z_{\infty} = \infty$.\\ 

\n  This equation belongs to family of confluent Heun equations (CHE) \\
\begin{equation}
\displaystyle{H_{\alpha, \beta, \gamma, \delta,\epsilon}u(z) =  z(z - 1)u''(z) + (\alpha z^{2} + \beta z  + \gamma)u'(z) + (\delta z + \epsilon) u(z) = 0}
\end{equation}

\n which has two regular singular points $z_{0} = 0$ and $z_{1} = 1$ and one irregular singular point $z_{\infty} = \infty$. \\

\n These equations are more complicated than the equations for classical special functions such as Bessel functions, hypergeometric functions, and confluent hypergeometric functions (see {\color{blue}[Betman et al.]},  {\color{blue}[Choun: [1-10)]]}) and  {\color{blue}[Abramovitz]}).\\

\n  Above equation includes, as a special (or confluent) case, the Mathieu equation, the equation for Coulomb spheroidal functions, and many other equations used in mathematical physics and applied mathematics. However, it cannot be solved in closed integral form, and analytic tools for investigation of its solutions are rather poor.\\

\n  It turns out that integral relations are useful for analytic studies of equation and similar objects. Integral relations for equations of the Heun class were discussed in  {\color{blue}[Erdelyi]} and {\color{blue}[Kasakov1 et al.]}. In  {\color{blue}[Kasakov2 et al.]} such integral relations were used in order to obtain Fredholm integral equations for eigenfunctions of the boundary problems associated with equations of the Heun class and in {\color{blue}[Kasakov1 et al.]}, integral relations between the eigenfunctions of different equations were established.\\

\n The aim result of this work is to specify the boundary conditions  of Schrodinger equations associated to ``Gribov-Intissar'' operators and to give an analysis of behavior of scattering amplitudes with respect the parameters.\\

\n As $ H_{\lambda', \mu, \lambda}$ and $H_{\mu, \lambda}$ are non-self-adjoint unbounded operators and that the space \\

\n  $\displaystyle{ \mathbb{B}_{0} = \{ \varphi \in \mathbb{B}; \varphi (0) = 0 \}}$ is invariant under their actions then the first idea is to show  that their respectively resolvent sets  are non empty and to show that their respectively resolvent is  compact.\\
 
\n In  {\color{blue}[Intissar2]}, it was showed that $ H_{\lambda', \mu, \lambda}^{-1}$ and $H_{ \mu, \lambda}^{-1}$  exist and they are compact using the compactness of the injection of their respective domains into Bargmann space and that their respective resolvent sets are non-empty.  \\

\n  Let $\psi \in \mathbb{B}_{0}$ and $\varphi \in D(H_{\lambda',\mu,\lambda})$ (respectively $\varphi \in D(H_{\mu,\lambda})$), we consider the equation  $H_{\lambda',\mu,\lambda}\varphi = \psi$  (respectively  $H_{\lambda',\mu,\lambda}\varphi = \psi$) which can be written under following form :\\
\begin{equation}
\displaystyle{(\lambda'z^{2} + i\lambda z)\varphi''(z) + (i\lambda z^{2}  + \mu z)\varphi'(z) = \psi(z)}
\end{equation}

\n respectively :\\
\begin{equation}
\displaystyle{ i\lambda z\varphi''(z) + (i\lambda z^{2}  + \mu z)\varphi'(z) = \psi(z)}
\end{equation}

\n Let $\psi \in \mathbb{B}_{0}$ and choosing the straight line connecting $-i\rho' , z \in \mathbb{C}$ parametrized  by \\

\n  $\displaystyle{\gamma : [0, 1] \rightarrow \mathbb{C}, \gamma(t) = -i\rho' + t(z + i\rho') \quad \quad (\gamma(0) = -i\rho' , \gamma(1) = z)}$\\

\n  If we define $\displaystyle{\int_{-i\rho'}^{z}\psi(\xi)d\xi:=\int_{\gamma}\psi(\xi)d\xi:= \int_{0}^{1}}\psi(\gamma(t))\gamma'(t)dt$ then our  equation can be transformed to following integral equation :\\
\begin{equation}
\displaystyle{\varphi(z) = \frac{1}{\lambda'}\int_{-i\rho'}^{z}e^{-i\rho'\eta}(\eta + i\rho')^{-(\delta +1)} [\int_{-i\rho'}^{\eta}e^{i\rho'\xi}(\xi + i\rho')^{\delta}\frac{\psi(\xi)}{\xi}d\xi]d\eta}
\end{equation}

\n In the same way we get on $]-i\infty, 0]$ an integral  equation with $\lambda' = 0$\\

\n However, the integral representation of $\varphi(z)$ in  the equations is hard to study in $\mathbb{C}$ for some existing results on eigenvalues and eigenfunctions of our operators. To overcome this difficulty, the study of integral equations associated to our equations are restricted on negative imaginary axis.Therefore led to put $\psi(y) = \varphi(-iy)$ ; $y \in [0, + \infty[$ to study the following problems :\\
\begin{equation}
\displaystyle{-\lambda y \psi_{\mu, \lambda}^{''}(y) + (\lambda y^{2} + \mu y)\psi_{\mu, \lambda}^{'}(y) = \sigma(\mu, \lambda)\psi_{\mu, \lambda}(y)}
\end{equation}
\begin{equation}
\displaystyle{(\lambda' y^{2} - \lambda y)\psi_{\lambda', \mu, \lambda}^{"}(y) + (\lambda y^{2} + \mu y)\psi_{\mu, \lambda}^{'}(y) = \sigma(\lambda', \mu, \lambda)\psi_{\lambda', \mu, \lambda}(y)}
\end{equation}

\n On \\
\begin{equation}
\displaystyle{  \mathbb{B}_{1} = \{\psi : [0, \infty[ \longrightarrow \mathbb{C}  \, entire \, ; \int_{0}^{\infty}\mid \psi(y)\mid^{2}e^{-\mid y \mid^{2}}dy < \infty \}}
\end{equation}

\n We note that above equations  are bi-confluent Heun equations and we observe that {\color{black}(5.17)} possesses the origin as regular singular point and $\infty$ as irregular singular point but {\color{black}(5.17)} possesses $ y = 0$ and $y = \displaystyle{\rho' = \frac{\lambda}{\lambda'}}$ as regular singular points  and $\infty$ as irregular singular point.\\

\n On negative imaginary axis an explicit inverse of $H_{\mu, \lambda}$  (respectively of $H_{\lambda', \mu, \lambda}$) is given by :\\

\n {\color{black}$\bullet$} For $ y \in [0, +\infty[$, we get:\\
\begin{equation}
\displaystyle{H_{\mu, \lambda}^{-1} \varphi(-iy) = \int_{0}^{\infty}\mathcal{N}_{\mu,\lambda}(y, s)\varphi(-is)ds}
\end{equation}

\n where  $\displaystyle{\mathcal{N}_{\mu,\lambda}(y, s) = \frac{1}{s}e^{-\frac{s^{2}}{2} - \frac{\mu}{\lambda}s}\int_{0}^{min(y, s)}e^{\frac{u^{2}}{2} +\frac{\mu}{\lambda}u}du}$\\

\n {\color{black}$\bullet$} The boundary conditions of the problem are given by:\\
\begin{equation}
\displaystyle{\frac{\partial u_{\mu, \lambda}}{\partial t}(t, y) -\lambda y \frac{\partial^{2} u_{\mu, \lambda}}{\partial y^{2}}(t, y) + (\lambda y^{2} + \mu y)\frac{\partial u_{\mu, \lambda}}{\partial y}(t, y) = 0}
\end{equation}
\n are \\
\begin{equation}
\displaystyle{ u_{\mu, \lambda}(t, 0) = 0 } \,and\,  \displaystyle{lim\, e^{-\frac{1}{2}(y - \frac{\mu}{\lambda})^{2}}\frac{\partial u_{\mu, \lambda}}{\partial y}(t, y) = 0}
\end{equation}

\n  as $y \longrightarrow + \infty$ for all $t$ \\

\n {\color{black}$\bullet$} For $ y \in [0, \rho^{'}]$, $\displaystyle{\rho' = \frac{\lambda}{\lambda'}}$ and $\displaystyle{\rho = \frac{\mu}{\lambda}}$, we get:\\ 
\begin{equation}
\displaystyle{H_{\lambda', \mu, \lambda}^{-1} \varphi(-iy) = \int_{0}^{\rho'}\mathcal{N}_{\lambda', \mu,\lambda}(y, s)\varphi(-is)ds}
\end{equation}

\n where $\displaystyle{\mathcal{N}_{\mu,\lambda}(y, s) = \frac{1}{\lambda s}e^{\rho' s}(1 -  \frac{s}{\rho'})^{\delta}\int_{0}^{min(y, s)}e^{-\rho' u }(1 - \frac{u}{\rho'})^{-(\rho + 1)} du}$\\

\n And we have the following properties\\

\n (i) The boundary conditions on regular singular points $y = 0$ and $y = \rho'$ of the problem :\\
\begin{equation}
\displaystyle{\frac{\partial u_{\mu, \lambda}}{\partial t}(t, y) + (\lambda' y^{2} - \lambda y)\psi_{\lambda', \mu, \lambda}^{"}(y) + (\lambda y^{2} + \mu y)\psi_{\mu, \lambda}^{'}(y) = 0}
\end{equation}

\n are \\
\begin{equation}
\displaystyle{ u_{\mu, \lambda}(t, 0) = 0 } \mbox{ and }  \displaystyle{ u_{\mu, \lambda}(t, y)} \mbox{\,is\, an\, analytic\, function\, around\, }y = \rho'
\end{equation}

\n (ii) The space of the restrictions of the elements of the domain of the operator $ H_{\lambda', \mu, \lambda}$ is included in \\
\begin{equation}
\displaystyle{\mathbb{B}_{2} = \{u \in L_{2}([0, \rho'], r(y)dy); H_{\lambda', \mu, \lambda}u \in L_{2}([0, \rho'], r(y)dy) , u(0) = 0\}}
\end{equation}

\n where the expression of $H_{\lambda', \mu, \lambda}u$ is given by\\
\begin{equation}
\displaystyle{H_{\lambda', \mu, \lambda}u = - \frac{1}{r(y)}\frac{\partial}{\partial y}[-\lambda'y(\rho' - y)r(y)\frac{\partial u}{\partial y}]} \, with\, \displaystyle{r(y) = \frac{1}{y}e^{\rho' y}\mid y - \rho'\mid^{\rho'(\rho' + \rho) - 1}}
\end{equation}

\n (iii) $\displaystyle{H_{\lambda', \mu, \lambda}}$ acting on $ L_{2}([0, \rho'], r(y)dy)$ with domain :\\

\n $\displaystyle{\mathbb{B}_{2} = \{u \in L_{2}([0, \rho'], r(y)dy); H_{\lambda', \mu, \lambda}u \in L_{2}([0, \rho'], r(y)dy) , u(0) = 0\}}$ \\

\n is symmetric. \\

\n We will show that $H_{\lambda', \mu, \lambda}$ has one real eigenvector $\varphi_{0} \in \mathbb{B}_{2}$ such that $\varphi_{0}  \geq 0$ (almost everywhere with respect to $r(y)dy$ on  $[0, \rho']$).\\

\n If $\sigma_{0}$ stands for the eigenvalue of $H_{\lambda', \mu, \lambda}$ associated with $\varphi_{0}$, the symmetry of $H_{\lambda', \mu, \lambda}$ implies that, for all $\varphi \in \mathbb{B}_{2}$ we have $\displaystyle{\Re e[(H_{\lambda', \mu, \lambda} - \sigma_{0}I)\varphi \geq 0}$. The last results motivates the following definition:\\
\begin{definition}
 {\bf (Local energy)}\\

\n  For any $\varphi \in \mathbb{B}_{2}$, the local energy is the  function $\mathcal{E} : [0, \rho'] \longrightarrow \overline{\mathbb{R}}$ defined by :\\

\n  $\displaystyle{\mathcal{E}_{\varphi}(y) = \frac{\Re e (H_{\lambda', \mu, \lambda}\varphi(y))}{\Re e(\varphi(y))}}$. \\
\end{definition}
\n Therefore from what precedes, we get\\

\n Let $\sigma_{0}$ be the smallest eigenvalue of  $H_{\lambda', \mu, \lambda}$ then we have\\

\n ($ \alpha$) For $\forall$ $\varphi \in \mathbb{B}_{2}$ such that $\displaystyle{\Re e(\varphi) \geq 0 , inf_{[0, \rho']}(\mathcal{E}_{\varphi}) \leq \sigma_{0} \leq sup_{[0, \rho']}(\mathcal{E}_{\varphi}) }$. \\

\n From above property, we deduce that:\\

\n  ($ \beta $) Let $\sigma_{0}$ be the smallest eigenvalue of  $H_{\lambda', \mu, \lambda}$  acting on Bargmann space then we have for\\ $\forall$ $\varphi \in D(H_{\lambda', \mu, \lambda}^{*})$ such that  $\Re e(\varphi) \geq 0 $,  $\displaystyle{inf_{[0, \rho']}\frac{[\Re e(H_{\lambda', \mu, \lambda}^{*}\varphi)}{\varphi}] \leq \sigma_{0} \leq sup_{[0, \rho']}\frac{[\Re e(H_{\lambda', \mu, \lambda}^{*}\varphi)}{\varphi}] }$. \\

\n By using Kato's inequalities, Thirring (1979) in {\color{blue}[Thirring]} have obtained the following theorem on Schrodinger operator acting on $L^{2}(\mathbb{R}^{n})$ (for this topic see also {\color{blue}[Duffin]} (1947), {\color{blue}[Barnsley]} (1978) and {\color{blue}[Schmutz]} (1985))\\
\begin{theorem}
 (Thirring)\\

\n Let $\mathbb{H} = - \Delta + \mathbb{V}$ be a Schrodinger operator acting on $L^{2}(\mathbb{R}^{n})$ having an eigenvalue {\bf below} the essential spectrum. Then the lowest eigenvalue $\epsilon_{0}$ of $\mathbb{H}$ is such that for any strictly positive $\varphi \in D(\mathbb{H})$,\\

\n $\displaystyle{inf_{\mathbb{R}^{n}}(\mathbb{V} - \frac{\Delta \varphi}{\varphi}) \leq \epsilon_{0} \leq sup_{\mathbb{R}^{n}}(\mathbb{V} - \frac{\Delta \varphi}{\varphi})}$.\\
\end{theorem}
\n  In Gribov's Reggeon calculus, the Regge-poles are not the only singularities of the amplitude. There are also branch points which correspond to the exchange of several Reggeons. A Regge pole can be interpreted as corresponding to a single scattering. Regge cuts-multiple scatterings of hardons'$^{s }$ constituents.\\

\n We recall that the transition of a closed system of particles from an initial state $ \mid \varphi >$ to a final state $ \mid  \psi >$ is described in quantum theory by the $\mathbb{S}$ matrix (called scattering matrix) $\displaystyle{| \psi > = \mathbb{S} | \phi >}$ and the matrix elements of the $\mathbb{S}$  $\displaystyle{\mathbb{S}_{\psi, \varphi} = < \psi \mid \mathbb{S} \mid  \varphi >}$ can be represented in the form:\\
\begin{equation}
\displaystyle{\mathbb{S}_{\psi, \varphi} = \delta_{\psi, \varphi} + C\mathbb{T}_{\psi, \varphi}}
\end{equation}

\n where $\displaystyle{ \delta_{\psi, \varphi}  = 1}$  if the state does not change $(\mid \psi >  = \mid \varphi >)$ i.e. no interaction, $C$ denote the conservation of
energy and momentum and $\displaystyle{\mathbb{T}_{\psi, \varphi}}$ is called the transition (scattering) amplitude from the state $ \mid \varphi >$ to a final state $ \mid  \psi >$. We can consult the following references on the subject which summarizes the actual knowledge of Reggeon theory ( {\color{blue}[Gribov2]},  { {\color{blue}[Baker et al.]}, {\color{blue}[Kaidalov]} and {\color{blue}[Boreskov et al.]}).\\

\section{Generalized diagonalisation  of  semigroup $\displaystyle{e^{-tH_{\lambda'', \lambda', \mu, \lambda}}}$, $t > 0$}

\subsection{ Diagonalization of semigroup $e^{-tH_{\lambda'}}$,  $ t > 0$ where  $H_{\lambda'} = \lambda'A^{*^{2}}A^{2} + \mu A^{*}A + i\lambda (A^{*}(A + A^{*})A, \, \lambda' > 0$}

\n In this under-section,  we consider the  following Cauchy problem  :\\
\begin{equation}
u'(t) = H_{\lambda'}u(t) ; u(0) = \phi \in B
\end{equation}
\n Let $\{\phi_{k}\}; k=1, 2,....$ are the eigenfunctions of  $H_{\lambda'}$ associated to eigenvalues $\{\sigma_{k}; k=1, 2,....\}$.\\

\n what conditions can be imposed on $H_{\lambda'}$ to write $u (t)$ as a limit of partial sums of expression $\displaystyle{\sum_{k=1}^{+\infty}C_{k}e^{-\sigma_{k}t}\phi_{k}}$ ?\\

\n Let $\lambda' \neq 0$, we will apply a fine result of Lidskii  {\color{blue}[Lidskii]} to show the existence of a total system of eigenfunctions of  $H_{\lambda'}$.\\
\begin{proposition}
 ( {\color{blue}[Intissar2]})\\

\n  Let $\lambda' > 0$ and $\lambda'^{2} \leq \mu\lambda' + \lambda^{2}$, we consider the Cauchy's problem :\\

\n $u'(t) = H_{\lambda'}u(t) ; t > 0 $ and $u(0) = \phi \in B$ \\

\n $\displaystyle{H_{\lambda'} = (\lambda' z^{2} + i\lambda z)\frac{\partial^{2}}{\partial z^{2}} + (i \lambda z^{2} +\mu z)\frac{\partial }{\partial z}}$ acting on Bargmann space \\

\n $\displaystyle{\mathbb{B} = \{\varphi : \mathbb{C} \longrightarrow \mathbb{C} \, entire \, ; \int_{\mathbb{C}}\mid \varphi(z) \mid^{2} e^{-\mid z \mid^{2}}dxdy < + \infty \}}$\\

\n Then we have  $u(t) = \displaystyle{\sum_{k=1}^{+\infty}\frac{< \phi,\phi_{k}^{*} > }{< \phi_{k},\phi_{k}^{*} >}e^{-\sigma_{k}t}\phi_{k}}$  where \\

\n $ \phi_{k} \in D(H_{\lambda'})$ and $\phi_{k}^{*} \in D(H_{\lambda'}^{*}) = D(H_{\lambda'}) = D(S)$\\

\n $H_{\lambda'}\phi_{k} = \sigma_{k}\phi_{k}$ et $H_{\lambda'}^{*}\phi_{k}^{*} = \sigma_{k}\phi_{k}^{*}$\\
\end{proposition}
\n The proof of this proposition is based on a thin Lidskii theorem which will be given below in simplified form.\\
\begin{theorem}
( {\color{blue}[Lidskii]}, theorem 3, p:208)\\

\n Let $T$  be a linear operator with dense domain  $D(T)$ in Hilbert space  $\mathcal{H}$  compact resolvent.\\

\n We suppose that the eigenvalues of $T$ are simple and  $0 \in \rho(T)$ ($\rho(T)$ is the resolvent set. of $T$).\\

\n Let \\
\begin{equation}
\left \{ \begin{array} {c} u'(t) + Tu(t) = 0 ; t > 0  \\
\quad\\
and \quad \quad \quad \quad \quad \quad \quad \quad\quad\\
\quad\\
u(0) = \phi \in \mathbb{B} \quad \quad \quad \quad.  \quad\\\
\end{array} \right.
\end{equation}

\n If $T$ satisfies the following two conditions\\

\n (i) $\exists \quad p ; 0 < p < 1$ and $\displaystyle{\sum_{k=1}^{+\infty}\mid \alpha_{k}\mid^{p}}$ converge\\

\n where $\{\alpha_{k}\}; k=1, 2, .....$ are the eigenvalues of $\sqrt{T^{*^{-1}}T^{-1}}$.\\

\n (ii) $\exists \quad \epsilon > 0 ; -\frac{\pi}{2} + \epsilon \leq < arg <T\phi, \phi> \leq \frac{\pi}{2} - \epsilon \quad \forall \phi \in D(T)$\\

\n Then \\

\n (1) The problem (6.2)  admits an unique solution.\\

\n (2) The solution $u(t)$ of Cauchy's problem (6.2) is a limit of partial sums of expression\\

\n $\displaystyle{\sum_{k=1}^{+\infty}\frac{< \phi,\phi_{k}^{*} > }{< \phi_{k},\phi_{k}^{*} >}e^{-\beta_{k}t}\phi_{k}}$ o$\grave{u}$ $ \phi_{k} \in D(T)$ and $\phi_{k}^{*} \in D(T^{*})$\\

\n and\\

\n $T\phi_{k} = \beta_{k}\phi_{k}$ and $T^{*}\phi_{k}^{*} = \overline{\beta}_{k}\phi_{k}^{*}$\\
\end{theorem}

\n Let  $\beta_{0} > 0$ and  $H_{\lambda'} =  H_{1} + iH_{2}$ where $H_{1} = \lambda' S + \mu A^{*}A + \beta_{0}I$  and  $H_{2} = \lambda A^{*}(A + A^{*})A$ with $S = A^{*^{2}}A^{2}$\\

\n Then the verification of hypotheses of Lidskii's theorem is based on following lemma\\
\begin{lemma}

\n  Let $\lambda' > 0$ and $H_{\lambda'} = H_{1} + iH_{2}$, then we have\\

\n (i) $\exists \quad C > 0 ; \mid < H_{2}\phi, \phi >\mid \leq C \mid < H_{1}\phi, \phi >\mid \quad \forall \phi \in D(S)$.\\

\n (ii) the operator  $(I + H_{2}H_{1}^{-1})$ is invertible.\\

\n (iii) $\exists \quad \epsilon > 0 ; -\frac{\pi}{2} + \epsilon \leq <Arg <H_{\lambda'}\phi, \phi> \leq \frac{\pi}{2} - \epsilon \quad \forall \phi \in D(H_{\lambda'})$\\

\n (iv) the series $\displaystyle{\sum_{k=1}^{+\infty}\mid \alpha_{k}\mid^{p}}$ converges $\quad \forall p > \frac{1}{2}$\\

\n where $\{\alpha_{k}\}; k=1, 2, .....$ are the eigenvalues  $\sqrt{H_{\lambda'}^{*^{-1}}H_{\lambda'}^{-1}}$, (this operator is said to belong to Carleman's class $\mathcal{C}_{p}$)\\
\end{lemma}

\n {\bf Proof}\\

\n (i) and (ii) follow from above section. \\

\n (iii) follow from (i).\\

\n (iv) As $H_{\lambda'}^{-1} =  H_{1}^{-1}(I + H_{2}H_{1}^{-1})^{-1}$ then we deduce that  $(I + H_{2}H_{1}^{-1})^{-1}$ is bounded by a constant $C$ and $H_{1}^{-1} \in \mathcal{C}_{p} \quad \forall p > \frac{1}{2}$ because its eigenvalues are $\gamma_{k} = \frac{1}{\lambda'k(k-1) + \mu k + \beta_{0}}\approx \frac{1}{k^{2}}$.\\

\n Let $\{\alpha_{k}\}$ be the eigenvalues of $\sqrt{H_{\lambda'}^{*^{-1}}H_{\lambda'}^{-1}}$,it is well known (by applying the minimax theorem, see {\color{blue} [Gohberg2 et al]}, p: 27)  there exist  $C > 0$ such that  $ \alpha_{k} \leq C \gamma_{k}$ in particular $\displaystyle{\sum_{k=1}^{+\infty} \alpha_{k}^{p} < + \infty \quad \forall p > \frac{1}{2}.}$\\

\n Now, as our operator checks the hypotheses of Lidskii's theorem then we deduce the proposition 6.1.\\

\n $\bullet$ For $\lambda' = 0$, above Lidskii's theorem is not applied then the question of diagonalisation of semigroup associated to  $H = \mu A^{*}A + i\lambda A^{*}(A + A^{*})A$  is open.\\

\n  Nevertheless we have the following \\

\subsection{Asymptotic behavior of $e^{-tH}$, $ t \rightarrow  + \infty$;$ H = \mu A^{*}A + i\lambda A^{*}(A + A^{*})A; \mu > 0$}

\n  We consider the Cauchy problem :\\
\begin{equation}
\left \{ \begin{array} {c} u'(t) = -Hu(t) , t > 0 \\
\quad\\
\n and \quad\quad  \quad\quad  \quad\quad  \quad\quad\\
\quad\\
 \n u(0) = \phi \in B \quad\quad  \quad\quad\\
 \end{array} \right.
\end{equation}
\n Then we have\\
\begin{proposition}

\n (i) For $\mu > 0$, the semigroup $e^{-tH}$ associated to $H$ is compact in particular we have\\

\n (ii)  \begin{equation}
\sigma (e^{-tH}) = e^{-t\sigma(H)} \cup \{0\} \end{equation}

\n  (iii) Let $\sigma_{0}$ and $\sigma_{1}$ are respectively the smallest and the second eigenvalue of the operator $H$.\\

\n Then\\

\n  $\mid\mid e^{-tH}\mid\mid = e^{-\sigma_{0}t} + O(e^{-\epsilon t})\quad \forall \epsilon < \sigma_{1}$.\\

\n (iv) $< e^{-tH}\phi, \psi > = <\phi, \phi_{0} > <\phi_{0} , \psi >e^{-\sigma_{0}t} + O(e^{-\sigma_{0}t})\quad \forall \phi \in \mathbb{B}, \quad \forall \psi \in \mathbb{B}$ where $\phi_{0}$  is the eigenfunction of  $H$ associated to $\sigma_{0}$.\\
\end{proposition}
\n {\bf Proof}\\

\n (i) This result was proven in {\color{blue}[intissar2]} and (ii) is a classical consequence of compact semigroups see  for example {\color{blue}[Pazy]}.\\

\n (iii) Let $\sigma_{0}$ be the smallest eigenvalue of $H$. Since it is simple then there exist an eigenfunction $\phi_{0} \neq 0$ associated to $\sigma_{0}$, which generates an space  $\mathbb{F}_{0}$ of one dimension.\\

\n From now, we take $\mid\mid \phi_{0}\mid\mid = 1$,  then there exist $\mathbb{F}_{1} \subset \mathbb{B}$ such that\\

\n $\mathbb{B} = \mathbb{F}_{0} \oplus \mathbb{F}_{1}$, $\phi_{0} \, \notin \, \mathbb{F}_{1}$ and $e^{-tH}\mathbb{F}_{1}\subset \mathbb{F}_{1}.$\\

\n Let $P$ be the projection of  $B$ on $ \mathbb{F}_{0}$ and  $Q = I- P$ be the projection of $\mathbb{B}$ on $ \mathbb{F}_{1}$, then we can write\\
\begin{equation}
\phi = P\phi + Q\phi = c_{1}\phi + Q\phi, \quad \forall \phi \in \mathbb{B} \, \mbox{and} \, c_{1} \, \mbox{ is a constant }
\end{equation}

\n in particular\\
\begin{equation}
e^{-tH}\phi = c_{1}e^{-\sigma_{0}t}\phi + e^{-tH}Q\phi
\end{equation}

\n Let $r(T)$ be the spectral ray of an op\'erateur $T$, since $e^{-tH}$ is compact,  we can use a classical result  (see {\color{blue}[Kato]}) to obtain \\
\begin{equation}
r(e^{-tH}{_{\mid \mathbb{F}_{1}}}) = e^{-\sigma_{1}t}.
\end{equation}

\n $\sigma_{1}$ is the second eigenvalue of $H$.\\

\n Let's put it now, $t = n + s$ where $n$ is entire part of $t$  then  $0 \leq s \leq 1$ and we deduce that\\

\n  $\displaystyle{\mid\mid e^{-tH}{_{\mid \mathbb{F}_{1}}}\mid\mid = \mid\mid e^{-nH}e^{-sH}{_{\mid \mathbb{F}_{1}}}\mid\mid \leq \mid\mid e^{-nH}{_{\mid \mathbb{F}_{1}}}\mid\mid}$.\\

\n As  $\displaystyle{lim \mid\mid e^{-nH}{_{\mid \mathbb{F}_{1}}}\mid\mid^{\frac{1}{n}} = e^{-\sigma_{1}}}$ as $n \longrightarrow +\infty$  then we  deduce that \\

\n  $\displaystyle{\mid\mid e^{-nH}{_{\mid \mathbb{F}_{1}}}\mid\mid^{\frac{1}{n}} = e^{-\sigma_{1}}}$ +  $\displaystyle{\epsilon_{n}; \epsilon_{n} \rightarrow 0}$ as $n \rightarrow +\infty$.\\

\n and\\

\n  $\displaystyle{\mid\mid e^{-nH}{_{\mid \mathbb{F}_{1}}}\mid\mid = (e^{-\sigma_{1}} + \epsilon_{n})^{n}}$\\

\n  $\displaystyle{ = e^{-n\sigma_{1}}(1 + e^{\sigma_{1}}\epsilon_{n})^{n}}$\\

\n  $\displaystyle{ =  e^{-n\sigma_{1}}(1 + \omega_{n})^{n}; \omega_{n}}$\\

\n  $\displaystyle{ = e^{\sigma_{1}}\epsilon_{n}}$\\

\n Let $\epsilon < \sigma_{1}$, then\\

\n $\displaystyle{e^{-n\sigma_{1}}(1 + \omega_{n})^{n} = e^{-\epsilon t + \epsilon t-n\sigma_{1}}(1 + \omega_{n})^{n}}$\\ 

\n  $\displaystyle{ = e^{-\epsilon t + n\epsilon + s\epsilon -n\sigma_{1}}(1 + \omega_{n})^{n}}$ \\

\n $\displaystyle{= e^{-\epsilon t}e^{n(\epsilon-\sigma_{1})}e^{s\epsilon} (1 + \omega_{n})^{n}}.$\\

\n and\\

\n $\displaystyle{\mid\mid e^{-tH}{_{\mid \mathbb{F}_{1}}}\mid\mid  \leq \mid\mid e^{-nH}{_{\mid \mathbb{F}_{1}}}\mid\mid = e^{-\epsilon t}e^{n(\epsilon-\sigma_{1})}e^{s\epsilon} (1 + \omega_{n})^{n}}$.\\

\n it follows that\\

\n $\displaystyle{lim \quad Log (e^{\epsilon t}e^{-s\epsilon}\mid\mid e^{-tH}{_{\mid \mathbb{F}_{1}}}\mid\mid) = -\infty} $ as $t \rightarrow +\infty$\\

\n and \\

\n  $\displaystyle{lim \quad e^{\epsilon t}\mid\mid e^{-tH}{_{\mid \mathbb{F}_{1}}}\mid\mid = 0 }$ as $t \rightarrow +\infty$.\\

\n (iv) Just write  $\displaystyle{ < e^{-tH}\phi, \psi > = <e^{-tH}P\phi, \psi > + < e^{-tH}Q\phi, \psi >}$.\\

\n and since  $\displaystyle{\mid < e^{-tH}Q\phi, \psi > \mid \leq c_{1}e^{-\epsilon t}}$ with $\epsilon < \sigma_{0}$, it follows that :\\

\n  $\displaystyle{\mid < e^{-tH}\phi, \psi >\mid = Ce^{-\sigma_{0}t} + O(e^{-\sigma_{0}t}) \quad \forall \phi \in B, \quad \forall \phi \in \mathbb{B}}.$\\

\n where  $\displaystyle{C = < \phi, \phi_{0} >< \phi_{0}, \psi >}$.\\

\subsection{Generalized diagonalization of semigroups $e^{-t H_{\lambda''}}$ and $e^{-t\sqrt{H}_{\lambda''}}$, $ t \geq 0$ , where $H_{\lambda"} = \lambda'' A^{*^{3}}A^{3} + \lambda' A^{*^{2}}A^{2} + \mu A^{*}A + i \lambda A^{*}(A+ A^{*})A$, $\lambda'' > 0$.}

\n In this sub-section we prove that the system of generalized eigenvectors of the operator $H_{\lambda"} = \lambda'' A^{*^{3}}A^{3} + \lambda'A^{*^{2}}A^{2} + \mu A^{*}A + i \lambda A^{*}(A+ A^{*})A$, $\lambda'' > 0$ is an unconditional basis in Bargmann space $\mathbb{B}$. \\

\n We also give a generalized diagonalization in the Abel sense of the semigroups associated to this operator and to its square root, respectively. Finally, some open problems are pointed out.\\

\n {\bf $\bullet$ (A) Some preliminary results and completeness of system of generalized eigenvectors of the operator $H_{\lambda''}$}\\

\n In the following, we adopt the notations,\\

\n $U\phi(z) =  A^{*}A \phi(z)$,  with domain $D(U) = \{\phi \in \mathbb{B}; U\phi \in \mathbb{B}\}$ (Harmonic Oscillator).\\

\n $S\phi(z) =  A^{*^{2}}A^{2}\phi(z)$,  with domain $D(S) = \{\phi \in \mathbb{B}; S\phi \in \mathbb{B}\}$\\

\n $\hat{G} = G + I$\\

\n $G\phi(z) = A^{*^{3}}A^{3} \phi(z)$,  with domain $D(G) = \{\phi \in \mathbb{B}; G\phi \in \mathbb{B}\}$\\

\n $H_{\lambda''} = H_{\lambda'', \lambda', \mu, \lambda}$\\

\n $H_{\lambda'} = H_{0, \lambda', \mu, \lambda}$\\

\n $H_{\mu} = H_{0, 0, \mu, \lambda}$\\

\n For $\lambda'' \neq 0$, we first begin to collect the basic results for the operators we defined previously.\\

\n 1. In Bargmann space, the adjoint of the derivation operator $A$ is the multiplication operator $A^{*}$.\\

\n 2. The injection from domain $D(A)$ of $A$ into the Bargmann space $\mathbb{B}$ is compact.\\

\n 3. \begin{equation}\mid\mid \phi \mid\mid \leq \mid\mid A^{*}\phi \mid\mid ,  \phi \in D(A^{*})\end{equation}

\n 4. $A^{*^{3}}A^{3} , A^{*^{2}}A^{2} $ and $A^{*}A $ are self-adjoint operators with compact resolvent.\\

\n 5. $H_{\lambda''} $ is maximal accretive operator for $\lambda'' > 0$, $\lambda' \geq 0$ and $\mu \geq 0$.\\

\n 6. For every $\epsilon > 0$ there exists $C_{\epsilon} > 0$ such that \\

\n \begin{equation}\displaystyle{\mid \mid H_{\lambda'}\phi \mid\mid \leq \epsilon \mid\mid G \phi \mid\mid + C_{\epsilon}\mid\mid \phi \mid\mid,  \, \forall \, \phi \in D(G)}\end{equation}

\n 7. \begin{equation}\displaystyle{D_{max}(H_{\lambda''}) = D_{min}(H_{\lambda''}) = D(G) , \lambda'' \neq 0}\end{equation} 

\n 8. According to the perturbation theory of semigroups {\color{blue} [Kato] } and {\color{blue}[Pazy]}, $- H_{\lambda''}$ generates an analytic semigroup $\displaystyle{e^{-tH_{\lambda''}}}$.\\

\n Now we recall some useful definitions :\\
\begin{definition}

\n  Let $L$ and $B$ be linear operators in a Hilbert space $\mathcal{H}$. We say that $B$ is $p$-subordinate $(p \in [0, 1])$  to $L$  if $D(L) \subset D(B)$. and there exists a strictly positive constant $C$ such that \\
\begin{equation}
\displaystyle{\mid\mid B \phi \mid\mid \leq C\mid\mid L \phi \mid\mid^{p}\mid\mid  \phi \mid\mid^{1-p}} \, \mbox{ for every} \, \phi \in D(L)
\end{equation}

\n  For $p =1$, we say that $B$ is subordinate to $L$.\\
\end{definition}
\begin{definition}

\n  Let $L$ be a linear operator in a Hilbert space $\mathcal{H}$. The operator $B$ is said to be $L$-compact with order$ p \in [0, 1]$ if $D(L) \subset D(B)$
and for any $\epsilon > 0$, there exists a constant $C_{\epsilon} > 0$ such that \\
\begin{equation}
\displaystyle{\mid\mid B \phi \mid\mid \leq \epsilon \mid\mid L \phi \mid\mid^{p}\mid\mid  \phi \mid\mid^{1-p} + C_{\epsilon}\mid\mid \phi \mid\mid}  \mbox{for every}\, \phi \in D(L)
\end{equation}
\n The operator $B$ is called $L$-compact, if $B$ is $L$-compact with unit order.\\
\end{definition}
\n From these definitions, we deduce the following results:\\

\n 1. Let $L$ be an operator in $\mathcal{H}$ with a dense domain $D(L)$. and at least one regular point $\lambda$.\\

\n We suppose that:\\

\n {\color{black}($\alpha$)} $D(L)) \subset D(B)$.\\

\n  {\color{black}($\beta$)}  For any $\epsilon > 0$, there exists a constant $C_{\epsilon} > 0$ such that  $\displaystyle{\mid\mid B \phi \mid\mid \leq \epsilon \mid\mid L \phi \mid\mid + C_{\epsilon} \mid\mid  \phi \mid\mid}$  for every $\phi \in D(L)$\\

\n {\color{black} ($\gamma$)} $\displaystyle{(L - \lambda I)^{-1}}$ is compact.\\

\n Then $\displaystyle{B(L - \lambda I)^{-1}}$ is a compact. operator\\

\n 2. Let $L$ be an operator in $\mathcal{H}$ with a dense domain $D(L)$. and at least one regular point $\lambda$.\\

\n We suppose that:\\

\n  {\color{black}($\alpha$)} $ D(L) \subset D(B)$.\\

\n  {\color{black} ($\beta$)}  $\displaystyle{B(L - \lambda I)^{-1}}$ is a compact. operator\\

\n Then for any $\epsilon > 0$ , there exists a constant $C_{\epsilon} > 0$ such that $\displaystyle{\mid\mid B \phi \mid\mid \leq \epsilon \mid\mid L \phi \mid\mid + C_{\epsilon} \mid\mid  \phi \mid\mid}$  for every $\phi \in D(L)$\\
\begin{definition}

\n {\color{black}$\bullet$} A sequence $\displaystyle{\{V_{n}\}_{n=1}^{\infty}}$  of subspaces of a Hilbert space $\mathcal{H}$ is called a {\color{black}basis of subspaces}. if any vector $f$ belonging to $\mathcal{H}$ can be {\color{black} uniquely} represented as a series $\displaystyle{\phi = \sum_{ n= 1}^{\infty}\phi_{n}}$ such that $ \phi_{n} \in V_{n}$.\\

\n  {\color{black}$\bullet$} A basis of subspaces is said to be unconditional if it remains a basis for $\mathcal{H}$  under any permutation of the subspaces appearing in it, i.e., if the preceding series converges unconditionally for any $\phi \in \mathcal{H}$.\\
\end{definition}
\begin{definition}

\n A linearly independent sequence $\displaystyle{\{\phi_{n}\}_{n=1}^{\infty}}$  is called an {\color{black}unconditional basis with parentheses} for $\mathcal{H}$ if there exists a
subsequence $\displaystyle{\{n_{k}\}_{k=1}^{\infty}}$. of the positive integers such that the subspaces spanned by the vectors $\displaystyle{\{\phi_{n}\}_{n_{k-1}}^{n_{k} - 1}}$ form an unconditional basis for $\mathcal{H}$.\\
\end{definition}
\n Now we will prove  that the system of generalized eigenvectors of the Gribov operator $H _{\lambda''}$ is an unconditional basis in the Bargmann space $\mathbb{B}$. The analysis is essentially based on the results obtained by Markus in {\color{blue}[Markus] }and by Yacubov and Mamedov in {\color{blue}[Yacubov et al]}, which we can apply to the operator of Reggeon field theory.\\

\n We begin by given some useful lemmas : \\
\begin{lemma}

\n $H_{\mu}$ is $\frac{1}{2}$-subordinate to $\hat{G}$.\\
\end{lemma}
\n {\bf Proof}\\

\n Let $\displaystyle{\hat{U} = \mu A^{*}A}$ then  $\displaystyle{\hat{U}}$ is a self-adjoint operator with compact resolvent in the Bargmann space $\mathbb{B}$, and $A^{*}A$ is a generator of an analytic semigroup.  Let $\displaystyle{\hat{U}\phi = \sum_{k=1}^{\infty} \lambda_{k} < \phi , e_{k} > e_{k}}$ be its spectral decomposition, where $\lambda_{k} = \mu k$ is the $kth$ eigenvalue of $\displaystyle{\hat{U}}$ associated to the eigenvector $\displaystyle{e_{k}(z) = \frac{z^{k}}{\sqrt{k!}}}$.\\

\n For $\beta > 0$, we define the operator $\displaystyle{\hat{U}^{\beta}}$ by\\
\begin{equation}
\displaystyle{\hat{U}^{\beta}\phi = \sum_{k=1}^{\infty} \lambda_{k}^{\beta} < \phi , e_{k} > e_{k}}
\end{equation}
\n with domain  $\displaystyle{D(\hat{U}^{\beta}) = \{\phi \in \mathbb{B}  ;  \sum_{k=1}^{\infty} \mid \lambda_{k}\mid^{2\beta} \mid < \phi , e_{k} >\mid^{2}  <  \infty \} }$. \\

\n From $\displaystyle{D(\hat{G}) = D(U^{3})}$ and $\displaystyle{\mid\mid \hat{G}\phi \mid\mid \sim \mid\mid U^{3}\phi \mid\mid }$, it is easy to deduce that \\
\begin{equation}
\displaystyle{D(\hat{G}^{\frac{3}{2}})  \subset D(H_{\mu})}
\end{equation} 

\n and\\

\n  There exists $C_{\mu, \lambda} > 0$  such that\\
\begin{equation}
\displaystyle{\mid\mid H_{\mu}\phi \mid\mid \leq C_{\mu, \lambda} \mid\mid U^{\frac{3}{2}}\phi \mid\mid }\mbox{ for any}  \displaystyle{\phi \in D(U^{\frac{3}{2}})}. 
\end{equation}. 

\n Now as $\displaystyle{\mid\mid U^{\frac{3}{2}}\phi \mid\mid^{2} = < U^{\frac{3}{2}}\phi , U^{\frac{3}{2}}\phi >  = < U^{3}\phi , \phi > \leq \mid\mid U^{3}\phi \mid\mid \mid\mid \phi \mid\mid}$ we deduce that \\
\begin{equation}
\displaystyle{\mid\mid H_{\mu}\phi \mid\mid \leq C_{\mu, \lambda} \mid\mid U^{3}\phi \mid\mid^{\frac{1}{2}}\mid\mid \phi  \mid\mid^{\frac{1}{2} }} \mbox{ for any}  \displaystyle{\phi \in D(U^{3})}
\end{equation}

\n and \\
\begin{equation}
\displaystyle{\mid\mid H_{\mu}\phi \mid\mid \leq C_{\mu, \lambda} \mid\mid \hat{G}\phi \mid\mid^{\frac{1}{2}}\mid\mid \phi  \mid\mid^{\frac{1}{2} }}\mbox{ for any}  \displaystyle{\phi \in D(G)}
\end{equation}
\begin{lemma}

\n Let $\displaystyle{\{e_{k}(z) = \frac{z^{k}}{\sqrt{k!}}\}_{k \in \mathbb{N}}}$  be the usuall basis of Bargmann space.\\

\n Let $(\lambda_{k}), k =1, 2, ...$ be the eigenvalues of $G$. Then for each $\lambda \neq \lambda_{k}$, we have \\
\begin{equation}
\displaystyle{(G - \lambda I)^{-1}\phi = \sum_{k=1}^{\infty}\frac{1}{\lambda_{k} -  \lambda} < \phi, e_{k} > e_{k}}
\end{equation}

\n Moreover, if $\Im m \lambda \neq 0$ and $\lambda$ belongs to a ray with origin zero and of angle $\theta$ with $\theta \neq 0$ and $\theta \neq \pi$ then we have:\\
\begin{equation}
\displaystyle{\mid\mid (G - \lambda I)^{-1}\mid\mid \leq \frac{1}{\mid \Im m \lambda \mid} = \frac{c(\theta)}{\mid \lambda \mid}}
\end{equation}
\end{lemma}
\n {\bf Proof}\\

\n This follows immediately from the fact that G is a self-adjoint operator with compact resolvent.  \hfill { } $\square$\\
\begin{lemma}

\n There exists a sequence of circles $\displaystyle{C(0, r_{k}), k = 1, 2, .... }$ with radii $r_{k}$  going to infinity such that \\

\n $\displaystyle{\mid\mid (\hat{G} - \lambda I)^{-1}\mid\mid \leq \frac{2}{\mid \lambda \mid^{\frac{2}{\beta}}}}$ for any $\beta \geq 3$ and $\mid \lambda \mid = r_{k}$.\\
\end{lemma}
\n {\bf Proof}\\

\n First, we remark that the eigenvalues of the operator $\hat{G}$ are given by $\lambda_{k} = k(k-1)(k-2)$  and that $\lambda_{k+1} - \lambda_{k} \geq k^{2}$ for $ k \geq 2$.\\

\n We set $\displaystyle{ r_{k} = \frac{\lambda_{k+1} + \lambda_{k}}{2} = \lambda_{k} + \frac{\lambda_{k+1} - \lambda_{k}}{2}}$ then from the fact that $\hat{G}$ is a self-adjoint operator  and the equality  $\displaystyle{\mid\mid (\hat{G} - \lambda I)^{-1} \mid\mid = \frac{1}{d(\lambda , \sigma(\hat{G})}}$ where $\lambda \in \rho(\hat{G})$ and $d(\lambda , \sigma(\hat{G})$ denotes the distance between the point $\lambda$ and the spectrum of the operator $\hat{G}$ we get :\\
\begin{equation}
\displaystyle{\mid\mid (\hat{G} - \lambda I)^{-1} \mid\mid = \frac{1}{d(\lambda , \sigma(\hat{G})} \leq \frac{2}{\lambda_{k+1} - \lambda_{k}} \leq \frac{2}{k^{2}}}, \mbox{where} \mid \lambda \mid = r_{k}
\end{equation}

\n Now from $\displaystyle{ r_{k} = \frac{\lambda_{k+1} + \lambda_{k}}{2} = k^{3} - 2k^{2} + k \leq k^{3} \leq k^{\beta}}$ for any $\beta \geq 3$, we deduce that \\

\n $\displaystyle{ k^{-2} \leq (r_{k})^{-\frac{2}{\beta}}}$ and  $\displaystyle{\mid\mid (\hat{G} - \lambda I)^{-1} \mid\mid \leq 2(r_{k})^{-\frac{2}{\beta}} = \frac{2}{\mid\lambda\mid^{\frac{2}{\beta}}}}$ for any $\beta \geq 3$  \hfill { }$\square$\\
\begin{lemma}

\n Let $\beta$ which satisfies $ 3 \leq \beta < 4$ then \, $ \forall \, \epsilon > 0$ there exists  $C_{\epsilon} > 0$ such that \\

\n $\displaystyle{ \mid\mid H_{\mu}\phi \mid\mid \leq \epsilon \mid\mid \hat{G}\phi\mid\mid ^{\frac{2}{\beta}} \mid\mid \phi \mid\mid ^{1 - \frac{2}{\beta}}}$ $\forall \, \phi \in D(G)$.\\
\end{lemma}
\n {\bf Proof}\\

\n Let us recall an interpolation inequality. For $a, b, \alpha $, and $\gamma$ positive numbers such that $\displaystyle{0 \leq \alpha \leq \gamma }$ then \\
\begin{equation}
\displaystyle{a^{\alpha}b^{\gamma-\alpha} \leq a^{\gamma} + b^{\gamma}}
\end{equation}

\n Hence, for each $\epsilon_{1} > 0$, there exists $C_{\epsilon_{1}} > 0$ such that  \\

\n  $\displaystyle{a^{\alpha}b^{\gamma-\alpha} \leq \epsilon_{1}a^{\gamma} +C_{\epsilon_{1}} b^{\gamma}}$\\

\n Writing $\displaystyle{\mid\mid \hat{G}\phi \mid\mid^{\frac{1}{2}} \mid\mid \phi \mid\mid^{\frac{1}{2}}}$ as   $\displaystyle{\mid\mid \hat{G}\phi \mid\mid^{\frac{1}{2}} \mid\mid \phi \mid\mid^{\frac{1}{2}} =  \mid\mid \phi \mid\mid^{1- \frac{2}{\beta}}\mid\mid \hat{G}\phi \mid\mid^{\frac{1}{2}} \mid\mid \phi \mid\mid^{\frac{2}{\beta}- \frac{1}{2}}}$  and as $\beta \leq 4$ we can apply the foregoing interpolation inequality to $\displaystyle{\mid\mid \hat{G}\phi \mid\mid^{\frac{1}{2}}}$ and $\displaystyle{\mid\mid \phi \mid\mid^{\frac{2}{\beta}- \frac{1}{2}}}$ with $\alpha = \frac{1}{2}$ and $\gamma = \frac{2}{\beta}$ to deduce that\\
\begin{equation}
\displaystyle{\mid\mid \hat{G}\phi \mid\mid^{\frac{1}{2}} \mid\mid \phi \mid\mid^{\frac{2}{\beta}- \frac{1}{2}} \leq \epsilon_{1}\mid\mid \hat{G}\phi \mid\mid^{\frac{2}{\beta}} + \, C_{\epsilon_{1}}\mid\mid \phi \mid\mid^{\frac{2}{\beta}} } 
\end{equation} 
\begin{center}
for every $ \phi \in D(G)$\\
\end{center}
\n and\\
\begin{equation}
\displaystyle{\mid\mid \hat{G}\phi \mid\mid^{\frac{1}{2}} \mid\mid \phi \mid\mid^{\frac{1}{2}} \leq \epsilon_{1}\mid\mid \hat{G}\phi \mid\mid^{\frac{2}{\beta}}  \mid\mid \phi \mid\mid^{1 - \frac{2}{\beta}} + \, C_{\epsilon_{1}}\mid\mid \phi \mid\mid} 
\end{equation} 
\begin{center}
for every $ \phi \in D(G)$\\
\end{center}
\n Finally, from this last inequality and from Lemma 6.9, we obtain\\
\begin{equation}
\displaystyle{ \mid\mid H_{\mu} \phi \mid\mid \leq C_{\mu,\lambda}\epsilon_{1}\mid\mid \hat{G}\phi \mid\mid^{\frac{1}{2}} \mid\mid \phi \mid\mid^{1- \frac{2}{\beta}} + \quad C_{\mu,\lambda}C_{\epsilon_{1}}\mid\mid \phi \mid\mid }
\end{equation} 
\begin{center}
 for every $\phi \in D(G)$\\
\end{center}
\n That is, for all $\epsilon > 0$  there exists $C_{\epsilon} > 0 $ such that \\
\begin{equation}
\displaystyle{ \mid\mid H_{\mu} \phi \mid\mid \leq \epsilon \mid\mid \hat{G}\phi \mid\mid^{\frac{1}{2}} \mid\mid \phi \mid\mid^{1- \frac{2}{\beta}} + \quad C_{\epsilon}\mid\mid \phi \mid\mid } 
\end{equation}
\begin{center}
for every $ \phi \in D(G)$\\
\end{center}
\n . \hfill { } $\square$\\

\n Let us now recall two results due to Yacubov and Mamedov {\color{blue}[Yacubov et al] } and Markus {\color{blue}[Markus]} :\\
\begin{theorem}
 (Yakubov and Mamedov in {\color{blue}[Yakubov et al]})\\

\n  Let $T$  be a closed densely defined linear operator in a Hilbert $\mathcal{H}$ with compact resolvent and assume that there exist a sequence of circles $C(0, r_{k}), k = 1, . . . , $ with radii $r_{k}$ going to infinity, a constant $C > 0$, and an integer $m \geq -1$ such that \\
\begin{equation}
\displaystyle{ \mid\mid (T - \lambda I)^{-1} \mid\mid  \leq \mid \lambda \mid^{m}} \mbox{ for } \mid \lambda \mid = r_{k}
\end{equation}

\n  Then the spectrum of the operator $T$ is discrete and for any \, $\phi \in D(T)$, there exists a subsequence of partial sums of the series $\displaystyle{\sum_{k}^{ }\mathbb{P}_{k}\phi }$ converging to $\phi$ in the sense of $\mathcal{H}$ where  \\

\n $\displaystyle{\mathbb{P}_{k} = \frac{1}{2i\pi}\int_{C(0, r_{k})}(T - \lambda I)^{-1} d\lambda}$\\
\end{theorem}
\begin{theorem}
( {\color{blue}[Markus]})\\

\n  Let $T$ be a closed densely defined linear operator in a Hilbert $\mathcal{H}$ with compact resolvent, let $L$ be a {\color{black}normal operator}
with compact resolvent and its spectrum lying on a finite number of rays, and. assume that,\\
\begin{equation}
\displaystyle{\overline{lim}\frac{\mathcal{N}(r, L)}{r^{1-p}} < \infty }\, as \,r \longrightarrow \infty \mbox{ for some} \,\, p \in [0, 1)
\end{equation} \\

\n  where $\mathcal{N}(r, L)$ denotes the sum of the multiplicities of all eigenvalues of $L$ contained in the disk $\displaystyle{\{ \lambda \in \mathbb{C} ; \mid \lambda \mid \leq r\}}$. \\

\n If the operator $B =T- L $ is subordinate to $L$ with order $p$, then the system of root vectors of $T$ forms an {\color{black}unconditional basis with parentheses } in $\mathcal{H}$.\\
\end{theorem}
\n Theorems 6.13 and 6.14 are used in the proof of the next theorem, which forms the main part of this sub-section.\\
\begin{theorem}

\n  For $\lambda'' \neq 0$  and {\color{black} $\lambda' = 0$}, we have the following statements:\\

\n (i) The spectrum of the operator $H_{\lambda''}$ is discrete and for any \, $\phi \in D(H_{\lambda''}^{2})$  there exists a subsequence of partial sums of series $\displaystyle{\sum_{n}^{}\mathbb{P}_{n}\phi}$ converging to $\phi$ in $\mathbb{B}$\\

\n (ii) The system of generalized eigenvectors of the operator $H_{\lambda''}$ is an {\color{black} unconditional basis with parentheses} in $\mathbb{B}$.\\
\end{theorem}
\n {\bf Proof}\\

\n (i) Let $\lambda \in \rho(\hat{G})$. Due to the equality $\displaystyle{\hat{G}(\hat{G} - \lambda I)^{-1} = I + \lambda (\hat{G} - \lambda I)^{-1}}$ and from the inequality of Lemma 6.11, we obtain \\
\begin{equation}
\displaystyle{\mid\mid \hat{G}(\hat{G} - \lambda I)^{-1}\mid\mid \leq  1 + 2\mid \lambda\mid^{1 - \frac{2}{\beta}}}\mbox{ for any} \beta \geq 3\mbox{ and} \displaystyle{\mid \lambda \mid = r_{k} \longrightarrow +\infty}. 
\end{equation}

\n Therefore,  for any $\beta \geq 3$ and $\displaystyle{\mid \lambda = r_{k}}$ sufficiently large, we have:\\
\begin{equation}
\n \displaystyle{\mid\mid \hat{G}(\hat{G} - \lambda I)^{-1}\mid\mid \leq  3\mid \lambda\mid^{1 - \frac{2}{\beta}}}
\end{equation}
\n Now, it follows from Lemma 6.12 that for each $\epsilon > 0$, there exists $C_{\epsilon} > 0$ such that:\\
\begin{equation}
\displaystyle{\mid\mid H_{\mu}(\hat{G} - \lambda I)^{-1}\phi \mid\mid \leq \epsilon \mid\mid \hat{G}(\hat{G} - \lambda I)^{-1}\phi \mid\mid^{\frac{2}{\beta}} \mid\mid (\hat{G} - \lambda I)^{-1}\phi \mid\mid^{1 - \frac{2}{\beta}} + C_{\epsilon} \mid\mid (\hat{G} - \lambda I)^{-1}\phi \mid\mid}
\end{equation}
\n From (6.29), (6.30) and lemma 6.11, it follows that \\

\n $\displaystyle{\mid\mid H_{\mu}(\hat{G} - \lambda I)^{-1}\phi \mid\mid \leq 3\epsilon \mid\mid \phi \mid\mid + 2C_{\epsilon}\mid \lambda \mid^{-\frac{2}{\beta}} = (3\epsilon  + 2C_{\epsilon}\mid \lambda \mid^{-\frac{2}{\beta}} )\mid\mid \phi \mid\mid}$\\

\n Hence, for $\epsilon < \frac{1}{3}$ and $\displaystyle{\mid \lambda \mid = r_{k} \longrightarrow +\infty}$ we get :\\
\begin{equation}
\displaystyle{\mid\mid H_{\mu}(\hat{G} - \lambda I)^{-1}\phi \mid\mid \leq q < 1.}
\end{equation}

\n Then, using the Neuman identity $\displaystyle{(\hat{G} + H_{\mu} - \lambda I)^{-1} = (\hat{G}  - \lambda I)^{-1}\sum_{k=1}^{\infty}[H_{\mu}(\hat{G} - \lambda I)^{-1}]^{k}}$,\\

\n we obtain:\\
\begin{equation}
\displaystyle{\mid\mid (\hat{G} + H_{\mu} - \lambda I)^{-1}\mid\mid \leq  C_{\mu}\mid\mid(\hat{G}  - \lambda I)^{-1}\mid\mid \leq \hat{C}_{\mu} \mid \lambda \mid^{-\frac{2}{\beta}}}
\end{equation}
\begin{center}
as $\mid \lambda \mid = r_{k} \longrightarrow \infty$.
\end{center}

\n and \\
\begin{equation}
\displaystyle{\mid\mid (\lambda''\hat{G} + H_{\mu} - \lambda I)^{-1}\mid\mid \leq  C_{\lambda''}\mid\mid(\hat{G}  - \lambda I)^{-1}\mid\mid \leq \hat{C}_{\lambda''} \mid \lambda \mid^{-\frac{2}{\beta}}}
\end{equation}
\begin{center}
 as $\mid \lambda \mid = r_{k} \longrightarrow \infty$. 
 \end{center}

\n i.e.,\\
\begin{equation}
\displaystyle{\mid\mid ( H_{\lambda''} - \lambda I)^{-1}\mid\mid \leq \hat{C}_{\lambda''} \mid \lambda \mid^{-\frac{2}{\beta}}}
\end{equation}
\begin{center}
 as $\mid \lambda \mid = r_{k} \longrightarrow \infty$. 
 \end{center}

\n From the compactness of $\displaystyle{ (\hat{G} - \lambda I)^{-1}}$ and the previous Neuman identity,it follows that the resolvent of the operator $H_{\lambda''}$ is compact in $\mathbb{B}$. Therefore, due to estimate  (6.34) , the operator $H_{\lambda''}$ satisfies the assumptions of Theorem 6.13 and the first assertion of the theorem is proved.\\

\n (ii) Let $\displaystyle{\mathcal{N}(C(0, r_{k} ., \hat{G})}$. be the sum of the multiplicities of the eigenvalues of the operator $G$ which are included in the circle $C(0, r_{k})$ . with radius $\displaystyle{r_{k} = \frac{k(k - 1)(2k -1)}{2}}$. Then, for $\beta \geq 3$ we get :\\
\begin{equation}
\displaystyle{\overline{lim}\, \frac{\mathcal{N}(C(0, r_{k} ., \hat{G})}{r_{k}^{1- \frac{2}{\beta}}} = lim \, k. k^{3(\frac{2}{\beta} - 1)} \leq 1 }
\end{equation}
\begin{center}
 as $ k \longrightarrow +\infty$.
 \end{center}

\n Consequently, the second assertion of the theorem follows from theorem 6.14. \hfill { } $\square$\\
\begin{remark}

\n (1) By adding the term $\lambda'' A^{*^{3}}A^{3}$ to $H_{\lambda'}$, we obtain a strong diagonalization of $H_{\lambda''}$;\\

\n (2) For $\lambda'' = 0$ and $\lambda' \neq 0$ the results of the preceding theorem remain open, although we hope to obtain a similar diagonalization.\\

\n (3) For $\lambda'' = 0$ and $\lambda' = 0$  and $\mu \neq 0$ notice that $H_{\mu}$ is very far from normal. Not only its self-adjoint and skew-adjoint parts do not commute
but there is no inclusion in either way between their domains or with the domain of their commutator.\\
\end{remark}
\n {\bf {\color{black} $\bullet$ (B) Generalized diagonalization of the semigroup $\displaystyle{e^{-tH_{\lambda''}^{\frac{1}{2}}}, t > 0}$}}\\

\n The aim of the end of this sub-section is to study the semigroup $\displaystyle{e^{-tH_{\lambda''}^{\frac{1}{2}}}, t > 0}$ on Bargmann space $\mathbb{B}$. In particular, we prove that the series  $\displaystyle{\sum_{n=1}^{+\infty}e^{-tH_{\lambda''}^{\frac{1}{2}}}\mathbb{P}_{n}}$ converges strongly to $\displaystyle{e^{-tH_{\lambda''}^{\frac{1}{2}}}}$.\\

\n We begin by recalling the summability method in Abel sense and Lidskii's theorem to  give a generalized diagonalization of semigroup associated to $H_{\lambda''}$.\\

\n Let $T$  be an operator with a dense domain and a discrete spectrum in a Hilbert space $\mathcal{H}$. In the following, $\{\mathcal{H}_{n}\}$ $n = 1, 2, . . . ,$ are used to denote
the system of all root subspaces of the operator $T$  which are associated to the eigenvalues numbered in increasing order and repeated according multiplicities. $\{\phi_{n}\}$, $n =1, 2, . . . $, denote the system of root vectors of the operator $T$, obtained by consecutive numbering of the basis of the sub-spaces $\{\mathcal{H}_{n}\}$ $n = 1, 2, . . . ,$, composed from the Jordan sequences. \\

\n The Fourier coefficients with respect to the system $\{\phi_{n}\}$, $n =1, 2, . . . $ are defined for any vector $\phi $ by $\displaystyle{c_{n} = \frac{< \phi , \phi_{n}^{*} >}{< \phi_{n}, \phi_{n}^{*} >}}$,  where the system $\{\phi_{n}^{*}\}$, $n =1, 2, . . . $, denote the system of root vectors of the operator $T^{*}$. \\

\n Even if the system $\{\phi_{n}\}$, $n =1, 2, . . . $ is  is complete, it is well known that it is not possible to say anything about the convergence of the Fourier series $\displaystyle{\sum_{n = 1}^{\infty}c_{n}\phi_{n}}$ to $\phi$.\\

\n Let $\alpha$ be a strictly positive real number, we introduce the following polynomials with respect to the real parameter $t$,\\
\begin{equation}
\displaystyle{P_{m}^{\alpha}(\frac{1}{\xi}, t) = \frac{\xi^{-\alpha}}{m!}\frac{\partial^{m}}{\partial \xi^{m}}e^{-t \xi^{-\alpha}}} ( m = 0, 1, 2, ....
\end{equation}
\n and we compare the series $\displaystyle{\sum_{n=1}^{\infty}c_{n}\phi_{n}}$ with the series  $\displaystyle{\sum_{n=1}^{\infty}c_{n}(t)\phi_{n}}$ whose  coefficients are defined in the following way:\\

\n {\color{black} $\bullet$ } If $\phi_{n}$ is an eigenvector of the operator $T$ associated to the eigenvalue $\lambda_{n}$ {\color{black} without} a Jordan chain, we set :\\
\begin{equation}
\displaystyle{c_{n}(t) = e^{-t\lambda_{n}^{\alpha}}c_{n}}\, \mbox{ where} \,\, \displaystyle{c_{n} = \frac{< \phi , \phi_{n}^{*} >}{< \phi_{n}, \phi_{n}^{*} >}}
\end{equation}
\n and\\

\n {\color{black} $\bullet$ } If $\{ \phi_{n}, \phi_{n}^{1}, \phi_{n}^{2}, ......, \phi_{n}^{s}\} , s \geq 0$ is a Jordan chain corresponding to the eigenvalue $\lambda_{n}$, then we set :\\
\begin{equation}
\displaystyle{c_{n+j}(t) = e^{-t\lambda_{n}^{\alpha}}\sum_{m=0}^{s-j}P_{m}^{\alpha}(\lambda_{n}, t), 0 \leq j \leq s}
\end{equation}
\begin{definition}

\n We say that the series $\displaystyle{\sum_{n=1}^{\infty}c_{n}\phi_{n}}$  is summable by Abel's method of {\color{black}order $\alpha$} if there exists a sequence of integers $m_{0}, m_{1}, ...., m_{n}$ with $m_{0}= 1$ such that, for any $t > 0$, the series $\displaystyle{ u(t) = \sum_{m_{n-1}}^{m_{n} -1}c_{n}(t)\phi_{n}}$ converges and $\displaystyle{ lim \, u(t) = \phi}$ as $t \longrightarrow 0^{+}$.\\
\end{definition}

\n Now, we need the following definition before stating Lidskii's theorem which played an important role in the spectral study of our operators.\\
\begin{definition}

\n Let $K$ be a compact operator in the Hilbert space $\mathcal{H}$. $K$ is said to belong to the Carleman-class $\mathcal{C}_{p}, p > 0$, with order $p$, if the
series $\displaystyle{\sum_{n=1}^{+\infty}s_{n}^{p}(K)}$. converges, where $s_{n}(K)$. are the eigenvalues of the operator $\displaystyle{\sqrt{K^{*}K}}$.\\

\n  In the particular case $p = 2$, $\mathcal{C}_{2}$ is exactly the Hilbert-Schmidt class and for $p = 1$, $\mathcal{C}_{1}$ is said class of nuclear operators or trace operators.\\
\end{definition}
\n For a systematic treatment of the operators of Carleman-class, we refer to the Gohberg and Krein's book {\color{blue}[Gohberg et al]}.\\
\begin{theorem}
  (Lidskii's Theorem {\color{blue} [Lidskii]})\\

\n Let $K$ be a compact operator acting on a Hilbert space $\mathcal{H}$. Suppose that the series $\displaystyle{\sum_{n = 1}^{\infty}s_{n}^{p}(K)}$ converges for some $p$, where $s_{n}^{p}(K)$ are the eigenvalues of the operator $\displaystyle{\sqrt{K^{*}K}}$  and suppose that the values of the quadratic form $< K\phi; \phi >$ lie in a sector $\mathbb{S}_{\rho}$ the complex $z$-plane defined by \\
\begin{equation}
\displaystyle{\mathbb{S}_{\rho} = \{ z \in \mathbb{C}; \mid arg(z) \mid  \leq \frac{\pi}{2\rho} \}} \mbox{ with} \rho > max (p, \frac{1}{2})
\end{equation}
\n Then for every $\phi$ in the range of the operator $K$, the corresponding Fourier series with respect to the system of root vectors of $K$ is summable by Abel's
method of order $\alpha$ to the vector $\phi$  where $ p \leq \alpha < \rho $ .\\
\end{theorem}
\begin{lemma}

\n For $\lambda'' \neq 0$, the resolvent of $H_{\lambda''}$ belongs to Carleman-class $\mathcal{C}_{p}$ for any $p > \frac{1}{3}$.\\
\end{lemma}
\n {\bf Proof}\\

\n As the eigenvalues of the self-adjoint operator $G$ are $\lambda_{n} = n(n-1)(n-2)$ then the resolvent of $G$ belongs to Carleman-class $\mathcal{C}_{p}$ for any $p > \frac{1}{3}$.\\

\n Now, as $\displaystyle{\lambda'' G + H_{\mu} -\sigma I)^{-1} = (\lambda'' G  -\sigma I)^{-1}\sum_{k = 0}^{\infty}[-H_{\mu}\lambda'' G  -\sigma I)^{-1}]^{k}}$, by applying the minimax theorem, it follows that the resolvent operator of $H_{\lambda''}$ belongs to Carleman-class  $\mathcal{C}_{p}$ for any $p > \frac{1}{3}$. \hfill { } $\square$\\
\begin{lemma}

\n For $\lambda'' > 0$ and $\mu > 0$,  there exist $\beta_{0} > 0$ and $\delta > 0$ such that the values of the quadratic form  $ < (H_{\lambda''} + \beta_{0}I)\phi , \phi >$ lie in the sector of complex $z$-plane,\\
\begin{equation}
\displaystyle{\mathbb{S}_{\delta} = \{ z \in \mathbb{C};  - \frac{\pi}{2} + \delta \leq arg(z)   \leq\frac{\pi}{2} - \delta  \}} \mbox{  for any} \phi \in D(H_{\lambda''} ).
\end{equation}
\end{lemma}
\n {\bf Proof}\\

\n We remark that\\

\n $\displaystyle{\mid < i\lambda A^{*}(A + A^{*})A\phi , \phi > \mid  \leq \mid \lambda \mid \mid <  (A + A^{*})A\phi , A\phi > \mid }$\\

\n  $\displaystyle{ \leq \mid \lambda \mid [\mid < A^{2}\phi , \phi > \mid + \mid < A\phi , A^{2}\phi > \mid ]}$\\

\n  $\displaystyle{ \leq 2\mid \lambda \mid. \mid\mid A^{2}\phi  \mid\mid. \mid\mid A \phi  \mid\mid}$\\

\n Using the estimate $\displaystyle{ \mid\mid  \phi  \mid\mid \leq \mid\mid A \phi  \mid\mid}$, we get \\

\n $\displaystyle{\mid < i\lambda A^{*}(A + A^{*})A\phi , \phi > \mid  \leq 2\mid \lambda \mid. \mid\mid A^{3}\phi  \mid\mid. \mid\mid A \phi  \mid\mid + \beta_{0}\mid\mid \phi \mid\mid^{2}}$\\

\n  $\displaystyle{ \leq \mid \lambda \mid [ \mid\mid A^{3}\phi  \mid\mid^{2} +  \mid\mid A \phi  \mid\mid^{2} ]  + \beta_{0}\mid\mid \phi \mid\mid^{2}}$\\

\n  $\displaystyle{ \leq  C < (\lambda'' G + \mu A^{*}A + \beta_{0}I)\phi , \phi > }$.\\

\n This last inequality implies that\\
\begin{equation}
-\displaystyle{tan(\frac{\Im m< H_{\lambda''} \phi, \phi >}{\Re e< (H_{\lambda''} +\beta_{0}I) \phi, \phi >})} \mbox{ is bounded}.
\end{equation}

\n Then, there exists $\delta > 0$ such that\\
\begin{equation}
\displaystyle{- \frac{\pi}{2} + \delta \leq arg < (H_{\lambda''} + \beta_{0}I) \phi , \phi > \leq  \frac{\pi}{2} - \delta}
\end{equation}
\begin{center}
\n for any $\phi \in D(H_{\lambda''})$ 
\end{center}
\begin{theorem}

\n For $\lambda'' > 0$ and $\mu > 0$, we have the following statements:\\
\n (i) The system of generalized eigenvectors of the Gribov operator $H_{\lambda''}$ is  complete in Bargmann space $\mathbb{B}$.\\

\n (ii) The corresponding Fourier series with respect to the system of generalized eigenvectors of the Gribov operator $H_{\lambda''}$ is summable by Abel's 
method of {\color{black}unit order } and the series $\displaystyle{\sum_{n = 1}^{\infty}e^{-t H_{\lambda''}}\mathbb{P}_{n}}$  converges strongly to  $\displaystyle{e^{-t H_{\lambda''}}}$ for any $t > 0.$\\
\end{theorem}
\n {\bf Proof}\\

\n By means of Lemmas 6.20 and 6.21, the ``Gribov-Intissar'' operator $H_{\lambda''}$ satisfies the assumptions of Lidskii's Theorem ; therefore, the earlier theorem follows directly from the Lidskii's theorem. \\

\n In particular, if we set $\displaystyle{u(t) = e^{-t H_{\lambda''}}\phi}$, then $u(t)$ is the solution of the Cauchy problem:\\

\begin{equation}
\left \{ \begin{array} {c} \displaystyle{u'(t) = - H_{\lambda''}u(t)} \\
\quad\\
\displaystyle{u(0) = \phi} \quad\quad\quad \quad\quad  \\
\end{array}\right.
\end{equation}
and the series $\displaystyle{\sum_{n = 1}^{\infty}e^{-t H_{\lambda''}}\mathbb{P}_{n}}$  converges strongly to  $\displaystyle{e^{-t H_{\lambda''}}}$ for any $t > 0.$ \hfill{ } $\square$\\

\n For the study of the semigroup $\displaystyle{e^{-t H_{\lambda''}^{\frac{1}{2}}}}$ on Bargmann space, we need the following theorems.\\
\begin{theorem}
( {\color{blue}[Kato]})\\

\n Let $T$ be a maximal accretive operator acting in a Hilbert space $\mathcal{H}$. Then there exists an unique maximal accretive square root $\sqrt{T}$ of $T$ such that $(\sqrt{T})^{2} = T$ and $\sqrt{T}$ has the following additional properties:\\

\n (1) $\sqrt{T}$ is $m$-sectorial with the numerical range contained in the sector :\\

\n $\displaystyle{\mathbb{S} = \{ z \in \mathbb{C} ; \mid arg z \mid \leq \frac{\pi}{4} \}}$\\

\n (2) $D(T)$  is a core of $\sqrt{T}$ .\\

\n (3) $T$ has a compact resolvent if and only if $\sqrt{T}$ has.\\
\end{theorem}
\n From the previous theorem and Krein's theorem 20, Chap. I, see {\color{blue}[Krein]}, one can define a square-root $\sqrt{H_{\lambda''}}$ such that $-\sqrt{H_{\lambda''}}$ generates an analytic semigroup $\displaystyle{e^{-t H_{\lambda''}^{\frac{1}{2}}}}$. \\

\n The function $\displaystyle{u(t) = e^{-t H_{\lambda''}^{\frac{1}{2}}}\phi}$ is a solution of {\color{black}the abstract elliptic problem } :\\
\begin{equation}
\left \{ \begin{array} {c} \displaystyle{u''(t) =  H_{\lambda''}u(t)} \\
\quad\\
\displaystyle{u(0) = \phi} \quad\quad\quad \quad\quad  \\
\end{array}\right.
\end{equation}
\begin{theorem}
( {\color{blue}[Macaev et al , Intissar3 {\color{black}(2001)}]})\\

\n  Let $K$ be a compact operator acting on a Hilbert space $\mathcal{H}$. We suppose that:\\

\n (i) $\displaystyle{\mid  Arg < K\phi, \phi > \mid \leq \frac{\pi}{2}}$ for any $\phi \in \mathcal{H}$.\\

\n (ii) $\displaystyle{K \in  \mathcal{C}_{p}}$.\\

\n Then $\displaystyle{{\color{black}\sqrt{K}} \in  \mathcal{C}_{2p}}$.\\
\end{theorem}
\n As the operator $H_{\lambda''}$ belongs to Carleman-class $ \mathcal{C}_{p}$  for any $p > \frac{1}{3}$  it follows from the preceding theorem that $\sqrt{H_{\lambda''}}$ belongs to Carleman-class $ \mathcal{C}_{p}$  for any $p > \frac{2}{3}$. Thus, by Lidskii's theorem, we get the following result:\\
\begin{theorem}

\n The series $\displaystyle{\sum_{n=1}^{\infty}e^{-t H_{\lambda''}^{\frac{1}{2}}}\mathbb{P}_{n}}$ converges strongly to $\displaystyle{e^{-t H_{\lambda''}^{\frac{1}{2}}}}$ for any $t > 0$. In particular, if we set $\displaystyle{u(t) =  e^{-t H_{\lambda''}^{\frac{1}{2}}}\phi}$ with $\phi \in D(G^{\frac{3}{2}})$, then $u(t)$ is the solution of the second-order problem (6.44).\\
\end{theorem}
\begin{remark}

\n For $\lambda'' = 0$ and $\lambda' \neq 0$, the resolvent of $H_{\lambda''}$ belongs Carleman-class $ \mathcal{C}_{p}$  for any $p > \frac{1}{2}$ see {\bf{\color{blue}[Aimar2 et al] }} or  {\bf{\color{blue}[Intissar4 {\color{black}(1994)}]}}. and we can apply the Lidskii's theorem to give a generalized diagonalization of the semigroup  $\displaystyle{e^{-t H_{\lambda''}}}$ for any $t > 0$. But the problem of diagonalization of  $\displaystyle{e^{-t H_{\lambda''}^{\frac{1}{2}}}}$ is open.\\

\n  For $\lambda'' = 0$ and $\lambda' = 0$ and $\mu \neq 0$, the resolvent of $ H_{\mu}$ belongs to Carleman-class $\mathcal{C}_{1+ \epsilon}$ for any $\epsilon > 0$ see {\bf{\color{blue}[Aimar3 et al] }} or {\bf{\color{blue}[Aimar4 et al] }}.. However, the completeness of the system of generalized eigenvectors of $H_{\mu}$ and the diagonalization of the semigroup $\displaystyle{e^{-t H_{\mu}}}$ are open problems.\\
\end{remark}
\section{Regularized trace formula associated to ``magic Gribov-Intissar'' operator on Bargmann space }

\n  In this section, we obtain a regularized trace formula for magic Gribov operator\\ $ H = \lambda{''}G + H_{\mu,\lambda}$ acting on Bargmann space where $$G = A^{*3}A^{3} \quad \quad and \quad \quad  H_{\mu,\lambda} = \mu A^{*}A + i\lambda A^{*}( A + A^{*})A$$ Here $A$ and $A^{*}$ are the standard Bose annihilation and creation operators and in Reggeon field theory, the real parameters $\lambda{''}$ is the magic coupling of Pomeron, $\mu$  is Pomeron intercept, $\lambda$ is the triple coupling of Pomeron and $i^{2} = -1$.\\
\n By applying some abstract results of  Sadovnichii-Podolskii [traces of operators with relatively compact perturbations. Mat. Sb. 193 (2), (2002) 129-152], we give the number of corrections sufficient for the existence of finite formula of the trace of  concrete magic Gribov's operator.\\

 \subsection{ Introduction and main result of this section}

\n  As is known, the trace of a finite-dimensional matrix is the sum of all the eigenvalues. But in an infinite dimensional space, in general, ordinary differential operators do not have a finite trace.\\

\n  In 1953, Gelfand and Levitan considered the Sturm-Liouville operator\\

\n $\left\{\begin{array}[c]{l}-y''(x) + q(x)y(x) = \sigma y(x)\\ \quad \\ y'(0) = 0, y'(\pi) = 0\\

\n q(x) \in \mathcal{C}^{1} [0, \pi],\quad \displaystyle{\int_{0}^{\pi}q(x)dx} = 0 \\ \end{array}\right.\hfill { }  (*)$\\

\n and derived the formula\\

\n $\displaystyle{\sum_{n=1}^{\infty}(\sigma_{n} -\lambda_{n}) = \frac{1}{4}(q(0) + q(\pi))}$ $\hfill { }  (**)$\\

\n where $\sigma_{n}$ are the eigenvalues of the above operator and $\lambda_{n} = n^{2}$ are the eigenvalues of the same operator with $q(x)=0$.\\

\n The proof of this regularized trace formula for the Sturm-Liouville operator can been found in {\bf{\color{blue}[Gelfand et al]}}\\

\n The same regularized trace formula for the same problem was obtained with different method by Dikii {\bf{\color{blue}[Dikii1] }}.\\

\n  For the scalar Sturm-Liouville problems, there is an enormous literature (see for example {\bf{\color{blue}[Dikii2] }} or {\bf{\color{blue} [Sadovnichii1 et al]}}) on estimates of large eigenvalues and regularized trace formulae which may often be computed explicitly in terms of the coefficients of operators and boundary conditions.\\

\n After these studies, several mathematicians were interested in developing regularized trace formulae for different differential operators. According Sadovnichii and  Podolskii, these formulae gave rise to a large and very important theory, which started from the investigation of specific operators and further embraced the analysis of regularized traces of discrete operators in general form.\\

\n Among the results of Sadovnichii and  Podolskii established for abstract operators, we can recall that following :\\

\n Let $A_{0}$ be a self-adjoint positive discrete operator of domain $D(A_{0})$ acting in a Hilbert space, we denote by $\{\lambda_{n}\}$ its eigenvalues arranged in ascending order, $\{\phi_{n}\}$ is an orthonormal basis formed by the eigenvectors of $A_{0}$ and $R_{0}(\sigma)$ is its resolvent . By $B$ we denote the perturbing
operator, by $\{\sigma_{n}\}$ we denote the eigenvalues of the operator $A_{0} + B$ numbered in increasing order of their real parts, and $R(\sigma)$ stands for its resolvent.\\
Also, assume that $BA_{0}^{-\delta}$, $\delta > 0$ is a compact operator belonging to some finite-order
Schatten von Neumann class, i.e. the set of compact operators whose singular numbers form a convergent series $\displaystyle{\sum_{n=1}^{\infty}s_{n}^{p}}$
for some $p > 0$ is traditionally denoted by $\mathcal{C}_{p}$.\\

\n For operators $A_{0}$ and $B$ in {\bf{\color{blue} [Sadovnichii2 et al]}}, the following theorem is proved.\\
\begin{theorem}
(Sadovnichii-Podol'skii {\bf{\color{blue} [Sadovnichii2 et al]}})\\

\n Consider operator $A_{0}$ and $B$ be such that $A_{0}^{-1} \in C_{1}$ and $D(A_{0}) \subset D(B)$, and  suppose that there exist $\delta \in [0, 1)$ such that the operator $BA_{0}^{-\delta}$ can be continued to a bounded operator. Further, suppose that there exists $\omega \in [0, 1)$ such that $ \omega + \delta < 1$ and  $A_{0}^{-(1-\delta -\omega)}$ is a trace class operator, i.e. in $\mathcal{C}_{1}$. Then, there exist a subsequence of natural numbers $\{n_{m}\}_{m=1}^{\infty}$ and a subsequence of contours $\Gamma_{m} \subset \mathbb{C}$, that  for $\omega \geq  \frac{\delta}{l}$ the following relation holds:\\
\begin{equation}
 \lim \limits_{m \rightarrow \infty}(\displaystyle{\sum_{j=1}^{n_{m}}(\sigma_{j} - \lambda_{j}) +\frac{1}{2\pi i}\int_{\Gamma_{m}}\sum_{k=1}^{l}\frac{(-1)^{k-1}}{k} Tr((BR_{0}(\sigma))^{k}d\sigma) = 0}, 
 \end{equation}
\n In particular, for $l = 1$ we have \\
\begin{equation}
 \lim \limits_{m \rightarrow \infty} \displaystyle{\sum_{j=1}^{n_{m}}(\sigma_{j} - \lambda_{j} -  <B\phi_{j}, \phi_{j}>) = 0}
 \end{equation}
\end{theorem}
\begin{remark}

\n  1) This theorem has been successfully applied to concrete ordinary differential operators as well as to partial differential operators, we can see some examples given in {\bf{\color{blue} [Sadovnichii2 et al]}} {\bf{\color{blue} [Sadovnichii5 et al]}}).\\

\n 2) We can found in {\bf{\color{blue} [Sadovnichii3 et al]}} or in {\bf{\color{blue} [Sadovnichii4 et al]}}, an excellent survey dedicated by Sadovnichii and Podolskii to the history of the state of the art in the theory of regularized traces of linear differential operators with discrete spectrum and a detailed list of publications related to the present aspect.\\
\end{remark}
\n Usually, quantum Hamiltonians are constructed as self-adjoint operators; for certain situations,however, non-self-adjoint Hamiltonians are also of importance. In particular, the reggeon field theory (as invented by V. Gribov {\bf{\color{blue} [Gribov1]}}) for the high energy behaviour of soft processes is governed by the magic non-self-adjoint Gribov operator\\
\begin{equation}
H_{\lambda'',\lambda',\mu,\lambda} = \lambda{''}A^{*3}A^{3}+\lambda{'}A^{*2}A^{2} + \mu A^{*}A +i\lambda A^{*}(A + A^{*})A 
\end{equation}
\n where $a$ and its adjoint $a^{*}$ are annihilation and creation operators, respectively, satisfying the canonical commutation relation $[A, A^{*}] = I$.\\

\n In the case $\lambda{''} = 0$, Ando and Zerner in {\bf{\color{blue} [Ando et al]}}, Aimar in   {\bf{\color{blue} [Aimar1 et al to Aimar7 et al]}} and Intissar in {\color{blue} [Intissar1 to intissar 20]} have given a complete spectral theory for the operator $H_{0,\lambda{'},\mu,\lambda}$. In particular, one can consult the list of spectral properties of this operator summarized in  {\color{blue}[Intissar4 et al]} (2019).\\

\n  In the case $\lambda{''} \neq  0$ and $\lambda{'} \neq 0$, some assumptions of Sadovnichii-Podlskii's theorem recalled above are not verified for the operator $H_{\lambda{''},\lambda{'},\mu,\lambda}$.\\

\n In the case $\lambda{''} \neq  0$ and $\lambda{'} = 0$, Aimar et al in {\bf{\color{blue} [Aimar6 et al]}} have given some spectral properties of magic Gribov operator $H_{\lambda{''},0,\mu,\lambda}$, which is more regular than $H_{0,\lambda{'},\mu,\lambda}$.\\

\n In this section, we continue the spectral study of $H_{\lambda{''},0,\mu,\lambda}$ and we apply the results of Sadovnichii-Podlskii cited above to obtain a regularized trace formula of this operator.\\

\n We adopt the following notations: \\

\n $ H = H_{\lambda{''},0,\mu,\lambda} = \lambda{''}G + H_{\mu,\lambda}$ where $G = A^{*3}A^{3}$ \quad and \quad $H_{\mu,\lambda} = \mu A^{*}A + i\lambda A^{*}( A + A^{*})A$ \\ 

\n Here $A$ and $A^{*}$ are the standard Bose annihilation and creation operators and $\lambda{''}$, $\mu$, $\lambda$ are real parameters and $i^{2} = -1$.\\

\n It is convenient to regard the above operators as acting on Bargmann space $\mathbb{B}$ {\bf{\color{blue} [Bargmann1]}}.\\

\n $\mathbb{B}$ is defined as a subspace of the space O($\mathbb{C}$) of holomorphic functions on $\mathbb{C}$\, given by\\
\begin{equation}
 \mathbb{B} = \{\phi\in O(\mathbb{C}) ; < \phi, \phi > < \infty \} 
\end{equation}
\n where the paring \\
\begin{equation}
<\phi,\psi> = \displaystyle{\frac{1}{\pi}\int_{\mathbb{C}}}\displaystyle{\phi(z)\overline{\psi(z)}e^{-\mid z\mid^{2}}dxdy}\,\,\,  \forall\, \phi,\psi \in O(\mathbb{C})
\end{equation}
\n  and $dxdy$ is Lebesgue measure on $\mathbb{C}$.\\

\n The Bargmann space $\mathbb{B}$ with $\mid\mid \phi \mid\mid = \sqrt{<\phi, \phi>}$ is a Hilbert space and $\displaystyle{e_{n}(z) = \frac{z^n}{\sqrt{n!}}};\\ n = 0, 1, ....$ is an orthonormal basis in $\mathbb{B}$.\\

\n In this representation, the standard Bose annihilation and creation operators are defined by\\

\n $\displaystyle{A\phi(z) = \phi^{'}(z)}$ with  maximal  domain $\displaystyle{D(A) = \{\phi \in \mathbb{B} \,  \mbox{ such  that} \,  A\phi \in \mathbb{B}\}}$ \\

\n  and\\

\n $\displaystyle{A^{*}\phi(z) = z\phi(z)}$ \, \mbox{ with maximal domain}\, 
$\displaystyle{D(A^{*}) = \{\phi \in \mathbb{B}  \,  \mbox{ such  that} \,  A^{*}\phi \in \mathbb{B}\}}$ \\

\n Accordingly, for the operator  $ H := H_{\lambda^{''}, 0,\mu,\lambda}$ we have\\
\begin{equation}
\left\{\begin{array}[c]{l}H \phi(z)= \lambda^{''} z^{3}\phi^{'''}(z)+i\lambda z\phi^{''}(z) + (i\lambda z^{2} + \mu z)\phi^{'}(z) \quad \quad \quad  \quad \quad \quad\\ \quad\\ \mbox{with \ maximal domain } \\
\quad\\
D(H_{max}) = \{\phi \in \mathbb{B} ; H\phi \in\mathbb{B}\} \\ \end{array}\right.
\end{equation}
\begin{remark}

\n  Tomin, in an interesting article {\bf{\color{blue}[Tonin]}}, have derived several formulas of Gelfand-Levitan type for the first regularized trace of discrete operators under various conditions convenient for verification. But, for $\mu \neq 0 $ and $\lambda \neq 0 $ the operator $H_{\mu,\lambda} = H_{0, 0,\mu,\lambda}$ is not in the classes of perturbing linear operators considered by him.\\
 
\n Notice that in this case the operator $H_{\mu,\lambda}$ is very far from normal and not only its self-adjoint and skew-adjoint parts do not commute but there is no inclusion in either way between their domains or with the domain of their commutator.\\

\n It may be noted also that if  $\mu = 0 $ and $\lambda \neq 0 $ the spectrum of $H_{\mu,\lambda}$ is $ \sigma(H_{\mu,\lambda}) = \mathbb{C}.$\\
\end{remark}
\n Let us begin by reviewing the most important properties of the $ H := H_{\lambda^{''},0,\mu,\lambda}$\\

\n i) Let {\bf Pol} be the space of polynomials, then {\bf Pol} is dense in $\mathcal{B}$.\\

\n ii) We define $H_{min}$ as the closure of restriction operator $ H_{\mid_{ Pol}}$ on the polynomials acting in Bargmann space, i.e.\\
\begin{equation}
\left\{\begin{array}[c]{l}H_{min}\phi= \quad \lambda^{''} z^{3}\phi^{'''}(z)+i\lambda z\phi^{''}  + (i\lambda z^{2} + \mu z)\phi^{'}\\
\quad\\
\mbox{with \ maximal domain } \\
\quad\\
\n D(H_{min}) = \{\phi \in \mathcal{B} ; \, \exists p_{n} \in  Pol\,  and \, \psi \in \mathcal{B} \, ; p_{n}\rightarrow \phi \mbox{ and }\,\,  Hp_{n}\rightarrow \psi\} \end{array}\right.
\end{equation}

\n iii) $D(H_{min}) = D(H_{max}) = D(G)$ \\

\n  In Bargmann space, it may be noted also that\\

\n  iv) The operator $ G = A^{*3}A^{3} $ is positive, self adjoint operator.\\

\n  v) The functions $\displaystyle{e_{n}(z) = \frac{z^n}{\sqrt{n!}}}$ are  the orthonormal eigenvectors of $G$ corresponding to the eigenvalues $\lambda_{n} = n(n-1)(n-2)$ for $n \geq 3 $ and $\lambda_{n} = 0 $ for $  n \in \{0, 1,2\}$\\

\n  vi) Let $\widetilde{G} = G + I$ then  $ <\widetilde{G}\phi, \phi> \geq <\phi, \phi>$  $\forall \phi \in D(G)$\\

\n vii) $G $ could be replaced by $\widetilde{G}$ or $G + \sigma I$ with a scalar $\sigma$ without changing the nature of the problem that we will study.\\

\n viii) in {\bf{\color{blue}[Aimar6 et al]}} (see theorem 3.3, p: 595), Aimar et al have shown that the spectrum of the magic Gribov operator is discrete and that the system of generalized eigenvectors of this operator is an unconditional basis in Bargmann space $\mathbb{B}$.\\

\n The goal of this section consists in establishing news spectral properties of this operator and the number of corrections sufficient for the existence of finite formula of the regularized trace.\\

\n Then the main results to which is aimed this paper can be stated as follows for the magic Gribov operator \\
\begin{theorem}

\n  Let $\mathbb{B} $ be the Bargmann space, $ H = \lambda{''}G + H_{\mu,\lambda}$ acting on $\mathbb{B}$ where $ G = A^{*3}A^{3}$ and $ H_{\mu,\lambda} = \mu A^{*}A + i\lambda A^{*}(A + A^{*})A$ , $A$ and $A^{*}$ are the standard Bose annihilation and creation operators.\\

\n Then there exists an increasing sequence of radius $r_{m}$ such that $r_{m} \rightarrow \infty$ as $ m \rightarrow \infty$  and\\
\begin{equation}
\displaystyle{\lim\limits_{m \rightarrow} (\sum_{n=0}^{m}(\sigma_{n} - \lambda{''}\lambda_{n}) + \frac{1}{2i\pi}\int_{\gamma_{m}} Tr[\sum_{k=1}^{4}\frac{(-1)^{k-1}}{k}[H_{\mu,\lambda}(\lambda{''}G - \sigma I)^{-1}]^{k}]d\sigma) = 0} 
\end{equation}
\n  Where\\

 \n - $\sigma_{n}$ are the eigenvalues of the operator $ H = \lambda{''}G + H_{\mu,\lambda}$\\
 
\n  - $\lambda_{n} = n(n-1)(n-2)$ are the eigenvalues of the operator $ G $\\

 \n - $ (\lambda{''}G - \sigma I)^{-1}$ is the resolvent of the operator $\lambda{''}G$\\
 
\n  and\\
 
\n  - $\gamma_{m}$ is the circle of radius $r_{m}$ centered at zero in complex plane.\\
\end{theorem}
\n  This section is organized as follows. In following subsection 7.2, we improve some basic results of the operator $G$ and some key results to apply theorem 7..1. In subsection 7.3, we show that all assumptions of theorem 7.1 are fulfilled for the operators $ A_{0} = \lambda'' (G + I) $ and $B = H_{\mu,\lambda}$ with $\delta = \frac{1}{2}$, $\forall \omega \in [0, \frac{1}{6})$ and for $\frac{1}{8} \leq \omega < \frac{1}{6}$, we  deduce the main result of this Note. This work is concluded by noting that if $\lambda' \neq 0$ and $\lambda \neq 0$, the existence of finite formula of the regularized trace for the operator  $H_{\lambda'',\lambda',\mu,\lambda}$ defined by (7.3) is an open problem.\\

\subsection{Some auxiliary results}

\n  We begin this under-section by improving some basic results of  {\color{blue}[Aimar6 et al ]} on the operator $G$ and by recalling some useful definitions.\\

\n  For the discreteness of spectrum of the operator $ \tilde{G} = G + I$, it suffices to use the following Rellich's theorem (see\n   {\color{blue}[Krein]}, p. 386)\\
\begin{theorem}

\n  Let $B$ be a self-adjoint operator in $\mathbb{B}$ satisfying $<B\phi, \phi> \geq <\phi, \phi>, \phi \in D(B)$, where $D(B)$ is a domain of $B$.\\

\n Then, the spectrum of $B$ is discrete if and only if the set of all vectors $\phi \in D(B)$, satisfying $ <B\phi, \phi > \leq 1 $ is a precompact set.\\
\end{theorem}
\n In  {\color{blue}[Aimar6 et al ]}, it was shown the following basic spectral properties on the operator $G$\\
\begin{lemma}

\n  1) The operator $G$ has a compact resolvent.\\

\n 2) Let  $\lambda_{n} = n(n-1)(n-2)$ for $ n \geq 0 $ the eigenvalues of $G$ associated to eigenvectors $\displaystyle{e_{n}(z) = \frac{z^{n}}{\sqrt{n!}}}$. Then for each $\sigma \in \mathbb{C}$ such that $\sigma \neq \lambda_{n}$, we have\\
\begin{equation}
(G - \sigma I)^{-1}\phi = \displaystyle{\sum_{n=0}^{\infty}\frac{1}{\lambda_{n} - \sigma}<\phi, e_{n}>e_{n}}.
\end{equation}
\n Moreover, if $Im \sigma \neq 0$ and if $\sigma$ belongs to a ray with origin zero and of angle $\theta$ with $\theta \neq 0 $ and $\theta \neq \pi $ , we have \\
\begin{equation}
\mid\mid(G - \sigma I)^{-1}\mid\mid \leq \frac{1}{Im \sigma} = \frac{c(\theta)}{\mid \sigma \mid}
\end{equation}
\n 3) There exists a sequence of circles $C(0, r_{n}) , n =1, 2, . . . ,$ with radii $r_{n}$ going to infinity such that\\
\begin{equation}
\mid\mid (G - \sigma I)^{-1}\mid\mid \leq \frac{2}{\mid\sigma\mid^{2/\beta}} 
\end{equation}
\n for any $\beta \geq 3$ and $ \mid \sigma \mid =  r_{n}$ where $ r_{n} = \frac{\lambda_{n} + \lambda_{n+1}}{2}$\\
\end{lemma}
\n {\bf Proof}\\

\n  1) It is well known that the injection from $D(A)$ into the Bargmann space $\mathbb{B}$ is compact (see for example (iii) of lemma 3.6) and as the injection from  $D(A^{*3}A^{3})$ into $D(A)$ is continuous, then the injection from  $D(A^{*3}A^{3})$ into Bargmann space is also compact. \\

\n Classically the operators of the form  $ (A^{*3}A^{3} + I) $ are invertible. Then the resolvent set of $G$ is not void. Consequently the self-adjoint operator $G$ has compact resolvent and this proves again the discreteness of its spectrum.\\

\n  For the properties 2) and 3) of this lemma see the lemmas 3.1 and 3.2 in  {\color{blue}[Aimar6 et al ]}.\hfill { } $\square$\\

\begin{proposition}

\n  1) The resolvent of the operator $G$ belongs to the class Carlemann $ \mathcal{C}_{p}$,  $\forall p > \frac{1}{3}$.
In particulary the operator $G$ belongs to class of operators with trace resolvent.\\

\n 2) Let $\lambda_{m} = m(m-1)(m-2)$ then \\
\begin{equation}
\forall m \geq 2, \lambda_{m+1} - \lambda_{m} \geq \frac{3}{2}m^{2}
\end{equation}
\n 3) Let $\lambda_{n} = n(n-1)(n-2)$ for $ n = 2, 3, ....$ and $ \beta > \frac{1}{3}$ . Then there exists an increasing sequence of positive numbers $\tilde{r}_{m} = \frac{\lambda_{m}^{\beta} + \lambda_{m+1}^{\beta}}{2}$ such that $\tilde{r}_{m} \rightarrow \infty $ as $ m \rightarrow \infty $ and\\
\begin{equation}
 \displaystyle{\sum_{n=3}^{+\infty}\frac{1}{\mid \lambda_{n}^{\beta} - \tilde{r}_{m}\mid} \leq C}
\end{equation}
\n where $C$ is a positive number.\\
\end{proposition}
\n {\bf Proof}\\

\n 1) As $\lambda_{n} \sim n^{3}$ then  the series of term general $\frac{1}{\lambda_{n}^{p}}$ converges for all $p > \frac{1}{3}$, i.e. the resolvent of the operator $G$ belongs to Carleman's class  $C_{p}$,  $\forall p > \frac{1}{3}$.\\

\n In particulary the operator $G$ belongs to class of operators with trace resolvent.\\

\n 2) We note that $\lambda_{m+1} - \lambda_{m} = 3m(m-1)$ then $\lambda_{m+1} - \lambda_{m} \geq \frac{3}{2}m^{2}$.\\

\n 3) Let $m \geq 3$ and $\tilde{r}_{m} = \frac{\lambda_{m}^{\beta} + \lambda_{m+1}^{\beta}}{2}$ with $\lambda_{m}^{\beta} = m^{\beta}(m-1)^{\beta}(m-2)^{\beta}$ and $\beta > \frac{1}{3}$. Then \\ 

\n $ \displaystyle{\sum_{n=3}^{+\infty}\frac{1}{\mid \lambda_{n}^{\beta} - \tilde{r}_{m}\mid}}$ = $ \displaystyle{\sum_{n=3}^{m}\frac{1}{\tilde{r}_{m} - \lambda_{n}^{\beta}}}$ + $ \displaystyle{\sum_{n=m+1}^{+\infty}\frac{1}{\lambda_{n}^{\beta} - \tilde{r}_{m}}}$\\

\n Let $ n-m = k$ then \\

\n $\displaystyle{\sum_{n=m+1}^{+\infty}\frac{1}{\lambda_{n}^{\beta} - \tilde{r}_{m}}}$ = $\displaystyle{\sum_{k =1}^{+\infty}\frac{1}{\lambda_{m + k}^{\beta} - \tilde{r}_{m}}}$ = $\displaystyle{\sum_{k =1}^{+\infty}\frac{2}{2\lambda_{m + k}^{\beta} - \lambda_{m}^{\beta} - \lambda_{m + 1}^{\beta}}}$\\

\n Now to estimate $\lambda_{m + k}^{\beta} - \lambda_{m}^{\beta}$, we consider the function \\
\begin{equation}
f_{\beta}(x) = x^{\beta}(x-1)^{\beta}(x-2)^{\beta} \mbox{ for } x \geq 3
\end{equation}
\n It follows that\\

\n $f^{'}_{\beta}(x) = \beta f_{\beta - 1}(x)(3x^{2} - 6x + 2)$ \\

\n and\\

\n $f^{''}_{\beta}(x) = \beta f_{\beta - 2}(x){\Large[ }(3 (3\beta -1) x^{4} - 12(3\beta -1) x^{3} + 6 (8\beta -3)x^{2} - 12 (2\beta -1)x + 4 (\beta -1){\Large]}$\\

\n For $3\beta (3\beta -1) \geq 0$ in particulary for $ \beta > \frac{1}{3}$ and sufficiently large $x$ , it follows that  $f^{''}_{\beta}(x)$ take positive values for sufficiently large $x$ ($ x > x_{min} \geq 3$) then it follows that $f^{'}_{\beta}(x)$ increases and in particulary $f^{'}_{\beta}(x) < f^{'}_{\beta}(x + h)$ for all $ h > 0$  and for $x > x_{min}$ therefore $ f_{\beta}(x_{min} + h) - f_{\beta}(x_{min}) \leq f_{\beta}(x + h) - f_{\beta}(x)$ for $ x \geq x_{min}$.\\

\n Thus there exists a number $m_{0}$ depending only on $\beta$ such that, for $ m > m_{0} \geq 3$ we have :\\

\n $\lambda_{m + k}^{\beta} - \lambda_{m}^{\beta} \geq \lambda_{m_{0} + k}^{\beta} - \lambda_{m_{0}}^{\beta}$\\

\n By noting that $\lambda_{m_{0} + k}^{\beta} - \lambda_{m_{0}}^{\beta}\sim k^{3\beta}$ and \\

\n $\displaystyle{\sum_{k =1}^{+\infty}\frac{2}{2\lambda_{m + k}^{\beta} - \lambda_{m}^{\beta} - \lambda_{m + 1}^{\beta}}}$ = $\displaystyle{\sum_{k =1}^{+\infty}\frac{2}{\lambda_{m + k}^{\beta} - \lambda_{m}^{\beta} + \lambda_{m + k}^{\beta} - \lambda_{m + 1}^{\beta}}}$\\

\n $\displaystyle{ = \frac{2}{\lambda_{m + 1}^{\beta} - \lambda_{m}^{\beta}}}$ + $\displaystyle{\sum_{k =2}^{+\infty}\frac{2}{\lambda_{m + k}^{\beta} - \lambda_{m}^{\beta} + \lambda_{m + k}^{\beta} - \lambda_{m + 1}^{\beta}}}$\\

\n $\displaystyle{ \leq \frac{2}{\lambda_{m + 1}^{\beta} - \lambda_{m}^{\beta}}}$ + $\displaystyle{\sum_{k =2}^{+\infty}\frac{2}{\lambda_{m + k}^{\beta} - \lambda_{m}^{\beta}}}$ + $\displaystyle{\sum_{k=2}^{+\infty}\frac{2}{\lambda_{m + k}^{\beta} - \lambda_{m + 1}^{\beta}}}$ \\

\n  $\displaystyle{ \leq \sum_{k =1}^{+\infty}\frac{2}{\lambda_{m + k}^{\beta} - \lambda_{m}^{\beta}}}$ + $\displaystyle{\sum_{k =2}^{+\infty}\frac{2}{\lambda_{m + k}^{\beta} - \lambda_{m + 1}^{\beta}}}$ \\

\n $\leq C_{1}$ $\displaystyle{\sum_{k =1}^{+\infty}\frac{1}{k^{3\beta}} = C}$ ($C_{1}$ does not depend on $m$) \hfill { } $\square$\\
\begin{lemma}

\n  The operator $ H_{\mu,\lambda}( G - \sigma I)^{-1}$ is nuclear on Bargmann space where $\sigma$ belongs a resolvent set of the operator $G$.\\
\end{lemma}

\n {\bf Proof}\\

\n  Consider $\phi(z) = \displaystyle{\sum_{n=0}^{\infty}\phi_{n}e_{n}(z)}$ in Bargmann space $\mathbb{B}$, The matrix of the operator $H_{\mu,\lambda}$ in basis  ${e_{n}(z)}$ has the form \\
\begin{equation}
(H_{\mu,\lambda}\phi)_{n} = \alpha_{n-1}\phi_{n-1} + q_{n}\phi_{n} +\alpha_{n}\phi_{n+1}, n\geq 1 
\end{equation}
\n with $(H_{\mu,\lambda}\phi)_{0} = 0$\\

\n where\\

\n $q_{n} = \mu n$, and $\alpha_{n} = i\lambda n\sqrt{n+1}$, ( $\mu$ and $\lambda$ are real numbers and $i^{2} = -1$).\\

\n It is complex symmetric tri-diagonale matrix (but not Hermitian!) of the form \\
\begin{equation}
\left\{\begin{array}[c]{l}H_{\mu,\lambda} = (h_{m,n})_{m,n=0}^{\infty}\quad with \quad the \quad elements\\ h_{nn} = \mu n\cr h_{n,n+1}=h_{n+1,n}= i\lambda n\sqrt{n+1}; n = 0,1,2, \\ and\\
h_{mn} = 0 \quad for \quad \mid m - n \mid > 1  \\ \end{array}\right.
\end{equation}
\n Then\\

\n  All elements of the matrix $H_{\mu,\lambda}$ have order $O(n^{\frac{3}{2}})$ as $n \rightarrow \infty$ and as all elements of the matrix $(G -\sigma I)^{-1}$ have order $O(n^{-3})$ as $n \rightarrow \infty$ then the elements of the matrix $ H_{\mu,\lambda}(G -\sigma I)^{-1}$ have order $O(n^{\frac{-3}{2}}) $ as $ n \rightarrow \infty$ therefore $ H_{\mu,\lambda}(G -\sigma I)^{-1}$ is a nuclear operator.\hfill { } $\square$\\
\begin{lemma}

\n 1) Let be $\omega \in \mathbb{R} $ such that $ \omega < \frac{1}{6} $ then the operator $\tilde{G}^{-(\frac{1}{2} - \omega)}$ is a trace class operator on the Hilbert space $E$.\\

\n 2) Let be $\delta \in \mathbb{R} $ such that $ \frac{1}{2} \leq \delta < \frac{2}{3} $ then the operator $H_{\mu,\lambda}\tilde{G}^{-\delta}$ is bounded and the operator $\tilde{G}^{\delta}(\tilde{G} - \sigma I)^{-1}$ is nuclear; $\sigma  \in \rho (\tilde{G})$.\\

\n 3) Let be $\delta \in \mathbb{R}$ such that $ \frac{5}{6} < \delta \leq 1$, the operator $H_{\mu,\lambda}\tilde{G}^{-\delta}$ is nuclear and the operator $\tilde{G}^{\delta}(\tilde{G} - \sigma I)^{-1}$ is bounded; $\sigma  \in \rho (\tilde{G})$.\\
\end{lemma}
\n {\bf Proof }\\

\n 1)  The matrix of the operator  $\tilde{G}^{-(\frac{1}{2} - \omega)}$ in the base $e_{n}(z)$ is diagonal and its elements have order  $O(n^{-3(1/2 - \omega)}$ as $ n \rightarrow \infty $ then $\tilde{G}^{-(\frac{1}{2} - \omega)}$ $\forall \omega < \frac{1}{6}$ is a trace class operator on the Hilbert space $\mathbb{B}$.\\

\n  2)$H_{\mu,\lambda}\tilde{G}^{-\delta}$ in the base $e_{n}(z)$ is tridioagonal and its elements have order $O(n^{\frac{3}{2}-3\delta})$ as $ n \rightarrow \infty $ then it easy to see that,if $ \frac{1}{2} \leq \delta $ then we obtain the property 2)\\

\n  3) From 2) if $ \frac{5}{6} < \delta \leq 1$, the operator $H_{\mu,\lambda}\tilde{G}^{-\delta}$ is nuclear and the operator $\tilde{G}^{\delta}(\tilde{G} - \sigma I)^{-1}$ is bounded where $\sigma $ belongs to $\rho(\tilde{G})$ resolvent set of $\tilde{G}$. \hfill { } $\square$\\
\begin{proposition}
 \n  All assumptions of theorem 1.6.1 are fulfilled for the operators $A_{0} = \lambda''\tilde{G}$ and $B = H_{\mu, \lambda}$ in $E$ with $\delta = \frac{1}{2}$ $\forall \omega < \frac{1}{6})$. The condition $\omega \geq \frac{\delta}{l} = \frac{1}{2l}$ is satisfied $\forall l > 3$ then for $l = 4$ and $\frac{1}{8}\leq \omega < \frac{1}{6}$, we get that there exist a subsequence of naturel numbers $\{n_{m}\}_{m =1}^{\infty}$ and a sequence of contours $\Gamma_{m} \subset \mathbb{C}$ that the formula\\
\begin{equation}
\displaystyle{\lim\limits_{m \rightarrow \infty} (\sum_{k=0}^{n_{m}}(\sigma_{k} - \lambda{''}\lambda_{k}) + \frac{1}{2i\pi}\int_{\Gamma_{m}} \sum_{j=1}^{4}\frac{(-1)^{j-1}}{j}Tr[[H_{\mu,\lambda}(\lambda{''}G - \sigma I)^{-1}]^{j}]d\sigma) = 0}
\end{equation}
\n   is true\\
\end{proposition}
\n   {\bf Proof}\\

\n  The operator $BA_{0}^{-\frac{1}{2}} $ is bounded then take $\delta = \frac{1}{2}$ in Theorem 7.1\\
 By above Lemma,  assertion 1) implies that assumption on $A_{0}^{-(1 - \delta - \omega)}$ of Theorem 7.1 is satisfied for $\delta  =\frac{1}{2}$ and $\forall \omega < \frac{1}{6}$ .\\
 Hence, taking  $\frac{1}{8}\leq \omega < \frac{1}{6}$ and $l > 3$ of assertion 2), we deduce that assumption $\omega > \frac{\delta}{l}$ of Theorem 7..1 is
satisfied. All the assumptions of Theorem 7.1 are satisfied. Hence formula in this proposition holds. \hfill { } $\square$\\
\begin{remark}

\n 1) the choice to use $\displaystyle{f_{\beta}(x)  = x^{\beta}(x - 1)^{\beta}(x - 2)^{\beta}}$ for $x \geq 3$ (see (2.7)) in this work instead of using $f_{1}(x)$  permits us to consider the function of order $m$, $\displaystyle{f_{\beta}(x)  = x^{\beta}(x - 1)^{\beta}... (x - m+1)^{\beta}}$ for $x \geq m$, and to give (in another paper) the number corrections sufficient for existence of finite formula of the regularized trace of the operator\\
\begin{equation}
\displaystyle{H = A^{*n}A^{n} + \sum_{i+j \leq m}c_{i,j}A^{*^{i}}A^{^{j}} }
\end{equation}
\n \n  where $c_{i,j}$ are complex numbers. If $m < 2(n -1)$, we can found that the value of $l$ is $\displaystyle{[\frac{m}{2n - m - 2}] + 1}$ (the brackets denote the integer part).\\

\n 2) The following of this section is devoted to proof of regularized formula  for $n_{m} = m$. The key point is the following theorem which gives an estimate of $\mid \mid H_{\mu,\lambda}(\tilde{G} - \sigma)^{-1} \mid\mid_{1}$.\\
\end{remark}
\begin{theorem}
 (estimatation of $\mid \mid H_{\mu,\lambda}(\tilde{G} - \sigma)^{-1} \mid\mid_{1}$)\\

\n For all $\alpha$ ; $\displaystyle{ 0 \leq \alpha < \frac{1}{6} }$, then there exists a sequence of numbers $\displaystyle{\tilde{r}_{m} = [\frac{1}{2}(\lambda_{m}^{\frac{1}{2} - \alpha} + \lambda_{m+1}^{\frac{1}{2} - \alpha}]^{\frac{1}{\frac{1}{2} - \alpha}}}$ such that\\

\n $\displaystyle{\lambda_{m} < \tilde{r}_{m} < \lambda_{m+1}}$ and $\displaystyle{\mid \mid H_{\mu,\lambda}(G - \sigma I)^{-1}\mid\mid_{1} = o(\tilde{r}_{m})^{-\alpha}}$  where  $\displaystyle{\mid\mid . \mid\mid_{1}}$ is the nuclear norm.\\
\end{theorem}
\n {\bf Proof }\\

\n To estimate $\mid \mid H_{\mu,\lambda}(G - \sigma I)^{-1}\mid\mid_{1}$, we note that\\
$\mid \mid H_{\mu,\lambda}(\tilde{G} - \sigma I)^{-1}\mid\mid_{1} \leq \mid \mid H_{\mu,\lambda}\tilde{G}^{-\frac{1}{2}}\mid\mid.\mid \mid \tilde{G}^{\frac{1}{2}}(\tilde{G} - \sigma I)^{-1}\mid\mid_{1} = \mid \mid H_{\mu,\lambda}\tilde{G}^{-\frac{1}{2}}\mid\mid \displaystyle{\sum_{n=1}^{\infty}\frac{\lambda_{n}^{\frac{1}{2}}}{\mid \lambda_{n} - \sigma\mid}}$\\

\n  $\displaystyle{\leq\mid \mid H_{\mu,\lambda}\tilde{G}^{-\frac{1}{2}}\mid\mid \sum_{n=1}^{\infty}\frac{\lambda_{n}^{\frac{1}{2}}}{\mid \lambda_{n} - \mid\sigma\mid\mid}}$\\

\n  By using the inequality $$\mid \frac{a^{\frac{1}{2}}b^{\alpha}(a^{\frac{1}{2} -\alpha} - b^{\frac{1}{2} -\alpha})}{a - b}\mid \leq 1 $$ we get\\

\n  $\mid \mid H_{\mu,\lambda}(\tilde{G} - \sigma I)^{-1}\mid\mid_{1} \leq \mid \mid H_{\mu,\lambda}G^{-\frac{1}{2}}\mid\mid $ $\displaystyle{\sum_{n=1}^{\infty}\frac{\lambda_{n}^{\frac{1}{2}}}{ \lambda_{n}^{\frac{1}{2}}\mid \sigma\mid^{\alpha}\mid \lambda_{n}^{\frac{1}{2}-\alpha} - \mid\sigma\mid^{\frac{1}{2}-\alpha}\mid}}$\\

 \n $=\frac{\mid \mid H_{\mu,\lambda}\tilde{G}^{-\frac{1}{2}}\mid\mid }{\mid\sigma\mid^{\alpha}} \displaystyle{\sum_{n=1}^{\infty}\frac{1}{\mid \lambda_{n}^{\frac{1}{2}-\alpha} - \mid\sigma\mid^{\frac{1}{2}-\alpha}\mid}}$\\

\n  From the relation $\lambda_{n}^{\frac{1}{2}-\alpha}\sim n^{3(\frac{1}{2}-\alpha)}$ then for $0 \leq \alpha < \frac{1}{6}$ the operator $\tilde{G}^{-(\frac{1}{2}-\alpha)}$  is nuclear and the series converges.\\

\n  By choosing $\tilde{r}_{m} = \displaystyle{[\frac{\lambda_{m}^{\frac{1}{2}-\alpha} + \lambda_{m+1}^{\frac{1}{2}-\alpha}}{2}]^{\frac{1}{\frac{1}{2}-\alpha}}}$, we deduce that\\
\begin{equation}
\displaystyle{\sum_{n=1}^{\infty}\frac{1}{\mid \lambda_{n}^{\frac{1}{2} - \alpha} - \tilde{r}_{m}^{\frac{1}{2} - \alpha}\mid}} \leq C
\end{equation}
\n   where $C$ does not depend on $m$\\

\n  The arbitrariness of the choice of $\alpha$ allows us to deduce that for $\mid \sigma \mid = \tilde{r}_{m}$ we have\\
\begin{equation}
\mid\mid H_{\mu,\lambda}(\tilde{G} - \sigma I)^{-1}\mid\mid_{1} = o(\frac{1}{\tilde{r}_{m}^{\alpha}}) \mbox{ as } m \rightarrow \infty 
  \end{equation}
\n In following subsection, we recall the key points of proof of theorem 7.1 applied to our operator to deduce the theorem 7.4 of this section.\\

\subsection{The key points of proof of Sadovnichii-Podol'skii's theorem applied to ``magic Gribov-Intissar'' operator}

\n Suppose that $\gamma_{m}$ are the circles of raddii $\{\tilde{r}_{m}\}$ centered at zero. As  $\mid\mid H_{\mu,\lambda}(\lambda^{''}\tilde{G} - \sigma I)^{-1}\mid\mid < 1$(see theorem 3.3  {\color{blue}[Aimar6 et al ]}) and $ H_{\mu,\lambda}(\lambda{''}\tilde{G} - \sigma I)^{-1}$ is nuclear, then the perturbation determinant \\
\begin{equation}
\displaystyle{D_{H_{\lambda'',0,\mu,\lambda}/\lambda^{''}\tilde{G}}(\sigma) = det[(H_{\lambda'',0,\mu,\lambda} - \sigma I)(\lambda^{''}\tilde{G} - \sigma I)^{-1}] = det[I + H_{\mu,\lambda}(\lambda^{''}G - \sigma I)^{-1}].}
\end{equation}
is defined.\\

\n and\\
\begin{equation}
\displaystyle{Tr[(\lambda^{''}\tilde{G} - \sigma I)^{-1} - (H_{\lambda'',0,\mu,\lambda} - \sigma I)^{-1}] = \frac{d}{d\sigma}(ln D_{H_{\lambda'',0,\mu,\lambda}/\lambda^{''}\tilde{G}}(\sigma))}
\end{equation}
\begin{equation}
\displaystyle{ln(D_{H_{\lambda'',0,\mu,\lambda}/\lambda^{''}\tilde{G}}(\sigma) = Tr(ln[I + H_{\mu,\lambda}(\lambda{''}\tilde{G} - \sigma I)^{-1}}]
\end{equation}
\n Most of the results on infinite determinants of Hilbert space operators can be founded in ( {\color{blue}[Gohberg1 et al ]}, {\bf {\color{blue}[Gohberg2 et al ]}},  {\color{blue}[Reed et al]},  {\color{blue}[Simon1]} or  {\color{blue}[Simon2]})\\

\n We denote the eigenvalues  of $H$ by $\{\sigma_{n}\}$ and we continue to denote  the eigenvalues  of $\lambda{''}\tilde{G}$ by $\{\lambda_{n}\}$ then we have\\
\begin{equation}
\displaystyle{\sum_{n=1}^{m}(\sigma_{n}-\lambda_{n})  = -\frac{1}{2i\pi}\int_{\gamma_{m}}\sigma Tr((\lambda{''}\tilde{G} + H_{\mu,\lambda} - \sigma I)^{-1} - (\lambda{''}\tilde{G} - \sigma I)^{-1})d\sigma}
\end{equation}
\n As $\mid\mid  H_{\mu,\lambda}(\lambda{''}\tilde{G} - \sigma I)^{-1})\mid\mid < 1$ then the form of a power series for the logarithm is legitimate and we deduce that\\
\begin{equation}
\displaystyle{\sum_{n=1}^{m}(\sigma_{n}-\lambda_{n})  =  -\frac{1}{2i\pi}\int_{\gamma_{m}}[Tr(H_{\mu,\lambda}R_{\sigma}^{0}) + Tr(\displaystyle{\sum_{k=2}^{\infty }\frac{(-1)^{k-1}}{k}(H_{\mu,\lambda}R_{\sigma}^{0})^{k}})]d\sigma.}
\end{equation}
\n where $R_{\sigma}^{0} = (\lambda^{''}\tilde{G} -\sigma I)^{-1}$\\

\n Since $\displaystyle{H_{\mu,\lambda}R_{\sigma}^{0}}$ is nuclear operator, its trace can be calculated as the matrix trace in orthonormal basis $\displaystyle{e_{n}(z) = \frac{z^{n}}{\sqrt{n!}}, n= 0, 1, ...}$ then\\

\n $\displaystyle{\frac{1}{2i\pi}\int_{\gamma_{m}}Tr(H_{\mu,\lambda}R_{\sigma}^{0})d\sigma = - \sum_{n=0}^{m}<H_{\mu,\lambda}e_{n}, e_{n}> = -\sum_{n=1}^{m}n\mu}$  then\\
\begin{equation}
\displaystyle{\sum_{n=1}^{m}(\sigma_{n}-\lambda_{n} - n\mu)} = \displaystyle{-\frac{1}{2i\pi}\int_{\gamma_{m}}Tr(\displaystyle{\sum_{k=2}^{\infty }}\frac{(-1)^{k-1}}{k}(H_{\mu,\lambda}R_{\sigma}^{0})^{k})}d\sigma.
\end{equation}
\n Now let us estimate the terms of this series for $k \geq 3$. Let $\tilde{r}_{m} = \displaystyle{[\frac{\lambda_{m}^{\frac{1}{2}-\alpha} + \lambda_{m+1}^{\frac{1}{2}-\alpha}}{2}]^{\frac{1}{\frac{1}{2}-\alpha}}}$, we follows the techniques of proof in  {\color{blue}[Sadovnichii2 et al]} of Sadovnichii and  Podol'skii theorem 1.1 by taking $\delta = \frac{1}{2}$ then the following inequalities are valid\\

\n i) the norm of  the operator $H_{\mu,\lambda}R_{\sigma}^{0}$ is estimated as\\
\begin{equation}
\left \{ \begin{array}{c}\displaystyle{_{_{_{_{_{\sigma \in \gamma_{_{m}} }}}}}\!\!\!\!\!\!\!\!\!\!\!\!\!\! max \mid\mid H_{\mu,\lambda}(\tilde{G} -\sigma I)^{-1}\mid\mid \leq
\mid\mid H_{\mu,\lambda}\tilde{G}^{\frac{-1}{2}}\mid\mid _{_{_{_{_{\sigma \in \gamma_{_{m}} }}}}}\!\!\!\!\!\!\!\!\!\!\!\!\!\! max \quad (\quad _{_{_{_{_{n}}}}}\!\!\!\!\!\!\! max \frac{\lambda_{n}^{\frac{1}{2}}}{\mid \lambda_{n} - \sigma\mid})}\\

\displaystyle{ \leq c\quad _{_{_{_{_{n}}}}}\!\!\!\!\!\!\!\! max \quad \frac{\lambda_{n}^{\frac{1}{2}}}{\mid \lambda_{n} - \tilde{r}_{m}\mid} }\\

\displaystyle{\leq \frac{c}{\tilde{r}_{m}^{\frac{1}{2}} - \lambda_{m}^{\frac{1}{2}}}}\\

c > 0, c = \mbox{ const}.
\end{array} \right.
\end{equation}
\n ii) The estimate $\displaystyle{\mid Tr(H_{\mu,\lambda}R_{\sigma}^{0})^{k}\mid \leq \mid\mid H_{\mu,\lambda}R_{\sigma}^{0}\mid\mid^{k-1}\mid\mid H_{\mu,\lambda}R_{\sigma}^{0}\mid\mid_{1}}$  implies  \\
\begin{equation}
\displaystyle{\int_{\gamma_{m}}\mid Tr( H_{\mu,\lambda}R_{\sigma}^{0})^{k}\mid \mid d\sigma \mid \leq max_{_{\sigma \in \gamma_{m}}}\mid\mid H_{\mu,\lambda}R_{\sigma}^{0}\mid\mid_{1} \int_{\gamma_{m}}\mid\mid H_{\mu,\lambda}R_{\sigma}^{0}\mid\mid^{k-1} \mid d\sigma \mid}
\end{equation}
\n iii)
\begin{equation}
\displaystyle{\int_{\gamma_{m}}\mid\mid H_{\mu,\lambda}R_{\sigma}^{0}\mid\mid^{k-1} \mid d\sigma \mid \leq \frac{c^{k}\lambda_{m}^{\frac{1}{2}}}{\mid \lambda_{m}^{\frac{1}{2}} - \tilde{r}_{m}^{\frac{1}{2}}\mid^{k-2}}}
\end{equation}
\n iv) the remainder of the series for $ l \geq 3$ satisfies the relation:\\

\n  $\mid \displaystyle{-\frac{1}{2i\pi}\int_{\gamma_{m}}Tr(\displaystyle{\sum_{k=l}^{\infty }\frac{(-1)^{k-1}}{k}(H_{\mu,\lambda}R_{\sigma}^{0})^{k}})}d\sigma \mid \displaystyle{\leq \mid\mid H_{\mu,\lambda}R_{\tilde{r}_{m}}^{0}\mid\mid_{1}}.\displaystyle{\int_{\gamma_{m}}\sum_{k=l}^{\infty }\mid\mid H_{\mu,\lambda}R_{\sigma}^{0}\mid\mid^{k-1}  \mid d\sigma\mid}$\\

\n  $\displaystyle{\leq \mid\mid H_{\mu,\lambda}R_{\tilde{r}_{m}}^{0}\mid\mid_{1}\lambda_{m}^{\frac{1}{2}}} \displaystyle{\sum_{k=l}^{\infty}(\frac{c}{ \tilde{r}_{m}^{\frac{1}{2}} - \lambda_{m}^{\frac{1}{2}}})^{k-2}}$\\
 $\leq \displaystyle{C\frac{\tilde{r}_{m}^{\frac{1}{2}}\mid\mid H_{\mu,\lambda}R_{\tilde{r}_{m}}^{0}\mid\mid_{1}}{(\tilde{r}_{m}^{\frac{1}{2}} - \lambda_{m}^{\frac{1}{2}})^{l-2}}} $\\
\begin{equation}
\leq \displaystyle{C\frac{\tilde{r}_{m}^{\frac{1}{2} -\alpha(l-2)}\mid\mid H_{\mu,\lambda}R_{\tilde{r}_{m}}^{0}\mid\mid_{1}}{(\tilde{a}_{m}^{\frac{1}{2}} - \lambda_{m}^{\frac{1}{2}})^{l-2}}} = o\displaystyle{(\tilde{r}_{m}^{\frac{1}{2} - \alpha(l-1)})}, \quad \quad m \rightarrow +\infty \quad \quad \quad \quad
\end{equation}

\n Then for all $ l \geq 2$, we get\\
\begin{equation}
\displaystyle{\lim\limits_{m \rightarrow \infty}\sum_{n=0}^{m}(\sigma_{n} - \lambda_{n}) + \frac{1}{2i\pi}\int_{\gamma_{m}} Tr[\sum_{k=1}^{l}\frac{(-1)^{k-1}}{k}[H_{\mu,\lambda}(\lambda{''}\tilde{G} - \sigma I)^{-1}]^{k}]d\sigma = o(\tilde{r}_{m}^{\frac{1}{2} - \alpha l})}
\end{equation}

\n The right-hand side of the above equality tends to zero as $m \rightarrow +\infty$  provided that $ \frac{1}{2} - \alpha l \leq 0$ and as $\alpha$ is subject to the constraint $\frac{1}{8}\leq \alpha < \frac{1}{6}$ then  $l > 3$. By taking $\omega = \alpha$ and $l = 4$ then the formula of above proposition  is true for $n_{m} = m$, this completes the proof of theorem 7.4\\

\n This work is concluded  by noting that\\

\n (i) If we consider  the tri-diagonal matrices  \\
\begin{equation}
\left \{ \begin{array} {c}(\tilde{H}_{\mu,\lambda}\phi)_{n} = \tilde{q}_{n}\phi_{n} + z(\beta_{n-1}\phi_{n-1} +\beta_{n}\phi_{n+1}), n\geq 1  \quad \quad \quad   \quad \quad \quad\\
 (\tilde{H}_{\mu,\lambda}\phi)_{0} = 0   \quad \quad  \quad \quad \quad \quad \quad \quad  \quad \quad \quad \quad  \quad \quad \quad \quad \quad  \quad \quad   \quad \quad  \\
   where  \quad \quad   \quad \quad  \quad \quad \quad \quad \quad \quad \quad \quad  \quad \quad \quad \quad  \quad \quad \quad \quad  \quad \quad \quad \quad \\
\tilde{q}_{n} = \lambda^{''}\lambda_{n} + \mu n, \lambda_{n} = n(n-1)(n-2), \beta_{n} =  n\sqrt{n+1} \, \mbox{and} \, z = i\lambda
\end{array} \right.
\end{equation}
\n  then we observe that the diagonal matrix denoted by $D$ majories the off-diagonal one denoted by $B$ in the sense of the following condition\\

\begin{equation}
\mid \tilde{q}_{n}\mid \rightarrow \infty \, \mbox{and}\, \frac{\beta_{n}^{2}}{\mid \tilde{q}_{n}\tilde{q}_{n+1}\mid} \rightarrow 0\, \mbox{ as} \, n \rightarrow \infty
\end{equation}
\n under  above condition, well-know methods of Perturbation Theory  {\color{blue}[Kato]} give information about the spectra  $\sigma (D + zB)$ which is discrete and $\sigma(D + zB) = \{\sigma_{n}(z)\}_{1}^{\infty}$, where for each $n$ $\sigma_{n}(z)$ is analytic function at least for small $\mid z \mid$, i.e., in the disk $\mid z\mid < r_{n}$ for some $r_{n} > 0$ \\

 \n  we can introduce the regularized trace \\
\begin{equation}
tr(\sigma_{n}(z)) = \sum_{n=1}^{\infty}(\sigma_{n}(z) - \tilde{q}_{n} )
\end{equation}
 \n as an entire function therefore we can give an evaluation of this regularized trace by following the method of  {\color{blue}[Janas5 et al]}. \\

\n  (ii) In {\color{blue}[Aimar7 et al]}, we studied the trace of the semigroup $e^{-tH}$ where \\

\n $ H = H_{\lambda^{''},0,\mu,\lambda} = \lambda^{''}A^{*3}A^{3} + \mu A^{*}A +i\lambda A^{*}(A + A^{*})A$\\

\n  Using the estimates obtained in  {\color{blue}[Intissar13]} see also the next section which give an approximation of this semigroup by the unperturbed semigroup $\displaystyle{e^{-t\lambda^{''}A^{*3}A^{3}}}$ in nuclear norm, we have given an asymptotic expansion of this trace as $t \rightarrow 0^{+} $. This study is main topic of the following section.\\ 

\section{ Regularized trace formula of the semigroups associated to Gribov-Intissar's operators in Bargmann representation}

\n In above section, we had regularized the operator  $\mathbb{H}_{\mu,\lambda}$ by $\lambda''\mathbb{G}$ where  $\mathbb{G} = A^{*3} A^{3}$, i.e we had considered  $\mathbb{H}_{\lambda''} = \lambda'' \mathbb{G} + \mathbb{H}_{\mu,\lambda}$ where $\lambda''$ is the {\it magic coupling} of Pomeron. In this case, we had established an exact relation between the degree of subordination of the non-self-adjoint perturbation operator $\mathbb{H}_{\mu,\lambda}$ to the unperturbed operator $\mathbb{G}$ and the number of corrections necessary for the existence of finite formula of the regularized trace.\\

\n The goal of work in this section consists to study the trace of the semigroup $e^{-t\mathbb{}H_{\lambda''}}$, in particular to give an asymptotic expansion of this trace as $t \rightarrow 0^{+} $.\\

\subsection{ Preliminary introduction}

\n Usually, quantum Hamiltonians are constructed as self-adjoint operators; for certain situations, however, non-self-adjoint Hamiltonians are also of importance. In particular, the reggeon field theory (as invented (1967) by V. Gribov {\bf{\color{blue}[Gribov1]}}) for the high energy behaviour of soft processes is governed by the non-self-adjoint Gribov operator\\

\n $\mathbb{H}_{\lambda'',\lambda',\mu,\lambda} = \lambda''A^{*3}A^{3} + \lambda'A^{*2}A^{2} + \mu A^{*}A +i\lambda A^{*}(A + A^{*})A $\\

\n where $A$ and its adjoint $A^{*}$ are annihilation and creation operators, respectively, satisfying the canonical commutation relations $[A, A^{*}] :=  AA^{*} - A^{*}A = \mathbb{I}$.\\

\n It is convenient to regard the above operators as acting on Bargmann space $\mathbb{B}$ {\bf{\color{blue}[Bargmann1]}} :\\

\n $ \mathcal{B} = \{\phi: \mathbb{C} \rightarrow \mathbb{C} \, entire \, ;  \displaystyle{\int_{\mathbb{C}}}\displaystyle{\mid \phi(z)\mid^{2}e^{-\mid z \mid ^{2}} dxdy} < \infty \}$ \\

\n The Bargmann space $\mathbb{\mathbb{B}}$ with the paring :\\

\n $ < \phi, \psi > = \displaystyle{\int_{\mathbb{C}}}\displaystyle{\phi(z)\bar{\psi(z)}e^{-\mid z \mid ^{2}} dxdy}$\\

\n is a Hilbert space and $\displaystyle{e_{n}(z) = \frac{z^n}{\sqrt{n!}}; n = 0, 1, ....}$ is an orthonormal basis in $\mathbb{B}$.\\

\n In this representation, the standard Bose annihilation and creation operators are defined by\\

\n $\left\{
  \begin{array}{ c }
A\phi(z) = \phi^{'}(z) \quad \quad \quad \quad \quad \quad \quad \quad\\
\quad\\
\mbox{with  maximal  domain }\quad \quad \quad \quad\\
\quad\\
D(A) = \{\phi \in \mathbb{B} ; \quad A\phi \in \mathbb{B}\}\quad \quad\\
\end{array}\right.$\\
\quad\\
and\\
\quad\\
\n $\left\{
  \begin{array}{ c }
A^{*}\phi(z) = z \phi(z) \quad \quad \quad \quad \quad \quad \quad \quad\\
\quad\\
\mbox{with  maximal  domain }\quad \quad \quad \quad\\
\quad\\
D(A^{*}) = \{\phi \in \mathbb{B} ; \quad A^{*}\phi \in \mathbb{B}\}\quad \quad\\
\end{array}\right.$\\
\quad\\

\n Notice that $D(A) = D(A^{*})$ and $D(A)\hookrightarrow \mathbb{B}$ is compact.\\

\n It has been established in {\bf{\color{blue}[Intissar14]}} that $ \mathbb{T}_{p} = A^{*^{p}}A^{p+1}$, $p \in \mathbb{N}$ is a chaotic operator and in  {\color{blue}[Intissar2 et al]}, some sufficient conditions have been given on the weight sequence of a weighted chaotic shift operator $\mathbb{T}$ on a Hilbert space such that $\mathbb{T} + \mathbb{T}^{*}$ to be chaotic (in the sense of Devaney) where $\mathbb{T}^{*}$ is its adjoint. In particular, in  {\color{blue}[Intissar2 et al]}, it has been shown  that $\mathbb{T}_{1} + \mathbb{T}_{1}^{*} = A^{*}(A + A^{*})A =:\mathbb{H}_{I}$ is chaotic operator. In Bargmann representation, there not exists semigroup generated by the operator $\mathbb{H}_{I}$, because $\sigma(\mathbb{H}_{I}) = \mathbb{C}$ where $\sigma(\mathbb{H}_{I})$ is the spectrum of $\mathbb{H}_{I}$.\\

\n Notice also that \\

\n For $\mathbb{H}_{\mu,\lambda} =  \mu A^{*}A + i\lambda A^{*}( A + A^{*})A $ with domain $D(H_{\mu, \lambda} = \{\phi \in \mathbb{B}\,  \mbox{such that} \, H_{\mu,\lambda}\phi \in \mathbb{B}\}$, we have:\\

\n i) For $\mu \neq 0$ and $\lambda \neq 0$,  $\mathbb{H}_{\mu,\lambda}$ is very far from normal and not only its self-adjoint and skew-adjoint parts do not commute but there is no inclusion in either way between their domains or with the domain of their commutator.\\

\n ii) For $\mu > 0$ and $\lambda \in \mathbb{R}$, the resolvent of $\mathbb{H}_{\mu,\lambda}$ is compact and $e^{-t\mathbb{H}_{\mu,\lambda}}$ is compact (see {\bf{\color{blue}[Intissar2]}}).\\

\n iii) For $\mu > 0$ and $\lambda \in \mathbb{R}$, it has been established in {\color{blue}[Aimar3 et al]} and  {\color{blue}[Aimar4 et al]} that the resolvent of $\mathbb{H}_{\mu,\lambda}$ belongs to the class $\mathcal{C}_{1 + \epsilon} \quad \forall$ $\epsilon > 0$.\\

\n We recalling that a compact operator $\mathbb{K}$ acting on a  complex Hilbert space $\mathcal{H}$ belongs to the Carlemen class $\mathcal{C}_{p}$ of order $p$ if $\displaystyle{\sum_{n=1}^{\infty} s_{n}^{p} < \infty}$, where $s_{n}$ are $s$-numbers of operator $\mathbb{K}$ i.e, the eigenvalues of the operator $\sqrt{\mathbb{K}^{*}\mathbb{K}}$.\\

\n In particular, the operator $\mathbb{K}$ is called nuclear operator if $\mathbb{K} \in \mathcal{C}_{1}$ and Hilbert-Shmidt operator if $\mathbb{K} \in \mathcal{C}_{2}$\\

\n For $ p \geq 1$ the value $\displaystyle{(\sum_{n=1}^{\infty} s_{n}^{p})^{\frac{1}{p}}}$ is a norm denoted by $\mid\mid . \mid\mid_{p} $ and for $ p = 1$ it is called nuclear norm or trace norm.\\

\n We can consult {\bf{\color{blue}[Gogberg2 et al]}} for a systematic study of operators of Carleman class  $\mathcal{C}_{p}$ of order $p$.\\

\n Let $ \mathbb{H}_{\lambda'} = \lambda'\mathbb{S} +  \mathbb{H}_{\mu,\lambda}$ where $\mathbb{S} = A^{*^{2}}A^{2}$, this operator is more regular that $\mathbb{H}_{\mu,\lambda}$, its semigroup $e^{-t\mathbb{H}_{\lambda'}}$ is analytic and it has been established in {\bf{\color{blue}[Intissar 13]}} that the convergence of the usual Trotter product formula for $\mathbb{H}_{\lambda'}$ is of classical type and can be lifted to trace-norm convergence.\\

\n Morever, there exist $t_{0} > 0 $ and $C > 0$ such that :\\

\n $\mid\mid (e^{-\frac{t}{n}\lambda'\mathbb{S}} e^{-\frac{t}{n}\mathbb{H}_{\mu,\lambda}})^{n} - e^{-t\mathbb{H}_{\lambda'}}\mid\mid_{1} \leq$ $C\frac{log n}{n}$ , $ n = 2, 3, ..... $ $\forall$ $t \geq t_{0}$.\\

\n Now, if $\mathbb{H}_{\mu,\lambda}$ is regularized by $\lambda''\mathbb{G}$ where $\mathbb{G} = A^{*^{3}}A^{3}$ where $\lambda''$ is the magic coupling of Pomeron, we can consider:\\
\begin{equation}
 \mathbb{H}_{\lambda^{''}} = \lambda''A^{*^{3}}A^{3} + \mu A^{*}A + i\lambda A^{*}(A  + A^{*})A
\end{equation}
\n i.e\\

\n $ \left\{\begin{array}{ c }
\mathbb{H}_{\lambda^{''}}\phi(z)= \lambda^{''} z^{3}\phi^{'''}(z)+i\lambda z\phi^{''}(z) + (i\lambda z^{2} + \mu z)\phi^{'}(z)\quad \quad \\
\quad\\
\mbox{with maximal  domain }\quad\quad \quad \quad \quad\quad \quad \quad \quad \quad \quad\quad \quad \quad \quad\\
\quad\\
D(\mathbb{H}_{\lambda^{''}}) = \{\phi \in \mathbb{B}; \quad \mathbb{H}_{\lambda^{''}}\phi \in \mathbb{B}\}\quad \quad \quad \quad \quad \quad \quad \quad \quad \quad\\
\end{array}\right.$
\quad\\

\n  In  {\color{blue}[Aimar6 et al]} (see theorem 3.3, p. 595), it was shown that the spectrum of $H_{\lambda''}$ is discrete and that the system of generalized eigenvectors of this operator is an unconditional basis in Bargmann space $\mathcal{B}$.\\

\n Recently in {\color{blue}[Intissar 19]}, it was established a regularized trace formula for $\mathbb{H}_{\lambda''}$.\\

\n More precisely, it was shown the following result:\\
\begin{theorem}
(see above section)\\

\n  Let $\mathbb{\mathbb{B}}$ be the Bargmann space and $\mathbb{H}_{\lambda^{''}} = \lambda^{''}\mathbb{G} + \mathbb{H}_{\mu,\lambda}$ acting on $\mathbb{\mathbb{B}}$\\

\n  where\\

\n - $ \mathbb{G} = A^{*3}A^{3}$ and $ \mathbb{H}_{\mu,\lambda} = \mu A^{*}A + i\lambda A^{*}(A + A^{*})A$\\

\n - $A$ and $A^{*}$ are the standard Bose annihilation and creation operators satisfying \\

\n the commutation relation $[A , A^{*}] = \mathbb{I}$.\\

\n Then \\

\n There exists an increasing sequence of radius $r_{n}$ such that $r_{n} \rightarrow \infty$ as $ n \rightarrow \infty$ \\

\n and \\

\n  $\displaystyle{Lim \sum_{k=0}^{n}(\sigma_{k} - \lambda''\lambda_{k}) =}$\\
\begin{equation}
 \displaystyle{\lim\limits_{n \rightarrow \infty}\frac{1}{2i\pi}\int_{\gamma_{n}} tr[\sum_{j=1}^{4}\frac{(-1)^{j-1}}{j}[\mathbb{H}_{\mu,\lambda}(\lambda{''}\mathbb{G} -\sigma \mathbb{I})^{-1}]^{j}]d\sigma}
\end{equation}

\n where\\

\n -  $tr(\mathbb{A})$ denote the trace of operator $\mathbb{A}$.\\

\n  - $\sigma_{k}$ are the eigenvalues of the operator $ \mathbb{H}_{\lambda^{''}} = \lambda{''}\mathbb{G} + \mathbb{H}_{\mu,\lambda}$\\

\n  - $\lambda_{k} = k(k-1)(k-2)$ are the eigenvalues of the operator of $ \mathbb{G} $ associated to eigenvectors $e_{k}(z)$\\

\n  - $ (\lambda{''}\mathbb{G} - \sigma \mathbb{I})^{-1}$  is the resolvent of the operator $\lambda{''}\mathbb{G}$ \\

\n  and\\

\n  - $\gamma_{n}$ is the circle of radius $r_{n}$ centered at zero in complex plane.\\
\end{theorem}

\n The goal of this section consists to study the trace of the semigroup $e^{-t\mathbb{H}_{\lambda''}}$, in particular, to \\

\n give an asymptotic expansion of this trace as $t \rightarrow 0^{+} $.\\

\n Our procedure consists to prove that\\

\n  i) The semigroup $e^{-t\mathbb{G}}$ generated by the operator $\mathbb{G}$ is analytic and nuclear (Gibbs analytic semigroup).\\

\n  ii) \begin{equation} \left \{ \begin{array} {c} \forall \epsilon > 0, \exists C_{\epsilon} > 0  ; \mid\mid \mathbb{H}_{\mu,\lambda}\phi\mid\mid \leq \epsilon\mid\mid \mathbb{G}\phi\mid\mid + C_{\epsilon}\mid\mid\phi\mid\mid \forall \phi \in D(\mathbb{G})\\
\quad\\
\mbox{where} \, D(\mathbb{G}) = \{\phi \in \mathbb{B}, \mathbb{G}\phi \in \mathbb{B}\}
\end{array} \right.
\end{equation}

\n  iii) The operator $\mathbb{H}_{\mu,\lambda}\mathbb{G}^{-\delta}$ is bounded \quad $\forall$ $\delta \geq \frac{1}{2}$.\\

\n  Now, with the aid of the results of Angelscu et al {\bf{\color{blue}[Angelscu et al ]}} or of Zagrebnov {\color{blue}[Zagrebnov1, Zagrebnov2 ]} with Ginibre-Gruber inequality {\color{blue}[Ginibre]}, it easy to prove that the series of general term $S_{k}(t)_{k\in \mathbb{N}}$ defined by:\\

 \n $S_{0}(t)\phi = e^{-t\lambda''\mathbb{G}}\phi$\\

\n and\\

\n $S_{k+1}(t)\phi = - \displaystyle{ \int_{0}^{t}e^{-(t-s)\lambda''\mathbb{G}} \mathbb{H}_{\mu,\lambda} S_{k}(t)\phi ds}$\\

\n converges to $e^{-t\mathbb{H}_{\lambda''}}$ (nuclear norm).\\

\n By using the properties i), ii) and iii) we get:\\
\begin{equation}
\n {\small \displaystyle{\mid\mid e^{-t\mathbb{H}_{\lambda''}} - e^{-t\lambda''\mathbb{G}} \mid\mid_{1} = t\mid\mid e^{-t\lambda''\mathbb{G}} \mathbb{H}_{\mu,\lambda}\mid\mid_{1} + } \displaystyle{\mid\mid (\lambda''\mathbb{G})^{\delta} e^{-\frac{t}{3}\lambda''\mathbb{G} }\mid\mid_{1}O(t^{2})}; \quad t \rightarrow 0^{+} }
\end{equation}
\n To establish the above results, we  begin by given :\\

\n - In  subsection 8.2,  some spectral properties of semigroups $e^{-t\lambda''\mathbb{G}}$ and $e^{-t(\lambda''\mathbb{G} + \mathbb{H}_{\mu,\lambda})} $ in $\mathcal{C}_{p}$.\\

\n  - In subsection 8.3, we give the proof of the above formula (8.4) for the trace of $e^{-t\mathbb{H}_{\lambda''}}$ as $t \rightarrow 0^{+} $.\\

\subsection{Some spectral properties of semigroups $e^{-t\lambda''\mathbb{G}}$ and $e^{-t(\lambda''\mathbb{G} + \mathbb{H}_{\mu,\lambda})}$ in  Carleman  spaces $\mathcal{C}_{p}$ }

\n We begin by given some elementary spectral properties of $\mathbb{G}$ and $\mathbb{H}_{\lambda{''}}$ :\\
\begin{lemma}

\n 1) The operator $\mathbb{G}$ is self adjoint with compact resolvent.\\

\n 2) The eigenvalues of $\mathbb{G}$ are $\lambda_{n} = n(n-1)(n-2)$ for $ n \geq 0 $ associated to eigenvectors $\displaystyle{e_{n}(z) = \frac{z^{n}}{\sqrt{n!}}}$.\\

\n 3) $Lim \mid\mid \mathbb{G}e^{-t\mathbb{G}}\mid\mid = \frac{1}{e}$ as $ t \rightarrow 0.$\\

\n 4) the resolvent of $\mathbb{G}$ belongs to $ \mathcal{C}_{p}$ \quad $\forall p > \frac{1}{3}$.\\

\n 5) $ e^{-t\mathbb{G}}$ is nuclear semigroup and $\mid\mid e^{-t\mathbb{G}}\mid\mid_{1} \leq Ct^{-\frac{1}{3}}.$\\

\n where the constant $C$ does not depend on $t$\\

\n 6) $ e^{-t\mathbb{G}} \in \mathcal{C}_{p} $ \quad $\forall p > 0$.\\
\end{lemma}
\n {\bf proof}\\

\n 1) It is evident that $\mathbb{G}$ is self-adjoint and its resolvent is compact follows easily from the following Rellich theorem : \\
\begin{theorem}
- (see {\color{blue}{\bf[Fr\'echet]}}, p. 386)\\

\n  Let $\mathbb{B}$ be a self adjoint operator in a complex Hilbert space $\mathcal{H}$ such that \\ $< \mathbb{B}\phi, \phi> \quad \geq\quad <\phi, \phi>, \phi \in D(\mathbb{B})$, where $D(\mathbb{B}) = \{\phi \in \mathcal{H}; \mathbb{B}\phi \in \mathcal{H}\}$. \\

\n Then $\mathbb{B}$ is discrete if and only if $\{\phi \in D(\mathbb{B})$; $ < \mathbb{B}\phi, \phi) \leq 1 \}$ is pre-compact.\\
\end{theorem}
\n that the operator $\mathbb{G}$ is discrete. \hfill { } $\square$\\

\n Also, we can prove the above result by using the following observation: \\

\n Since $D(A)$ is compactly embedded in Bargmann space $\mathbb{B}$ and $D(A^{*3}A^{3}) \hookrightarrow D(A)$ is continuous then $D(A^{*3}A^{3})$ is compactly embedded in Bargmann space $\mathbb{B}$.\\

\n Since $ (A^{*3}A^{3} + \mathbb{I}) $ is invertible  then the operator $\mathbb{G}$ is discrete.\\

\n 2) It is obvious.\\

\n 3) As $\mathbb{G}e^{-t\mathbb{G}}e_{n} = n(n-1)(n-2)e^{-tn(n-1)(n-2)}e_{n}$ we get\\

\n  $\mid\mid \mathbb{G}e^{-t\mathbb{G}}\mid\mid  = \frac{1}{t}\quad_{_{n}}\!\!\!\!\!Sup\quad tn(n-1)(n-2)e^{-tn(n-1)(n-2)} = \frac{1}{e}$ this implies \\
\begin{equation}
\n \lim\limits_{t \rightarrow 0} \mid\mid \mathbb{G}e^{-t\mathbb{G}}\mid\mid = \frac{1}{e}
\end{equation}
\n 4) As $\mathbb{G}$ is self adjoint and its eigenvalues are :\\
\begin{equation}
\lambda_{n} = n(n-1)(n-2) \sim n^{3}
\end{equation}
\n this implies that the series of general term  $\frac{1}{n^{3p}}$ converges $\forall p > \frac{1}{3}$ and consequently, the resolvent of $\mathbb{G}$ belongs to Carleman class $ \mathcal{C}_{p}$ $\forall p > \frac{1}{3}$.\\

\n 5) $e^{-t\mathbb{G}}$ is self adjoint and it is of trace class, because the series of general term $e^{-t\lambda_{n}}$ converges $\forall t > 0 $.\\

\n Now let $ x \in [0, +\infty[$ and if $t \in [0, +\infty[ $, consider the function $f(x) = e^{-tx^{3}}$ and its derivative $f'(x) = -3tx^{2}e^{-tx^{3}}$  which non positive then the function $f(x)$ is decreasing and we have \\

\n $\displaystyle{\sum_{n=1}^{\infty}e^{-tn^{3}} \leq \int_{0}^{\infty}e^{-tx^{3}}dx}$ \\

\n By a change of variable in the above integral, we obtain that\\

\n $\displaystyle{\int_{0}^{\infty}e^{-tx^{3}}dx = Ct^{\frac{-1}{3}}}$\\

\n where the constant $C$ does not depend on $t$\\

\n This implies that\\

\n $\mid\mid e^{-t\mathbb{G}} \mid\mid_{1} \leq \displaystyle{\sum_{n=1}^{\infty}e^{-t^{3}n} \leq  Ct^{\frac{-1}{3}}}$\\

\n 6) $e^{-t\mathbb{G}}$ is Carleman class $\mathcal{C}_{p}$ $\forall p > 0 $, because the series of general term $e^{-tp\lambda_{n}}$  converges $\forall t > 0$ and $ \forall p > 0$.\\
\begin{remark}

\n  We can derive the property 6) from the fact that $e^{-t\mathbb{G}} \in \mathcal{C}_{1}$. In fact, since $\mathcal{C}_{1}\subset \mathcal{C}_{p}$ then $e^{-t\mathbb{G}} \in \mathcal{C}_{p}$ $\forall p > 1$.\\

\n Now if $ p < 1$, we choose an integer $ n$ such that $\frac{1}{n} < p$ , then  $\mathcal{C}_{\frac{1}{n}} \subset \mathcal{C}_{p}$.\\

\n Since $e^{-\frac{t}{n}\mathbb{G}} \in \mathcal{C}_{1}$ $\forall$ $\tau = \frac{t}{n} > 0$ then $(e^{-\frac{t}{n}\mathbb{G}})^{n} \in \mathcal{C}_{\frac{1}{n}}$ \\

\n and we obtain that $ e^{-t\mathbb{G}} \in \mathcal{C}_{p} $ $\forall$ $ p > 0$.\\
\end{remark}
\n From the above remark, we are now ready to prove following lemma :\\
\begin{lemma}

\n  1) Let $\mathbb{T}(t)$ be a semigroup on a complex Hilbert space $\mathcal{H}$. We assume that $\mathbb{T}(t)$ is selfadjoint and there exists $p_{0} > 0$ such that $\mathbb{T}(t) \in \mathbb{C}_{p_{0}}$\quad $\forall$ $ t > 0$.\\

\n  Then $\mathbb{T}(t) \in \mathcal{C}_{p}$ \quad $\forall$ $ t > 0$, \quad $\forall$ $ p > 0$.\\

\n  2) Let $\mathbb{T}(t)$ be a semigroup on a complex Hilbert space $\mathcal{H}$. We assume that there exists $p_{0} > 0$ \\

\n such that $\mathbb{T}(t) \in \mathcal{C}_{p_{0}}$ \quad $\forall$ $ t > 0$. Then \\

\n $\mathbb{T}(t) \in \mathcal{C}_{p}$ \quad $\forall$ $ t > 0$, \quad $\forall$ $ p > 0$.\\
\end{lemma}
\n {\bf Proof}\\

\n 1) let $ t = \frac{\tau}{p_{0}} $ and  $\hat{\mathbb{T}}(\tau) = \mathbb{T}(\frac{\tau}{p_{0}})$ then $\mathbb{T}(t) \in \mathcal{C}_{p_{0}}$ if and only if $\hat{\mathbb{T}}(\tau) \in  \mathcal{C}_{1}$. Since  $\mathcal{C}_{1}\subset \mathcal{C}_{p}$ for all $ p > 1$, it follows that $\hat{\mathbb{T}}(\tau) \in \mathcal{C}_{p}$ for all $ p > 1$.\\

\n  Let $ 0 < p < 1$, Since  $\hat{\mathbb{T}}(\tau)$ is self adjoint, then for each fixed  $\tau > 0 $, there exist a positive
decreasing sequence  $s_{n} \in l_{1}$ and an orthonormal sequence $e_{n}$ such that $\hat{\mathbb{T}}(\tau) = \displaystyle{\sum_{n=0}^{\infty}s_{n}e_{n}\otimes e_{n}}$.\\

\n  Let $p = \frac{1}{\delta}$ with $\delta >1$  and $ n$ such that $ n > \delta$, then for fixed $\frac{t}{n}$ , there exist  a positive decreasing sequence $r_{k} \in l_{1}$ and an orthonormal sequence $e_{k}$ such tat $\mathbb{T}(\frac{t}{n}) = \displaystyle{\sum_{k=0}^{\infty}r_{k}e_{k}\otimes e_{k}}$.\\

\n  It follows that  $\mathbb{T}(t) = [\mathbb{T}(\frac{t}{n})] = \displaystyle{\sum_{k=0}^{\infty}r_{k}^{n}e_{k}\otimes e_{k}}$.\\

\n Since, $r_{k} \in l_{1}$ then $r_{k}^{n} \in l_{\frac{1}{n}} \subset l_{p}$,i.e.  $\mathbb{T}(t) \in \mathcal{C}_{p}$.\\

\n 2) If $ p > p_{0}$ then  $\mathbb{T}(t) \in \mathcal{C}_{p}$ for all $ t > 0$ and all $ p > p_{0}$ because $\mathcal{C}_{p_{0}} \subset \mathcal{C}_{p}$ for all $ p > p_{0}$.\\

\n If $ p < p_{0}$, we choose an integer $ n $ such that $\frac{p_{0}}{n} < p$, since $\mathbb{T}(t) \in \mathcal{C}_{p_{0}}$ for all $ t > 0$ then $\mathbb{T}(\frac{t}{n}) \in \mathcal{C}_{p_{0}}$\\

\n  and consequently $ \mathbb{T}(t) = [\mathbb{T}(\frac{t}{n}]^{n} \in \mathcal{C}_{\frac{p_{0}}{n}} \subset \mathcal{C}_{p}$. \hfill { } $\square$\\
\begin{remark}

\n  Let $\mathbb{T}$ be a compact operator, we assume that $\mathbb{T}$  is positif and $\mid\mid \mathbb{T} \mid\mid \leq 1$ then there exist a positive sequence $ 0 < s_{n} < 1$ and an orthonormal sequence $e_{n}$ such that $\mathbb{T} = \displaystyle{\sum_{n=0}^{\infty}s_{n}e_{n}\otimes e_{n}}$.\\

\n Let $\mathbb{T}(t) = \displaystyle{\sum_{n=0}^{\infty}s_{n}^{t}e_{n}\otimes e_{n}}$, if we choose $\displaystyle{s_{n} \in \bigcap_{p=0}^{\infty}l_{p}}$ where $l_{p}$ is the space of $p$-summable sequences, then $\mathbb{T}(t) \in \mathcal{C}_{p}$ for all $ t > 0$ and all $ p > 0$, it follows that $\mid\mid \mathbb{T}(t) \mid\mid_{p} = \displaystyle{(\sum_{n=0}^{\infty}s_{n}^{tp})^{\frac{1}{p}}}$\\

\n  For $s < t $ we have $s_{n}^{t} < s_{n}^{s}$ then $\mid\mid \mathbb{T}(t) \mid\mid_{p}  < \mid\mid \mathbb{T}(s) \mid\mid_{p}$.\\

\n  By using the Beppo-Levi's theorem, we deduce that  $\mid\mid \mathbb{T}(t) \mid\mid_{p} \rightarrow \infty $ as $ t \rightarrow 0$.\\
\end{remark}

\begin{proposition}

 \n Let $\mathbb{B}_{0} = \{\phi \in \mathbb{B}; \phi(0) = 0\}$, $\mathbb{P}_{0} = \{ p \in \mathbb{P}; p(0) = 0\}$  where $ \mathbb{P}$ is the space of polynomials\\

 \n and \\

\n  $\mathbb{H}_{\lambda''}^{min}$ with domain $D_{min}(\mathbb{H}_{\lambda''})$ is the closure of the restriction of $\mathbb{H}_{\lambda''}$ on $\mathbb{P}_{0}$.\\

\n  Then we have:\\

\n {\bf(a)} $\forall$ $\epsilon > 0 $, there exists $C_{\epsilon} > 0 $ such that :\\
\begin{equation}
\mid\mid \mathbb{H}_{\mu,\lambda}\phi\mid\mid \leq  \epsilon \mid\mid \mathbb{G}\phi\mid\mid + C_{\epsilon}\mid\mid \phi\mid\mid \, \forall \, \phi \in D(\mathbb{G})
\end{equation}

\n {\bf(b)} $\forall$ $\epsilon > 0 $, there exists $C_{\epsilon} > 0 $ such that :\\
\begin{equation}
\mid < \mathbb{H}_{\mu,\lambda}\phi, \phi >\mid \leq  \epsilon < \mathbb{G}\phi, \phi > + C_{\epsilon}\mid\mid \phi\mid\mid^{2}\, \forall\,  \phi \in D(\mathbb{G})
\end{equation}
\n {\bf(c)} For $\lambda'' > 0$ and $\forall$  $\epsilon $ ; $0 < \epsilon < \lambda'' $, there exists $C_{\epsilon} > 0 $ such that :\\
\begin{equation}
\Re e < \mathbb{H}_{\lambda''}\phi, \phi > \geq ( \lambda'' - \epsilon )< \mathbb{G}\phi, \phi > -C_{\epsilon}\mid\mid \phi\mid\mid^{2} \,\forall \, \phi \in D(\mathbb{G})
\end{equation}

\n in particular, the range of $ \mathbb{H}_{\lambda''}$ is closed.\\

\n {\bf(d)} For $\lambda^{''} \geq 0$ and $\mu > 0$, $\mathbb{H}_{\lambda{''}}$ is accretive and maximal.\\

\n {\bf (e)} $D_{max}(\mathbb{H}_{\lambda{''}}) = D_{min}(\mathbb{H}_{\lambda{''}}) = D(\mathbb{G})$.\\

\n {\bf (f)} $ - \mathbb{H}_{\lambda{''}}$ generates an analytic semigroup $e^{-t\mathbb{H}_{\lambda{''}}}$, $ t > 0$.\\

\n {\bf (g)} For $\lambda^{''} \neq 0$, the resolvent of $\mathbb{H}_{\lambda{''}}$ belongs to Carleman class $\mathcal{C}_{p}$ for all $ p > \frac{1}{3}$.\\
\end{proposition}
\n {\bf Proof}\\

\n a) Let $\phi \in \mathbb{B}_{0}$, we have $\phi(z) = \displaystyle{\sum_{n=1}^{\infty}a_{n}e_{n}(z)}$. Then \\

\n $\mathbb{H}_{\mu,\lambda}\phi(z) = \displaystyle{\sum_{n=1}^{\infty}[\mu na_{n} + i\lambda (n-1)\sqrt{n}a_{n-1} + i\lambda n\sqrt{n+1}a_{n+1}] e_{n}(z)}$ \\

\n and\\

\n $\mathbb{G}\phi(z) = \displaystyle{\sum_{n=1}^{\infty}n(n-1)(n-2)a_{n}e_{n}(z)}$\\

\n we remark that there exists $C > 0$ such that $\mid \mid \mathbb{H}_{\mu,\lambda}\phi\mid \mid^{2} \leq C \displaystyle{\sum_{n=1}^{\infty}n^{3} \mid a_{n} \mid^{2}}$\\

\n and\\

\n $\mid \mid \mathbb{G}\phi\mid \mid^{2} \geq \frac{1}{36} \displaystyle{\sum_{n=1}^{\infty}n^{6}\mid a_{n}\mid^{2}}$.\\

\n Now, by using the Young's inequality, we get:\\

\n $\forall \epsilon > 0$ , $k^{3} \leq \epsilon k^{6} + \frac{1}{\epsilon}$ $\forall k \in \mathbb{N}$.\\

\n this implies that :\\

\n $\forall$ $\epsilon > 0 $, there exists $C_{\epsilon} > 0 $ such that $\mid\mid \mathbb{H}_{\mu,\lambda}\phi\mid\mid \leq  \epsilon \mid\mid \mathbb{G}\phi\mid\mid + C_{\epsilon}\mid\mid \phi\mid\mid$ for all $\phi \in D(\mathbb{G})$.\\

\n b) $ < \mathbb{H}_{\mu,\lambda}\phi, \phi > = \mu \mid\mid A\phi\mid\mid^{2} + i\lambda< A^{2}\phi, A\phi > + i\lambda< A\phi, A^{2}\phi >$ for all $\phi \in D(\mathbb{H}_{\mu,\lambda})$.\\

\n Then\\

\n $\mid < \mathbb{H}_{\mu,\lambda}\phi, \phi >\mid \leq \mu \mid\mid A\phi\mid\mid^{2} + 2\mid \lambda \mid \mid \mid A\phi\mid\mid.\mid\mid A^{2}\phi \mid\mid$ pour tout $\phi \in D(\mathbb{H}_{\mu,\lambda})$.\\

\n With the aid of following inequalities:\\

\n i) $\forall$ $\epsilon_{1} > 0 $ $\mid \mid A\phi\mid\mid.\mid\mid A^{2}\phi \mid\mid \leq  \epsilon_{1}\mid\mid A^{2}\phi \mid\mid^{2} + \frac{1}{\epsilon_{1}}\mid\mid A\phi \mid\mid^{2}.$\\

\n ii) $\forall$ $\epsilon_{2} > 0$ there exists $C_{\epsilon_{2}} > 0 $ such that $\mid\mid A\phi \mid\mid^{2}\leq \epsilon_{2}\mid\mid A^{3}\phi \mid\mid^{2} + C_{\epsilon_{2}}\mid\mid \phi\mid\mid^{2}$\\

\n iii) $\mid\mid A^{2}\phi \mid\mid^{2} \leq \mid\mid A^{3}\phi \mid\mid^{2}$\\

\n we obtain:\\

\n $\forall$ $\epsilon > 0 $, there $C_{\epsilon} > 0 $ such that $\mid < \mathbb{H}_{\mu,\lambda}\phi, \phi >\mid \leq  \epsilon < \mathbb{G}\phi, \phi > + C_{\epsilon}\mid\mid \phi\mid\mid^{2}$ for all $\phi \in D(\mathbb{G})$.\\

\n c) Since $Re < \mathbb{H}_{\lambda''}\phi, \phi > = \lambda''< \mathbb{G}\phi, \phi > + < \mathbb{H}_{\mu,\lambda}\phi, \phi >$ and $ \lambda'' > 0$\\

\n Then \\

\n $Re < \mathbb{H}_{\lambda''}\phi, \phi > \geq \lambda''< \mathbb{G}\phi, \phi > - \mid < \mathbb{H}_{\mu,\lambda}\phi, \phi >\mid$.\\

\n By using the above property we get:\\

\n  $Re < \mathbb{H}_{\lambda''}\phi, \phi > \geq \lambda''< \mathbb{G}\phi, \phi > - \epsilon< \mathbb{G}\phi, \phi > - C_{\epsilon}\mid\mid \phi\mid\mid^{2} = (\lambda'' -\epsilon) < \mathbb{G}\phi, \phi > - C_{\epsilon}\mid\mid \phi\mid\mid^{2}$\\

\n We choose $ 0 < \epsilon < \lambda'' $ to deduce that\\

\n  $Re < \mathbb{H}_{\lambda''}\phi, \phi > \geq - C_{\epsilon}\mid\mid \phi\mid\mid^{2}$\\

\n Consequently the range of $ \mathbb{H}_{\lambda''}$ is closed.\\

\n d) Since $A^{*}(A + A^{*})A $ is symmetric operator then \\

\n $Re < \mathbb{H}_{\lambda''}\phi,\phi > = \lambda''\mid \mid a^{3}\phi\mid\mid^{2} + \mu\mid\mid a\phi\mid\mid^{2} \geq \mu\mid\mid \phi\mid\mid^{2}$, $\forall \phi \in D_{min}(\mathbb{H}_{\lambda''})$ \\

\n Now for $\lambda''\geq 0$ and  $\mu >0$ we deduce that :\\

\n $Re < \mathbb{H}_{\lambda''}\phi,\phi > \geq \mu\mid\mid \phi\mid\mid^{2}$,$ \forall \phi \in D_{min}(\mathbb{H}_{\lambda''})$ .\\

\n This inequality will be not verified if we kept constant functions in Bargmann space $\mathbb{B}$.\\

\n Now, we would like to show that there exists  $\beta_{0} \in \mathbb{R}$ ; $\mathbb{H}_{\lambda''} + \beta_{0}\mathbb{I}$ is invertible.\\

\n We rewrite $\mathbb{H}_{\lambda''}$ in the following form:\\

\n $\mathbb{H}_{\lambda''} = \lambda''( \mathbb{G} + \frac{1}{\lambda''}\mathbb{H}_{\mu,\lambda}$ et $\mathbb{G} + \beta \mathbb{I} + \frac{1}{\lambda''}\mathbb{H}_{\mu,\lambda} = [\mathbb{I} + \frac{1}{\lambda''}\mathbb{H}_{\mu,\lambda}(\mathbb{G} + \beta \mathbb{I} )^{-1}](\mathbb{G} + \beta \mathbb{I})$.\\

\n Using the property a) to get:\\

\n $ \mid\mid \frac{1}{\lambda''}\mathbb{H}_{\mu,\lambda}(\mathbb{G} + \beta \mathbb{I} )^{-1}\psi\mid\mid \leq \epsilon \mid\mid \mathbb{G} (\mathbb{G} + \beta \mathbb{I} )^{-1}\psi\mid\mid + C_{\epsilon}\mid\mid (\mathbb{G} + \beta \mathbb{I} )^{-1}\psi\mid\mid$\\

\n $\leq \epsilon \mid\mid (\mathbb{G} +\beta \mathbb{I} - \beta \mathbb{I}) (\mathbb{G} + \beta \mathbb{I} )^{-1}\psi\mid\mid + C_{\epsilon}\mid\mid (\mathbb{G} + \beta \mathbb{I} )^{-1}\psi\mid\mid$\\

\n $ \leq \epsilon \mid\mid \psi \mid\mid + (\epsilon \beta + C_{\epsilon})\mid\mid (\mathbb{G} + \beta \mathbb{I} )^{-1}\psi\mid\mid$\\

\n and as $\mid\mid (\mathbb{G} + \beta \mathbb{I} )^{-1}\mid\mid \leq \frac{1}{\beta}$ then \\

\n $ \mid\mid \frac{1}{\lambda''}\mathbb{H}_{\mu,\lambda}(\mathbb{G} + \beta \mathbb{I} )^{-1}\psi\mid\mid \leq (2\epsilon + \frac{C_{\epsilon}}{\beta})$. \\

\n Now, we choose $ 0 < \epsilon < \frac{1}{2}$ and $\beta > \frac{C_{\epsilon}}{1-2\epsilon}$ to obtain :\\

\n $ \mid\mid \frac{1}{\lambda''}\mathbb{H}_{\mu,\lambda}(\mathbb{G} + \beta \mathbb{I} )^{-1}\mid\mid < 1$.\\

\n this implies $\mathbb{H}_{\lambda''}^{min} + \beta \mathbb{I}$ is invertible.\\

\n e) We begin to show that  $D_{max}(\mathbb{H}_{\lambda{''}}) = D_{min}(\mathbb{H}_{\lambda{''}})$.\\

\n First,  $D_{min}(\mathbb{H}_{\lambda{''}}) \subset  D_{max}(\mathbb{H}_{\lambda{''}})$ is trivial.\\

\n To show that $D_{max}(\mathbb{H}_{\lambda{''}}) \subset D_{min}(\mathbb{H}_{\lambda{''}})$, $\phi \in D_{max}(\mathbb{H}_{\lambda{''}})$ then $ (\mathbb{H}_{\lambda{''}} + \beta \mathbb{I})\phi \in \mathbb{E}_{0}$ for all $\beta$.\\

\n Since there exists  $\beta_{0} \in \mathbb{R}$ such that $\mathbb{H}_{\lambda''}^{min} + \beta_{0} \mathbb{I}$ is invertible of $D_{min}(\mathbb{H}_{\lambda{''}})$ on $\mathbb{E}_{0}$, then there exists
$\phi_{1} \in D_{min}(\mathbb{H}_{\lambda{''}})$ such that :\\

\n $(\mathbb{H}_{\lambda''} + \beta_{0} \mathbb{I})\phi = (H_{\lambda''}^{min} + \beta_{0} \mathbb{I})\phi_{1} $, in particular, we have $(\mathbb{H}_{\lambda''} + \beta_{0} \mathbb{I})(\phi - \phi_{1}) = 0$.\\

\n To deduce that $\phi = \phi_{1}$, we need to verify that $Ker(\mathbb{H}_{\lambda''} + \beta_{0} \mathbb{I}) = \{0\}$, where $Ker$ denote the kernel of the operator $\mathbb{H}_{\lambda''} + \beta_{0} \mathbb{I}$\\

\n we recall that the range of $\mathbb{H}_{\lambda''}^{min} + \beta_{0} \mathbb{I}$ is closed and the formal adjoint of $\mathbb{H}_{\lambda''}$ is $\mathbb{\mathbb{H}}_{\lambda''}$ where we substitute $\lambda$ by its opposite.\\

\n  Now, since the adjoint of the minimal is formal adjoint of the maximal and $\mathbb{H}_{\lambda''}^{min} + \beta_{0} \mathbb{I}$ is invertible then $Ker (\mathbb{H}_{\lambda''} + \beta_{0} \mathbb{I}) = \{0\}$ this implies that $\phi = \phi_{1}$ and $\phi \in D_{min}(\mathbb{H}_{\lambda''})$.\\

\n From the inequality a) and the theorem 111 in book's Kato {\bf{\color{blue}[Kato]}}, we deduce $D_{max}(\mathbb{H}_{\lambda''}) = D(\mathbb{G})$.\\

\n f) From the inequality a) and the theorem 2.1 in book's Pazy {\bf{\color{blue}[Pazy]}}, we deduce that $ - \mathbb{H}_{\lambda{''}}$ generates analytic semigroup $e^{-t\mathbb{H}_{\lambda''}}$ $ t > 0$.\\

\n g) For $\lambda^{''} \neq 0$ the resolvent of $\mathbb{H}_{\lambda{''}}$ is Carleman class $\mathcal{}C_{p}$ for all $ p > \frac{1}{3}$, this property is the lemma 4.1 of {\bf{\color{blue}[Aimar6 et al]}}. This ends the proof of this proposition.\\

\n Now, the above properties d) and e) allow us to show the following theorem:\\
\begin{theorem}

\n $- \mathbb{H}_{\lambda{''}}$ generates a semigroup $e^{-t\mathbb{H}_{\lambda''}}$ of Carleman class $\mathcal{C}_{p}$ for all $ p > 0$ and all $ t > 0$.\\
\end{theorem}
\n {\bf Proof}\\

\n  Let $\mathbb{B}_{0} = \{\phi \in \mathbb{B} ; \phi(0) = 0\}$ then on  $\mathbb{B}_{0}$ we have:\\

\n  $\Re e< \mathbb{H}_{\lambda''}\phi, \phi> = \lambda{''}\mid\mid A^{3}\phi\mid\mid^{2} + \mu\mid\mid A\phi\mid\mid^{2} \geq \mu \mid\mid \phi \mid\mid^{2}$ \\

\n  for $ \lambda{''} \geq 0$ and $\mu > 0$. \\

\n  From this inequality we deduce that $0$ belongs to resolvent set $\rho(\mathbb{H}_{\lambda{''}})$ of the operator $\mathbb{H}_{\lambda{''}}$.\\

\n  Let $\mathbb{T}(t) = \displaystyle{\int_{0}^{t}e^{-s\mathbb{H}_{\lambda''}}}\phi ds$ then\\

\n  $\mathbb{T}(t) = \mathbb{H}_{\lambda''}^{-1}( \mathbb{I} - e^{-t\mathbb{H}_{\lambda{''}}})$ and as the resolvante of $\mathbb{H}_{\lambda{''}}$ is Carleman class $\mathcal{C}_{p}$ for all $ p > \frac{1}{3}$ and the operator $\mathbb{I} - e^{-t\mathbb{H}_{\lambda''}}$ is bounded then $\mathbb{T}(t)$ is Carleman class $\mathcal{C}_{p}$ for all $ p > \frac{1}{3}$ and with the aid of lemma 8.5, we end the proof. \hfill { } $\square$\\

\subsection{Asymptotic expansion of trace of $e^{-t\mathbb{H}_{\lambda''}}$ as $t \rightarrow 0^{+}$}

\n We put $S(s) = e^{-(t-s)\lambda''\mathbb{G}} e^{-s\mathbb{H}_{\lambda''}}$ then for $\phi \in D(\mathbb{G})$, the application $\phi \rightarrow S(s)\phi $ is differentiable and $S'(s)\phi = \lambda''\mathbb{G}e^{-(t-s)\lambda''G}   e^{-sH_{\lambda''}} - e^{-(t-s)\lambda''G}H_{\lambda''}e^{-sH_{\lambda''}} = -e^{-(t-s)\lambda''G}H_{\mu,\lambda}e^{-sH_{\lambda''}}$.\\

\n On $[0, t]$, we have $\displaystyle{ \int_{0}^{t}S'(s)\phi ds = S(t)\phi - S(0)\phi =} -\displaystyle{ \int_{0}^{t}e^{-(t-s)\lambda''G}H_{\mu,\lambda}e^{-sH_{\lambda''}}\phi ds}$ \\

\n This implies that $e^{-tH_{\lambda''}}\phi $ is solution of the following integral equation :\\
\begin{equation}
e^{-tH_{\lambda''}}\phi - e^{-t\lambda''G}\phi = -\displaystyle{ \int_{0}^{t}e^{-(t-s)\lambda''G}H_{\mu,\lambda}e^{-sH_{\lambda''}}\phi ds}
\end{equation}
\n or\\
\begin{equation}
e^{-tH_{\lambda''}}\phi - e^{-t\lambda''G}\phi = -\displaystyle{ \int_{0}^{t}N(t,s)e^{-sH_{\lambda''}}\phi ds} \mbox{ with} \, N(t,s) = e^{-(t-s)\lambda''G}H_{\mu,\lambda}
\end{equation}
\n It is well known that the solution of equation (8.10) can be obtained by successive approximation method:\\
\begin{equation}
 e^{-tH_{\lambda''}} = \displaystyle{\sum_{k=0}^{\infty}S_{k}(t)}
\end{equation}
\n where \\

\n $S_{0}(t)\phi = e^{-t\lambda''G}\phi$\\

\n and\\

\n $S_{k+1}(t)\phi = - \displaystyle{ \int_{0}^{t}e^{-(t-s)\lambda''G}H_{\mu,\lambda} S_{k}(t)\phi ds}$\\

\n the convergence of (8.12) is in operator norm.\\

\n  Notice that:\\

\n  $S_{1}(t)\phi = - \displaystyle{ \int_{0}^{t}e^{-(t-t_{1})\lambda''G}H_{\mu,\lambda}e^{-t_{1}\lambda''G}\phi dt_{1}}$\\

\n  $S_{2}(t)\phi = \displaystyle{ \int_{0}^{t}\displaystyle{ \int_{0}^{t_{1}}e^{-(t-t_{1})\lambda''G}H_{\mu,\lambda}e^{-(t_{1}-t_{2})\lambda''G}H_{\mu,\lambda}e^{-t_{2}\lambda''G}\phi dt_{2}dt_{1}}}$\\

\n $S_{3}(t)\phi = (-1)^{3} \displaystyle{ \int_{0}^{t}\displaystyle{ \int_{0}^{t_{1}}\displaystyle{ \int_{0}^{t_{2}}e^{-(t-t_{1})\lambda''G}H_{\mu,\lambda}e^{-(t_{1}-t_{2})\lambda''G}H_{\mu,\lambda}e^{-(t_{2}-t_{3})\lambda''G}H_{\mu,\lambda}e^{-t_{3}\lambda''G}\phi dt_{3}dt_{2}dt_{1}}}}$\\

\n $S_{k}(t)\phi = (-1)^{k} \displaystyle{ \int_{0}^{t}\displaystyle{ \int_{0}^{t_{1}}\displaystyle{ \int_{0}^{t_{2}} .................................}}}
\displaystyle{ \int_{0}^{t_{k-1}}}$\\

\n $e^{-(t-t_{1})\lambda''G}H_{\mu,\lambda}e^{-(t_{1}-t_{2})\lambda''G}H_{\mu,\lambda}e^{-(t_{2}-t_{3})\lambda''G}H_{\mu,\lambda}
............................................ $\\

\n $e^{-(t_{k-1}-t_{k})\lambda''G}H_{\mu,\lambda}e^{-t_{k}\lambda''G}\phi dt_{k}............................................ dt_{3}dt_{2}dt_{1}$.\\
\begin{lemma}

\n  The series $e^{-tH_{\lambda''}} = \displaystyle{\sum_{k = 0}^{\infty}S_{k}(t)}$ converges in trace norm.\\
\end{lemma}
\n {\bf Proof}\\

\n The convergence in trace norm is obtained by using the results of Angelescu-Nenciu-Bundaru, in particular their  proposition in  {\color{blue}[Angelescu]}  or the results of Zagrebnov in  {\bf {\color{blue}[Zagrebnov1, Zagrebnov2]}}, in particular the theorem 2.1 in  {\bf {\color{blue}[Zagrebnov2]}} with the aid of  Ginibre-Gruber's inequality in   {\color{blue}[Ginibre et al]}. \hfill { } $\square$\\

\n Now we are going to derive an asymptotic expansion of the trace of $e^{-tH_{\lambda''}}$ as $t\rightarrow 0^{+}$\\

\n For $ s \geq 0 $ we have\\

\n $e^{-sH_{\lambda''}}\phi = e^{-s\lambda''G}\phi -\displaystyle{ \int_{0}^{s}N(s,s_{1})e^{-s_{1}H_{\lambda''}}\phi ds_{1}}$ with $N(s,s_{1}) = e^{-(s-s_{1})\lambda''G}H_{\mu,\lambda}$\\

\n and for $ 0 \leq s \leq t$, we substitute the above expression of $e^{-sH_{\lambda''}}\phi$ in (8.11) to get  :\\

\n $e^{-tH_{\lambda''}}\phi - e^{-t\lambda''G}\phi = -\displaystyle{ \int_{0}^{t}N(t,s)[e^{-s\lambda''G}\phi -\displaystyle{ \int_{0}^{s}N(s,s_{1})e^{-s_{1}H_{\lambda''}}\phi ]ds_{1}ds}}$\\

\n $= -\displaystyle{ \int_{0}^{t}N(t,s)e^{-s\lambda''G}\phi ds}$ + $ \displaystyle{ \int_{0}^{t}\displaystyle{ \int_{0}^{s}N(t,s)N(s,s_{1})e^{-s_{1}H_{\lambda''}}\phi ds_{1}ds}}$\\

\n $= -\displaystyle{ \int_{0}^{t}e^{-(t-s)\lambda''G}H_{\mu,\lambda}e^{-s\lambda''G}\phi ds}$ + $\displaystyle{ \int_{0}^{t}\displaystyle{ \int_{0}^{s}e^{-(t-s)\lambda''G}H_{\mu,\lambda}e^{-(s-s_{1})\lambda''G}H_{\mu,\lambda}e^{-s_{1}H_{\lambda''}}\phi ds_{1}ds}}$\\

\n Then we have: \\

\n $e^{-tH_{\lambda''}}\phi - e^{-t\lambda''G}\phi = \displaystyle {\int_{0}^{t}e^{-(t-s)\lambda''G}H_{\mu,\lambda}e^{-s\lambda''G}\phi ds}$ + \\
\begin{equation}
\displaystyle { \int_{\Delta}e^{-(t-s)\lambda''G}H_{\mu,\lambda}e^{-(s-s_{1})\lambda''G}H_{\mu,\lambda}e^{-s_{1}H_{\lambda''}}\phi ds_{1}ds}
\end{equation}
\n where $\Delta$ is the triangle $\{(s_{1},s); 0 \leq s_{1} \leq s \leq t \}$.\\

\n The matrix associated to $H_{\mu,\lambda}$ in the basis $ \{e_{n}\}$ can be written in this form :\\
\begin{equation}
H_{\mu,\lambda} e_{n} = i\lambda(n-1)\sqrt{n}e_{n-1} + n\mu e_{n} + i\lambda n\sqrt{n+1}e_{n+1}
\end{equation}
\n - The  family of infinite matrices associated to $H_{\mu,\lambda}$ is tridiagonal of the form $J + i\lambda H^{(i)}$, where the matrix $J$ is diagonal with entries $J_{nn} := n\mu$ , and the matrix $H^{(i)}$ is off-diagonal, with nonzero entries $H_{n,n+1}^{(i)} = H_{n+1,n}^{(i)} := H_{n}^{(i)} = n\sqrt{n+1}$.\\

\n - The  family of infinite matrices associated to $\lambda''G$ is diagonal with entries $G_{nn} := \lambda''n(n-1)(n-2)$\\

\n Let the infinite matrix $ ^{t\!}H_{\mu,\lambda} $ be obtained from $H_{\mu,\lambda}$ by transposing of the elements and the infinite matrix
 $H_{\mu,\lambda}^{\bot}$ be obtained from $H_{\mu,\lambda}$ by transposing and by taking complex conjugates of the elements. Then observe that\\

\n i) $H_{\mu,\lambda}$ is symmetric complex matrix i.e. $ H_{\mu,\lambda} =\, ^{t\!}H_{\mu,\lambda}$.\\

\n ii) $ H_{\mu,\lambda} \neq  H_{\mu,\lambda}^{\bot}$ (The symbol $\bot$ represents Dirac Hermitian conjugation; that is, transpose and complex conjugate.)\\

\n iii) As $H_{n}^{(i)} = O(n^{\alpha})$ with $\alpha = \frac{3}{2} > 1$ then the standard perturbation theory is not applicable.\\

\n iv) As $G_{nn} = \lambda''n(n-1)(n-2)$ then $G_{nn} = O(n^{3})$ as $n\rightarrow \infty$\\

\n v) For other properties on the matrix associated to $H_{\mu,\lambda}$, we can consult the {\color{blue}[Intissar 20]}.\\

\n vi) from the above observations, we deduce that the operator $H_{\mu,\lambda}G^{-\delta}$ is bounded for all $ \delta \geq \frac{1}{2}$.\\

\n Now, we present the aim result of this work in following theorem:\\
\begin{theorem}

\n Let  $H_{\lambda''} = \lambda'' G + H_{\mu,\lambda}$ the Gribov's operator acting on Bargmann's space.\\

\n where\\

\n $ G = A^{*3} A^{3}$ and $ H_{\mu,\lambda} =\mu A^* A + i \lambda A^* (A + A^*)A$\\

\n $[A, A^{*}] = I$ and $(\lambda'', \mu,\lambda)$ are reel parameters and $i^{2} = -1$.\\

\n Then\\

\n $\mid\mid e^{-tH_{\lambda''}} - e^{-t\lambda''G} \mid\mid_{1} = t\mid\mid e^{-t\lambda''G} H_{\mu,\lambda}\mid\mid_{1} + \mid\mid (\lambda''G)^{\delta} e^{-\frac{t}{3}\lambda''G }\mid\mid_{1}O(t^{2})$.\\
\end{theorem}
\n {\bf Proof }\\

\n {\bf (a)} We begin by computing the trace of the operator:\\

\n $I_{1}(t) = \displaystyle {\int_{0}^{t}e^{-(t-s)\lambda''G}H_{\mu,\lambda}e^{-s\lambda''G}ds}$\\

\n We have\\

\n $\mid\mid I_{1}(t)\mid\mid_{1} = \displaystyle {\int_{0}^{t}\mid\mid e^{-(t-s)\lambda''G}H_{\mu,\lambda}e^{-s\lambda''G}\mid\mid_{1}ds}$\\

\n $= \displaystyle {\int_{0}^{t}\displaystyle {\sum_{n=1}^{\infty}<e^{-(t-s)\lambda''G}H_{\mu,\lambda}e^{-s\lambda''G}e_{n},e_{n}> ds}}$\\

\n $ = \displaystyle {\int_{0}^{t}\displaystyle {\sum_{n=1}^{\infty}<e^{-t\lambda''G}H_{\mu,\lambda}e^{-s\lambda_{n}}e_{n}, e^{s\lambda_{n}}e_{n}>ds}}$ because $e^{s\lambda''G}$ is self adjoint\\

\n $= \displaystyle {\int_{0}^{t}\displaystyle {\sum_{n=1}^{\infty}<e^{-t\lambda''G}H_{\mu,\lambda}e_{n}, e_{n}>ds}}$\\

\n $ = \displaystyle {\int_{0}^{t}\mid\mid e^{-t\lambda''G}H_{\mu,\lambda}\mid\mid_{1}ds}$\\

\n $ = t\mid\mid e^{-t\lambda''G}H_{\mu,\lambda}\mid\mid_{1}$\\

\n Then we deduce that:\\
\begin{equation}
\mid\mid I_{1}(t)\mid\mid_{1} = t\mid\mid e^{-t\lambda''G}H_{\mu,\lambda}\mid\mid_{1}
\end{equation}
\n {\bf (b)} We begin to recall the symmetry property of the norm in Carleman class $C_{p}$\\

\n The symmetry of the norm in $C_{p}$ means that\\
\begin{equation}
\mid\mid K_{1}K_{2}K_{3}\mid\mid_{p} \leq \mid\mid K_{1}\mid\mid.\mid\mid K_{2}\mid\mid_{p}\mid\mid K_{3}\mid\mid.
\end{equation}
\n for any bounded operators $K_{1}$ and $K_{3}$ and $K_{2}\in C_{p}$\\

\n Consider the trace of the operator \\

\n $I_{2}(t) = \displaystyle { \int_{\Delta}e^{-(t-s)\lambda''G}H_{\mu,\lambda}e^{-(s-s_{1})\lambda''G}H_{\mu,\lambda}e^{-s_{1}H_{\lambda''}}ds_{1}ds}$\\

\n and let $\delta \geq\frac{1}{2}$ such that $H_{\mu,\lambda}G^{-\delta}$ bounded, then \\

\n $\mid\mid I_{2}(t)\mid\mid_{1} = $

\n $\displaystyle { \int_{\Delta}\mid\mid e^{-(t-s)\lambda''G}H_{\mu,\lambda}G^{-\delta}[G^{\delta}e^{-(s-s_{1})\lambda''G}]H_{\mu,\lambda}G^{-\delta}[G^{\delta}e^{-s_{1}H_{\lambda''}}]\mid\mid_{1} ds_{1}ds}$\\

\n As $t$ can be written as sum of three positif numbers $t = (t-s) +(s-s_{1}) + s_{1}$. It follows that at least one of them is not less than $\frac{t}{3}$; suppose, for example, that $ s-s_{1} \geq \frac{t}{3}$\\

\n Then \\

\n $\mid\mid G^{\delta}e^{-(s-s_{1})\lambda''G}\mid\mid_{1} \leq \mid\mid G^{\delta}e^{-\frac{t}{3}\lambda''G}\mid\mid_{1}$\\

\n By using the inequality (8.16) we deduce that\\

\n $\mid\mid e^{-(t-s)\lambda''G}H_{\mu,\lambda}G^{-\delta}[G^{\delta}e^{-(s-s_{1})\lambda''G}]H_{\mu,\lambda}G^{-\delta}[G^{\delta}e^{-s_{1}H_{\lambda''}}]\mid\mid_{1}$
$ \leq \mid\mid e^{-(t-s)\lambda''G}H_{\mu,\lambda}G^{-\delta}\mid\mid . \mid\mid G^{\delta}e^{-(s-s_{1})\lambda''G}\mid\mid_{1}.
 \mid\mid H_{\mu,\lambda}G^{-\delta}[G^{\delta}e^{-s_{1}H_{\lambda''}}]\mid\mid$\\

\n $ \leq \mid\mid H_{\mu,\lambda}G^{-\delta} \mid\mid^{2}\mid\mid G^{\delta}e^{-\frac{t}{3}\lambda''G}\mid\mid_{1}$\\

\n Then we have\\

\n  $\mid\mid I_{2}(t)\mid\mid_{1} \leq \mid\mid H_{\mu,\lambda}G^{-\delta} \mid\mid^{2}\mid\mid G^{\delta}e^{-\frac{t}{3}\lambda''G}\mid\mid_{1}\displaystyle { \int_{\Delta}dsds_{1}}$\\

\n  $\leq \mid\mid H_{\mu,\lambda}G^{-\delta} \mid\mid^{2}\mid\mid G^{\delta}e^{-\frac{t}{3}\lambda''G}\mid\mid_{1}t^{2}$\\

\n It follows that\\
\begin{equation}
\mid\mid I_{2}(t)\mid\mid_{1} = \mid\mid G^{\delta}e^{-\frac{t}{3}\lambda''G}\mid\mid_{1}O(t^{2})
\end{equation}
\n and consequently we have\\
\begin{equation}
\mid\mid e^{-tH_{\lambda''}} - e^{-t\lambda''G} \mid\mid_{1} = t\mid\mid e^{-t\lambda''G} H_{\mu,\lambda}\mid\mid_{1} + \mid\mid (\lambda''G)^{\delta} e^{-\frac{t}{3}\lambda''G }\mid\mid_{1} O(t^{2}).
\end{equation}

\begin{center}
 {\bf {\color{black} {\large{\color{red} References}}}}\\
\end{center}

\n  {\color{blue}[Abramovitz]} Abramovitz, M. and Stegun, I. A., Handbook of Mathematical Functions, New York (1968)\\

\n  {\color{blue}[Aimar1 et al ]}  Aimar, M.-T.,  Intissar,  A. and J.-M. Paoli, M.-P. :. Densit\'e des vecteurs propres g\'en\'eralis\'es d'une
classe d'op\'erateurs non auto-adjoints a r\'esolvante compacte, C.R. Acad. Sci. Paris, t. 315, S\'er. I 1992..\\

\n  {\color{blue}[Aimar2 et al ]}  Aimar, M.-T.,  Intissar,  A. and J.-M. Paoli, M.-P. :. Densit\'e des vecteurs propres g\'en\'eralis\'es d'une
classe d'op\'erateurs compacts non auto-adjoints et applications, Commun. Math. Phys. 156, 1993., 169-177.\\

\n  {\color{blue}[Aimar3 et al ]}  Aimar, M.-T.,  Intissar,  A. and J.-M. Paoli, M.-P. :. Quelques propri\'et\'es de r\'egularit\'e de l'op\'erateur
de Gribov, C.R. Acad. Sci. Paris, t. 320, S\'er. I 1995..\\

\n {\color{blue}[Aimar4 et al ]}  Aimar, M.-T.,  Intissar,  A. and J.-M. Paoli, M.-P. :. Quelques nouvelles propri\'et\'es de r\'egularit\'e de
l'op\'erateur de Gribov, Commun. Math. Phys. 172, 1995., 461-466.\\

\n {\color{blue}[Aimar5 et al ]}  Aimar, M.-T.,  Intissar,  A. and J.-M. Paoli, M.-P. :. Crit$\grave{e}$res de compl\'etude des vecteurs propres g\'en\'eralis\'es d'une classe d'op\'erateurs non auto-adjoints compacts ou  a r\'esolvante compacte et applications, R. I.M.S. 32, 2. 1996..\\

\n  {\color{blue}[Aimar6 et al ]} Aimar, M.-T, Intissar, A and Jeribi, A. :. On an Unconditional Basis of Generalized Eigenvectors of the Nonself-adjoint Gribov Operator in Bargmann Space, Journal of Mathematical Analysis and Applications 231, 588-602 (1999).\\

\n  {\color{blue}[Aimar7 et al ]} Aimar, M.-T, Intissar, A and Intissar, J. K. :. On Regularized Trace Formula of Gribov Semigroup Generated by the Hamiltonian of Reggeon Field Theory in Bargmann Representation, Complex Analysis and Operator Theory, March 2018, Volume 12, Issue 3, pp 615-627\\

\n {\color{blue}[Alessandrini et al]}, Alessandrini, V., Amati, D. and R. Jengo, R. :. One-dimensional quantum theory of the pomeron, Nucl. Phys. B 108 (1976) 425-446.\\

\n  {\color{blue}[Amati et al]} Amati, D., Le Bellac, M., G.Marchesini, G. and Ciafaloni, M. : Reggeon field theory for $\alpha(0) > 1$, Nucl. Phys. B l12 (1976) 107.\\

\n {\color{blue}[Ando et al]} Ando, T. and  Zerner, M. :. Sur une valeur propre d'un op\'erateur, Commun. Math. Phys. 93, 1984..\\

\n  {\color{blue}[Angelescu et al]} Angelescu, N.,  Nenciu, G. and  Bundaru, M. :. On the perturbation of Gibbs semigroups, Comm. Math. Phys., 42 (1975).\\

\n  {\color{blue}[Baker et al]} Baker, M. and Ter-Martirosyan, K.A., Gribov's Reggeon Calculus: Its physical basis and implications, Phys. Rep. C 28 (1976)\\

\n  {\color{blue}[Bargmann1]}  Bargmann, V, On a Hilbert space of analytic functions and an associated integral transform I, Comm. Pure Appl. Math. Vol. 14, Issue 3 (1962) 187-214.\\

\n {\color{blue}[Bargmann2]} Bargmann, V, On a Hilbert space of analytic functions and an associated integral transform II, Comm. Pure Appl. Math., Vol. 20, Issue 1 (1967) 1-101.\\

\n {\color{blue}[Bargmann 3]} Bargmann, V. : Remarks on a Hilbert space of analytic functions. Proc. Acad. Sc. 48. (1962) 199-204\\

\n  {\color{blue}[Barnsley]} Barnsley, M., Lower bounds for quantum mechanical energy levels. J. Phys. A, 11(1):55?68, 1978.\\

\n   {\color{blue}[Beckermann]}  Beckermann, B. :. Complex Jacobi matrices , J. Compt. Appl. Math.,  2001, 127 (1/2), 17-65\\

\n  {\color{blue} [Bender]}  Bender, C.M. :. Introduction to PT -Symmetric Quantum Theory, arXiv:quant-ph/0501052v1 11 Jan 2005 .\\

\n  {\color{blue} [Berberian] } Berberian, S.K., Tensor product of Hilbert spaces, \\https : =www:ma:utexas:edu=mparc=c=14=14- 2:pdf (2013).\\

\n  {\color{blue}[Boreskov et al.]} Boreskov, K.G.,  Kaidalov, A.B. and  Kancheli, O.V. :. Strong interactions at high energies in the Reggeon approch, Physics of Atomic Nuclei,vol. 69, n 10,(2006) pp.1765-1780.\\

\n {\color{blue}[Bronzan et al]}  Bronzan, J.B., Shapiro, J. and Sugar, R.L.: Phys. Rev. D14 (1976) 618.\\

\n {\color{blue}[Brower et al]}  Brower, R., Furman,  M. and M. Moshe, Phys. Lett. B 76, 213 (1978); B. Harms, S. Jones, and C.-I Tan, Nucl. Phys. 171, 392 ,1980 and Phys. Lett. B 91, 291, 1980. \\

\n {\color{blue}[Ciafaloni et al]} Ciafaloni, M.,  LeBellac, M. and Rossi, G.C. : Nucl. Phys. B130 (1977) 388.\\

\n {\color{blue}[Ciafaloni-Onofri ]} Ciafaloni, M. and  Onofri, E. :. Path integral formulation of reggeon quantum mechanics,  Nucl. Phys. B 151 (1979) 118-146.\\

\n {\color{blue}[Choun: [1-10)]]} Choun, Y.S., Generalization of the three-term recurrence formula and its applications, arXiv:1303.0806\\

\n - The analytic solution for the power series expansion of Heun function, Ann. Phys. 338 (2013), 21-31, e-Print: arXiv:1303.0830.\\

\n - Asymptotic behavior of Heun function and its integral formalism, arXiv:1303.0876. \\

\n -The power series expansion of Mathieu function and its integral formalism,Int. J. Differ. Equ. Appl.,14(2) (2015), 81-99, e-Print: arXiv:1303.0820.\\

\n - Lame equation in the algebraic form, arXiv:1303.0873.\\

\n - Power series and integral forms of Lame equation in Weierstrass's form, arXiv:1303.0878.\\

\n - The generating functions of Lame equation inWeierstrass's form, arXiv:1303.0879.\\

\n - Analytic solution for grand confluent hypergeometric function,arXiv:1303.0813.\\

\n - The integral formalism and the generating function of grand confluent hypergeometric function, arXiv:1303.0819.\\

\n - Special functions and reversible three-term recurrence formula (R3TRF), \\arXiv:1310.7811.\\

\n - Complete polynomials using 3-term and reversible 3-term recurrence formulas (3TRF and R3TRF),arXiv:1405.3610\\

\n  {\color{blue}[Delabeare et al]}  Delabaere, E, and Pham, F, Eigenvalues of complex hamiltionians with PT symmetry I, II. Phys.Lett.A,250, (1998) 25-32.\\

\n  {\color{blue}[Dikii1]} Dikii, L A. :. About a formula of Gelfand-Levitan, Usp. Mat. Nauk 8(2), (1953), 119-123. \\

\n  {\color{blue}[Dikii2]} Dikii, L A. :. New method of computing approximate eigenvalues of the Sturm-Liouville problem, Dokl. Akad. Nauk SSSR 116, (1957), 12-14.\\

\n   {\color{blue}[Dorey et al]} Dorey, P., Dunning, C. and Tateo, R., Spectral equivalences, Bethe Ansatz equations, and reality properties in PT-symmetric quantum mechanics. J.Phys. A34, (2001) 5679-5704.\\

\n {\color{blue}[Duffin]} Duffin, R.J., Lower bounds for eigenvalues. Phys. Rev., 71: 827-828, 1947.\\

\n  {\color{blue}[Dunford et al ]}  Dunford, N. and Schwartz,  J. T. :. Linear Operators, Part II and III, Interscience, New York, 1963 and 1971.\\

\n   {\color{blue}[Eppens]} Eppens, M. :. Real time vs. Imaginary time, section 4.3, pp : 32-34)\\

\n  {\color{blue}[Erdelyi]} Erdellyi, A. :.Integral equations for Heun functions, Quart. J. Math., 13,107-112 (1942)\\

\n  {\color{blue}[Fr\'echet]} Fr\'echet, M. :. Sur les ensembles compacts de fonctions de carr\'es sommables, Acta. Litt. Sci. Szegd, 8, (1937) 116-126.\\

\n  {\color{blue}[Garcia et al]}  Garcia, S. R. and  Putinar, M. :. Complex symmetric operators and applications , Transactions of the American Mathematical Society, 
Volume 358, Number 3,  (2005) Pages 1285-1315.\\

\n   {\color{blue}[Gelfand et al]}   Gelfand, I. M. and Levitan, B. M. :. On a Simple Identity for the Characteristic Values of a Differential Operator of Second Order, Dokl. Akad. Nauk SSSR, vol. 88, (1953), pp. 593-596\\

\n   {\color{blue}[Ginibre et al]}  Ginibre, J. , Gruber, C. :.  Green functions of anisotropic Heisenberg model, Comm. Math. Phys. 11 (1969).\\

\n  {\color{blue}[Gohberg1 et al ]}   Gohberg, I. , Goldberg, S. and  Krupnik, N. :. Traces and determinants of linear operators, Birkhauser, 2000.\\

\n  {\color{blue}[Gohberg2 et al ]}  Gohberg, C.  and M. G. Krein, M. G. :. Introduction to the Theory of Linear Non-Self Adjoint Operators, Vol. 18, Am. Math. Soc., Providence, RI, (1969).\\

\n {\color{blue}[Grassberger et al]} P.Grassberger, P. Sundermeyer, K.: Reggeon field theory and markov processes, Physics Letters B, Volume 77, Issue 2, 31 July 1978, Pages 220-222\\

\n  {\color{blue}[Gribov1]}  Gribov, V. :. A reggeon diagram technique, Soviet Phys. JETP 26, no. 2, (1968), 414-423\\

\n  {\color{blue}[Gribov2]} Gribov, V. :. Strong Interactions of Hadrons at High Energies, Gribov Lectures on Theoretical Physics, Prepared by Y. L. Dokshitzer and J. Nyiri, Cambridge University Press (2009)\\

\n  {\color{blue}[Intissar1 et al]}  Intissar, A, Le Bellac, M. and Zerner, M. :. Properties of the Hamiltonian of reggeon field theory, Physics Lett. 113, B, 6. 1982., 487-489.\\

\n  {\color{blue}[Intissar2 et al]}  Intissar, A. and Intissar, J.K. :. On Chaoticity of the Sum of Chaotic Shifts with Their Adjoints in Hilbert Space and Applications to Some Weighted Shifts Acting on Some Fock-Bargmann Spaces, Complex Analysis and Operator Theory, March 2017, Volume 11, Issue 3, pp 491-505\\

\n  {\color{blue}[Intissar3 et al]}  Intissar, A. and Intissar, J.K.:. Calcul diff\'erentiel , Fondement et applications, cours et exercices avec solutions, Editions CEPADUES (2017)\\

\n  {\color{blue}[Intissar 4 et al]}  Intissar, A.  and Intissar, J.K. :. A Complete Spectral Analysis of Generalized Gribov-Intissar's Operator in Bargmann Space, Complex Analysis and Operator Theory, April 2019, Volume 13, Issue 3, pp 1481-1510\\

\n  {\color{blue}[Intissar 1]}  Intissar, A. :. Sur une propri\'et\'e spectrale d'un op\'erateur non sym\'etrique intervenant dans la th\'eorie de Regge, C.R. Acad. Sci. Paris T 294, 715-718 (1982)\\

 \n  {\color{blue}[Intissar 2]}  Intissar, A. :. Etude spectrale d'une famille d'op\'erateurs non-sym\'etriques intervenant dans la th\'eorie des champs de reggeons, Commun. Math. Phys. 113, 1987., 263-297.\\

\n  {\color{blue}[Intissar 3]}  Intissar, A. :. Diagonalisation d'op\'erateurs non auto-adjoints intervenant dans la th\'eorie des champs des reggeons de Gribov. C.R. Acad. Sci. Paris T: 304, N2 Ser I, 43-46 (1987)\\

\n  {\color{blue}[Intissar 4]}  Intissar, A. :. Quelques propri\'et\'es spectrales de l'hamiltonien de la th\'eorie des champs de reggeons, C.R. Acad. Sci. Paris T 304, No 3 Ser I, 63-66 (1987) \\

\n  {\color{blue}[Intissar 5]}  Intissar, A. :.  Sur une m\'ethode de perturbation C.R. Acad. Sci. Paris T 304, No 4 Serie I, 95-98 (1987)\\

\n  {\color{blue}[Intissar 6]}  Intissar, A. :. Quelques nouvelles propri\'et\'es spectrales de l'hamiltonien de la th\'eorie des champs de reggeons, C.R. Acad. Sci. Paris, t. 308, S\'er. I (1989), 209-214\\

 \n  {\color{blue}[Intissar 7]}  Intissar, A. :. Th\'eorie spectrale dans l'espace de Bargmann, Cours de DEA, Universit\'e de Besan\c con, 1989..\\

\n  {\color{blue}[Intissar 8]}  Intissar, A. :. 20 heures de cours de DEA, Quelques \'el\'ements de la th\'eorie des op\'erateurs suivant Kato  et quelques \'el\'ements de la
th\'eorie des serni-groupes suivant Pazy, 1989\\

\n  {\color{blue}[Intissar 9]}  Intissar, A. :. Diagonalisation au sens d'Abel de l'op\'erateur de Gribov dans l'espace de Bargmann, Publ. Math. Besancon, 1994..\\

 \n  {\color{blue}[Intissar 10]}  Intissar, A. :.  Analyse Fonctionnelle et Th\'eorie Spectrale pour les Op\'erateurs Compacts Non Auto-Adjoints, Editions C\'epadues, Toulouse, 1997.\\

 \n {\color{blue}[Intissar 11]}  Intissar, A. :. Analyse de Scattering d'un op\'erateur cubique de Heun dans l'espace de Bargmann, Commun. Math. Phys. 199, 1998., 243-256.\\

\n  {\color{blue}[Intissar 12] }  Intissar, A. :. Diagonalization of Non-selfadjoint Analytic Semigroups and Application to the Shape Memory Alloys Operator, Journal of Mathematical Analysis and Applications 257, 120, (2001).\\

 \n  {\color{blue}[Intissar 13] } Intissar, A. :. Approximation of the semigroup generated by the Hamiltonian of Reggeon field theory in Bargmann space, Journal of Mathematical Analysis and Applications, vol. 305, no. 2,(2005), pp. 669-689\\

\n  {\color{blue}[Intissar 14]}  Intissar, A. :. On a chaotic weighted Shift $\displaystyle{z^{p}\frac{\partial^{p+1}}{\partial z^{p+1}}}$ of order $p$ in Bargmann space,
Advances in Mathematical Physics, Article ID 471314, (2011).\\

\n  {\color{blue}[Intissar 15]}  Intissar, A. :.  An elementary construction on nonlinear coherent states associated to generalized Bargmann spaces, International Journal of Mathematics and Mathematical Sciences,(2010)\\

\n {\color{blue}[Intissar 16]}  Intissar, A. :.  A short note on the chaoticity of a weight shift on concrete orthonormal basis associated to some Fock-Bargmann space, Journal of Mathematical Physics, 55, 011502 (2014)\\

 \n  {\color{blue}[Intissar 17]}  Intissar, A. :. :. Spectral Analysis of Non-self-adjoint Jacobi-Gribov Operator and Asymptotic Analysis of Its Generalized Eigenvectors, Advances in Mathematics (China), Vol.44, No.3, May, (2015) doi: 10.11845/sxjz.2013117b\\

\n  {\color{blue}[Intissar 18]}  Intissar, A. :. On the chaoticity of some tensor product weighted backward shift operators acting on some tensor product Fock-Bargmann spaces Complex Analysis and Operator Theory 2016\\

\n {\color{blue}[Intissar 19] }  Intissar, A. :. Regularized trace of magic Gribov operator on Bargmann space, J. Math. Anal. Appl. 437 (2016), 59-70\\

\n  {\color{blue}[Intissar 20] } Intissar, A. :. On spectral approximation of unbounded Gribov?Intissar operators in Bargmann space. Adv. Math. (China) 46(1), 13?33 (2017)\\

\n  {\color{blue}[Janas1 et al]} Janas, J. and Malejki, M. :. Alternatives approches to asymptotic behaviour of eigenvalues of some unbounded Jacobi matrices, J. Comput. Appl. Math., 2007, 200(1)  342-356\\

\n {\color{blue}[Janas2 et al]} Janas, J. and Moszynski, M. :. Spectral properties of Jacobi matrices by asymptotic analysis; J. Approx. Theory, 2003,   120(2) 309-336.\\

\n  {\color{blue}[Janas3 et al]} Janas, J. and Naboko, S. :. Jacobi matrices with power-like weights-grouping in blocks approch,  J. Funct. Anal., 1999, 166(2), 218-243.\\

\n {\color{blue}[Janas4 et al]} Janas, J. ,  Naboko, S. and Stolz, .G. :. Spectral theory for a class of periodically perturbed unbounded Jacobi matrices : elementary methods, J. Comput. Appl. Math., 2004, 171(1)  265-276\\

\n  {\color{blue}[Janas5 et al]} Janas, J. and   Naboko, S. :. Multithreshold spectral phase transition examples in class of unbounded Jacobi matrices, recent Advances in Operator Theory, Operation Theory Adv. Appl.,vol. 124, Birkhauser-Verlag, Basel, 2001, pp.267-285)\\

\n {\color{blue}[Jengo]} Jengo, R. : Nucl. Phys. B108 (1976) 447.\\

\n {\color{blue}[Kasakov1 et al.]} Kazakov, A. Ya. and Slavyanov, S. Yu., Integral relations for Heun-class special functions, Theor. Math. Phys., 107, 733-739 (1996).\\

\n   {\color{blue}[Kasakov2 et al.]} Kazakov, A. Ya. and Slavyanov, S. Yu. Integral equations for special functions of Heun class, Meth. Appl. Anal., 3, 447-456 (1996).\\

\n  {\color{blue}[Kato]}  Kato, T. :.Perturbation Theory for Linear Operators, Springer-Verlag, BerlinrNew York, 1966.\\

\n  {\color{blue}[Kaidalov]} Kaidalov, A. B. :. Pomeranchuk singularity and high- energy hadronic interactions, Usp. Fiz. Nauk, 46 (2003) 1153\\

\n   {\color{blue}[Krein]} Krein, S. G. :. Linear Differential Equations in Banach Space, Am. Math. Soc., Providence, RI, 1971.\\

\n {\color{blue}[Lax]} Lax, P. : Trace formulas for the Schrodinger operator. Comm. Pure Appl. Math., 47, (1994), 503-512.\\

\n {\color{blue}[Levin1]} Levin, E. M., Everything about Reggeons, arXiv :hep-ph/9710546.\\

\n {\color{blue}[Levin2]} Levin, E. M. , : The Pomeron : yesterday, today and tomorrow, hepph/ 9503399 , Lec- tures given at 3rd Gleb Wataghin School on High Energy Phenomenology, Campinas, Brazil, 11-16 Jul (1994).\\

 \n {\color{blue}[Levitan et al]} Levitan, B. M. and  Sargsyan, I.S.: Sturm-Liouville and Dirac Operators. Kluwer, Dordrecht, 1991.\\

\n   {\color{blue}[Lidskii]} Lidskii,  V. B. :. Summability of series in the principal vectors of non self-adjoint operators, Amer. Math. Soc. Trans. 2. 40, 1964..\\

\n {\color{blue}[Lidskii et al]}  Lidskii, V. B. and  Sadovnichii, V. A. : Regularized sums of roots of a class of entire functions. Functional Anal. i Prilozen., 1, (1967), 52-59 (in Russian); English transl.: Funct. Anal. Appl.,1, (1967), 133-139.\\

\n  {\color{blue}[Macaev et al]} Macaev, V. I. and Ju. A. Palant, Ju. A. :.O Stepenjah ogranicennogo dissipativnogo operatora, Ukranian Math. J. 14, 3,  1962..\\

\n  {\color{blue}[Makin]} Makin, A. S. :. Trace Formulas for the Sturm-Liouville Operator with Regular Boundary Conditions. Dokl. Acad. Nauk, 416, (2007), 308-313 (in Russian); English transl.: Dokl. Math., 76, (2007), 702-707.\\

\n  {\color{blue}[Markus]}  Markus, A. S. :. Introduction to the Spectral Theory of Polynomial Operator Pencils, Amer. Math. Soc., Providence, RI, 1988.\\

\n {\color{blue}[Moszynski] ]} Moszynski, M. :. Spectral properties of some Jacobi matrices with double weights, J. Math. Anal. Appl., 2003, 280(2), 400-412\\

 \n  {\color{blue}[Naymark]} Naymark, M. A.  :. Linear Differential Operators. Nauka, M. 528 (1969)\\

\n  {\color{blue}[Pazy]} Pazy,  A. :. Semigroups of Linear Operators and Applications to Partial Differential Equations, Springer-Verlag, BerlinrNew York, 1983.\\

\n  {\color{blue}[Poghosyan]} Poghosyan, M., An introduction to Regge Field Theory, Wilhelm und Else Heraeus Physics Summer School ``Diffractive and electromagnetic processes at high energies'', Heidelberg, Germany, September 2-6, (2013)\\

\n  {\color{blue}[Reed et al]}   Reed , M. and Simon, B. :. Methods of modern mathematical physics I : Functional analysis, Academic Press, 1980.\\

\n  {\color{blue}[Roy] } Roy, S. M., : High energy theorems for strong interactions and their comparison with experimental data, Phys. Rep. 5C (1972) 125.\\

\n  {\color{blue}[Sadovnichii1 et al]}  Sadovnichii, V. A. and V.E. Podolskii,V. E. :. On the class of Sturm-Liouville operators and approximate calculation of first eigenvalues, Mat Sbornik. 189(1), (1998), 133-148\\

\n  {\color{blue}[Sadovnichii2 et al] } Sadovnichii, V. A. and V.E. Podolskii,V. E. :.Traces of operators with relatively compact perturbations. Mat. Sb. 193 (2), (2002) 129-152\\

\n  {\color{blue}[Sadovnichii3 et al]}  Sadovnichii, V. A. and V.E. Podolskii,V. E. :. Trace of operators. Uspech Math Nauk. 61(5), (2006), 89-156\\

\n {\color{blue}[Sadovnichii4 et al] } Sadovnichii, V. A. and V.E. Podolskii,V. E. :. Traces of Differential Operators, Differential Equations, Vol. 45, No. 4,(2009), pp. 477-493.\\

\n  {\color{blue}[Sadovnichii5 et al]}  Sadovnichii, V. A. and V.E. Podolskii,V. E. :. Regularized Traces of Discrete Operators, Proceedings of the Steklov Institute of Mathematics, Pleiades Publishing, Inc.Suppl. 2, (2006), pp. 161-177.\\

\n {\color{blue}[Sansuc et al}  Sansuc, J.J. and  V. Tkachenko, V. : Characterization of the periodic and antiperiodic spectra of nonselfadjoint Hillls operators. Oper. Theory Adv. Appl., 98, (1997), 216-224.\\

\n {\color{blue}[Schmutz]} Schmutz, M., The factorization method and ground state energy bounds. Phys. Lett. A, 108(4):195?196, 1985. \\

\n {\color{blue}[Segal 1]} Segal, I.E. : Lecture at the 1960 Summer Seminar in Applied Mathematic, Boulder, Colorado. (1960)\\

\n {\color{blue}[Segal 2]} Segal, I.E. : Mathematical Characterization of the Physical Vacuum for a linear Bos-Einstein Field. Illinois,   J. Math. 6. (1962) 500-523\\

\n {\color{blue}[Segal 3]} Segal, I.E. : The complex-wave representation of the free boson field in : ``Topics in Functional Analysis'' (I. Gohberg and M. Kac. Eds.) . Advances in Mathematics Supplementary studies 3. Academic Press. New York. London. (1978) 321-343\\

\n  {\color{blue}[Silva]} Silva, L. O. :. Spectral properties of Jacobi matrices with rapidly growing power-like weights, In : Operator Methods in Ordinary and Partial  Differential Equations (Albeverio, S., Elander, N., Everitt, W. N. and Kurasov, P. eds). Operator Theory:  Advances and Applications , vol.  132,  Basel : Birkhauser, 2002, 387-394\\

\n {\color{blue}[Simon1]} Simon, B. :. Notes on infinite determinants of Hilbert space operators, Advances in Mathematics 24 (1977), pp. 244-273.\\

\n {\color{blue}[Simon2]} Simon, B. :.Trace ideals and their applications, Mathematical Surveys and Monographs, Volume 120, AMS, 2nd Ed. , (2005).\\

\n  {\color{blue}[Thirring]} Thirring, W. :.  Quantum mechanics of atoms and molecules, volume 3 of A course in mathematical physics. Springer-Verlag, New York, 1979.\\

\n  {\color{blue}[Tonin] }  Tomin, N. G. :. Several Formulas for the First Regularized Trace of Discrete Operators, Mathematical Notes, vol. 70, no. 1, (2001), pp. 97-109.\\

\n {\color{blue}[Tur et al]} Tur, E. String Jacobi matrices with very fast decreasing weights, Integr. Oper. Theory, 2004, 50(1), 115-128.\\

\n  {\color{blue}[White]} White,A. :. CERN preprint TH 2445 (1978).\\

\n  {\color{blue}[Yakubov et al]} Yakubov, S. Ya and Mamedov,  K. S. :. On the multiple completeness of a system of eigenelements and associated elements of a polynomial operator pencil and on multiple expansions in that system, Funktsional. Anal. i Prilozhen. 91. 1975..\\

\n  {\color{blue}[Zagrebnov1]}   Zagrebnov, V. A. :. On the families of Gibbs semigroups, Commun. Math. Phys. 76 (1980) 269-276\\

\n  {\color{blue}[Zagrebnov2]}   Zagrebnov, V. A. : . Perturbations of Gibbs semigroups, Commun. Math. Phys. 120 (1989) 653-664\\

\n   {\color{blue} [zerner]}  Zerner, M. :. Quelques propri\'et\'es spectrales des op\'erateurs positifs, Journal of Functional Analysis  72, 381417 (1987)\\

\end{document}